\documentclass[a4paper,fleqn,usenatbib,useAMS]{mnras}

\usepackage{graphicx}	
\usepackage{amsmath}	
\usepackage{amssymb}	
\usepackage{multicol}        
\usepackage{bm}		
\usepackage{pdflscape}	

\newcommand{\cn}{$C_\mathrm{n}^2$} 
\newcommand{\comment}[1]{}
\newcommand{\as}{$^{\prime\prime}$} 
\newcommand{\red}[1]{\textcolor{red}{#1}}
\makeatletter
\AtBeginDocument{\let\red\@firstofone}
\makeatother

\usepackage[T1]{fontenc}
\usepackage{ae,aecompl}

\usepackage{newtxtext,newtxmath}

\title[Turbulence profiling with GCM at Paranal]{Atmospheric turbulence forecasting with a General Circulation Model for Cerro Paranal}

\author[J. Osborn]{J. Osborn\thanks{Contact e-mail: \href{mailto:james.osborn@durham.ac.uk}{james.osborn@durham.ac.uk}}
and M. Sarazin
\\
$^{1}$Centre for Advanced Instrumentation, Department of Physics, Durham University, South Road, Durham, DH1 3LE, UK\\
$^{2}$European Southern Observatory, Karl-Schwarzshild-Str.2, 85748 Garching bei Muenchen, Germany}
\begin{document}

\date{Accepted . Received ; in original form }

\pagerange{\pageref{firstpage}--\pageref{lastpage}} \pubyear{2014}

\maketitle

\label{firstpage}

\begin{abstract}
 \red{In addition to astro-meteorological parameters, such as seeing, coherence time and isoplanatic angle, the vertical profile of the Earth's atmospheric turbulence strength and velocity is important for instrument design, performance validation and monitoring, and observation scheduling and management. Here we compare these astro-meteorological parameters as well as the vertical profile itself from a forecast model based on a General Circulation Model from the European Centre for Median range Weather Forecasts and the stereo-SCIDAR, a high-sensitivity turbulence profiling instrument in regular operation at Paranal, Chile. The model is fast to process as no spatial nesting or data manipulation is performed. This speed enables the model to be reactive based on the most up to date forecasts. We find that the model is statistically consistent with measurements from stereo-SCIDAR. The correlation of the median turbulence profile from the model and the measurement is 0.98. We also find that the distributions of astro-meteorological parameters are consistent. We compare contemporaneous measurements and show that the free atmosphere seeing, isoplanatic angle and coherence time have correlation values of 0.64, 0.40 and 0.63 respectively. We show and compare the profile sequences from a large number of trial nights. We see that the model is able to forecast the evolution of dominating features. In addition to smart scheduling, ensuring that the most sensitive astronomical observations are scheduled for the optimum time, this model could enable remote site characterisation using a large archive of weather forecasts and could be used to optimise the performance of wide-field AO system.}
\end{abstract}


\section{Introduction}
The Earth's turbulent atmosphere degrades the image quality from astronomical telescopes. This is exacerbated as telescopes become larger. Adaptive Optics (AO) systems must be implemented in order to recover the spatial resolution by compensating for the phase aberration induced by the turbulence. In the current era of large 8-10~m class telescopes and the future 40~m extremely large telescopes it is of critical importance to have thorough knowledge of the vertical structure of the turbulence strength (for example, \citealp{Osborn2016d,Gendron2014, Basden2010, Vidal2010, Neichel2008}) and velocity (for example, \citealp{Osborn16c,Kulcsar2006,Paschall1993}. In addition, if this knowledge can be forecast in advance then this enables some significant benefits in operational efficiency of the modern observatory and, critically, will enable the most sensitive of observations to be scheduled, and executed in the optimum conditions \citep{Masciadri2013b}.

Here we present a turbulence model which uses parameters directly extracted from General Circulation Models (GCMs) such as the European Centre for Medium range Weather Forecasts (ECMWF), without any further manipulation. By processing the atmospheric parameters we can derive forecasts of the vertical turbulence profiles and astro-meteorological parameters, such as the coherence time and isoplanatic angle.

There are many applications for such a function, here we list a sample:
\begin{itemize}
\item{Site characterisation and selection, without the need for on-site instrumentation. This will be extremely useful for possible site identification and selection of potential new observatories as well as the characterisation of existing observatories without atmospheric monitoring instrumentation.}
\item{Night by night astronomical parameter forecasts / nowcasts without dedicated instrumentation. It is extremely useful to be able to monitor the atmospheric conditions during an observation, for example, for performance validation.}
\item{Dynamic scheduling based on astronomical parameter forecast, enabling the most sensitive experiments to be executed in the optimum conditions.}
\item{Instrument optimisation based on astronomical parameter forecast. For example, wide-field Adaptive Optics (AO) instrumentation requires a model of the atmosphere within the control system. If this can be built during the day, at least for a first estimate, then minimal on-sky time will be lost for AO calibration.}
\item{In addition to the field of astronomical instrumentation turbulence forecasting is essential to estimate the feasibility of optical communications with satellites for ground stations around the world.}
\end{itemize}

There have been several studies into forecasting of optical turbulence. \cite{Trinquet2007} introduced a model to convert the standard atmospheric parameters, such as wind velocity, pressure, humidity and temperature into $C_T^2$ profiles, from which $C_n^2$ profiles can be derived. \citeauthor{Trinquet2007} concentrated on validating the model with radiosonde measurements. However, \cite{Ye2011} used the model with the Global Forecast System (GFS) outputs to estimate the seeing and free atmosphere seeing at several sites round the world. This study was ambitious however the model from \citeauthor{Trinquet2007} contains an empirical weighting function which was defined at the Observatoire de Haute Provence in France and was therefore a limitation for the study of \citeauthor{Ye2011}.

\cite{Giordano2013} followed a similar line, using the statistical \citeauthor{Trinquet2007} model with a meso-scale model of the Weather Research and Forecasting model (WRF). This mesoscale approach provides a secondary simulation stage, enabling higher resolution inputs into the \citeauthor{Trinquet2007} model. 

Significant effort has been applied into the field of mesoscale numerical models for the forecast of optical turbulence profiles and atmospheric parameters, for example \cite{Masciadri2013b}. \citeauthor{Masciadri2013b} opted to develop an hydrodynamic model of the atmospheric turbulence enabling then to calculate the 3D map of the atmospheric turbulence in the model spatial range. The results so far are extremely promising. 

Another mesoscale model to be developed recently is the Mauna Kea Weather Centre mesoscale model which is operational at the Mauna Kea observatory, Hawaii \citep{Cherubini2008}. \citeauthor{Cherubini2008} follow a similar route to \citeauthor{masciadri2013a}, however, several differences are made in the physical modelling of the turbulence and a different GCM model is used as the input.

These mesoscale models require site specific calibration and employ further spatial nesting to increase the spatial resolution. This nesting improve the fidelity of the results at the sacrifice of processing time. The Earth's atmosphere is a dynamic system where changes can happen quickly. For the application of smart scheduling it is critical that the most up to date forecast can be used as an input in order to maximise the probability of successfully forecasting the required parameters. Minimal forecast calculation time is a requirement to enable rapid response to changing conditions.

Here, we pursue a low spatial and temporal resolution alternative, which can easily be applied anywhere in the world, \red{and ideally} free from any site specific calibration. An additional advantage of the GCM approach is that the reprocessing is minimal meaning that the forecast can be updated as soon as a new forecast is released ensuring that the latest model is being used. The processing time for the mesoscale model to reach thermodynamic equilibrium can be several hours (15 hours in the case of \cite{Masciadri2017}). This low resolution alternative can be processed in seconds \red{once the meteorological forecast has been received}.

Without a site specific calibration the GCM's limited spatial resolution will not be able to reproduce the atmospheric parameters which are influenced by the local topography. The challenging goal of accurately reproducing a highly localised model, for a particular telescope, for example, requires more sophisticated modelling capabilities, such as \cite{Masciadri2013b}. Even with a mesoscale model, the local ground layer turbulence is difficult to model. However, above the surface layer, in the free atmosphere ($>$1-2~km) where any local effects are negligible by definition, the GCM based turbulence model can provide a good reflection of the reality. This is the most critical part of the atmosphere which limits the performance of wide-field AO systems, for example. Also, integrated parameters such as the coherence time and isoplanatic angle are dominated by high altitude turbulence. This will be sufficient to estimate the field-of-view, PSF stability or overall performance of any particular instrument and therefore enable dynamic queue scheduling.

\red{As with all models, a structure coefficient to normalise the magnitude of the turbulence is required. This coefficient can be constant for the full profile or, as with most previous models, vary with altitude. The coefficient is used to parameterise the stability of the atmosphere. Here we normalise the model based on a subset of measurements at ESO Paranal. This normalisation process ensures the integrated turbulence strength is consistent but does not calibrate the structure of the turbulence in any way. We avoid a full calibration process (of the structure of the profile), as pursued in previous models in order to keep the model as general as possible with the the ultimate goal of applying it globally. It is not yet clear if such a normalisation process will be general enough to be applied globally, but in this work we concentrate on reporting the performance of the model at ESO Paranal.}

\comment{
Previously, the validation of these models has been difficult due to the lack of a precise turbulence profiler instrument. Some studies compare the models with profiles derived from balloon borne radiosonde or radar \citep{Trinquet2007}, but these profiles are also model based and not a direct measurement.
}

We compare the forecasts with the measurements from a high-precision optical turbulence profiler, stereo-SCIDAR, at Cerro Paranal, the site of the Very Large Telescope and 20~km from the site of the Extremely Large Telescope. This comparison allows us to easily validate the forecasts. We have previously shown that the wind velocity profile from these models correlates well with the turbulence velocity profiles from the stereo-SCIDAR, despite the low spatial resolution \citep{Osborn16c}, demonstrating that the model can be reliable used to forecast turbulence velocity. However, in order to fulfil the potential of the model we must also be able to forecast the strength as well as the velocity of the optical turbulence. Here we compare the optical turbulence profile and the derived astro-meteorological parameters from the GCM model with those extracted from stereo-SCIDAR. This work concentrates on validating the model at ESO Paranal, although further work to validate the model at other sites is required.

In section~\ref{sect:SCIDAR} we describe the stereo-SCIDAR instrument which is used for the validation of the forecasts. In section~\ref{sect:GCM} we describe the GCM models used in this work and in section~\ref{sect:model} we describe the turbulence model used in this publication. The results are in section~\ref{sect:results}.

\section{Stereo-SCIDAR}
\label{sect:SCIDAR}
Stereo-SCIDAR is a dedicated high-precision, high-sensitivity, high-altitude resolution optical turbulence profiler \citep{Shepherd13}. The stereo-SCIDAR is therefore an ideal instrument to use for comparison with the numerical forecast models. 

The instrument was developed as part of the canary AO demonstrator project \citep{Morris2014} and was installed on the 2.5~m Isaac Newton Telescope, La Palma for a total of 28 nights in 2014 and 2015. In addition, a version of the instrument has been in regular operation at ESO Cerro Paranal \citep{Derie2016} since April 2016 \citep{Osborn2018} (table~\ref{tab:scidarDataParanal}). These \red{83 nights of} data from Cerro Paranal, will be used for the validation of the turbulence forecasts.

\begin{table*}
\caption{ESO Paranal, Stereo-SCIDAR data volume}
\label{tab:scidarDataParanal}
\begin{tabular}{@{}clccc}
\hline
Year & Month & Days & Hours & Number of Profiles\\
\hline
2016 & April    & 26 - 29	& 18.43	& 607\\
	 & July		& 22 - 26	& 37.12	& 1143\\
     & October  & 30 - 31	& 10.65	& 301\\
     & November & 1 - 2		& 10.80	& 302\\
     & December & 10 - 12	& 11.62	& 308\\
2017 & March	& 7 - 9		& 16.46	& 469\\
	 & April	& 12 - 18	& 37.34	& 988\\
     & May		& 5 - 9		& 16.06	& 419\\
     & June		& 8 - 10	& 19.97	& 511\\
     & July		& 3 - 9		& 37.60	& 962\\
     & August	& 3 - 8		& 34.42 & 930\\
     & November & 4 - 9, 18 - 20, 29 - 30 & 45.63 & 1076\\
     & December & 1 - 6, 8 - 18 & 56.69 & 1483 \\
2018 & January	& 13 - 24   & 44.19	&	1192\\
\hline
Totals:	&		& 83	&	396.97 & 10691	\\
\hline
\end{tabular}
\end{table*}

\section{General Circulation Models}
\label{sect:GCM}
General circulation models (GCM) have been used to provide wind velocity profiles for previous astronomical studies (for example, \citealp{Hagelin2010,Osborn16c}). They have also been used as the input for mesoscale turbulence forecast models (for example, \citealp{Giordano2013,Masciadri2017}). In this study we use the European Centre for Medium-range Weather Forecasts (ECMWF) forecasts \footnote{https://www.ecmwf.int/en/forecasts/datasets/}. 

\subsection{ECMWF}
The ECMWF model is a non-hydrostatic model. The model is refreshed every 6 hours and provides a forecast for every hour. Two level models are produced, pressure level and model level. For the pressure level forecast, parameters are forecast at 1000, 950, 925, 900, 850, 800, 700, 600, 500, 400, 300, 250, 200, 150, 100, 70, 50, 30, 20, 10, 7, 5, 3, 2, 1 mbar. For the model levels, forecasts are provided at 137 levels. The altitude levels are hybrid, defined as lines of constant pressure above surface pressure. The altitude resolution is generally a couple of tens of metres near the ground a few kilometres above the tropopause.

Here, we use publicly available data from ECMWF from the ERA5 catalogue. Historical data is freely available up until 2 months in the past. The data has 0.3degree spatial resolution and is only available for the models produced at 06:00 and 18:00 UT, with forecasts for every hour up to 19 hours. Here, we use the best case data, i.e. data that was produced at most 11 hours before (for example, 06:00+11 hours). To extract the parameters for the site of Cerro Paranal in the 0.3 degree grid \red{(equivalent to approximately 30~km by 30~km grid)}, we linearly interpolate between the four nearest data points. 

\section{Turbulence model}
\label{sect:model}
Our aim is to validate a numerical model which will accept atmospheric parameter forecasts, such as wind velocity, temperature and pressure, and output a low resolution turbulence profile. The model should be valid globally and not require any site specific calibration. It should also be computationally easy and not require long processing times.

The Gladstone relation can be used to estimate the value of the optical turbulence refractive index structure constant, $C_\mathrm{n}^2$ given the temperature, pressure and temperature structure constant, $C_T^2$,
\begin{equation}
C_n^2 = \left(\frac{80\times10^{-6}P}{T^2}\right)^2 C_T^2.
\label{eq:Gladstone}
\end{equation}
Here, we use the modification introduced by \cite{Masciadri2017},
\begin{equation}
C_n^2 = \left(\frac{80\times10^{-6}P}{T\theta}\right)^2 C_T^2,
\label{eq:Gladstone2}
\end{equation}
where, 
\begin{equation}
\theta = T\left(\frac{P_0}{P}\right)^{R/c_p},
\end{equation}
is the potential temperature, $R/c_p=0.286$,$P_0=1000$~mbar. This modification is introduced as equation~\ref{eq:Gladstone} assumes that the atmosphere is in hydrostatic equilibrium and that the gradient of temperature follows the adiabatic approximation \citep{Masciadri2017}. However, in the free atmosphere the temperature gradient is not as high and the potential temperature should be used \citep{Tatarski1971}.

$C_T^2$ can be estimated with \citep{Tatarski1971},
\begin{equation}
C_T^2 = k L^{4/3}\left(\frac{\delta\bar{\theta}(z)}{\delta z}\right)^2,
\end{equation}
where $z$ is the altitude, $k$ depends on the stability of the atmosphere \citep{Masciadri2001} and $L$ is the scale of the largest energy input into the turbulent flow and can be defined as \citep{Masciadri2001},
\begin{equation}
L(z)=\sqrt{\frac{2E}{\frac{g}{\theta(z)}\frac{\delta\bar{\theta}(z)}{\delta z}}},
\end{equation}
where $E$ is the turbulent kinetic energy. Here, we use $E=S^2$, where $S$ is the vertical wind shear,
\begin{equation}
S=\left[\left(\frac{\delta u}{\delta z}\right)^2+\left(\frac{\delta v}{\delta z}\right)^2\right]^{1/2},
\end{equation}
$u$ and $v$ being the the two horizontal components of the wind velocity. 

Therefore, to estimate \cn we use the following,
\begin{equation}
C_n^2(z) = k \left(\frac{80\times10^{-6}P(z)}{T(z)\theta(z)}\right)^2 L(z)^{4/3}\left(\frac{\delta\bar{\theta}(z)}{\delta z}\right)^2.
\label{eq:Cn2}
\end{equation}

In some models $k$ is a function of altitude, $k(z)$, and is used to calibrate the model to a particular site. Higher values are used to amplify the turbulence in unstable atmosphere zones and lower values suppress the turbulence strength in stable zones. Here, we want to develop a single global model with no site specific calibration, \red{we therefore use a single value coefficient for the full atmosphere to avoid over calibrating for a single site. $k$ is calculated by calibrating the integral turbulence strength from the model with the 50\% of the stereo-SCIDAR data (selected at random), in this case $k$=6.0. The whole dataset (including the 50\% calibration data) is used for the validation in the following sections.} 

\comment{
For this application we are primarily interested in modelling turbulent zones so we choose a high value. \red{\cite{Trinquet2007} take $k$ to be 2.8.}
, $k=10$. Choosing a high value may lead to over-estimating the turbulence magnitude of low turbulence zones, however it should provide a better estimate of the unstable atmospheric regions that are most important for our study. In this way the model presented here is different to other work as it is completely free of any site specific calibration and altitude dependent correction factors.
}

\section{Results}
\label{sect:results}

\subsection{Global turbulence}
\begin{figure}
\centering
    	\includegraphics[width=0.5\textwidth]{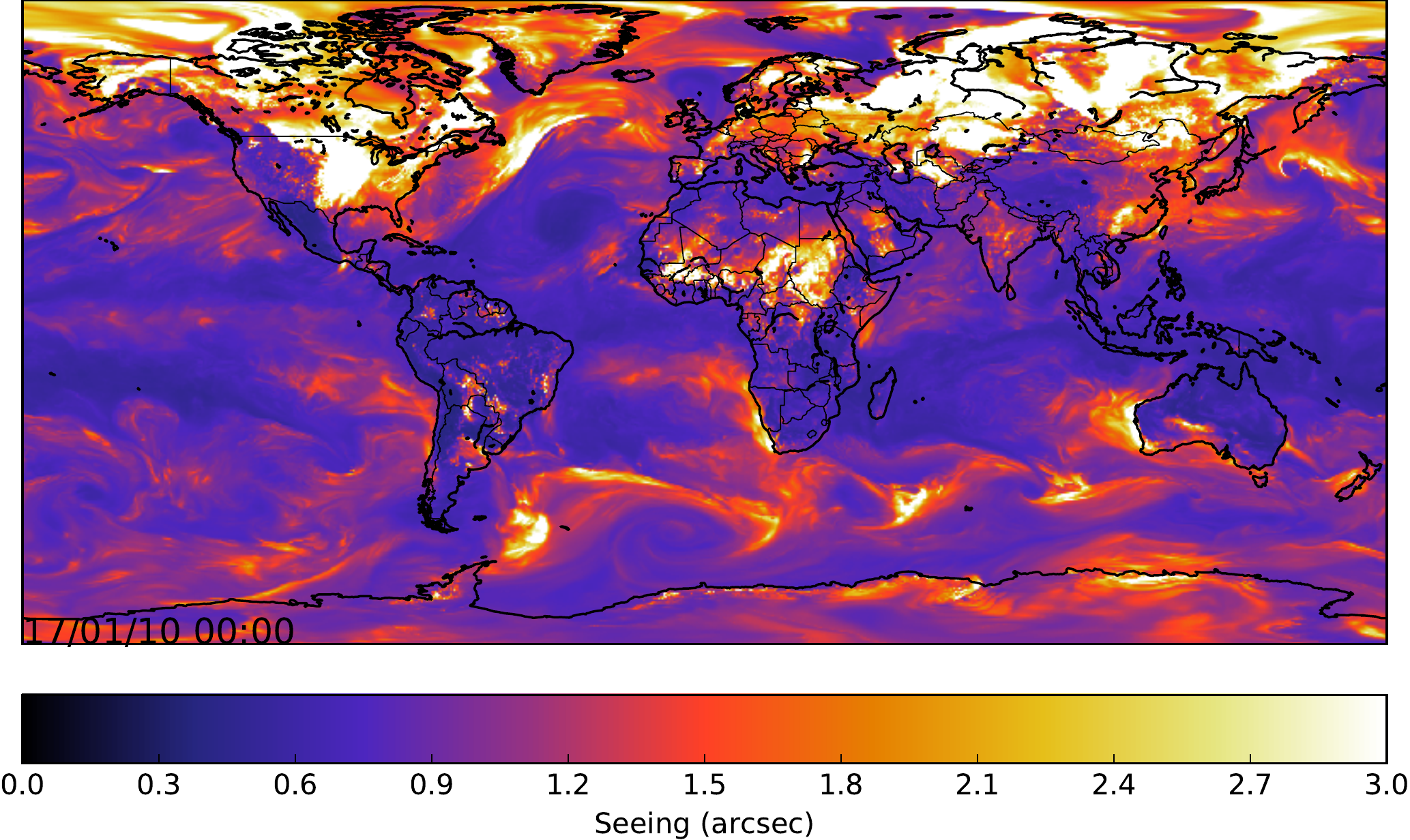}
\caption{Global free atmosphere seeing (integrated from 1~km above the model altitude) map derived from ecmwf gcm for the night of 10th January 2017.}
\label{fig:cn2Global}
\end{figure}

Using the model described in section~\ref{sect:model} we are able to estimate the full vertical profile of turbulence strength and velocity globally from general circulation model weather forecast data. As an example figure~\ref{fig:cn2Global} shows an example global of the integrated free atmosphere seeing (altitudes over 1~km above the grid altitude level).

\subsection{Comparisons}

To compare the results of the ECMWF model and the stereo-SCIDAR measurements, the ECMWF model is interpolated onto the same altitude grid as the stereo-SCIDAR (i.e. 250~m resolution from the observatory level to 25~km above the observatory level). We compare the atmospheric parameters derived from the ECMWF model forecast for the hour with the median stereo-SCIDAR measurement from 2.5 minutes either side of the hour, ie a 5 minute sampling period. The stereo-SCIDAR has the dome contribution automatically subtracted.

\comment{
To quantify the error associated with this, we compare all the stereo-SCIDAR measurements of atmospheric parameters with the mean for the hour. We find that the expected average bias and rmse for the stereo-SCIDAR dataset is 0.06\as and 0.26\as respectively for the seeing, 0.65~ms and 2.7~ms respectively for the coherence time and 0.09\as and 0.68\as respectively for the isoplanatic angle. These values are similar to the measured metrics suggesting that it is the best we can do for one hour temporal sampling given the inherent evolution of the atmosphere. The performance is limited by atmospheric variation and measurement errors rather than model fidelity. This may be improved by the greater temporal sampling made possible by a meso-scale model, however, the comparison with published results suggests that even the meso-scale model is not able to improve on the precision of the GCM model.
}

\subsection{Turbulence velocity}
The turbulence velocity from GCM models has been compared with stereo-SCIDAR and discussed previously, for example \cite{Osborn16c} (La Palma) and \cite{Osborn2018} (Paranal). Here, we again show the result comparisons for completion.

\begin{figure}
\centering
\includegraphics[width=0.5\textwidth]{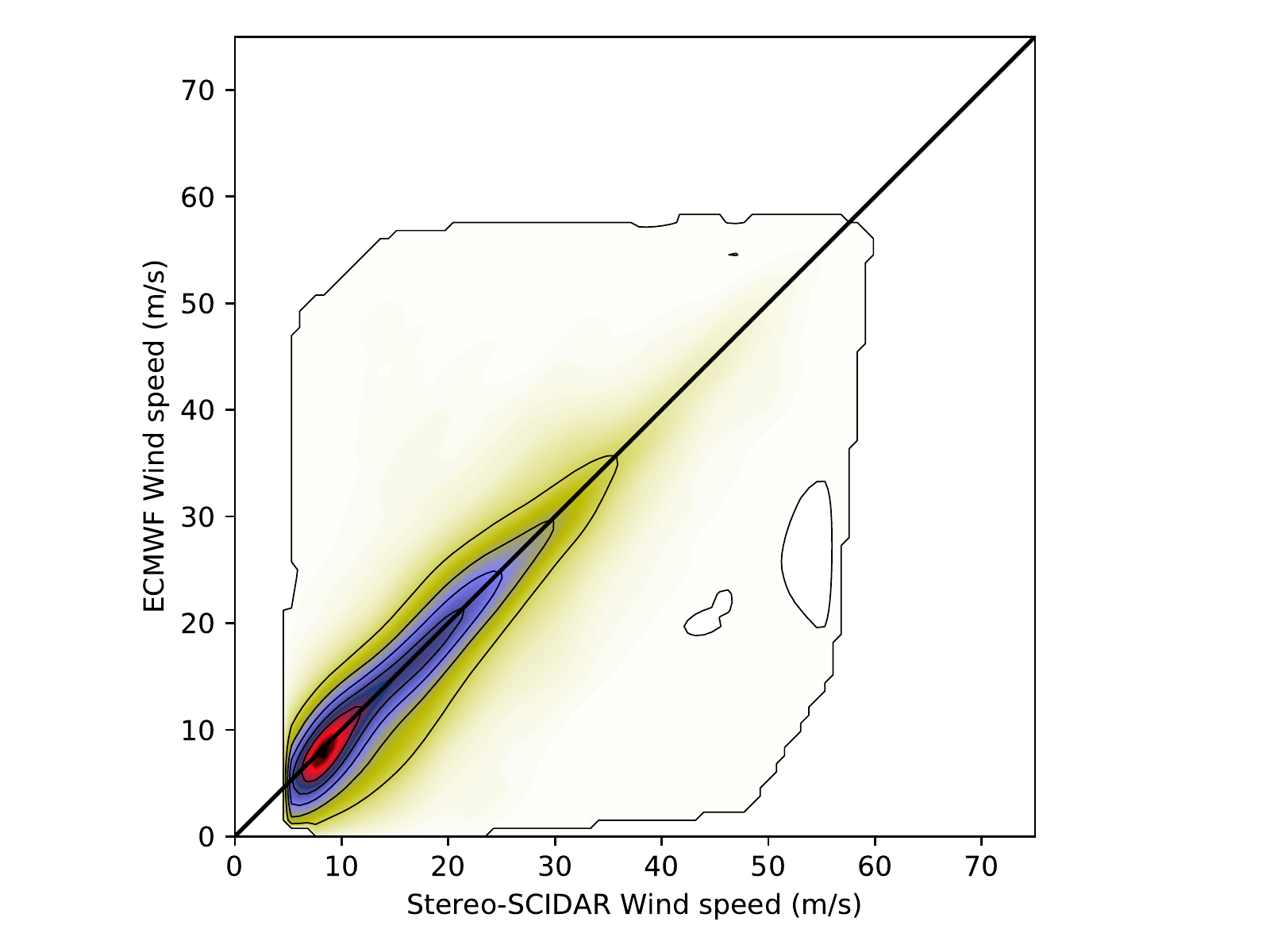}
\caption{Comparison of turbulence speed as measured by stereo-SCIDAR and the ECMWF model. The correlation is 0.81 with a bias and rmse of 0.22~m/s and 6.6~m/s respectively.}
\label{fig:speedComparison}
\end{figure}

\begin{figure}
\centering
    	\includegraphics[width=0.5\textwidth]{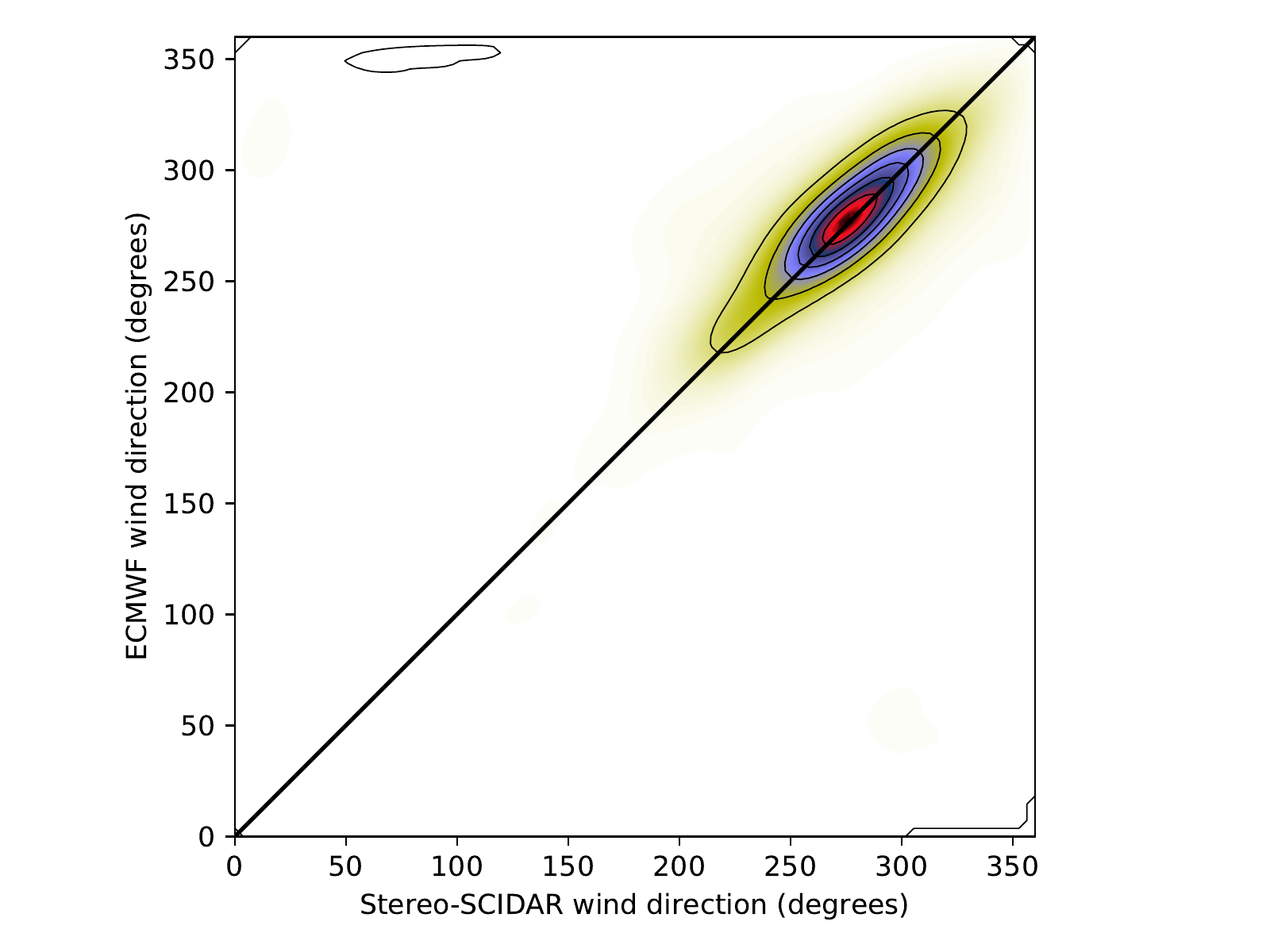}
\caption{Comparison of turbulence direction as measured by stereo-SCIDAR and the ECMWF model. The correlation is 0.73 with a bias and rmse of -0.85~degrees and 29.22~degrees respectively.}
\label{fig:dirComparison}
\end{figure}

Figure~\ref{fig:speedComparison} and figure~\ref{fig:dirComparison} shows the comparison between wind speed and direction from the stereo-SCIDAR and ECMWF for all altitudes. The correlation values of this comparison are 0.82 and 0.77 for wind speed and direction respectively. We see that the RMSE of the wind direction is large (29~degrees).The reason for this discrepancy is thought to be due to the large wind shear within turbulent zones in the free atmosphere at Paranal, resulting in a dispersion of velocity vectors for the turbulent zone. The model does not have sufficient vertical resolution to resolve the velocity dispersion that is measured by the stereo-SCIDAR.

\subsection{Ground layer turbulence}
Due to the limited spatial resolution of the GCM (0.3~degrees in this case which corresponds to approximately 30~km at the equator), the topological map is a coarse representation of reality. For example, ESO Paranal observatory is located on the summit of cerro Paranal at an altitude of 2635~m above sea level. However, due to the spatial averaging of the GCM the altitude of the grid corresponding to Paranal is at 926~m (grid altitude). This discrepancy causes a significant issue for the vertical profile of the atmospheric parameters. If we take the ground level to be the grid altitude  then high altitude layers will be offset in altitude by a corresponding amount, and if we instead take the ground to be the altitude of the observatory then the ground turbulence is missed. To rectify this problem, we propose to take the ground layer from the grid altitude (926~m) to 1~km above the grid altitude (1926~m) and add that into the levels corresponding to the observatory altitude (2635~m to 3635~m). This ground layer must be interpolated to take into account the different altitude resolution of the model at the grid altitude and the observatory altitude.

\subsection{Median Profiles}
\begin{figure*}
\centering
$\begin{array}{cc}
	\includegraphics[width=0.5\textwidth,trim={0cm 0 0cm 0}]{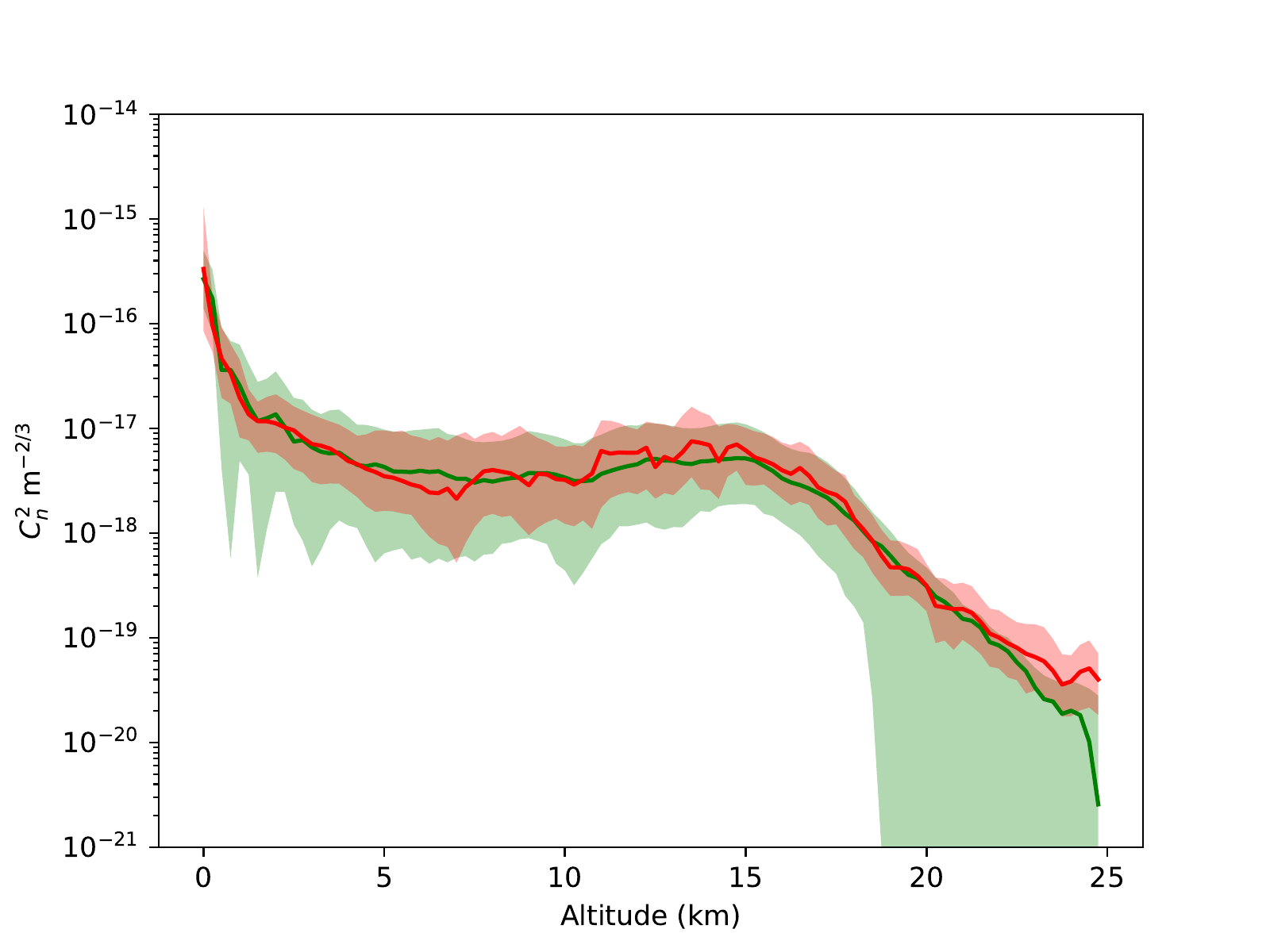} &
	\includegraphics[width=0.5\textwidth,trim={0cm 0 0cm 0}]{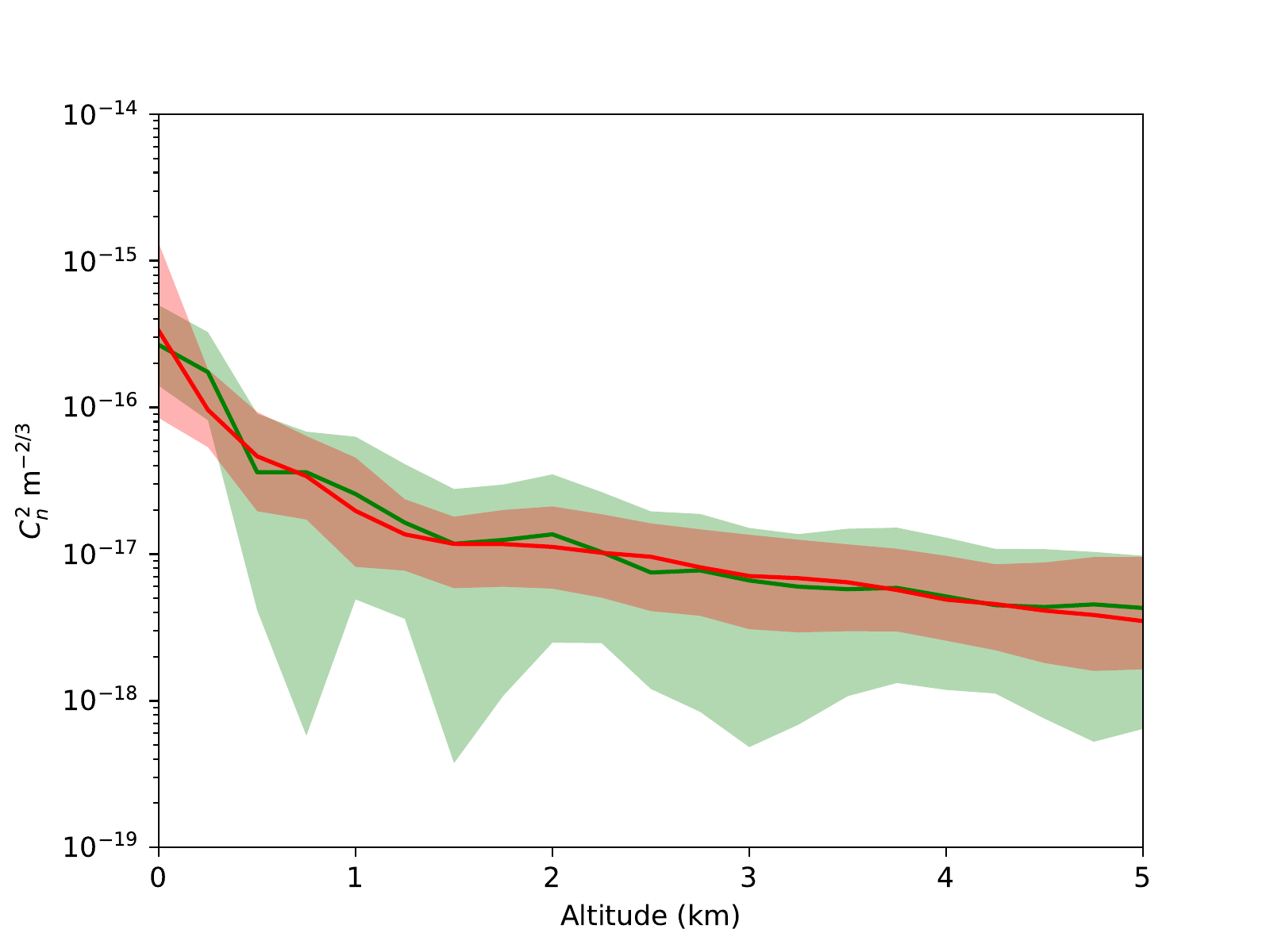}
\end{array}$
\caption{Median turbulence profile (left) with a magnified view of the first 5~km (right) from the stereo-SCIDAR (green) and from ECMWF (red) for ESO Paranal. The inter-quartile range is shown in the coloured region. The two curves are well correlated (0.98). The altitudes are from observatory level (2.6~km above sea level).}
\label{fig:medianProfileParanal}
\end{figure*}

An important and interesting application is to use numerical GCM forecasts to derive typical or median vertical turbulence profiles for astronomical observatory sites. In figure~\ref{fig:medianProfileParanal}  we show the median profiles from stereo-SCIDAR and from the ECMWF for ESO Paranal, Chile. The two profiles have high correlation (0.98). However, the model does estimate stronger turbulence in the first bin, close to the ground. It is thought that this could be due to the stereo-SCIDAR automatically subtracting the dome turbulence. It is unlikely for any model to be able to accurately estimate the turbulence very close to the telescope due to the local effects of the telescope itself. 

The model shows a narrower inter-quartile range than the measured data. This suggests that the model does not forecast the extreme of events. This is likely to be a manifestation of the limited altitude resolution. For example, a velocity shear within a resolution element may not be seen by the model. It is common in stereo-SCIDAR data to see wind shear of up to 20~degrees within a single turbulent layer of less then a few hundred metres in depth \citep{Osborn16c}. This might not be reproduced by the GCM.

Although, median profiles are not realistic typical profiles, and they will almost certainly never occur, they do give an estimate of the median strength of the turbulence at each altitude. This is useful for site selection, as well as instrument design and performance estimation for future instrumentation. 

\begin{figure*}
\centering
$\begin{array}{cc}
    	\includegraphics[width=0.5\textwidth,trim={2cm 0 1cm 0}]{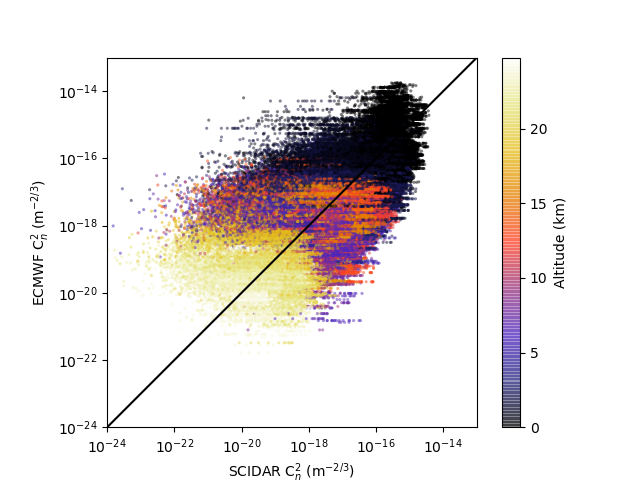}&
	\includegraphics[width=0.5\textwidth,trim={0cm 0 0cm 0}]{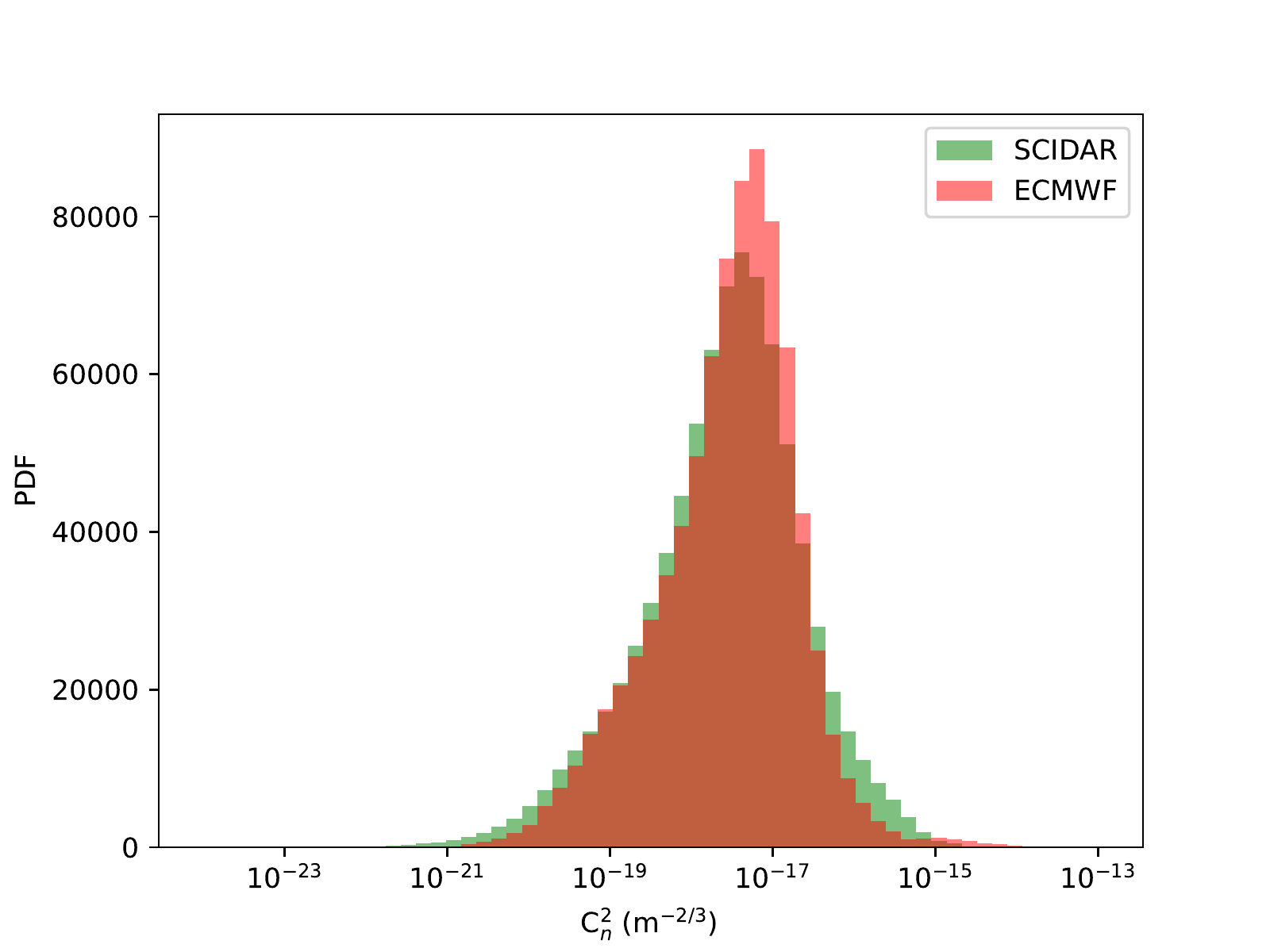}
\end{array}$
\caption{On the left is as catter of the refractive index structure parameter for every layer comparison. The colour indicates the altitude of the layer from observatory level. To make this comparison the ECMWF model turbulence profile was interpolated onto the same altitude grid as the stereo-SCIDAR profiles. The correlation coefficient is 0.63 with a bias of -8.07$\times10^{-18}$~m$^{-2/3}$ and rmse of 3.01$\times10^{-16}$~m$^{-2/3}$. The plot on the right is the histogram on turbulence strength values measured by stereo-SCIDAR (green) and the ECMWF model (red). The distributions are similar, showing the same range of values and the similar shape of distribution.}
\label{fig:cn2Comparison}
\end{figure*}

Figure~\ref{fig:cn2Comparison} shows the comparison of turbulence strength layer by layer for the stereo-SCIDAR data and the ECMWF model. The comparison appears linear with a small bias of -2.29$\times10^{-17}$~m$^{-2/3}$. We see that the model tends to estimate stronger turbulence at the ground and this could be due to the fact that the stereo-SCIDAR automatically subtracts the dome turbulence contribution. Also shown in figure~\ref{fig:cn2Comparison} is the comparison of the distribution of the turbulence strength values. We see that that distributions of the measured and modelled turbulence strength is similar. The first, second and third quartiles for the turbulence strength ($C_\mathrm{n}^2$) is $7.1\times10^{-19}$, $3.2\times10^{-18}$ and $1.1\times10^{-17}$ m$^{-2/3}$ respectively for stereo-SCIDAR and $8.9\times10^{-19}$, $3.7\times10^{-18}$ and $1.0\times10^{-17}$ m$^{-2/3}$ respectively for the model.

\comment{
cn2
7.096e-19 3.1572e-18 1.0856e-17
8.90323166541e-19 3.67753300138e-18 1.04719376289e-17
0.9554929771 0.907285599489
}

\subsection{Nightly Conditions}
\begin{figure*}
\centering
$\begin{array}{cc}
	\includegraphics[width=0.5\textwidth,trim={0cm 0 1cm 0}]{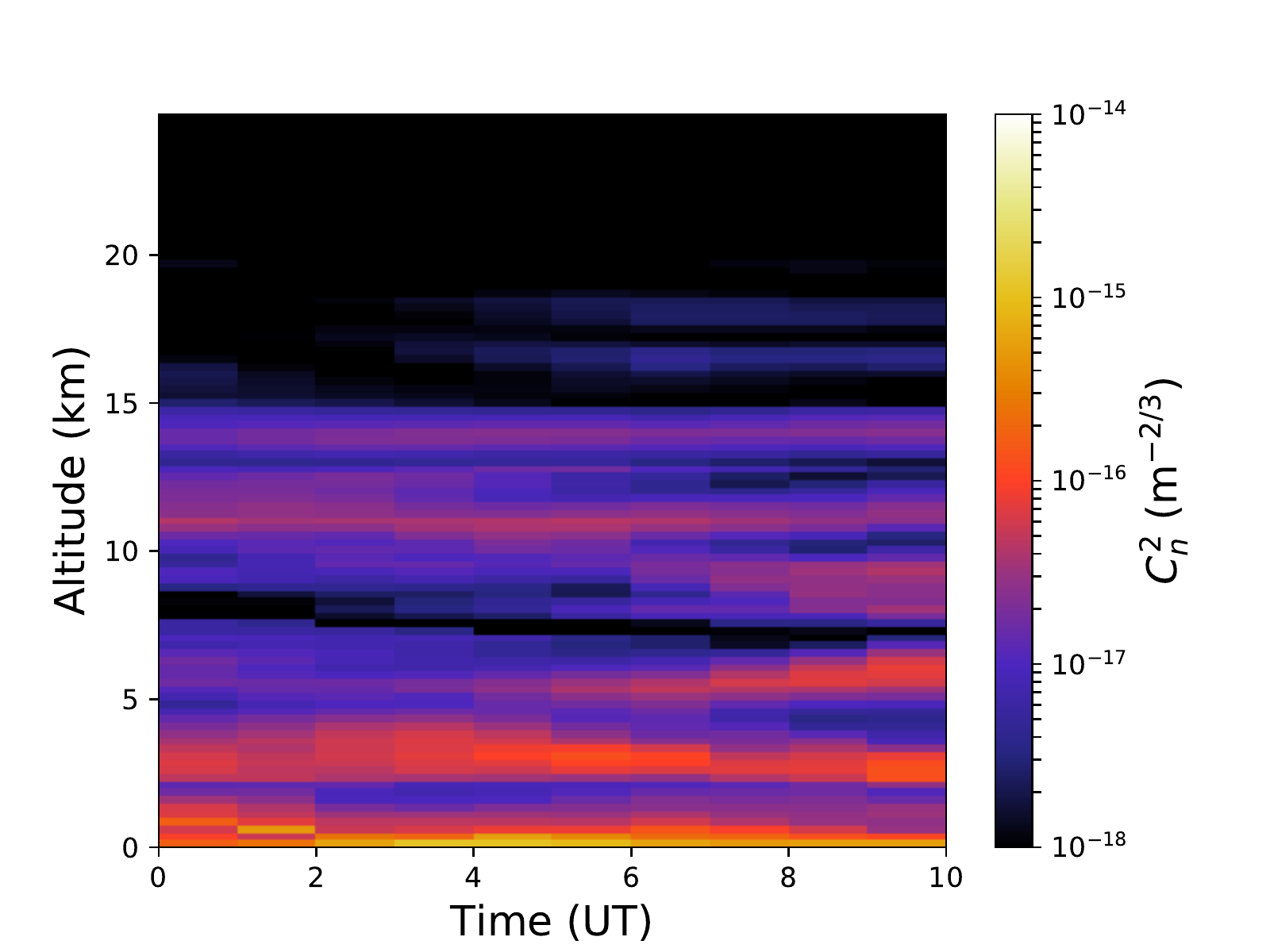}&
    \includegraphics[width=0.5\textwidth,trim={0cm 0 1cm 0}]{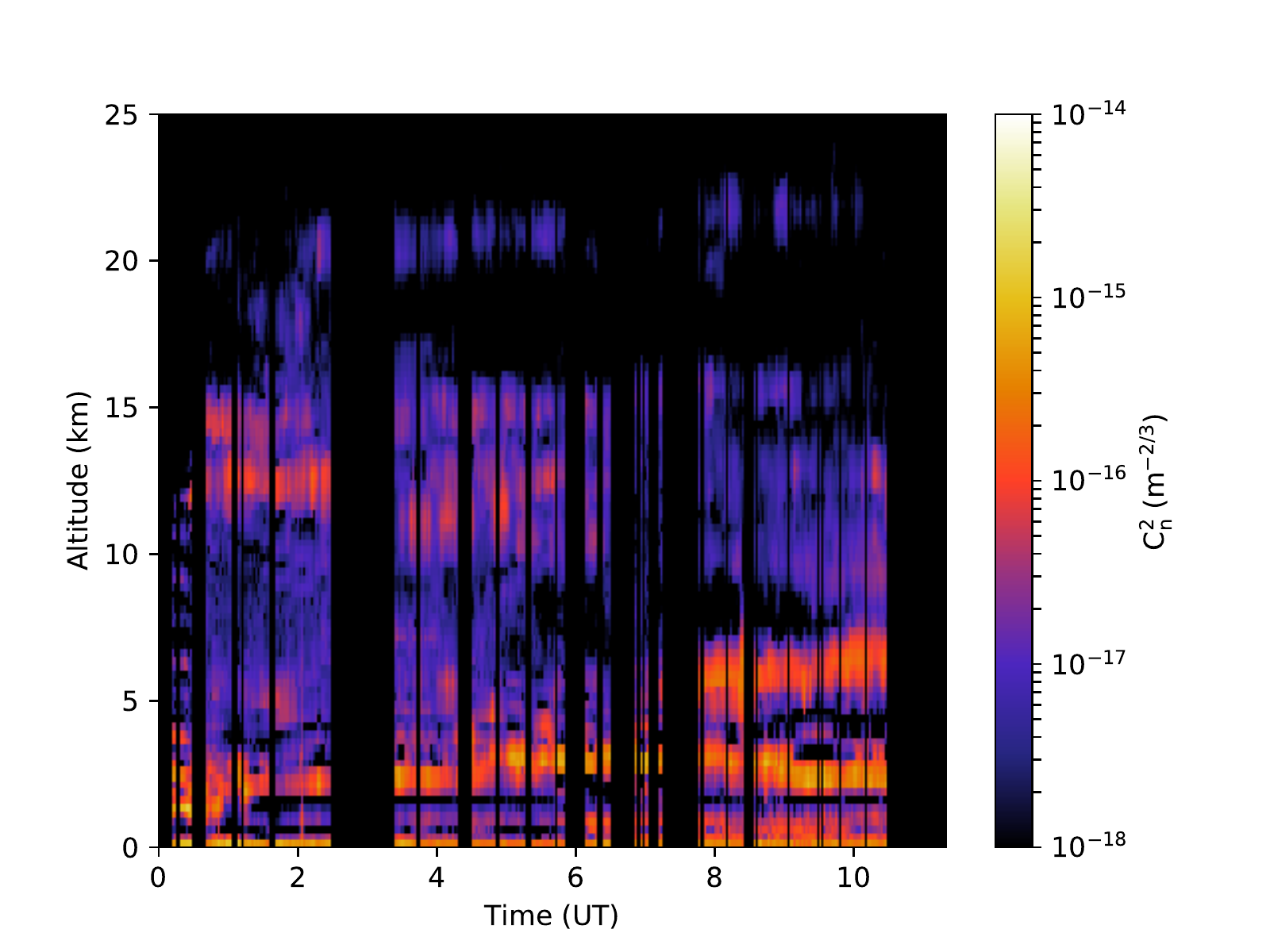}\\
    \mbox{ECMWF} & \mbox{SCIDAR}
 \end{array}$
\caption{Turbulence profile sequences for the night beginning 24th July 2016. The altitude shown is from observatory level. The measured stereo-SCIDAR turbulence profile sequence is shown on the right. Similarities can be seen between the two sequences. For example, a turbulent zone at approximately 2.5~km above observatory level can be seen to effectively split and diverge at approximately 06:00UT. }
\label{fig:ECMWFSequence}
\end{figure*}

Figure~\ref{fig:ECMWFSequence} shows an example of how the turbulence profiles evolves over a night for the model and for the stereo-SCIDAR measured profiles. Similar features can be seen on both, suggesting that the model can indeed forecast the dominant phenomena. A large sample of the profile sequences for the comparison nights are shown in appendix~\ref{ap:sequences}. This is a forecast and is therefore not guaranteed to perfectly reproduce the measurements. However, strong features, which dominate instrument performance limitations can be seen forecast on many nights.

\begin{figure}
\centering
	\includegraphics[width=0.45\textwidth]{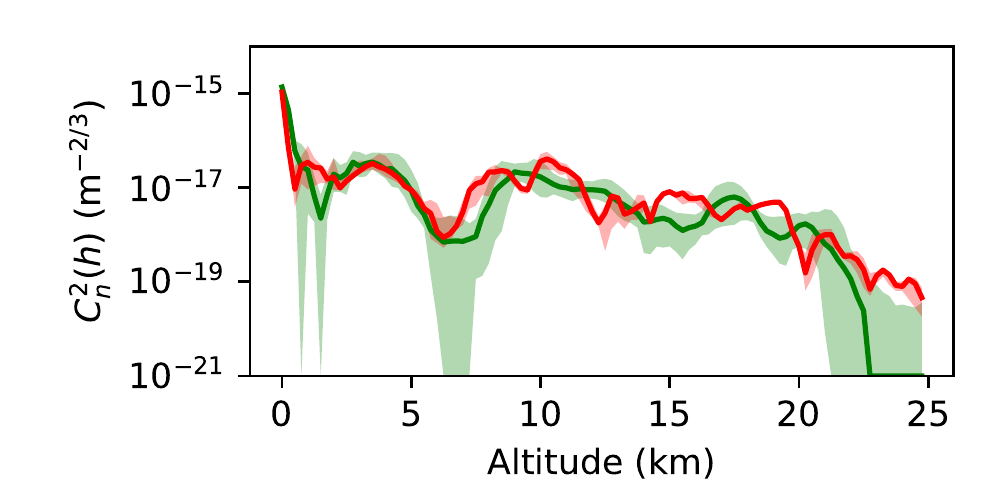}
\caption{Median turbulence profile sequences for the night beginning 24th July 2016. The altitude shown is from observatory level. The green curve is measurement from the stereo-SCIDAR and the red curve is the forecast from ECMWF data.}
\label{fig:ECMWFMedian}
\end{figure}

The median nightly measured and ECMWF forecast profiles for all of the stereo-SCIDAR nights at ESO Paranal are shown in \ref{sect:nightMedian}, an example is shown in figure~\ref{fig:ECMWFMedian}. Generally, the agreement is good. Large features which dominate the profile are seen in both. The shapes of the profiles differ from night to night demonstrating the versatility of the model, as well as the variability of the atmospheric turbulence structure.

\subsection{Astro-meteorological parameters comparisons}

The turbulence profile is certainly an important function for many applications. However, derived parameters such as the integrated seeing, free atmosphere seeing, coherence time and isoplanatic angle are also vitally important and will enable performance prediction and dynamic scheduling.

\red{Note that the model is normalised using 50\% of the stereo-SCIDAR data, we therefore expect the integrated seeing comparison to have a low bias}.

Figure~\ref{fig:paramsComparison} shows the comparison of the integrated seeing, the free atmosphere seeing ($h$>1~km), the ground layer seeing ($h$<1~km), the coherence time and isoplanatic angle. Table~\ref{tab:stats} presents the statistics of the parameter comparison. 
\begin{figure*}
\centering
$\begin{array}{cc}
	\includegraphics[width=0.45\textwidth,trim={1cm 0.3cm 1cm 0}]{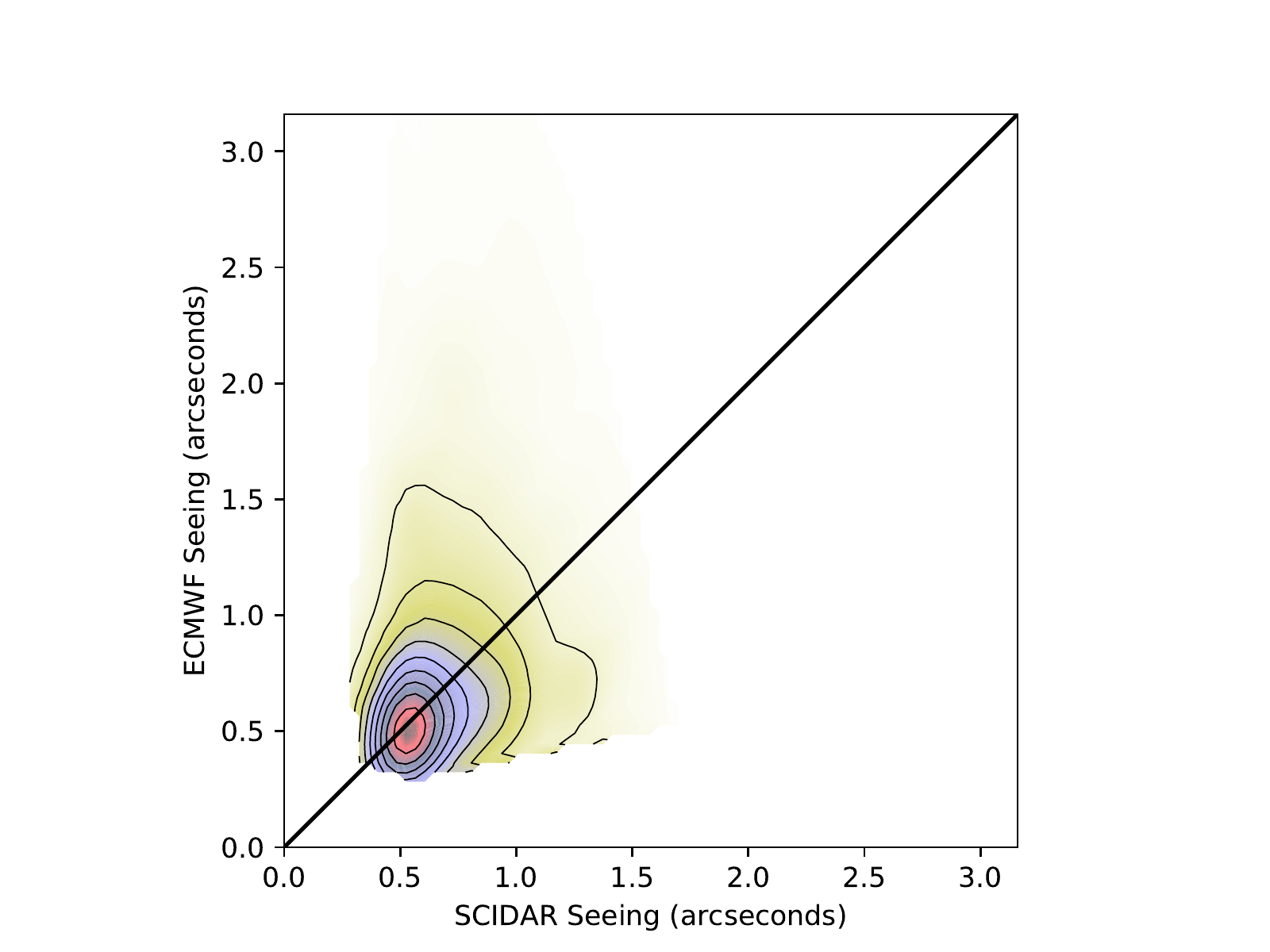} &
    	\includegraphics[width=0.45\textwidth,trim={1cm 0.3cm 1cm 0}]{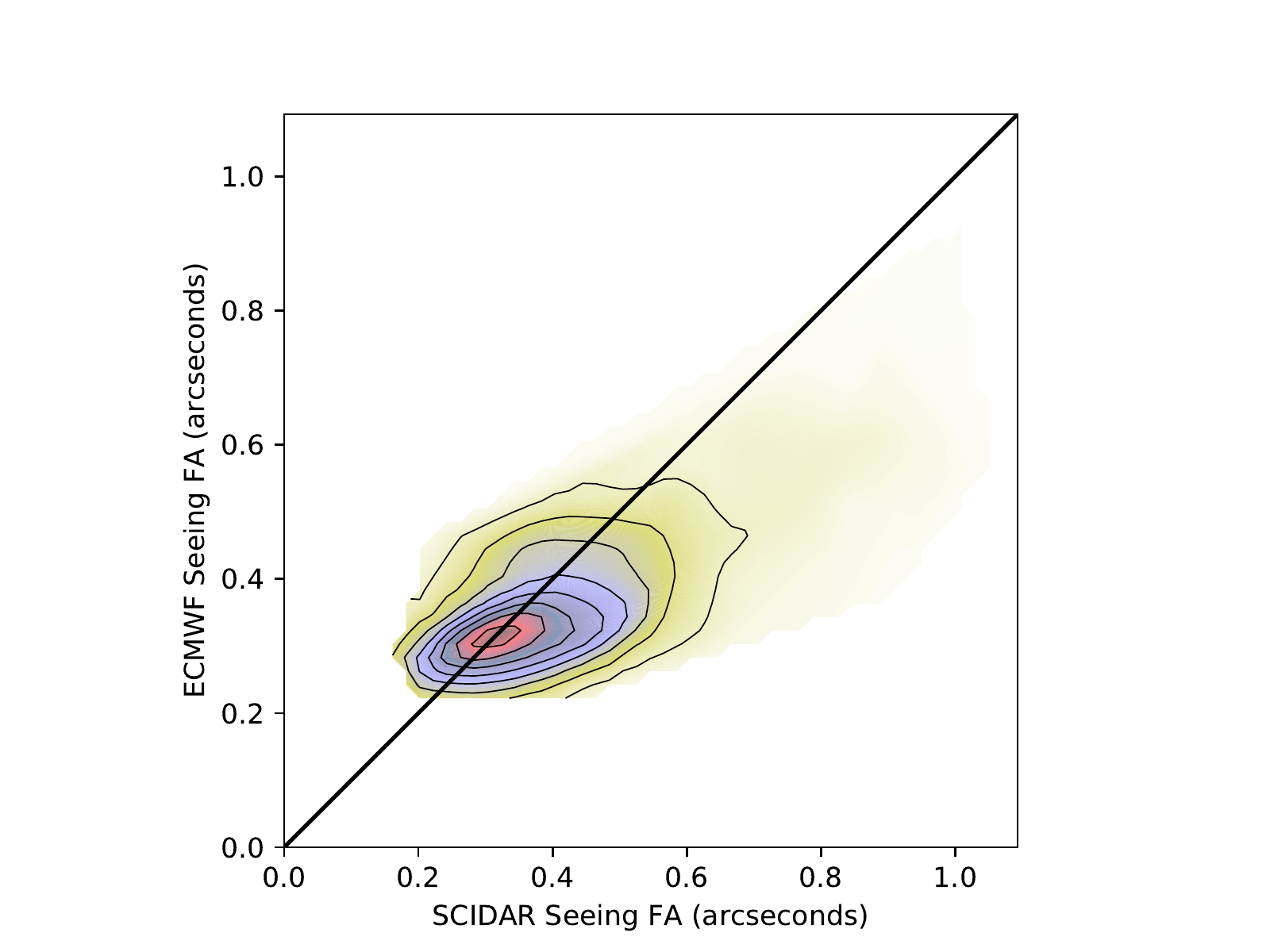} \\
	\mbox{(a)}& \mbox{(b)}\\
	\includegraphics[width=0.45\textwidth,trim={1cm 0 1cm 0cm}]{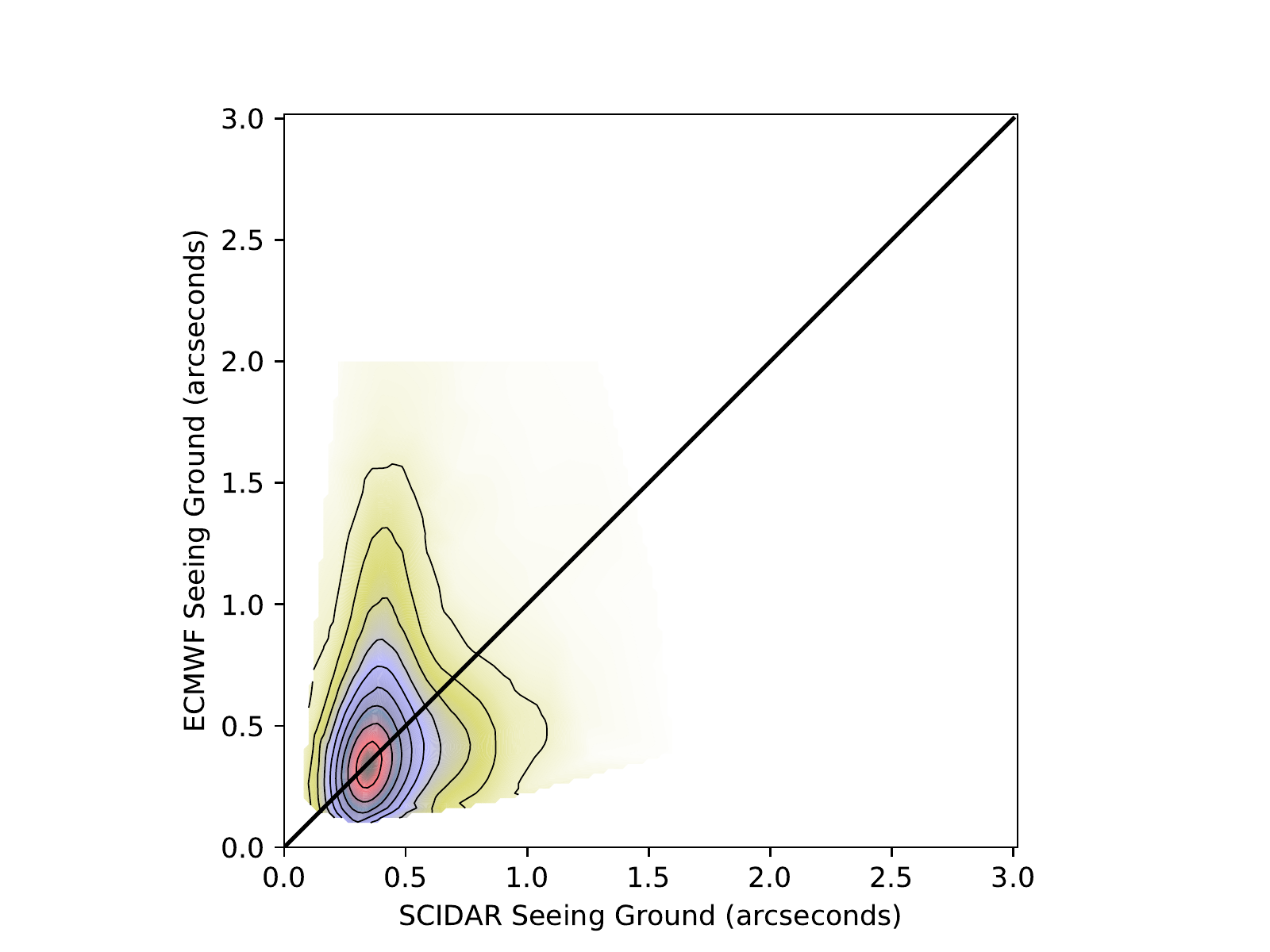} &
	\includegraphics[width=0.45\textwidth,trim={1cm 0 1cm 0cm}]{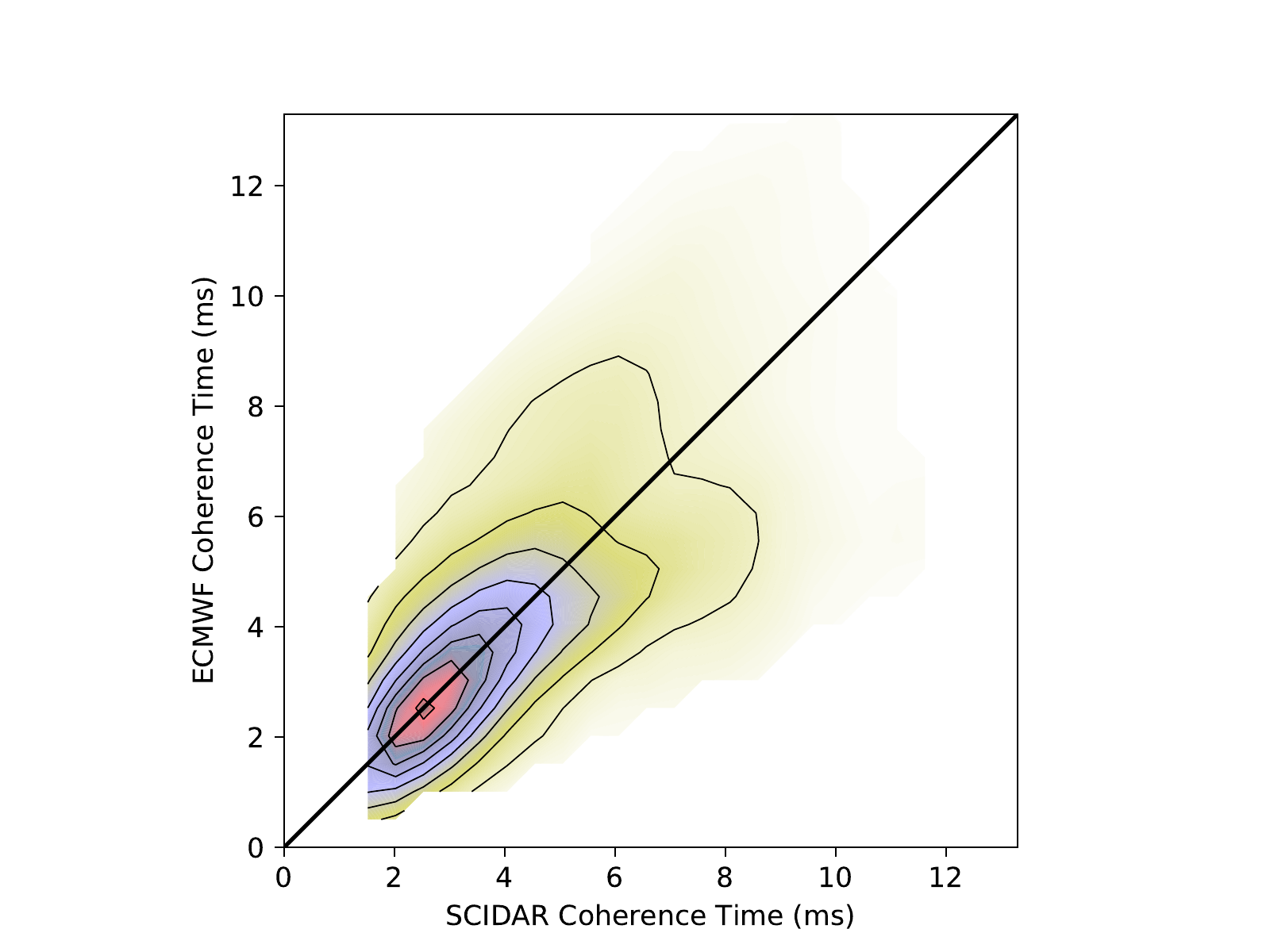} \\
	\mbox{(c)}& \mbox{(d)}\\
    	\includegraphics[width=0.45\textwidth,trim={1cm 0 1cm 0}]{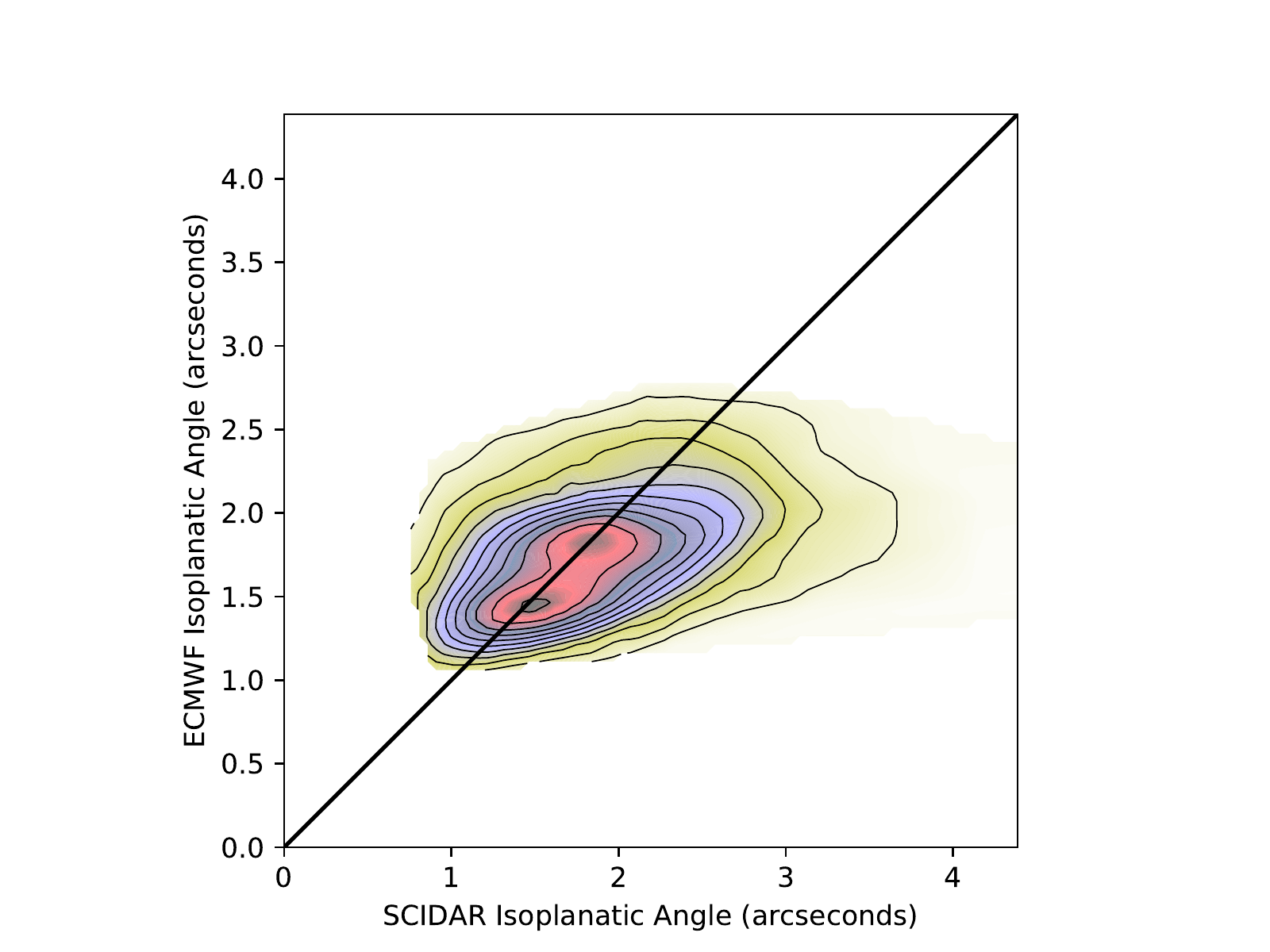}\\
	\mbox{(e)}
\end{array}$
\caption{\red{Comparisons of the atmospheric parameters estimated from ECMWF and measured by stereo-SCIDAR at ESO Paranal. (a) is the seeing, (b) the free atmosphere seeing, (c) the ground layer seeing, (d) the coherence time and (e) is the for the isoplanatic angle.}}
\label{fig:paramsComparison}
\end{figure*}
\comment{
\begin{figure*}
\centering
$\begin{array}{cc}
	\includegraphics[width=0.33\textwidth,trim={2cm 0 1cm 0}]{images/ecmwfScidarCoherenceTime_PAR.pdf} &
    	\includegraphics[width=0.33\textwidth,trim={2cm 0 1cm 0}]{images/ecmwfScidarIsoplanaticAngle_PAR.pdf}
\end{array}$
\caption{Comparisons of the coherence time (left) and isoplanatic angle (right) from the ECMWF model for ESO Paranal and the stereo-SCIDAR. The coherence time shows good agreement in fast conditions with more uncertainty in slower conditions. The model shows a smaller range of values for the isoplanatic angle than the measurements.}
\label{fig:paramsComparison}
\end{figure*}
}

\begin{table*}
\caption{Parameter comparison statistics}
\label{tab:stats}
\begin{tabular}{@{}llll}
\hline
Parameter & Correlation & Bias & RMSE\\
\hline
Seeing 				& 0.30 & -0.01$^{\prime\prime}$ 	& 0.31$^{\prime\prime}$ \\
Seeing (FA) 			& 0.64 & 0.08$^{\prime\prime}$ 	& 0.16$^{\prime\prime}$\\
Seeing (Ground)		& 0.24 & -0.05\as				& 0.33\as\\
$C_n^2$				& 0.63	& -1.13$\times10^{-18}$ m$^{-2/3}$	& $1.67\times10^{-16}$ m$^{-2/3}$\\
Coherence Time 		& 0.63 & -0.20 ms 				& 1.95 ms\\
Isoplanatic Angle		& 0.40 & 0.05$^{\prime\prime}$ 	& 0.62$^{\prime\prime}$\\
\hline
\end{tabular}
\end{table*}

Table~\ref{tab:parameterValues} compares the 1st and 3rd quartiles and the median values for the parameters derived from the ECMWF model and as measured by the stereo-SCIDAR. We see that the two techniques provide statistics within twice the standard deviation of the measurements over the 5 minute sampling period. The $\chi^2_\nu$ is the reduced $\chi^2$ parameter, where a value of 1 indicates a good fit between the measurement and the model.

\begin{figure*}
\centering
$\begin{array}{cc}
	\includegraphics[width=0.5\textwidth,trim={0cm 0 0cm 0}]{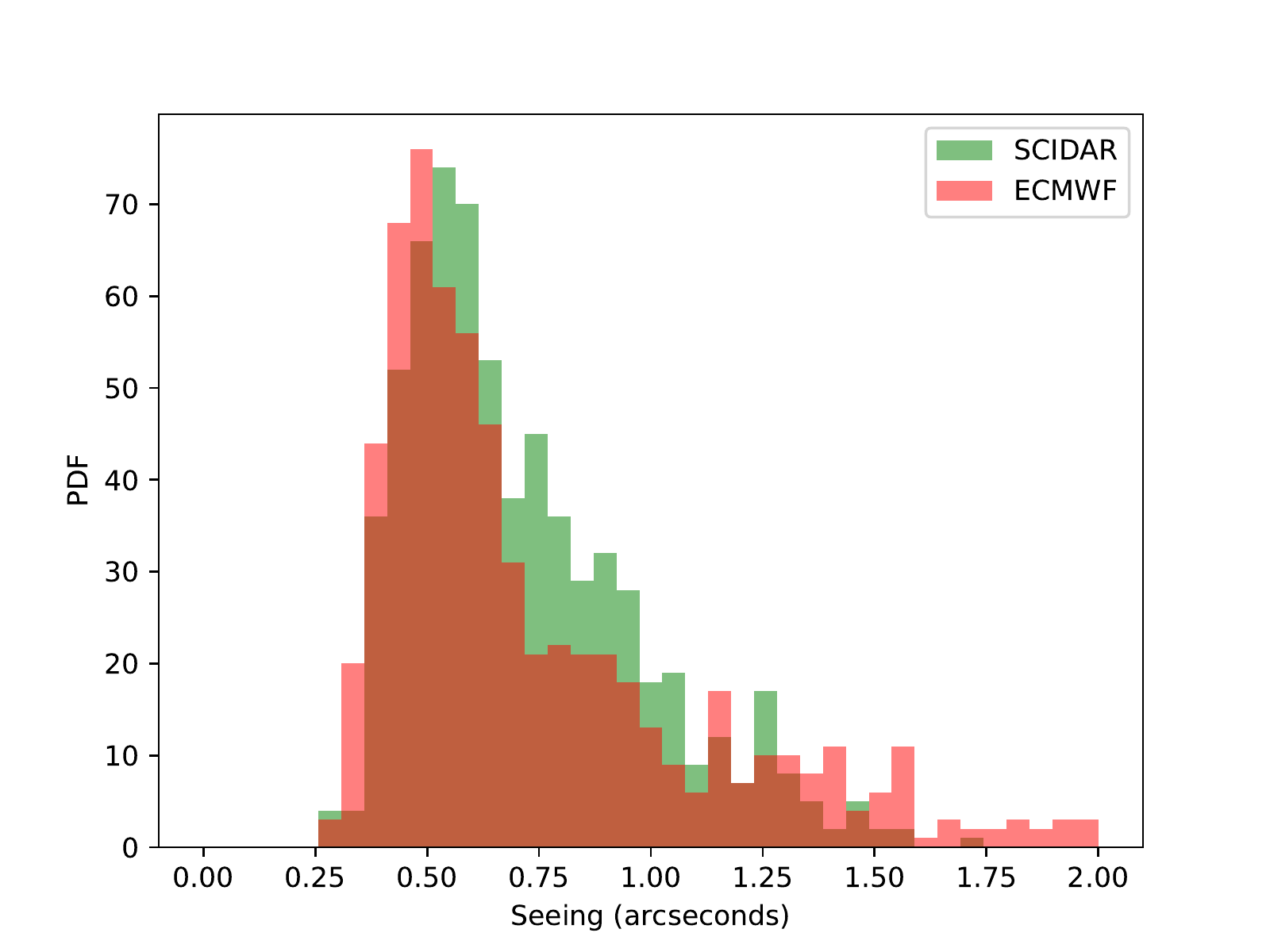} &
    	\includegraphics[width=0.5\textwidth,trim={0cm 0 0cm 0}]{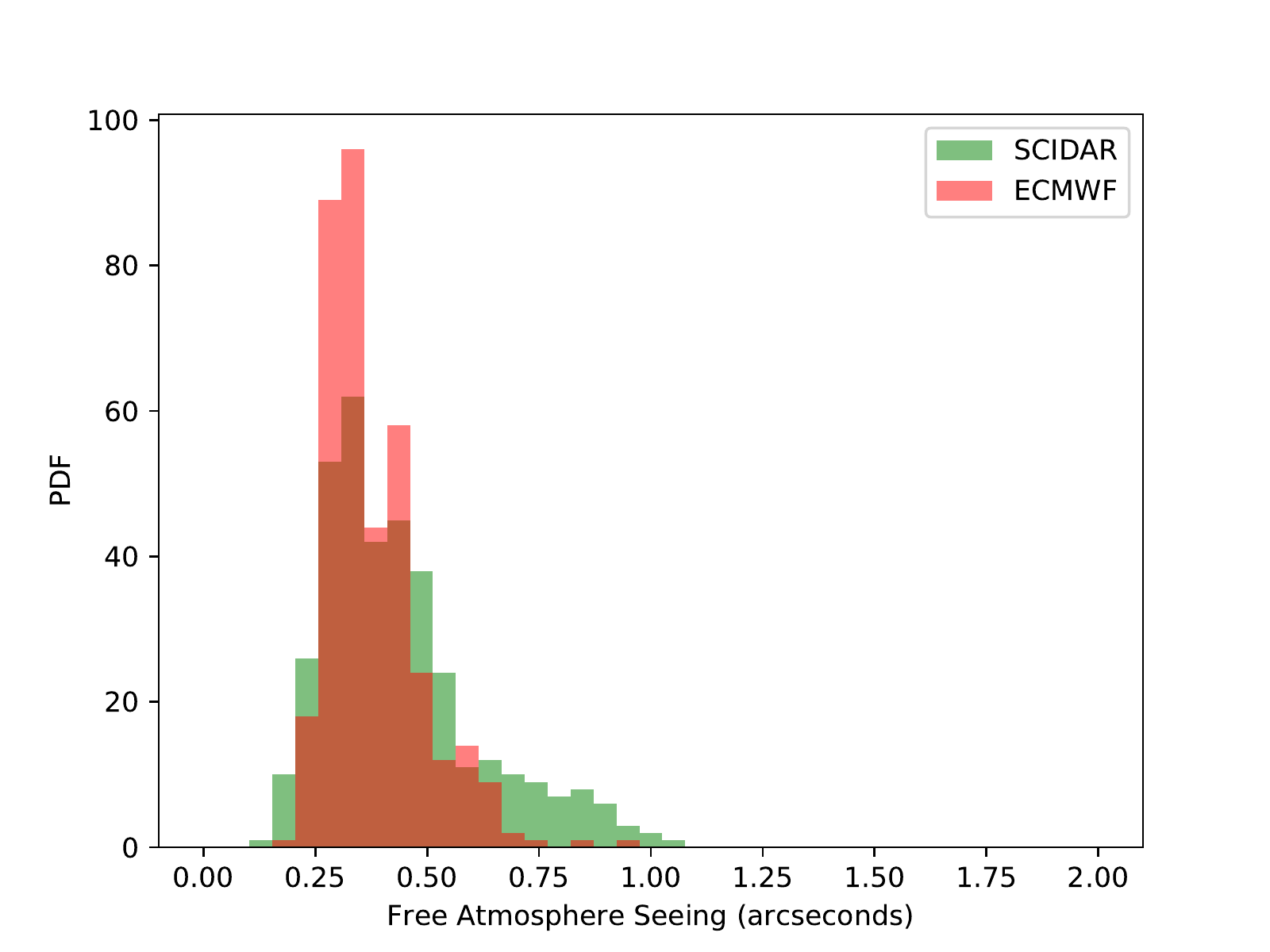}\\
	\mbox{(a)} & \mbox{(b)}\\
	\includegraphics[width=0.5\textwidth,trim={0cm 0 0cm 0}]{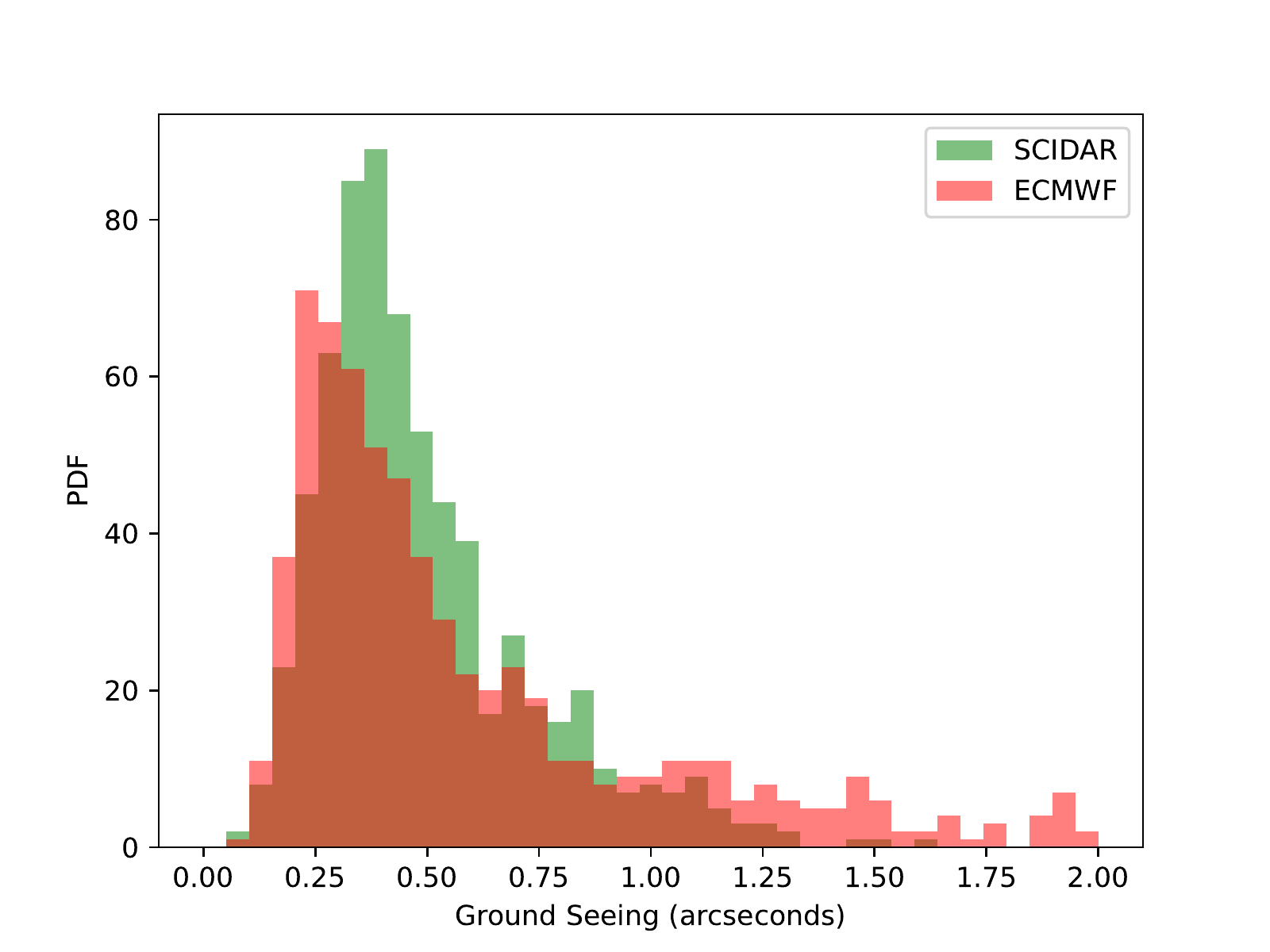}&
	\includegraphics[width=0.5\textwidth,trim={0cm 0 0cm 0}]{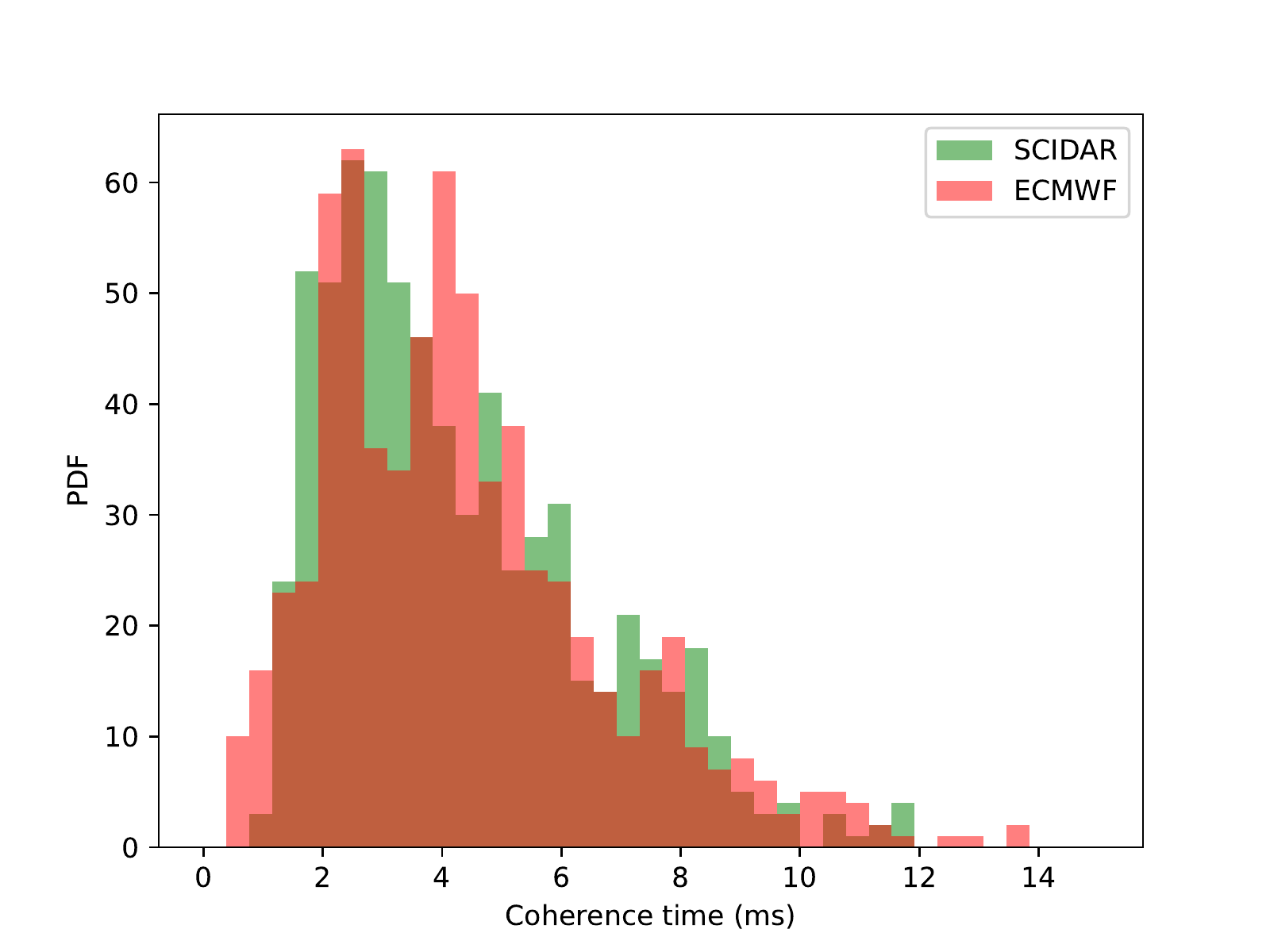} \\
	\mbox{(c)} & \mbox{(d)}\\
    	\includegraphics[width=0.5\textwidth,trim={0cm 0 0cm 0}]{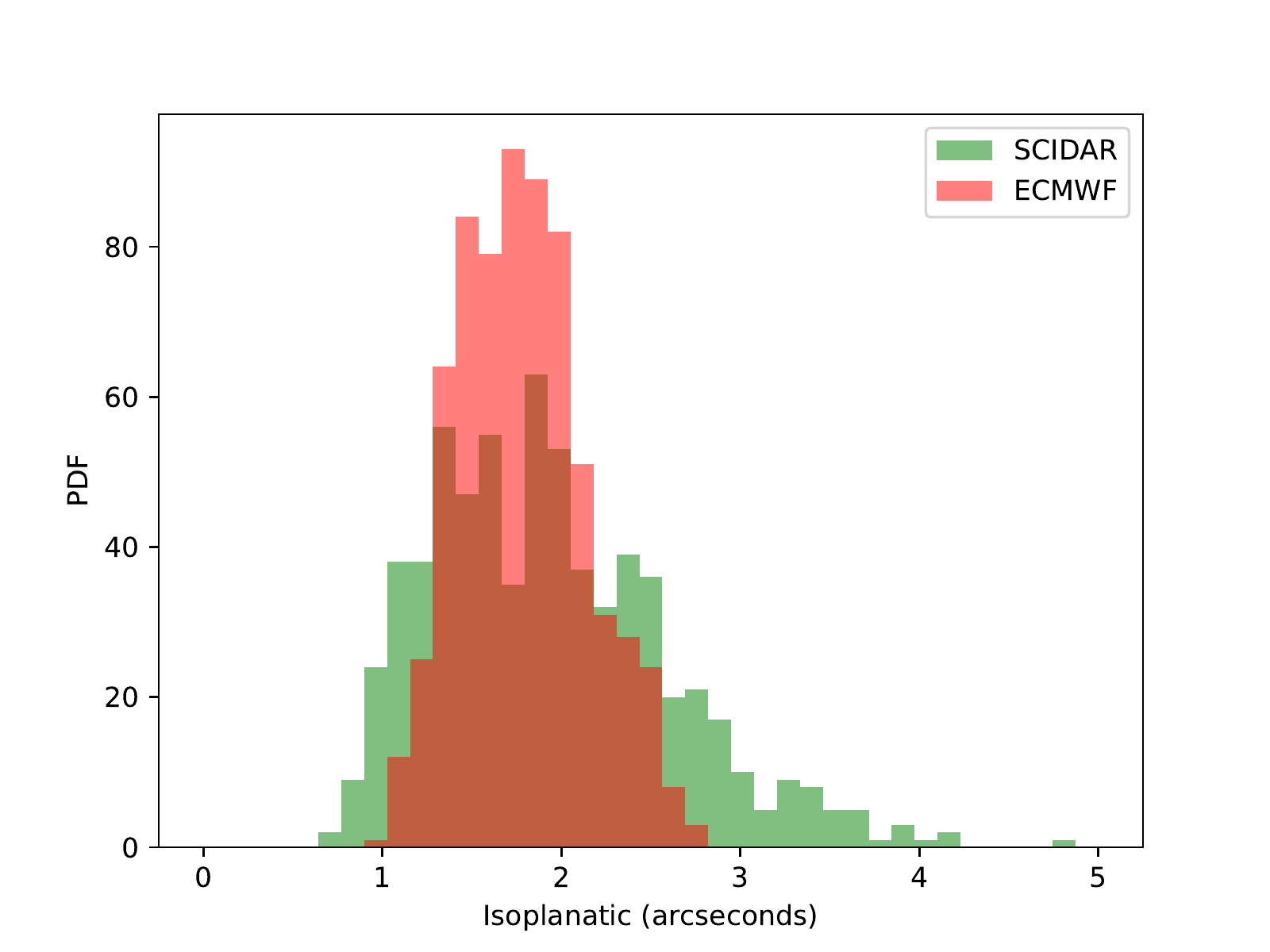}\\
	\mbox{(e)}
\end{array}$
\caption{\red{Comparisons of the distribution of atmospheric parameters estimated from ECMWF and measured by stereo-SCIDAR at ESO Paranal. (a) is for the total integrated seeing. It can be seen that in this case the ECMWF model and the stereo-SCIDAR measurements share a very similar distribution. (b) shows the comparison for the free atmosphere only ($h$>1~km above observatory level or 3.6~km above observatory level). In this case the the ECMWF model does not show the variability of the free atmosphere seeing that is measured by the stereo-SCIDAR, particularly for the more turbulent conditions. (c) shows the distribution of seeing values integrated up to 1~km. The distribution of the coherence time and the isoplanatic angle are shown in (d) and (e) respectively. The coherence time shows good agreement, however the isoplanatic angle from the ECMWF model does not show the same variability of values as the stereo-SCIDAR measurements.}}
\label{fig:seeingDist}
\end{figure*}

Figure~\ref{fig:seeingDist} shows the distributions of parameter values from the ECMWF model and the stereo-SCIDAR measurements. The distributions of integrated seeing and coherence time are similar, however the free atmosphere seeing and isoplanatic angle show a limited range of values compared to the measurements. In this case the bias is still low, suggesting that the median values of these parameters can be used if not the full distribution.

\begin{table*}
\caption{Parameter statistics}
\label{tab:parameterValues}
\begin{tabular}{@{}lllllllll}
\hline
Parameter &  \multicolumn{4}{c}{Stereo-SCIDAR} &\multicolumn{3}{c}{ECMWF} & $\chi^2_\nu$ \\ 
\hline
 				& Q1		& Median 	&Q3 & Standard Deviation& Q1		& Median 	& Q3	 &	\\
\hline
Seeing			&0.48\as	&0.61\as	&0.87\as 	& 0.07\as	&0.52\as	&0.62\as	&0.81\as&1.3				\\
Seeing (FA)		&0.30\as 	&0.34\as 	&0.44\as 	& 0.06\as	&0.31\as 	&0.41\as 	&0.55\as&2.9 			\\
Seeing (Ground)	&0.33\as	&0.42\as	&0.60\as	& 0.07\as	&0.33\as	&0.45\as	&0.65\as&1.9\\
Coherence Time	& 2.46 ms	& 3.97 ms	&5.67 ms	& 0.57 ms	&2.45 ms	& 3.61 ms	& 5.43 ms&1.7\\
Isoplanatic Angle	&1.48\as	&1.73\as 	&1.93\as 	&0.24\as	& 1.27\as	& 1.70\as	& 2.17\as	&4.8	\\
\hline
\end{tabular}
\end{table*}

\comment{
\begin{figure*}
\centering
$\begin{array}{cc}
	\includegraphics[width=0.33\textwidth,trim={0cm 0 0cm 0}]{images/coherenceHist.pdf} &
    	\includegraphics[width=0.33\textwidth,trim={0cm 0 0cm 0}]{images/isoHist.pdf}
\end{array}$
\caption{Comparisons of the distribution of the coherence time (left) and isoplanatic angle (right) from the ECMWF model for ESO Paranal and the stereo-SCIDAR. The coherence time shows good agreement, however the isoplanatic angle from the ECMWF model does not show the same variability of values as the stereo-SCIDAR measurements.}
\label{fig:paramsDist}
\end{figure*}
}

This GCM model has a low correlation when compared to the stereo-SCIDAR for the ground layer seeing (integrated from the ground to $h$=1~km). This is expected as the model cannot include the  effects of the local topology. This is obvious when examining the scatter of contemporaneous measurements, the correlation is 0.24. This low agreement near the ground also means that the total integrated seeing has a low contemporaneous agreement.  However, the distribution of ground layer seeing and integrated seeing is similar for both measured and modelled values, both have $\chi^2_\nu$ values less than 2. This suggests that the model does forecast an appropriate ground layer strength but the exact value depends on interaction with local structure. The good agreement of the parameter distributions means that this GCM model could be used statistically for site characterisation purposes. This will be particularly interesting if the model is validated at other locations.

The free atmosphere seeing shows a high correlation (0.64) of contemporaneous measurements with a low bias, but limited range of forecast values. This suggests that the model can be used to forecast the free atmosphere seeing although extreme values, particularly strong high altitude turbulence may be underestimated, as demonstrated by the relatively high $\chi^2_\nu$ of 2.9.

The coherence time shows a high correlation of contemporaneous measurements (0.63) and very similar parameter distributions, suggesting that the coherence time is a parameter which can be forecast by this model.

For the isoplanatic angle the correlation of contemporaneous values is low (0.40) and the inter-quartile range of the forecast parameters is limited compared to the measurements. However, the median value for the measurement and model is consistent, suggesting that the median isoplanatic angle can be recovered from forecast data although extreme events, particularly strong high altitude turbulence can be underestimated. This is consistent with the findings for the free atmosphere seeing and is likely to be caused by the limited altitude resolution of the GCM model.

\red{As this GCM model has coarse spatial resolution (0.3 degree) it is not possible to resolve the turbulence caused by local topography near the ground. However, the integrated parameters such as isoplanatic angle and coherence time tend to be dominated by high altitude turbulence where such fine resolution is not required. The bias on the integrated seeing is low, but it should be remembered that 50\% of the stereo-SCIDAR data is used to normalise the integrated turbulence strength of the model.}

\subsection{Discussion}

Comparisons with previous studies is difficult as the studies concentrate on different sites around the world, use different instruments for the validation and quote their results with different metrics. \red{It should be noted that the results presented here are for the best case of GCM. The data used is re-anaylsis data and we use the optimum (ie shortest) forecast time available.} Table~\ref{tab:comparisons} shows the bias and RMSE extracted from previous studies. The results from this study compare favourably with that of \citeauthor{Ye2011} suggesting that the model used here is more reliable as a GCM approach. We note the Mauna Kea Weather Centre mesoscale model quotes a seeing correlation coefficient of 0.18 and an RMSE of 0.32\as \cite{Cherubini2008}. The results from this study are also comparable to those of the mesoscale models of \citeauthor{Masciadri2017, Giordano2013}. This suggests that any potential gain from using the complicated mesoscale approach is possibly negated by the long lead times required for the processing.

Modelling the ground layer of the atmospheric turbulence was expected to be difficult with a GCM approach due to the low spatial resolution. However, we note that the contemporaneous comparison statistics for the integrated seeing is comparable to that of the more complicated mesoscale models.

\begin{table*}
\caption{Model Comparisons with the MASS-DIMM instrument from previous studies}
\label{tab:comparisons}
\begin{tabular}{@{}lllllllll}
\hline
Parameter				& \multicolumn{2}{c}{\cite{Masciadri2017}}	& \multicolumn{2}{c}{\cite{Giordano2013}}	& \multicolumn{2}{c}{\cite{Ye2011}}& \multicolumn{2}{c}{This Study}\\
\hline
		  			& Bias 				& RMSE				& Bias 	& RMSE	& Bias 	& RMSE 	& Bias	&RMSE\\
\hline
Seeing 				& -0.09 $^{\prime\prime}$ 	& 0.48$^{\prime\prime}$	& -0.17\as	& 0.58\as	& 0.3\as	&0.22\as	& -0.01\as	& 0.31\as\\
Seeing (FA)			& -					& -					& -0.12\as	& 0.4\as	& -		& -		& 0.08\as 	& 0.16\as\\
Coherence Time 		& -0.99 ms (summer)	& 1.90 ms	(summer)		& 0.63 ms	& 4.64 ms	& -		& -		& -0.20 ms& 1.95 ms\\
					& -1.28 ms (winter)		& 2.20 ms (winter)		&		&		&		&		&		&		\\
Isoplanatic Angle		& 0.20$^{\prime\prime}$ 	& 0.60$^{\prime\prime}$	& 0.34\as	& 0.73\as	& -		& -		& 0.05\as	& 0.62\as\\
\hline
\end{tabular}
\end{table*}

\section{Towards Operational Real-time Forecasts}
The study here was facilitated by low-resolution, publicly available, historical data directly from ECMWF. However, ESO does have an agreement with ECMWF to access current forecasts. Due to data volume, this data is not recorded and so could not be used for this study. However, this dataset has a  higher spatial resolution of 0.1 degree grid. As the computation of the turbulence profile is trivial, negligible computational time is required to process the data after the forecast is released. This will enable rapid reaction to changes in forecast as well as monitoring of conditions on the build up to the observation.

In order to be operationally useful a further study into the accuracy of the turbulence forecast as a function of the forecast time is required. In addition, this work concentrates on validating the model at ESO Paranal, although further work to validate the model at other sites is required.

\section{Conclusions}
We have developed an optical atmospheric turbulence forecast based on global GCM forecast data. \red{This model is extremely fast to calculate, this low processing time removes any latency between a forecast being released and a useable forecast being computed, enabling rapid reaction to changing forecast conditions.}

\red{The ultimate goal is to use the model globally without any site specific calibration. However, the model does require a stability coefficient. In this case we normalise the integrated turbulence strength using 50\% of the stereo-SCIDAR data. It is not clear whether this normalisation is valid globally and so in this work we concentrate on reporting the performance at ESO Paranal.}

\comment{The model does not require a site-specific calibration and is extremely fast to calculate. Alternative complicated meso-scale models must be calibrated for each season and every site individually. In addition to low processing time removes any latency between a forecast being released and a useable forecast being computed, enabling rapid reaction to changing forecast conditions.}

We have shown that a GCM forecast can be used to forecast the vertical profile of the Earth's atmospheric turbulence. From this forecast astro-meteorological parameters, such as the seeing, coherence time and isoplanatic angle can be calculated. We use the model level (137 vertical levels) forecast from the European Centre for Medium range Weather Forecasts (ECMWF) and compare with measurements from the dedicated optical turbulence profiling instrument stereo-SCIDAR. 

In addition to the integrated parameters, the model provides an estimate of the vertical turbulence profile of the optical turbulence strength and velocity.  The median vertical profile of the atmospheric turbulence from the stereo-SCIDAR data is well correlated with the model (0.98). We have shown comparisons of the stereo-SCIDAR and ECMWF model for the sequence of turbulence profiles as well as the nightly median profiles. From these comparisons we see that the model does forecast strong turbulent features. It is these strong features that will limit the performance and are therefore most important for astronomical observations. 

We find that the integrated free atmosphere turbulence profile can be forecast with a correlation of 0.64. The coherence time is also estimated with a high correlation (0.63) and low bias and rmse. However, the isoplanatic angle has a lower correlation (0.40). This suggest that the model is good at forecasting the wind velocity, and hence the good estimate of the coherence time, however errors in the high altitude turbulence strength forecast is amplified by the isoplanatic angle calculation, leading to a poorer estimate. The integrated seeing has a correlation of 0.30. The bias is low (-0.01$^{\prime\prime}$), however the rmse is large (0.31\as), as expected, as the model does not include the turbulence inducing topography on the ground. The two techniques provide statistics within twice the standard deviation of the measurements over the 5 minute sampling period.

For astronomical applications, where the telescope is located in a dome, the ground layer of the optical turbulence becomes less important as no current model will be able to forecast the dynamics inside the dome which can be caused by the interaction of many complicated mechanisms (such as internal heat sources and the interaction of external air at the dome interface). An update to the model to account for low altitude and dome induced turbulence would increase the usefulness of the solution.

In addition, to contemporaneous comparison, we also show that the distribution of astro-meteorological parameters form the model agrees with those from stereo-SCIDAR. This includes parameters such as ground layer seeing and integrated seeing. This is particularly interesting as it suggests that although the model can not currently forecast the exact magnitude of the turbulence near the ground, statistically the value is consistent with the measured distribution. Therefore, if validated elsewhere, this none-calibrated, global model can be used to calculate site statistics from archived GCM weather forecast data, unlocking the potential of a vast dataset.

The GCM model here shows comparable results with more computationally intensive mesoscale models suggesting that, currently, there is no advantage in using a complicated mesoscale model to extract atmospheric parameters such as seeing (free atmosphere and total integrated), coherence time and isoplanatic angle.

\section*{Acknowledgments}
FP7/2013-2016: The research leading to these results has received funding from the European Community's Seventh Framework Programme (FP7/2013-2016) under grant agreement number 312430 (OPTICON). Horizon 2020: This project has received funding from the European Union's Horizon 2020 research and innovation programme under grant agreement No 730890. This material reflects only the authors views and the Commission is not liable for any use that may be made of the information contained therein. This work was supported by the Science and Technology Funding Council (UK) (ST/L00075X/1). We also acknowledge ECMWF for access to the weather forecast data through the MARS access system. This research made use of python including numpy and scipy \citep{numpy}, matplotlib \citep{matplotlib} and Astropy, a community-developed core Python package for Astronomy \citep{astropy}. We also made use of the python AO utility library `AOtools'.

\bibliographystyle{mnras}
\bibliography{library}


\appendix
\section{Nightly profiles}
\label{ap:sequences}
\begin{figure*}
\centering
$\begin{array}{cccc}
	\includegraphics[width=0.23\textwidth,trim={2cm 0 1cm 0}]{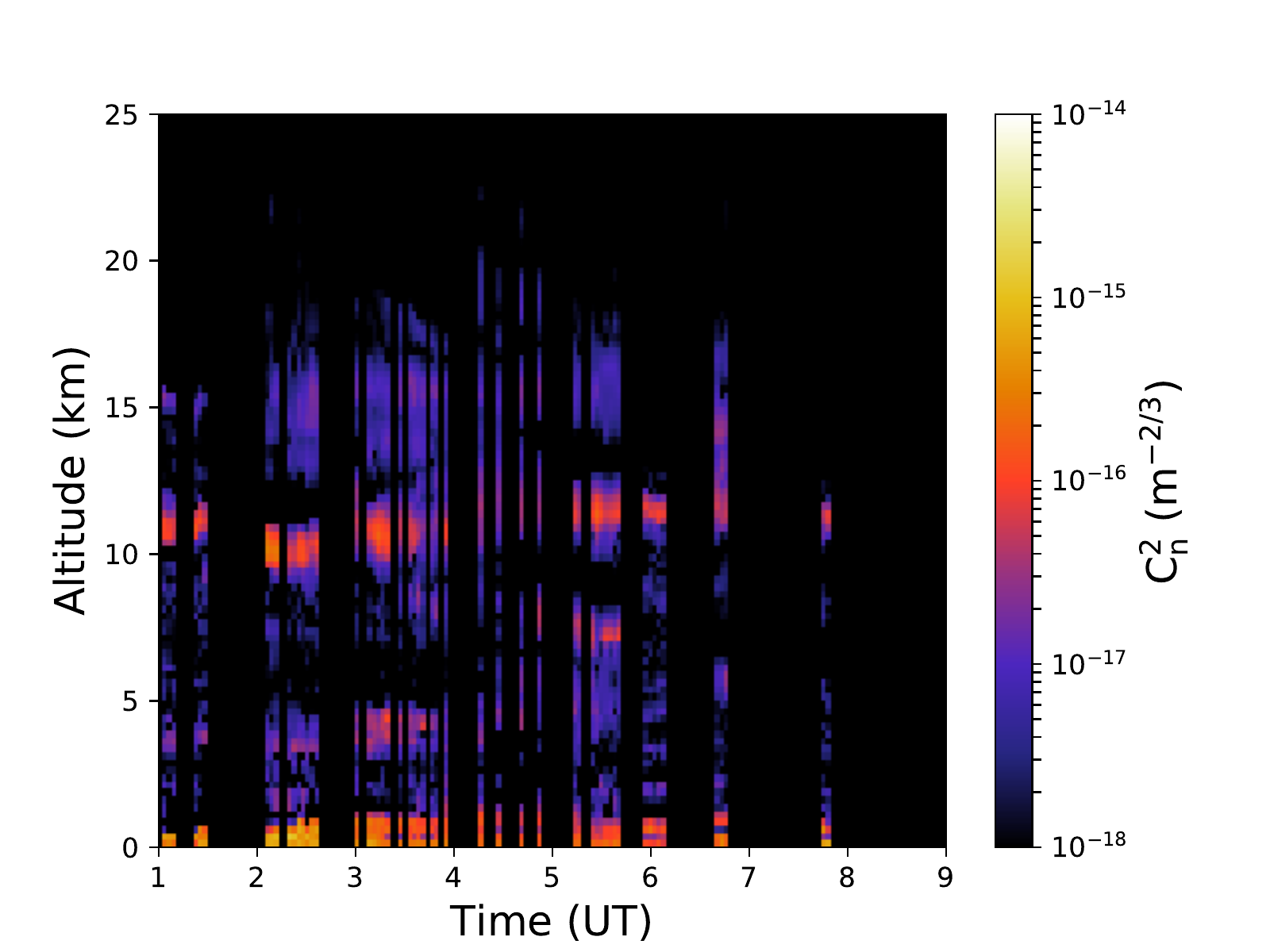} &
    	\includegraphics[width=0.23\textwidth,trim={2cm 0 1cm 0}]{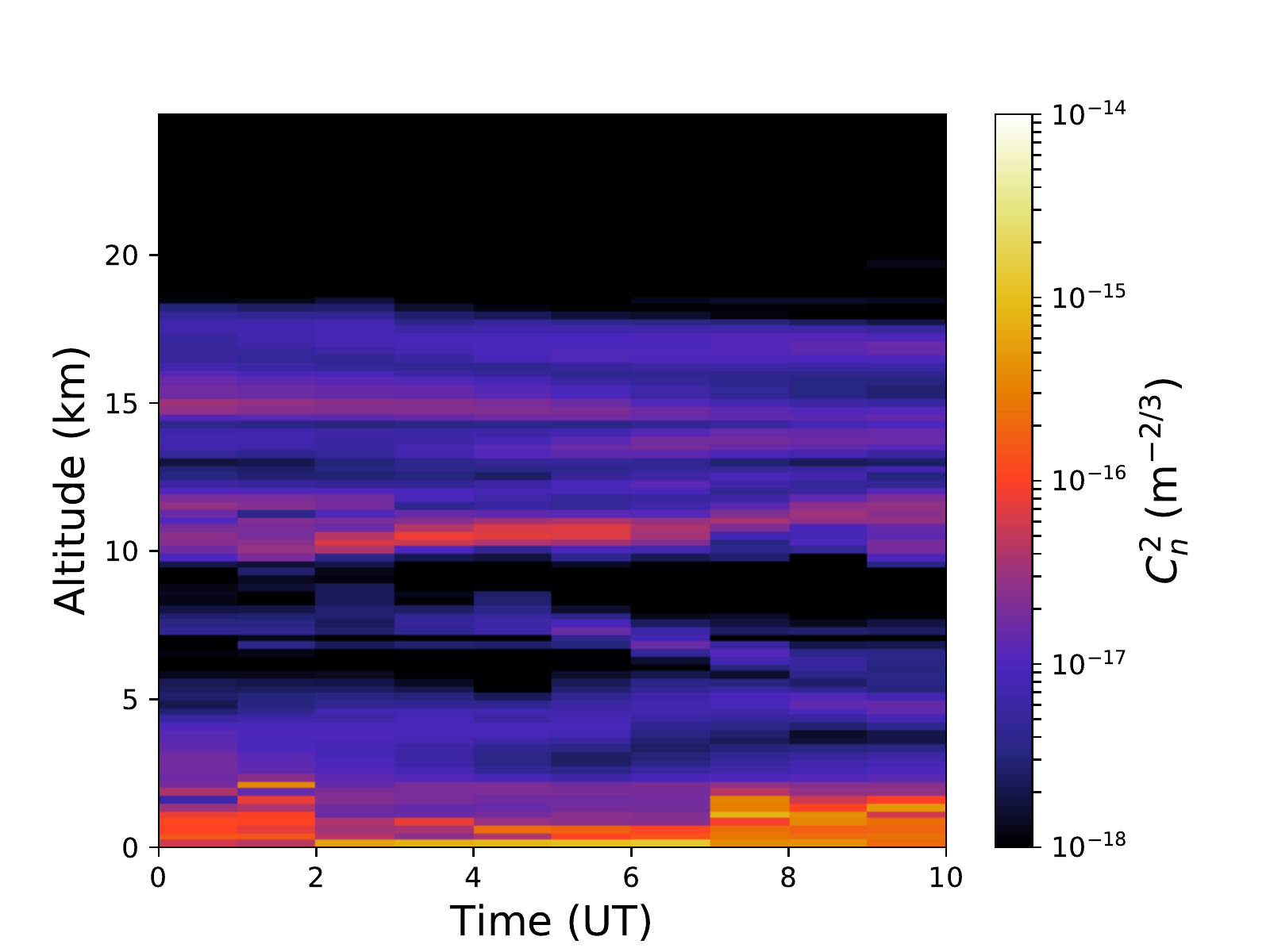} &
    	\includegraphics[width=0.23\textwidth,trim={2cm 0 1cm 0}]{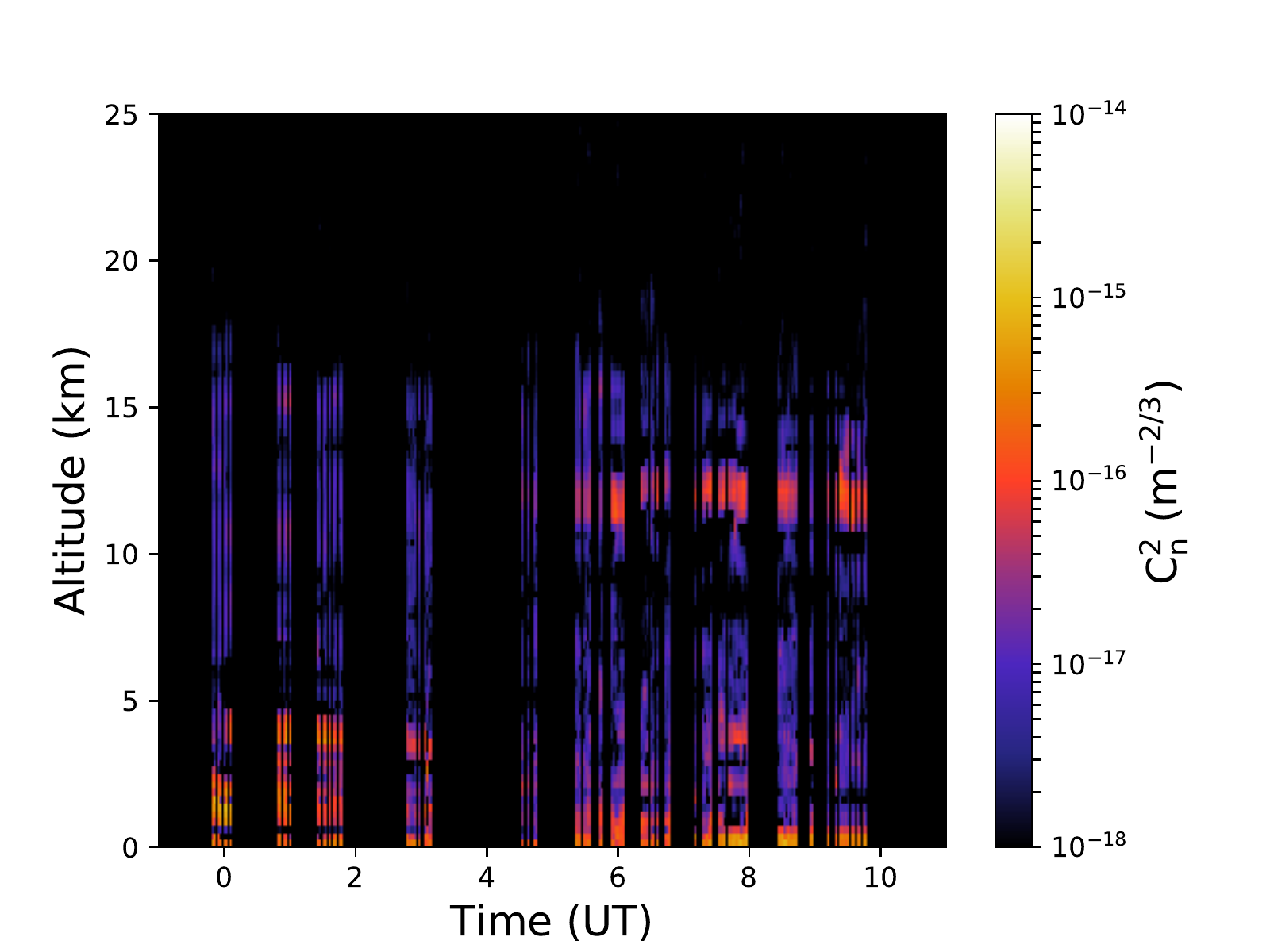} &
    	\includegraphics[width=0.23\textwidth,trim={2cm 0 1cm 0}]{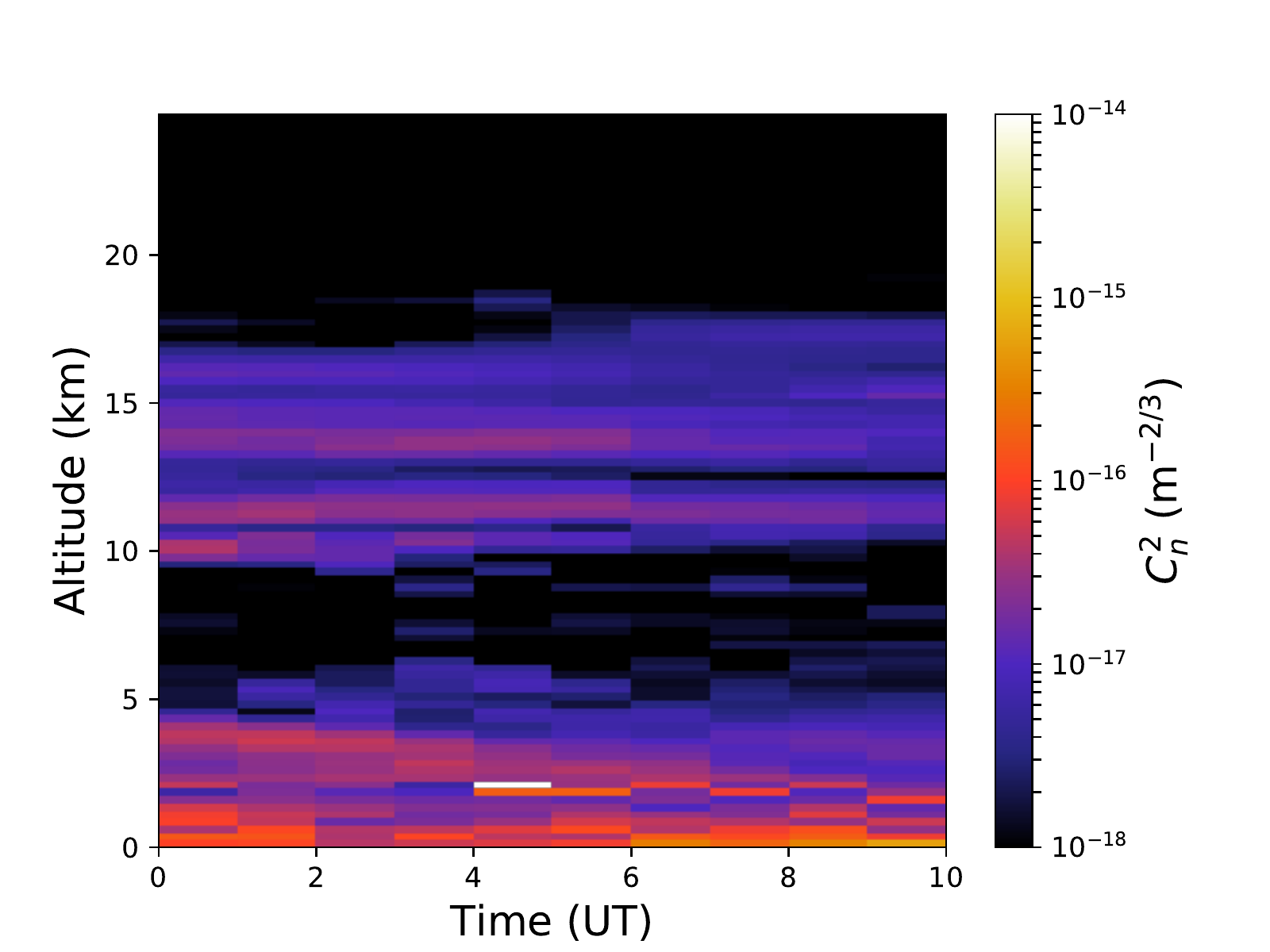} \\
	\includegraphics[width=0.23\textwidth,trim={2cm 0 1cm 0}]{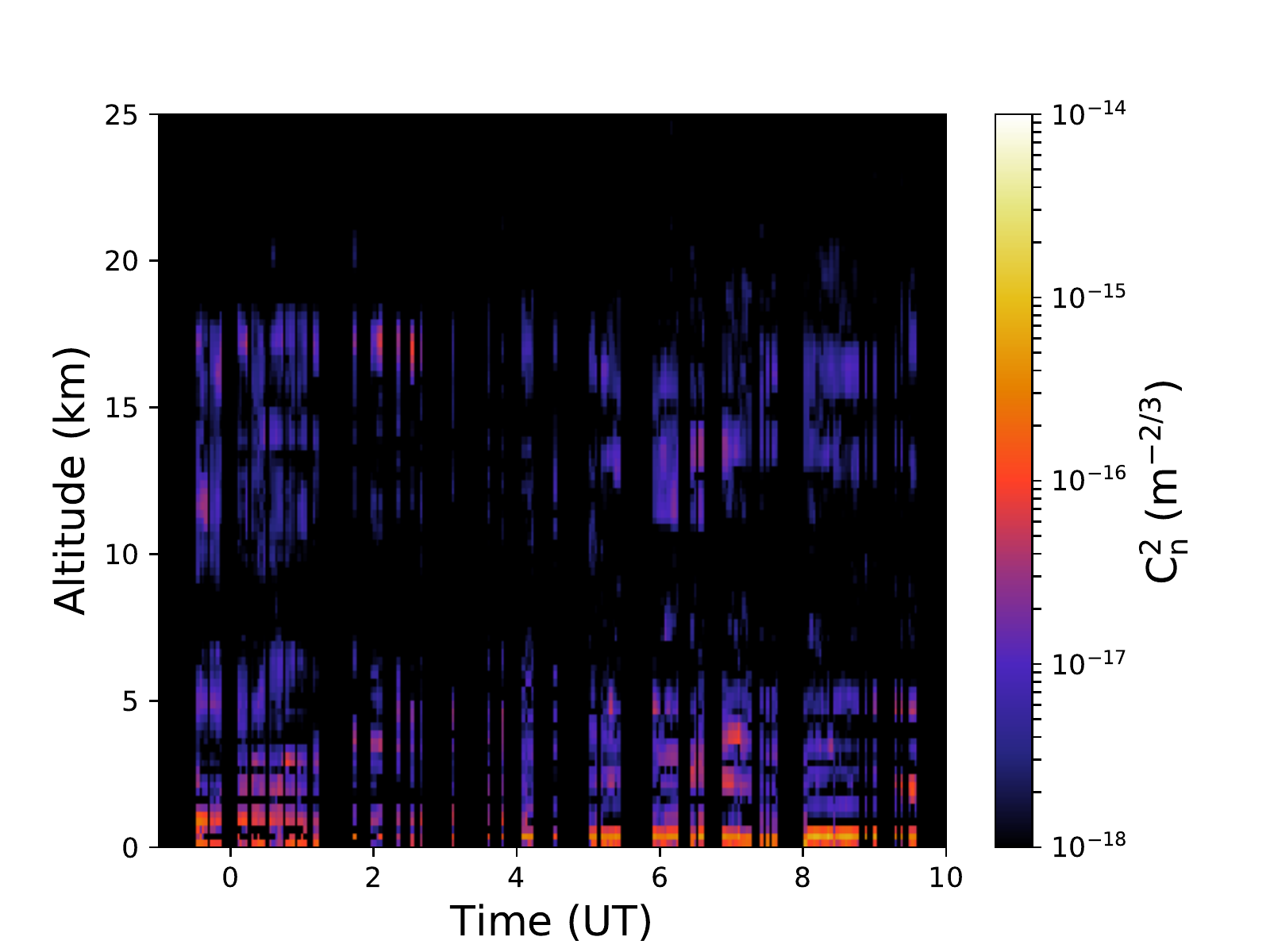} &
    	\includegraphics[width=0.23\textwidth,trim={2cm 0 1cm 0}]{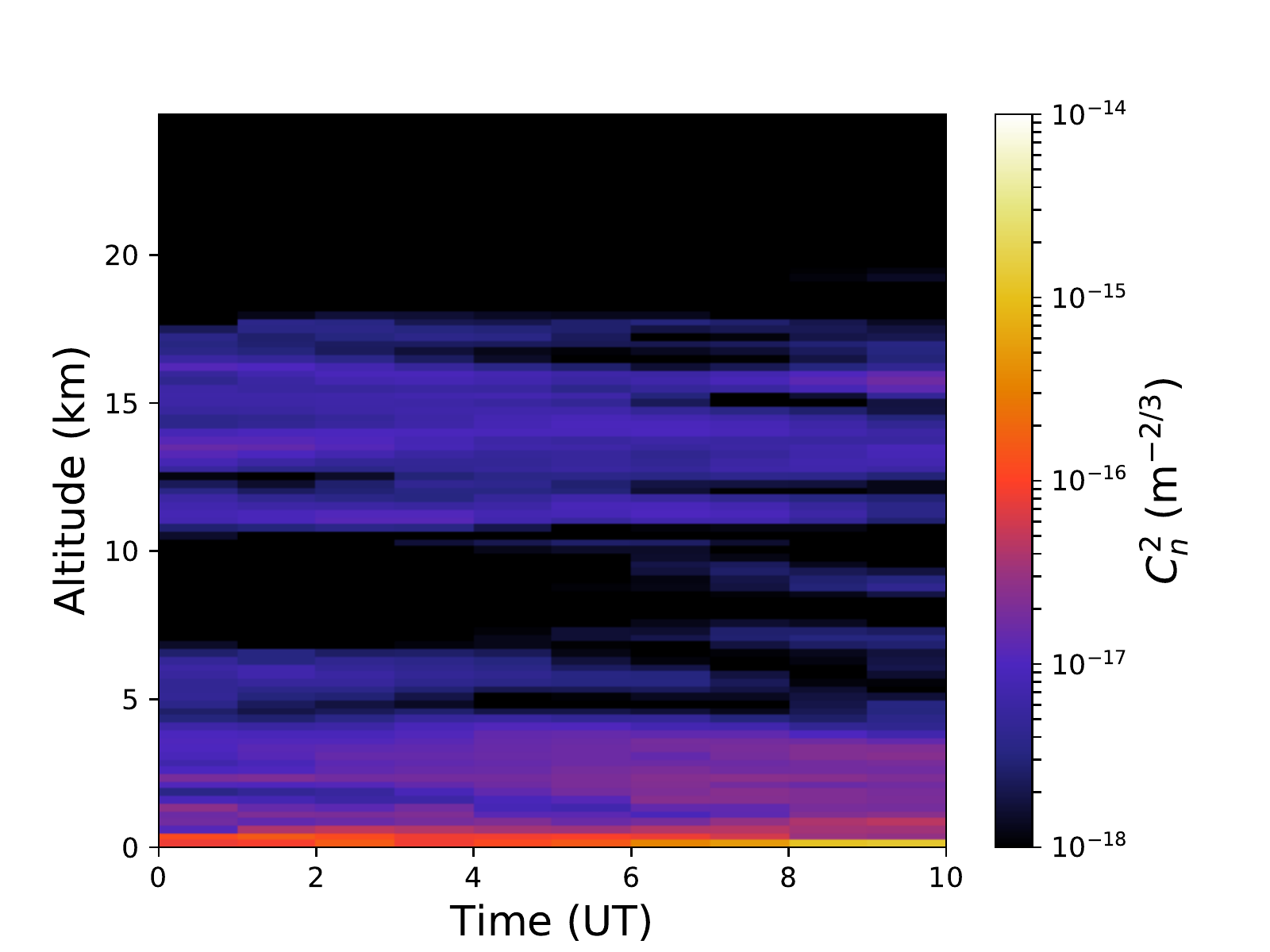} &
    	\includegraphics[width=0.23\textwidth,trim={2cm 0 1cm 0}]{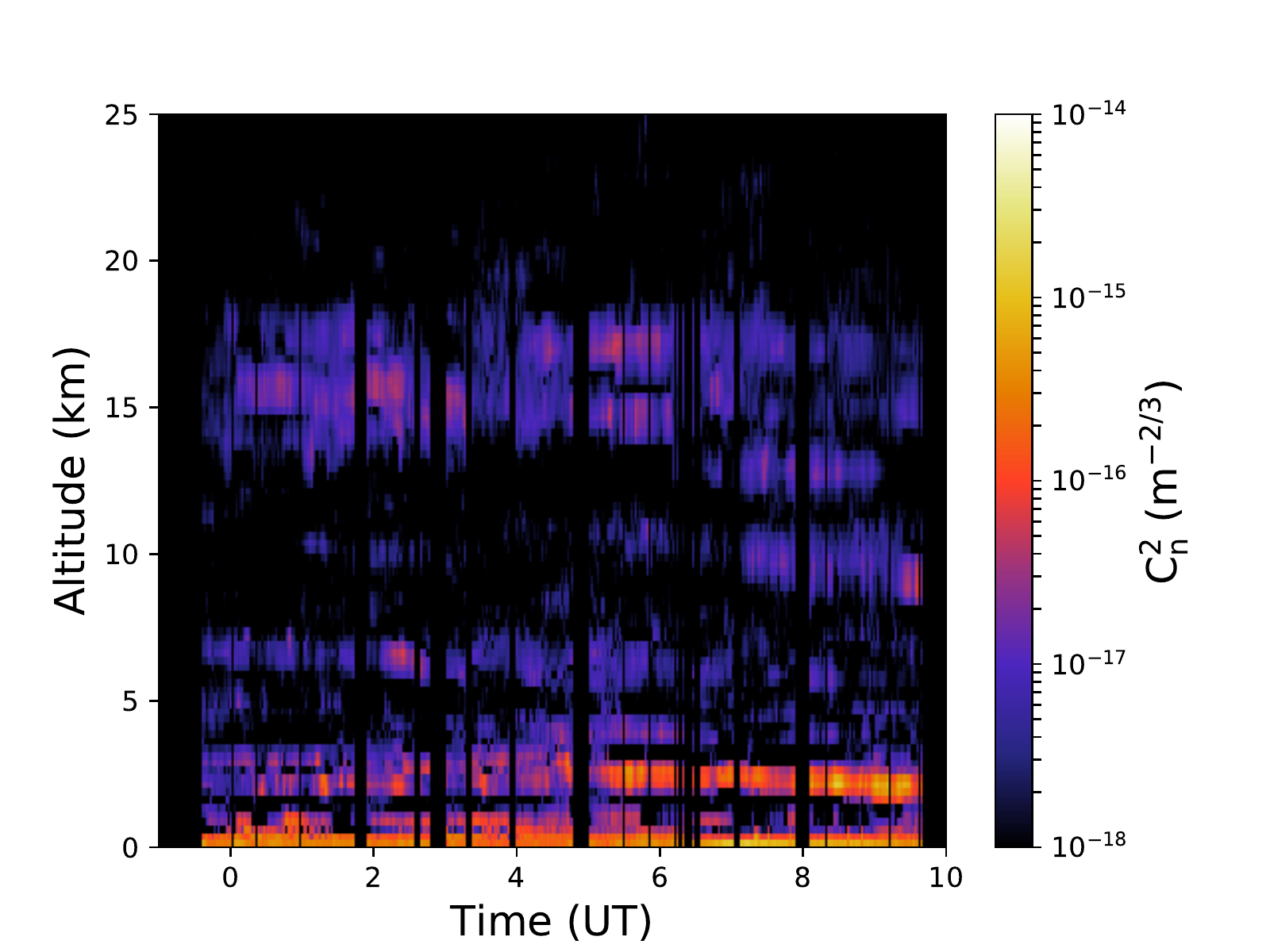} &
    	\includegraphics[width=0.23\textwidth,trim={2cm 0 1cm 0}]{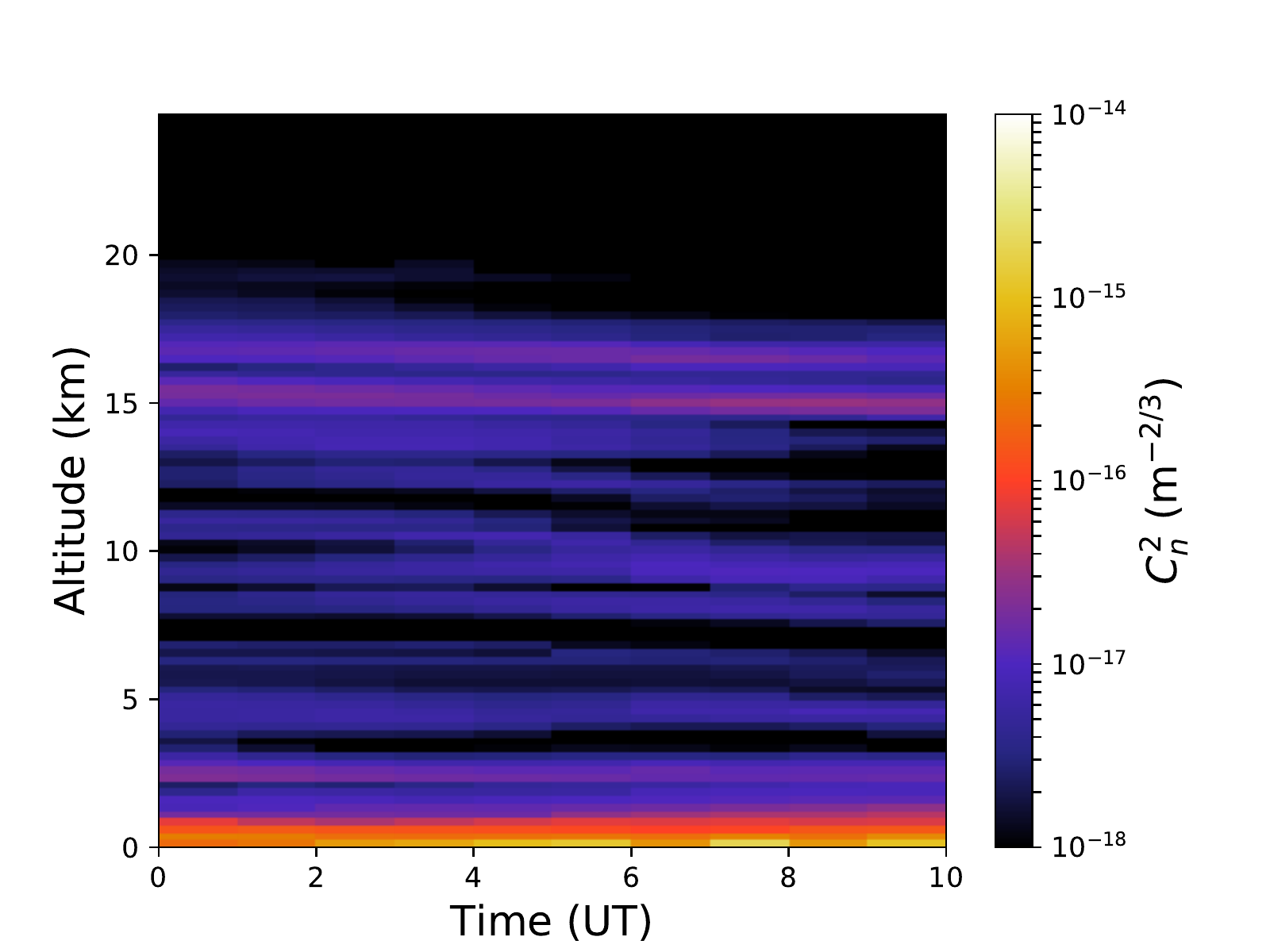} \\
	
	\includegraphics[width=0.23\textwidth,trim={2cm 0 1cm 0}]{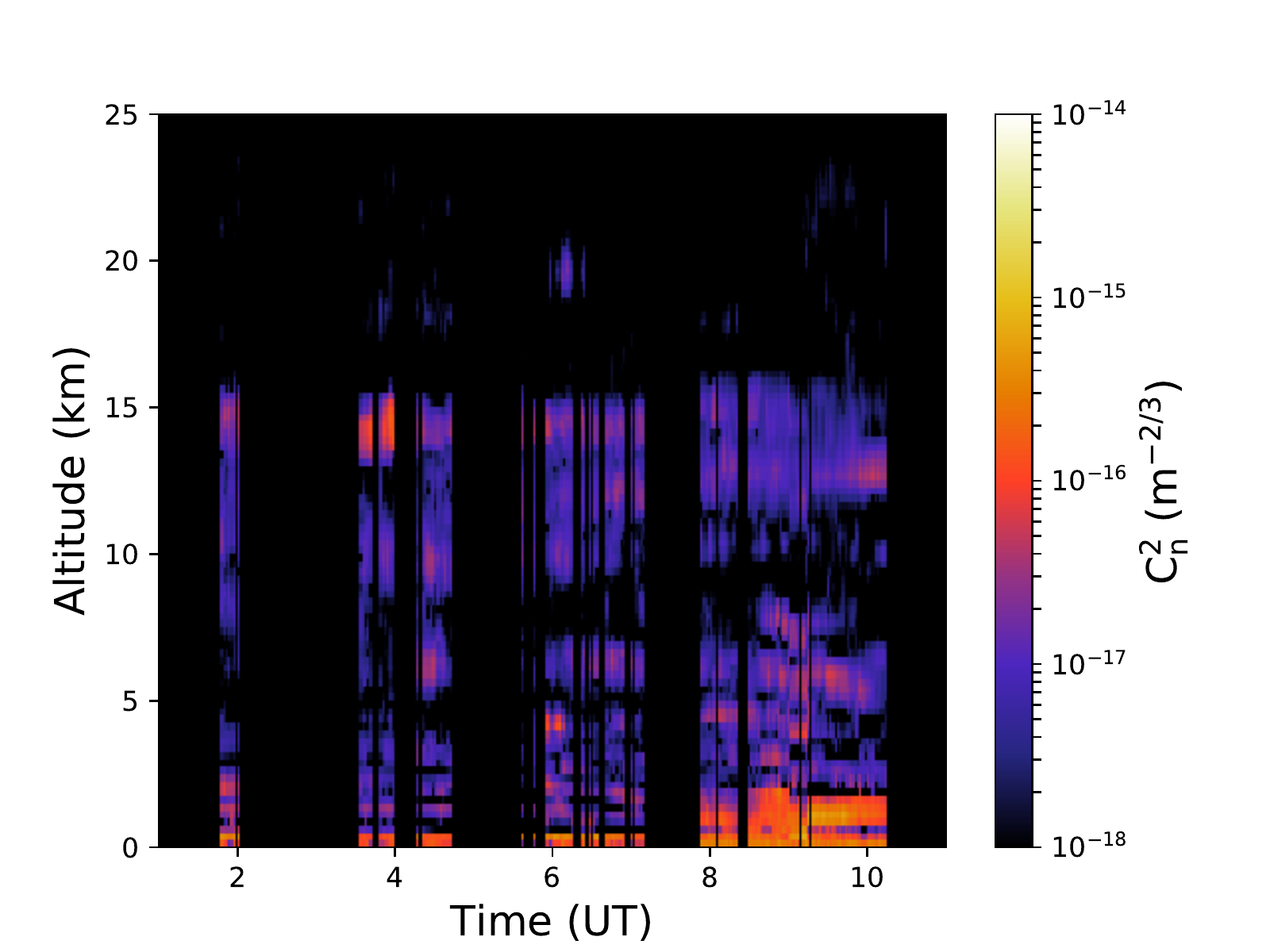} &
    	\includegraphics[width=0.23\textwidth,trim={2cm 0 1cm 0}]{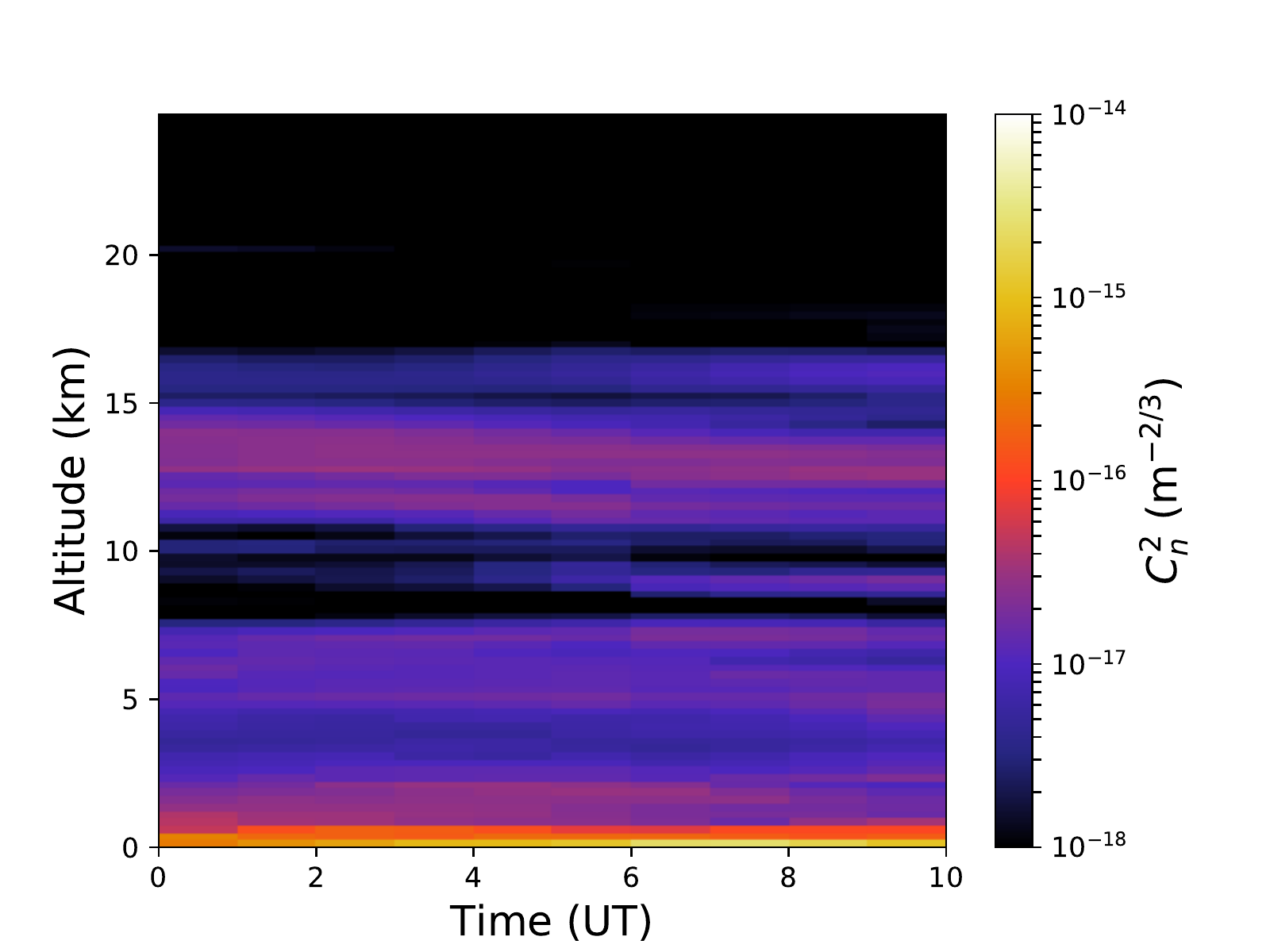} &
    	\includegraphics[width=0.23\textwidth,trim={2cm 0 1cm 0}]{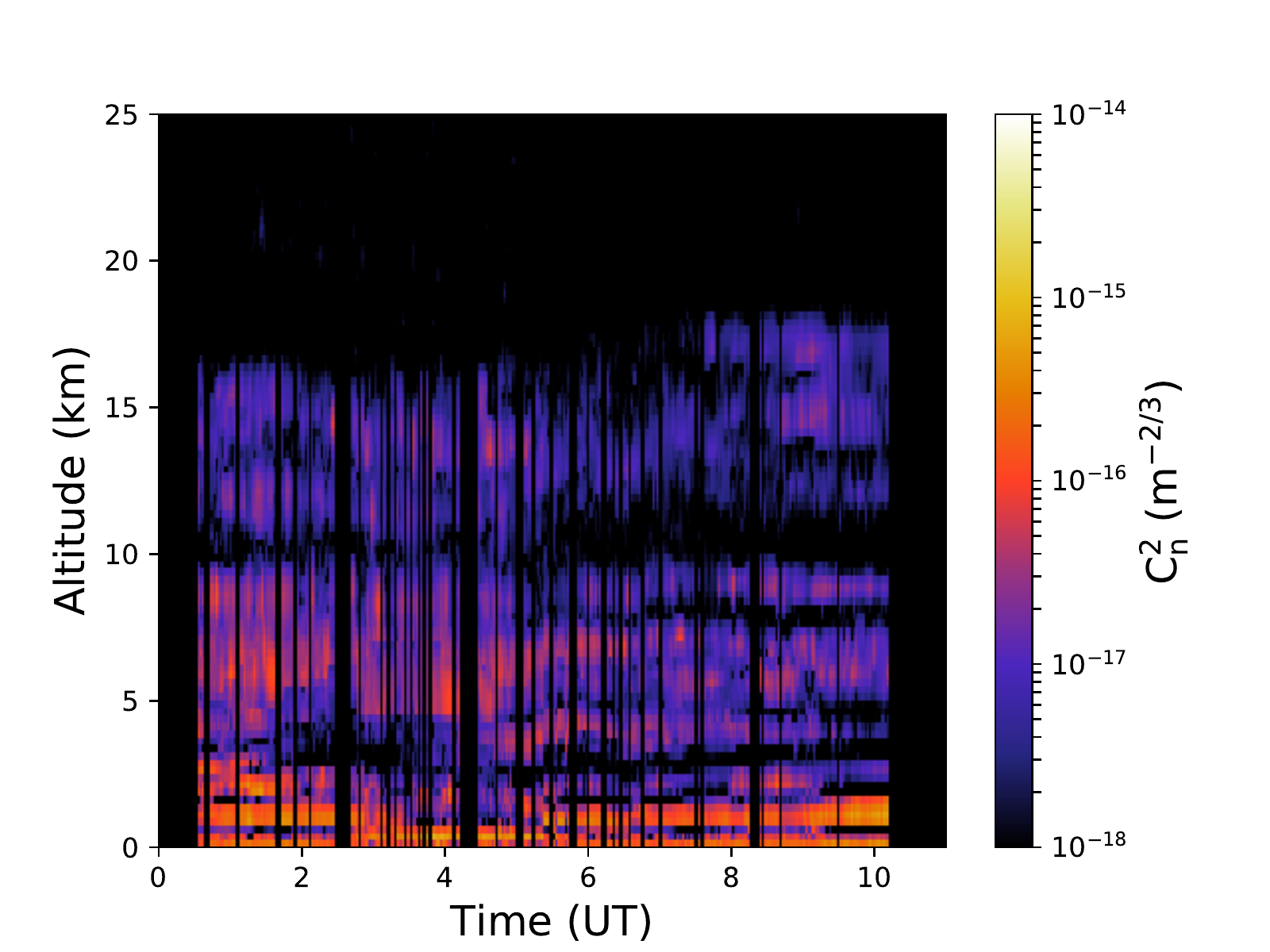} &
    	\includegraphics[width=0.23\textwidth,trim={2cm 0 1cm 0}]{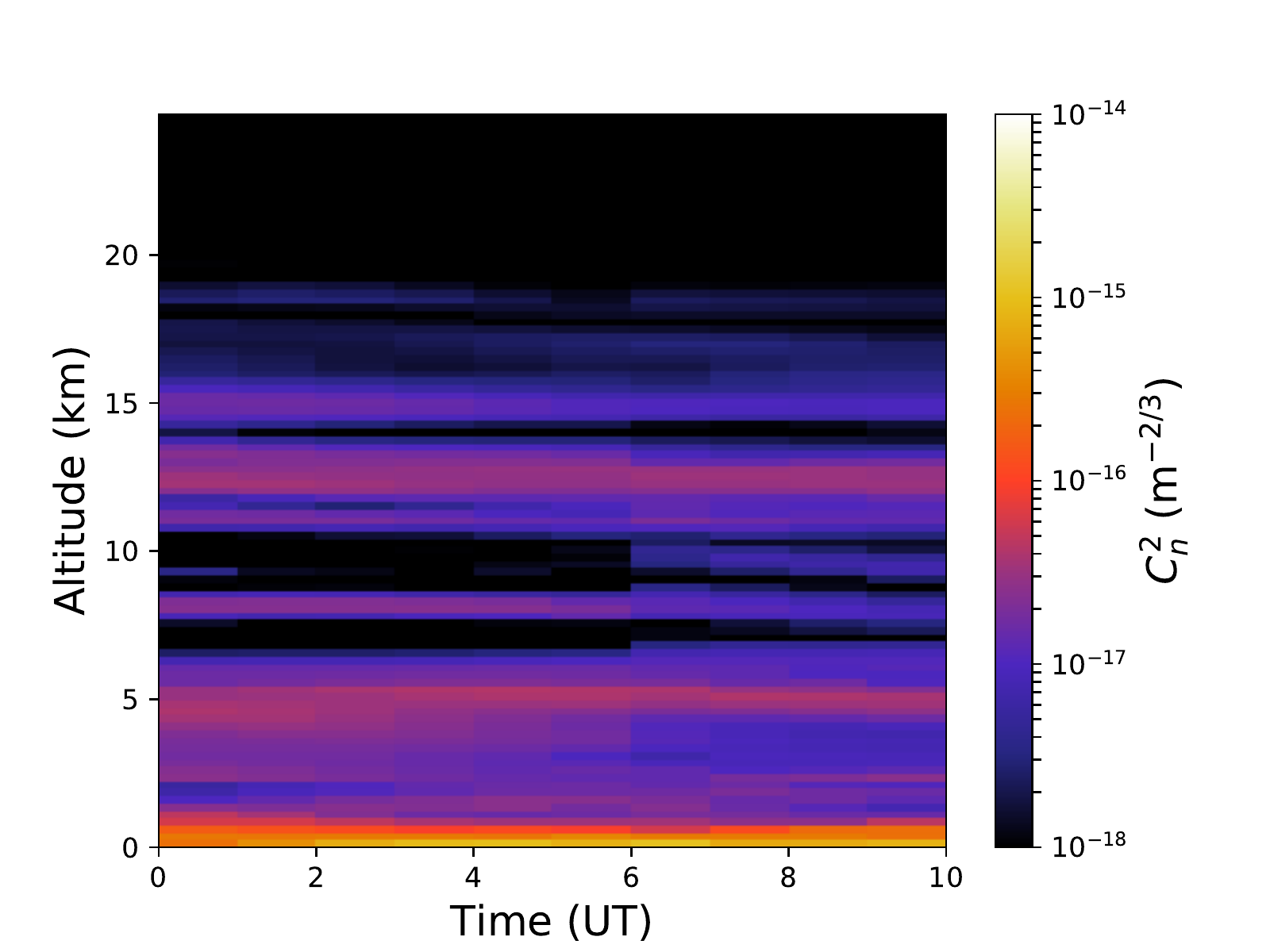} \\
	\includegraphics[width=0.23\textwidth,trim={2cm 0 1cm 0}]{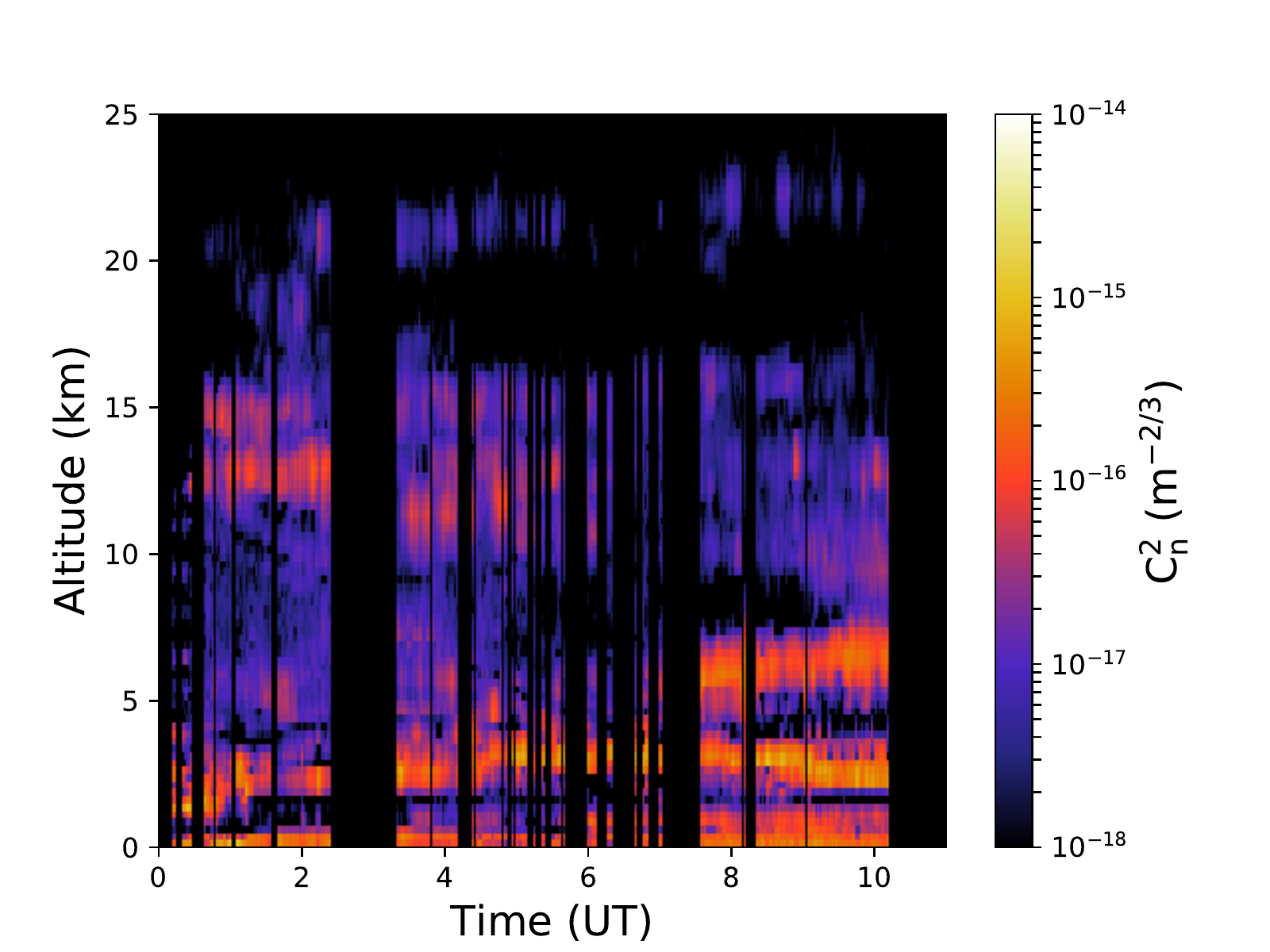} &
    	\includegraphics[width=0.23\textwidth,trim={2cm 0 1cm 0}]{images/20160725_23_ECMWFSequence_PAR.pdf} &
    	\includegraphics[width=0.23\textwidth,trim={2cm 0 1cm 0}]{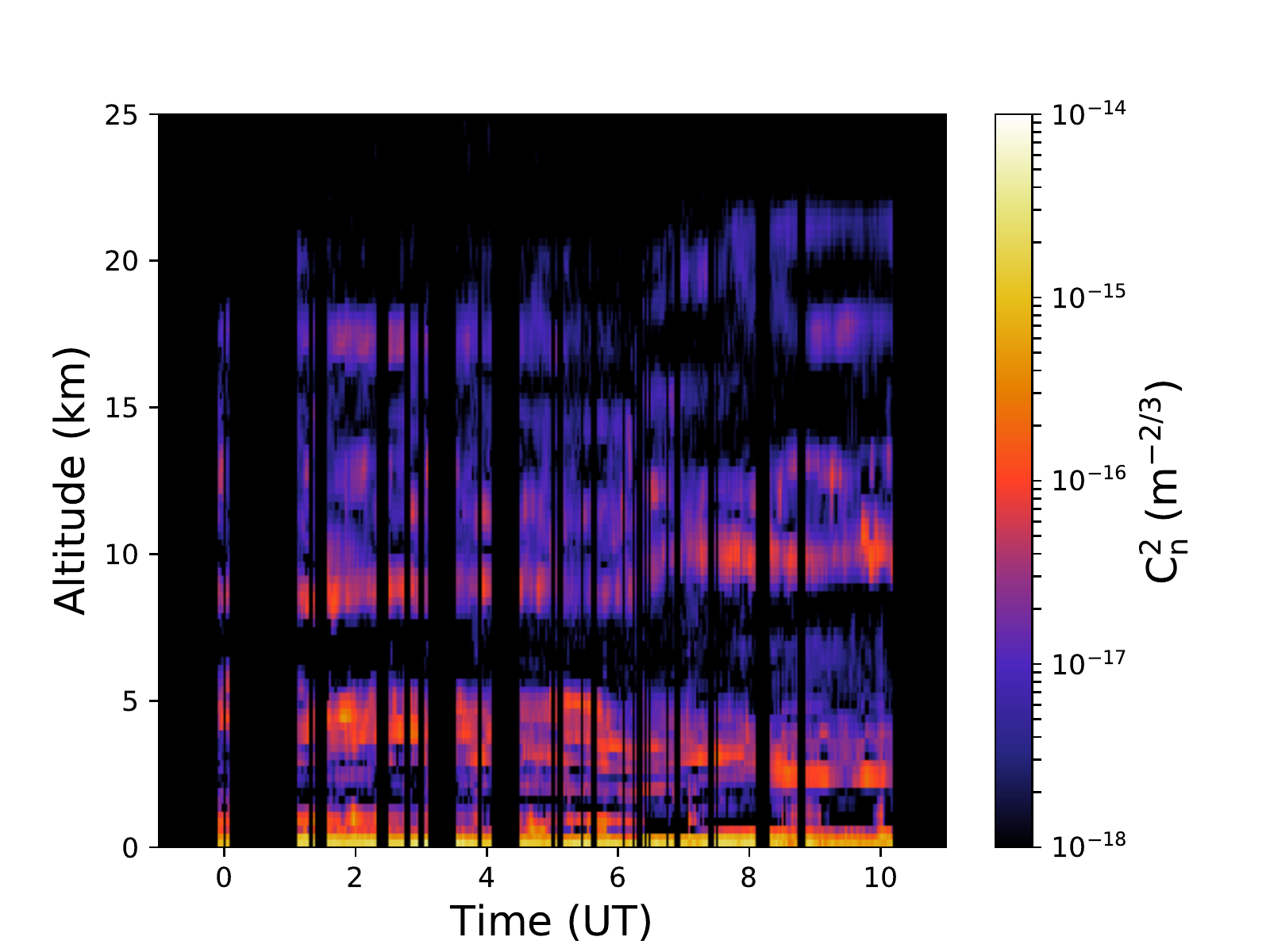} &
    	\includegraphics[width=0.23\textwidth,trim={2cm 0 1cm 0}]{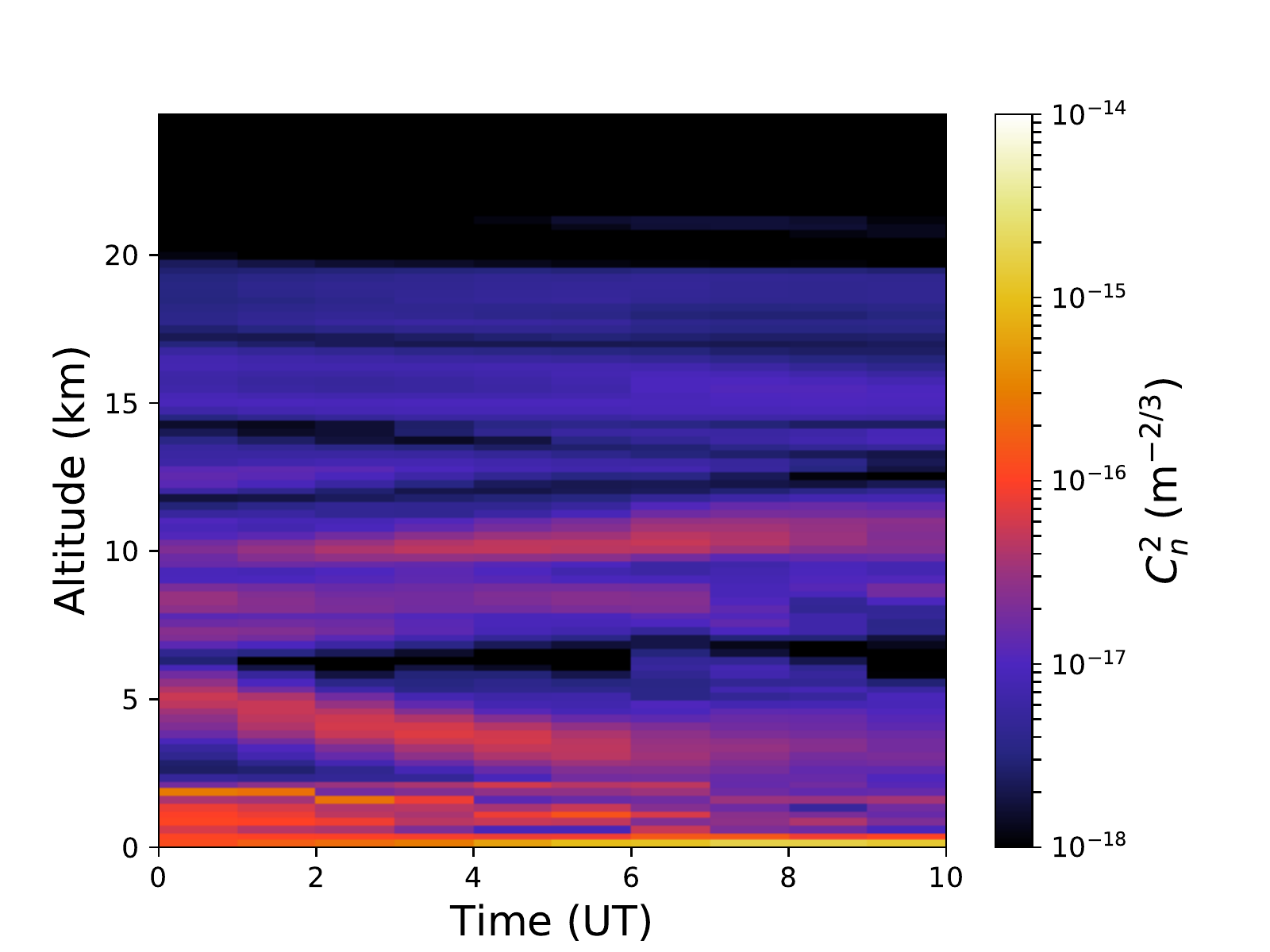} \\

    	\includegraphics[width=0.23\textwidth,trim={2cm 0 1cm 0}]{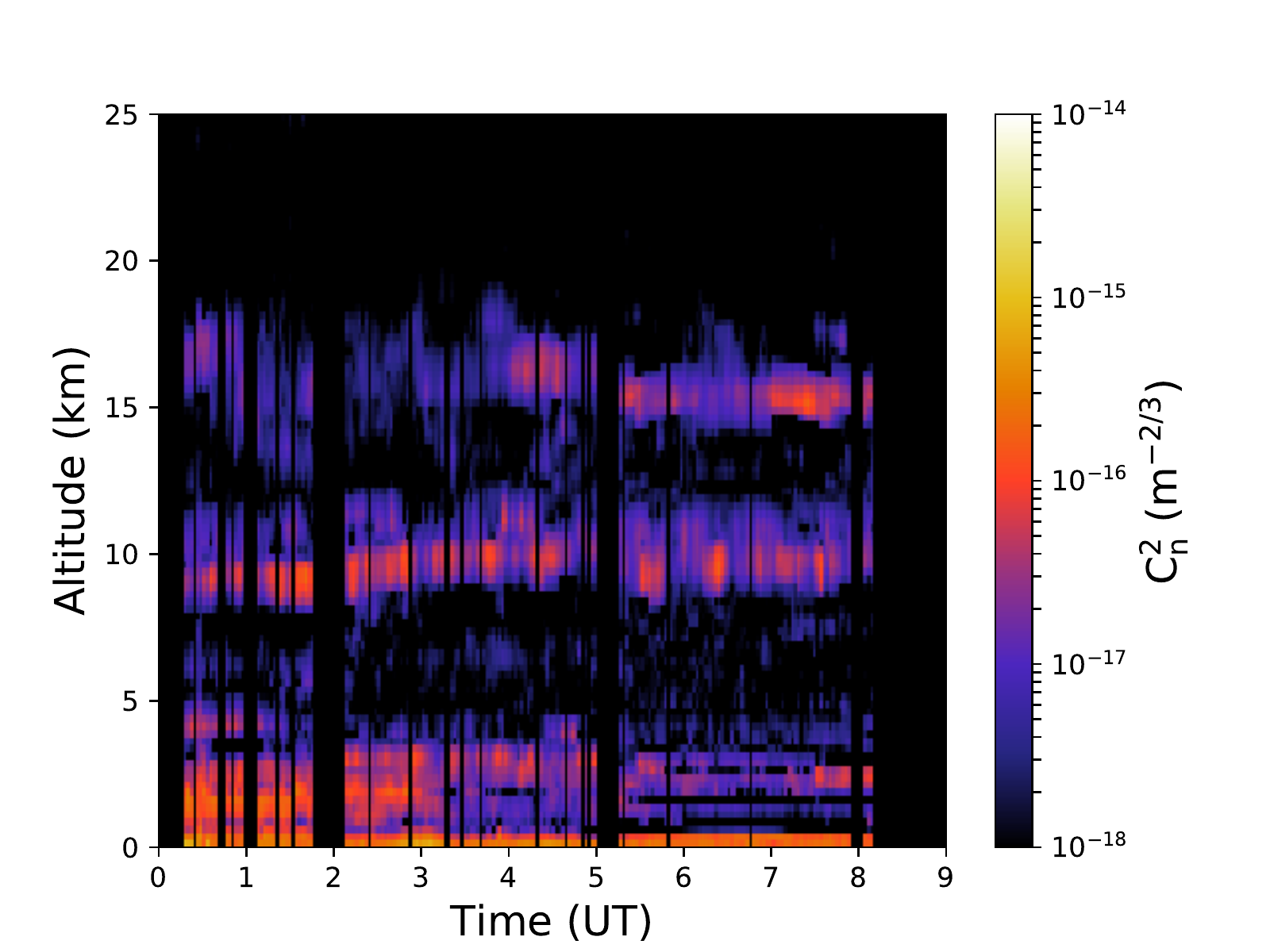} &
    	\includegraphics[width=0.23\textwidth,trim={2cm 0 1cm 0}]{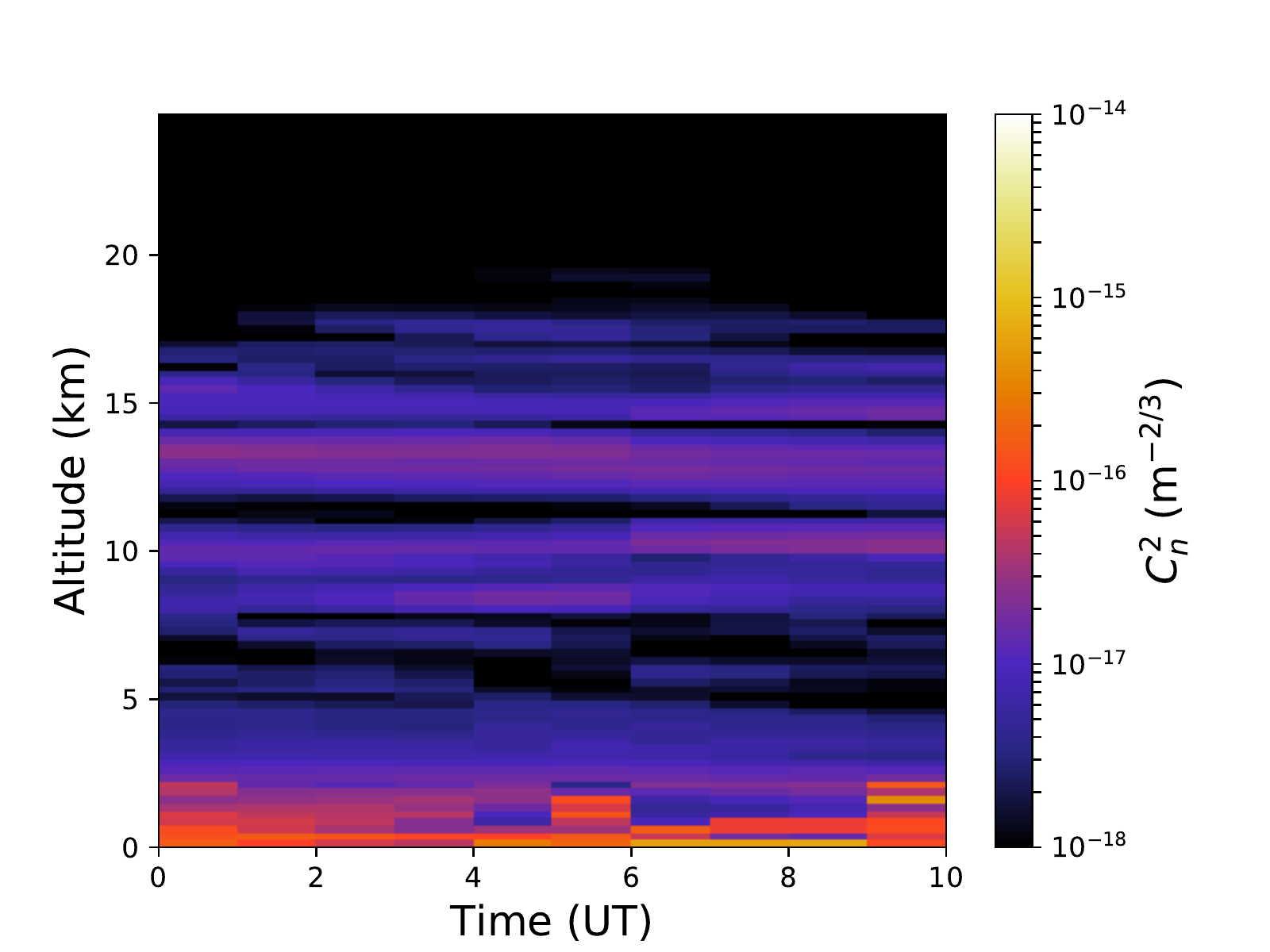} &
	\includegraphics[width=0.23\textwidth,trim={2cm 0 1cm 0}]{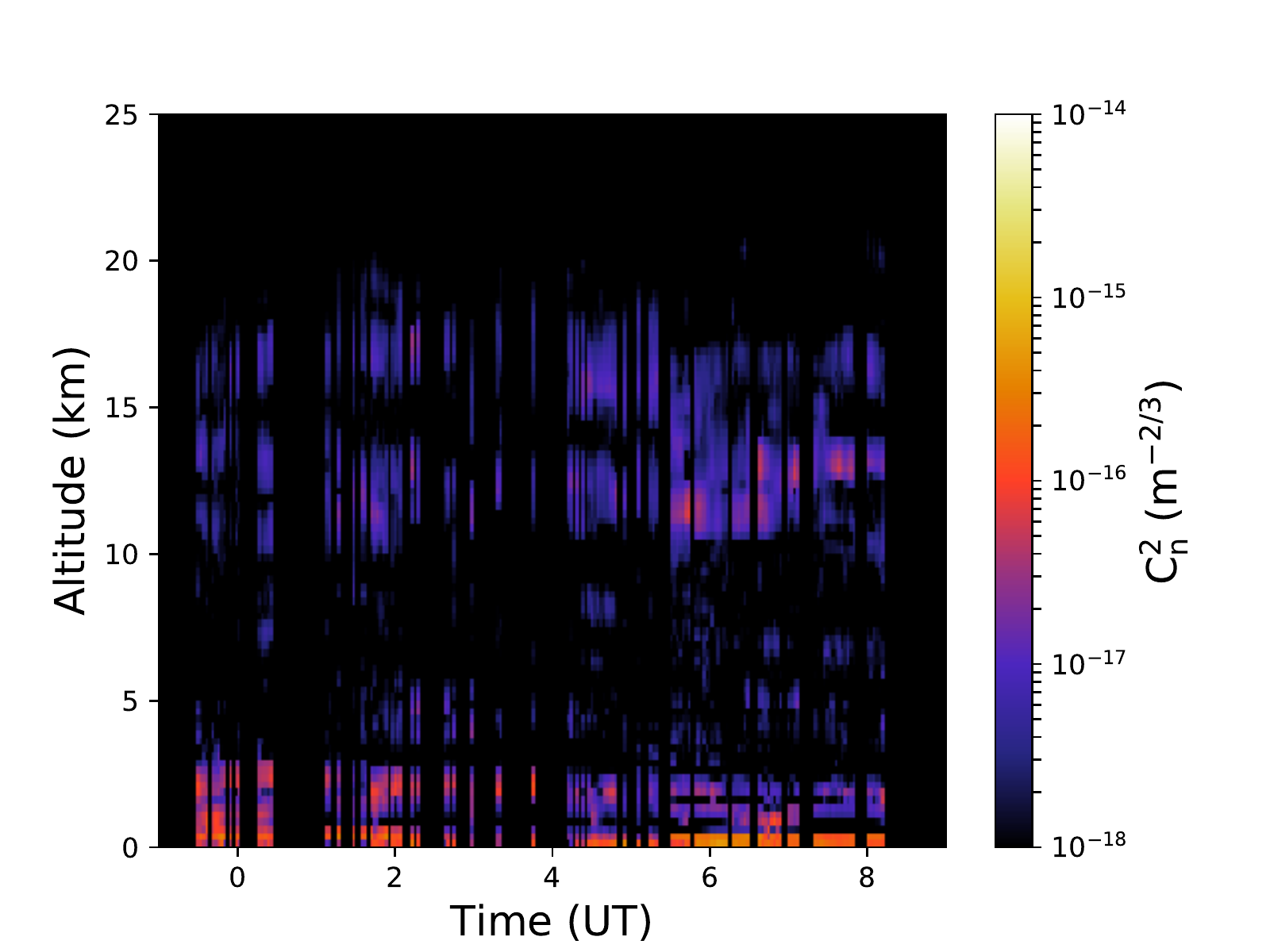} &
    	\includegraphics[width=0.23\textwidth,trim={2cm 0 1cm 0}]{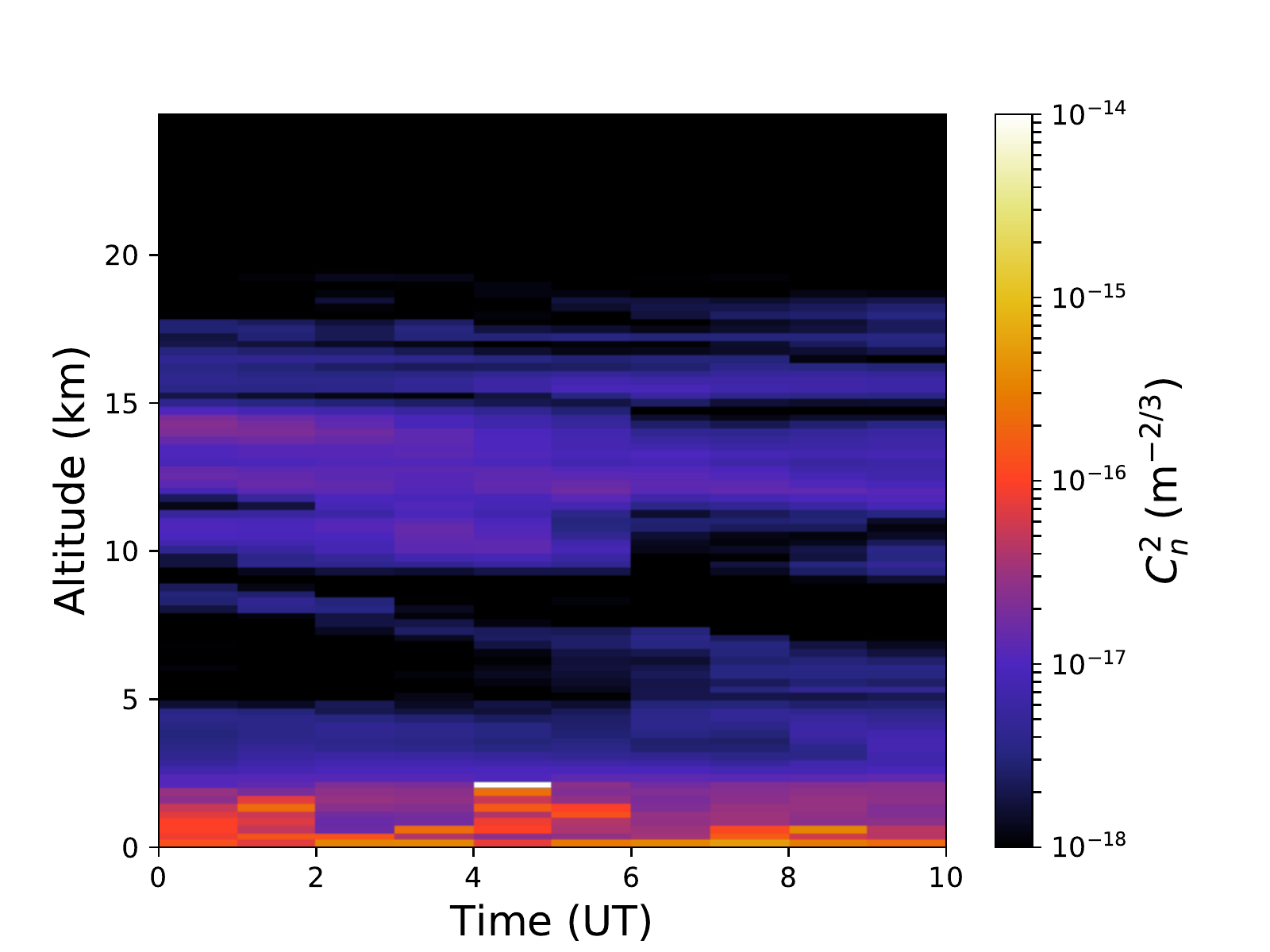} \\
    	\includegraphics[width=0.23\textwidth,trim={2cm 0 1cm 0}]{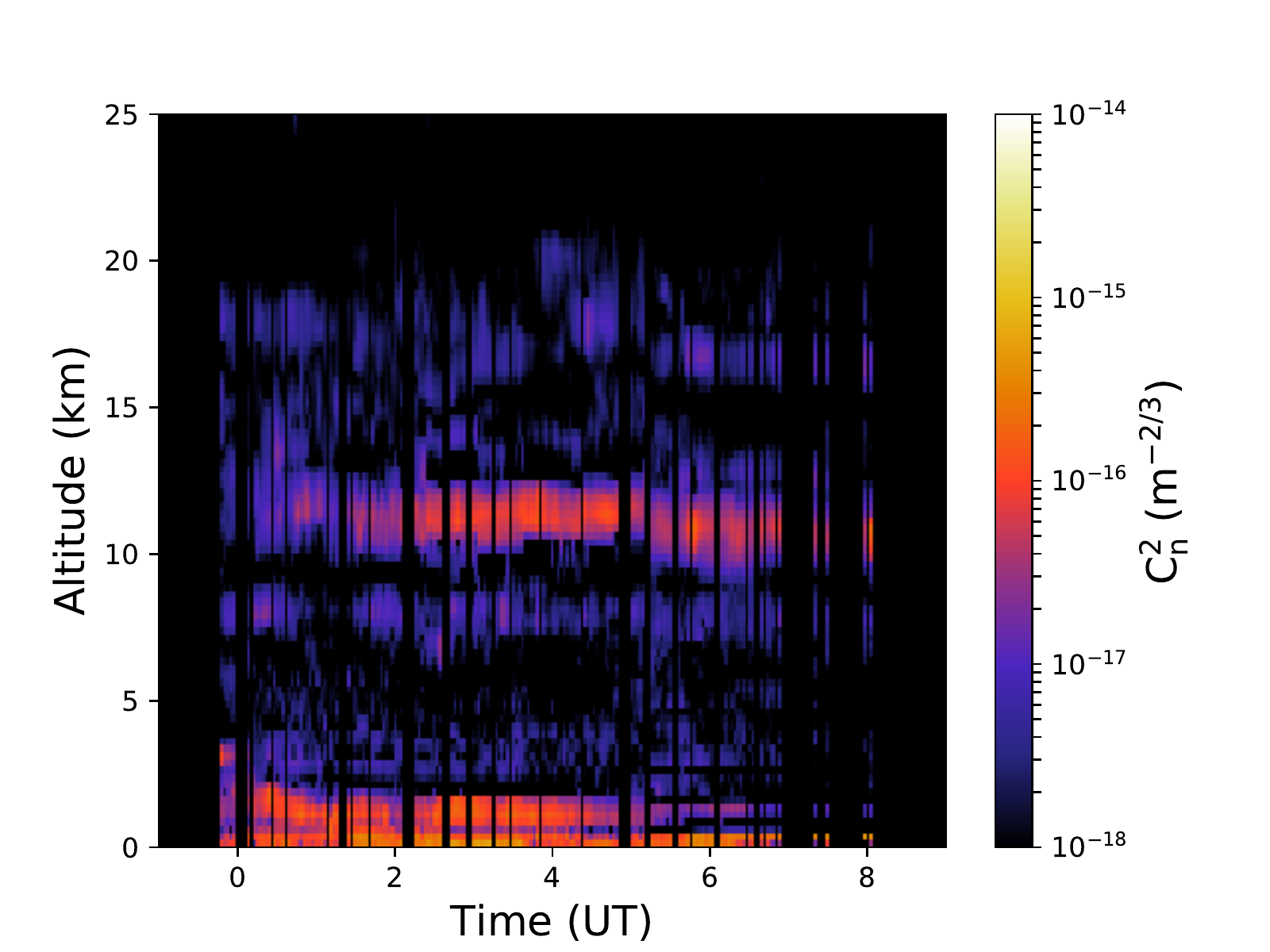} &
    	\includegraphics[width=0.23\textwidth,trim={2cm 0 1cm 0}]{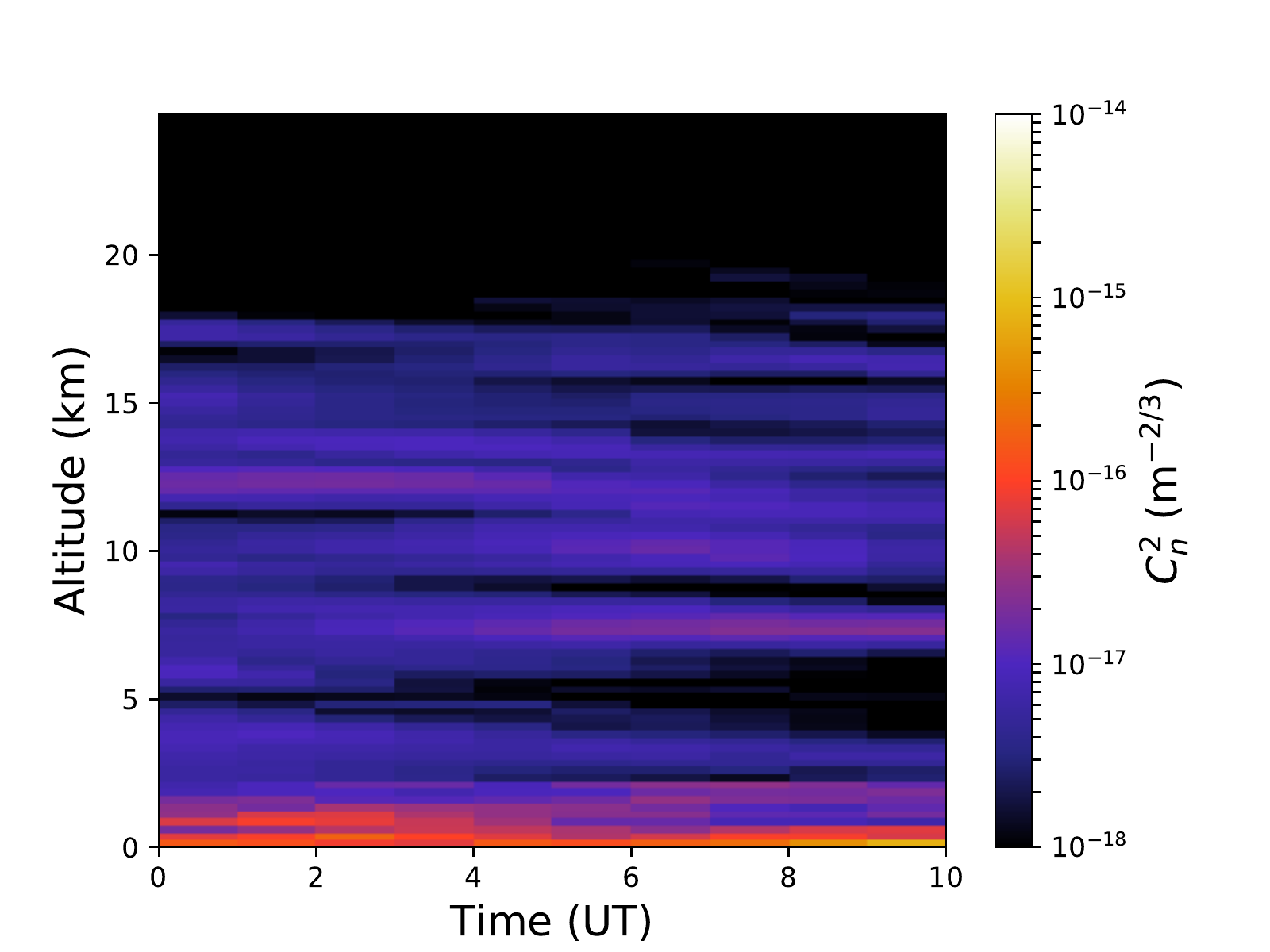} &
	\includegraphics[width=0.23\textwidth,trim={2cm 0 1cm 0}]{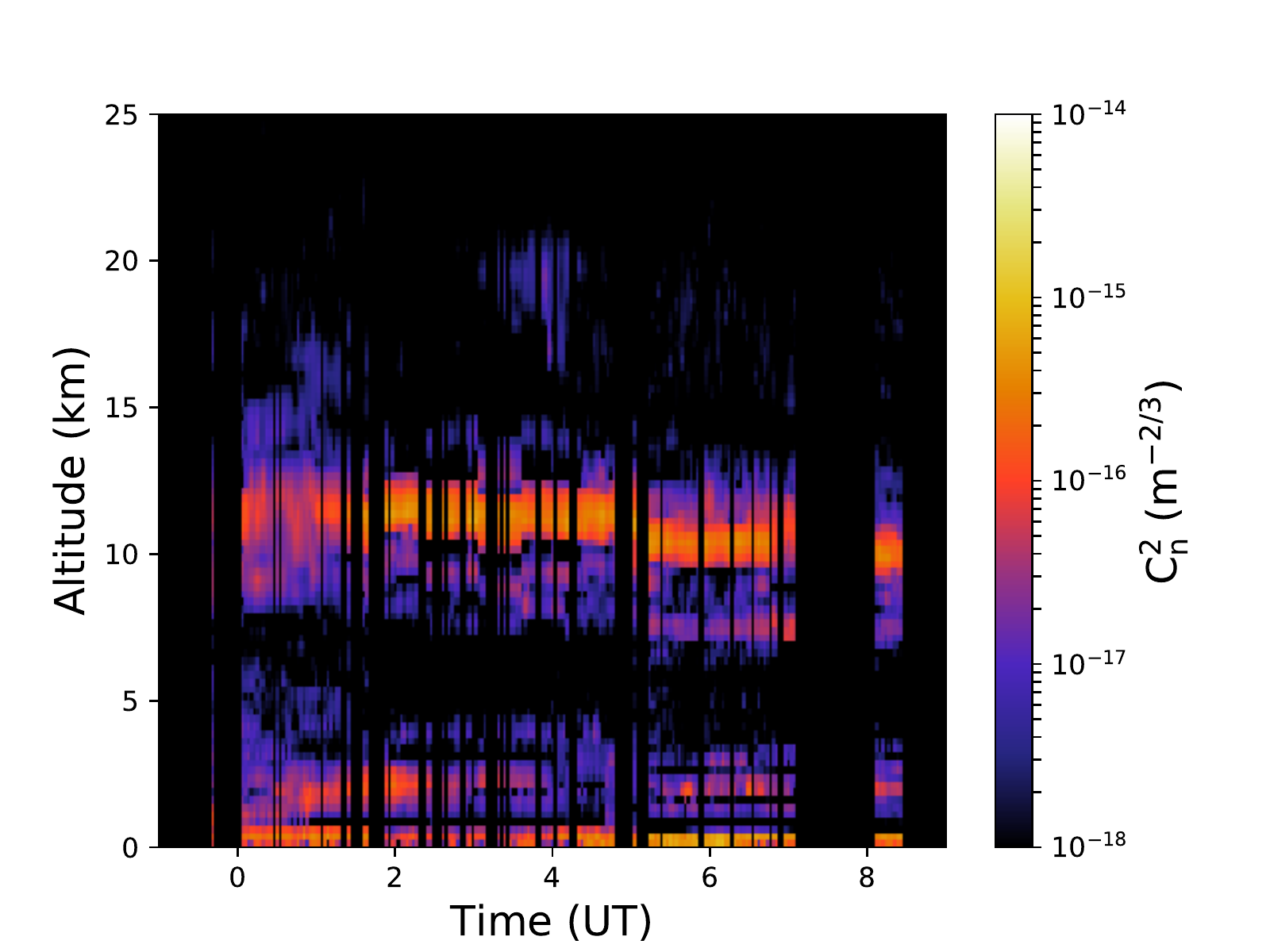} &
    	\includegraphics[width=0.23\textwidth,trim={2cm 0 1cm 0}]{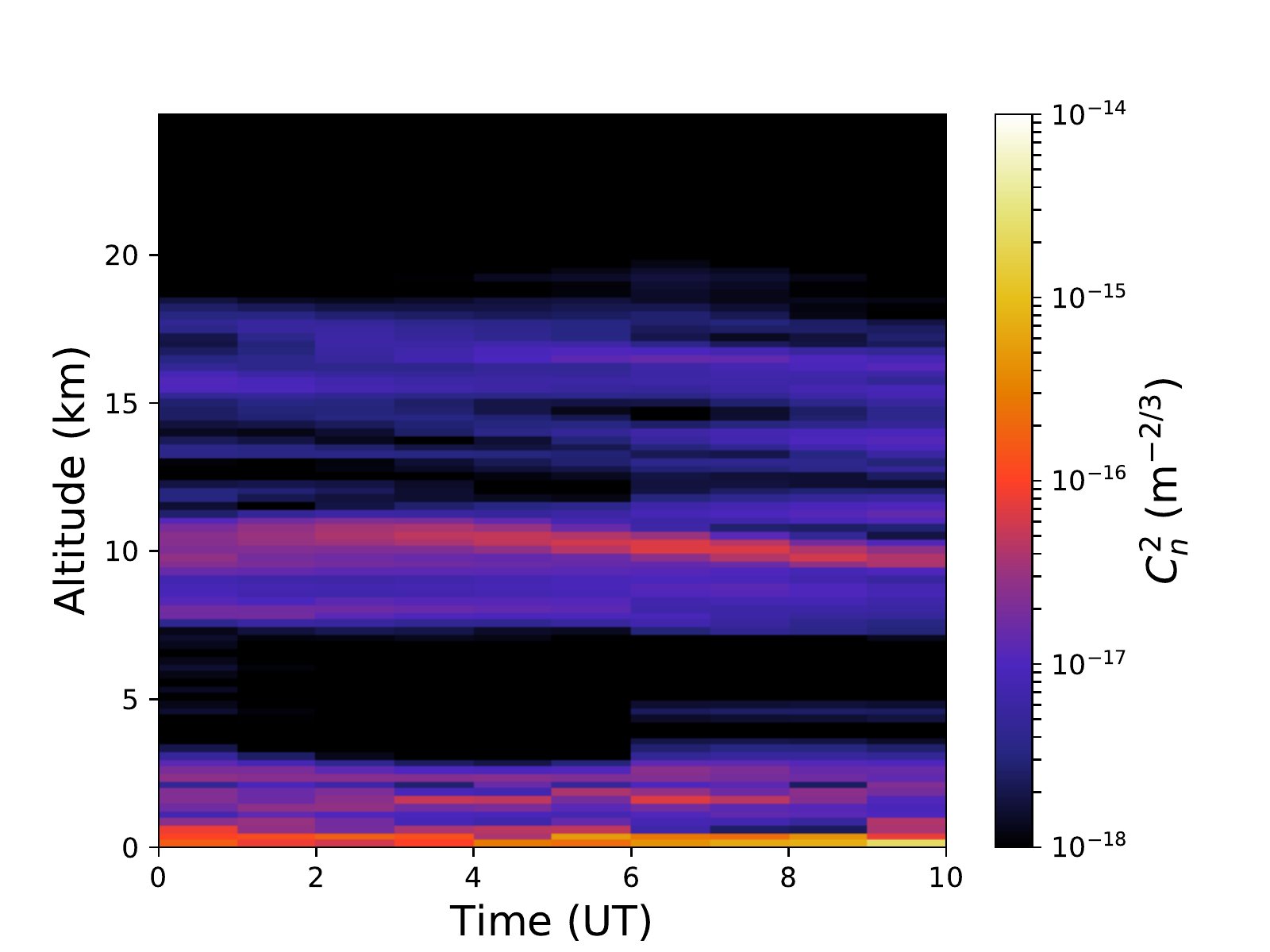} \\
\end{array}$
\caption{Example vertical profiles as measured by the stereo-SCIDAR (green) and estimated by the ECMWF GCM model (red). The profiles shown are the median for an individual night of observation. The coloured region shows the interquartile range. These profiles are from the nights beginning 26th - 29th April, 22nd-25th July, 30th-31st October and 1st-2nd November 2016.}
\label{fig:seqProfiles1}
\end{figure*}

\begin{figure*}
\centering
$\begin{array}{cccc}	
	
    	\includegraphics[width=0.23\textwidth,trim={2cm 0 1cm 0}]{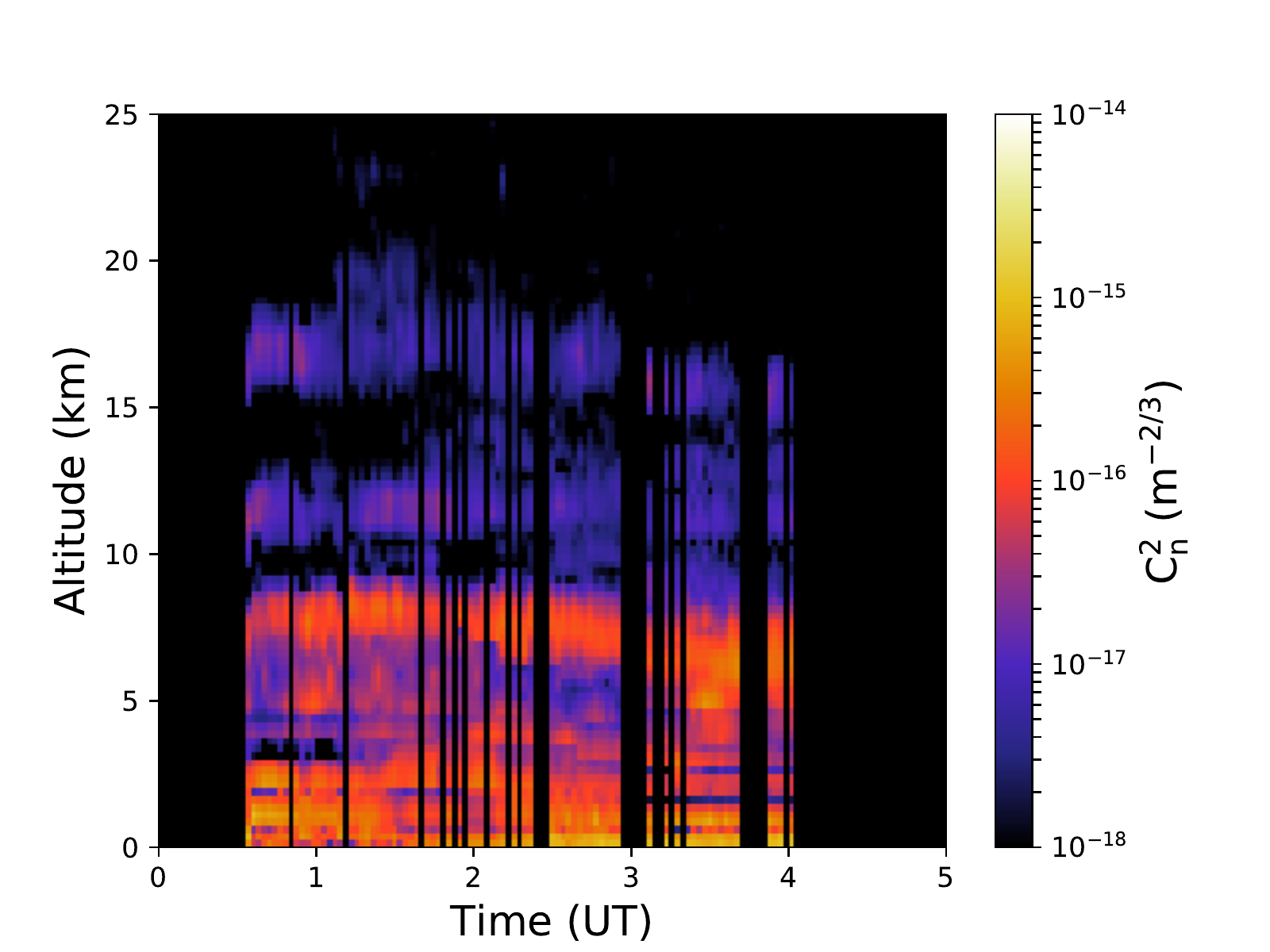} &
    	\includegraphics[width=0.23\textwidth,trim={2cm 0 1cm 0}]{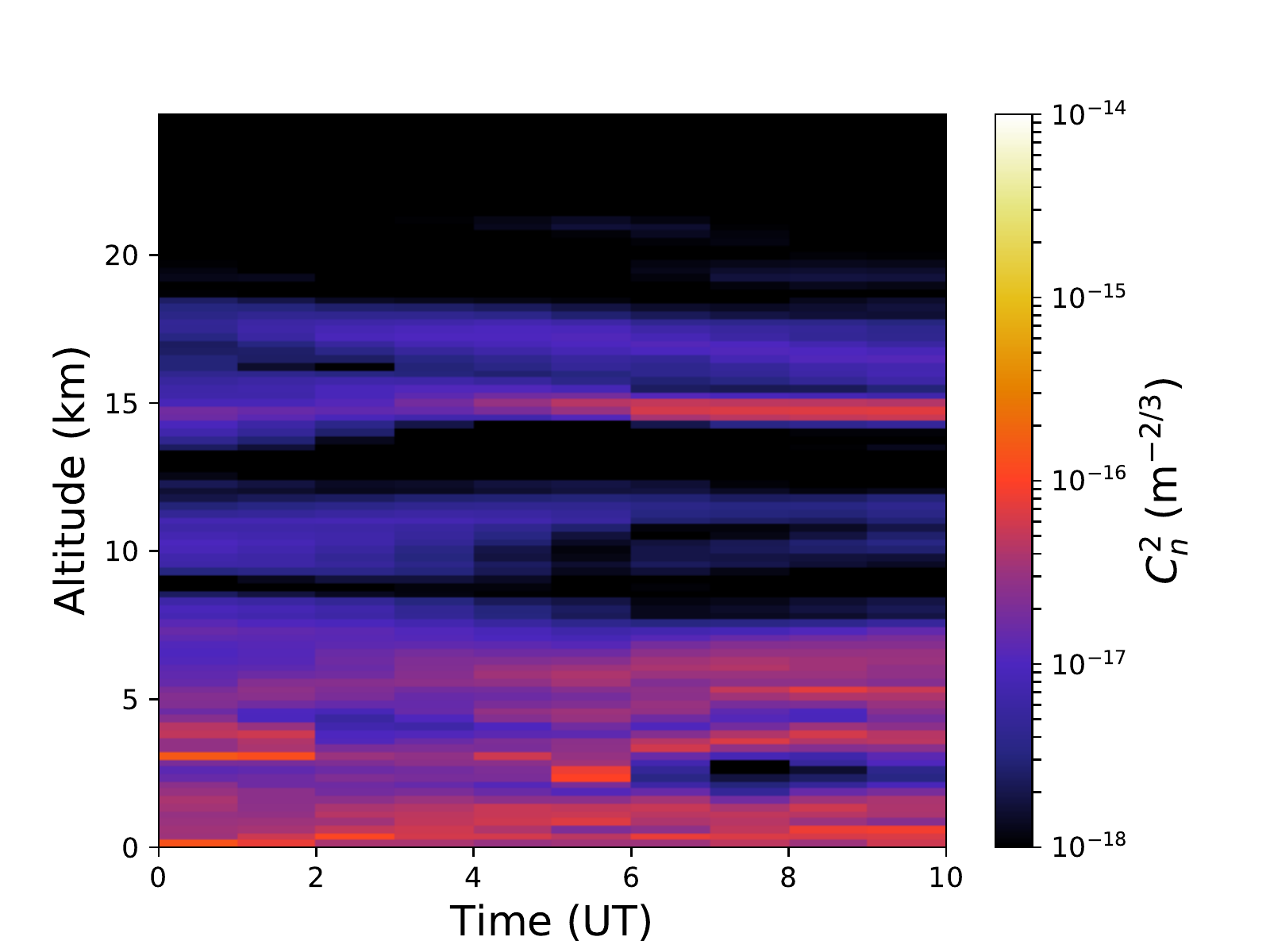} &
	\includegraphics[width=0.23\textwidth,trim={2cm 0 1cm 0}]{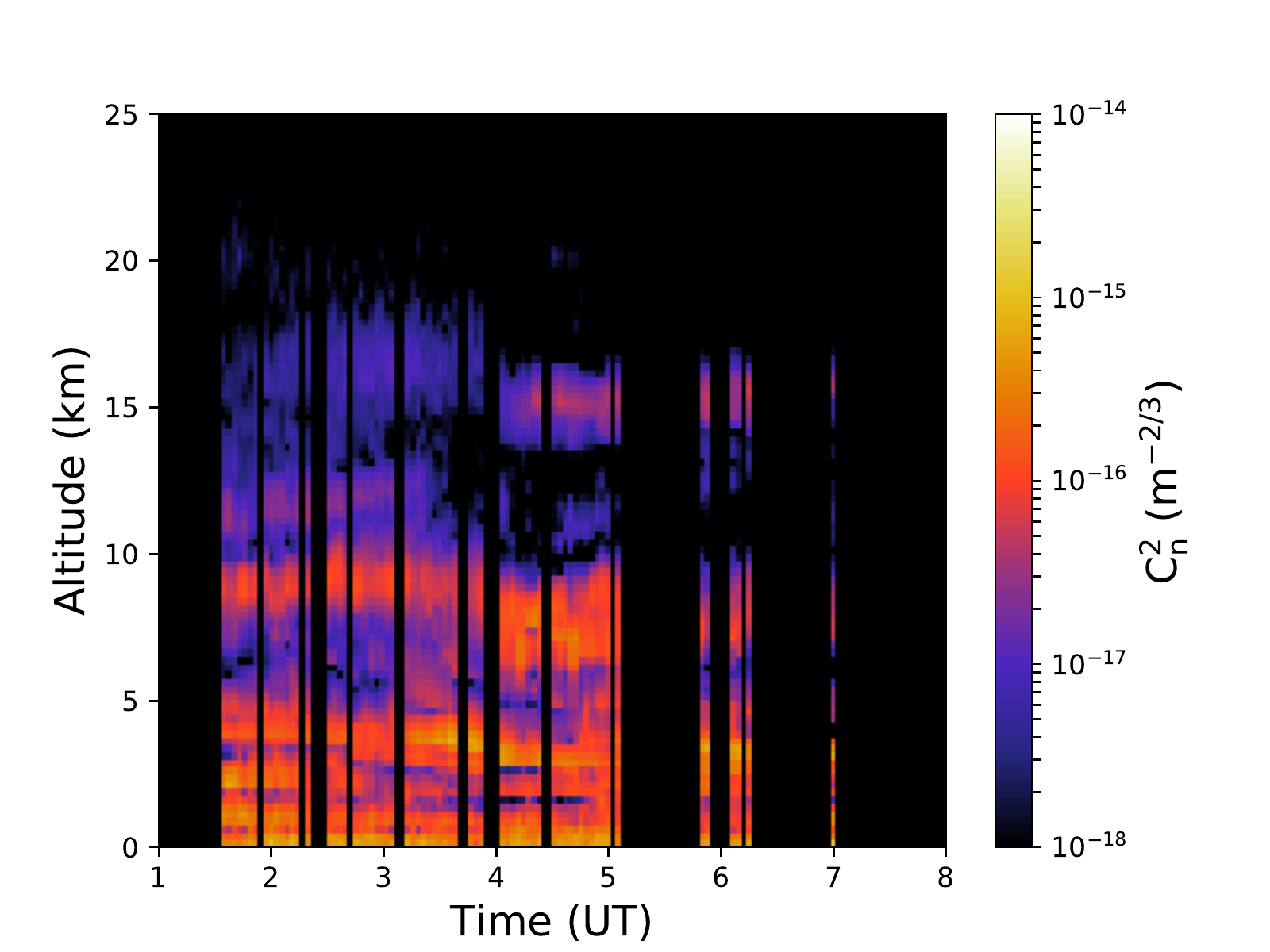} &
    	\includegraphics[width=0.23\textwidth,trim={2cm 0 1cm 0}]{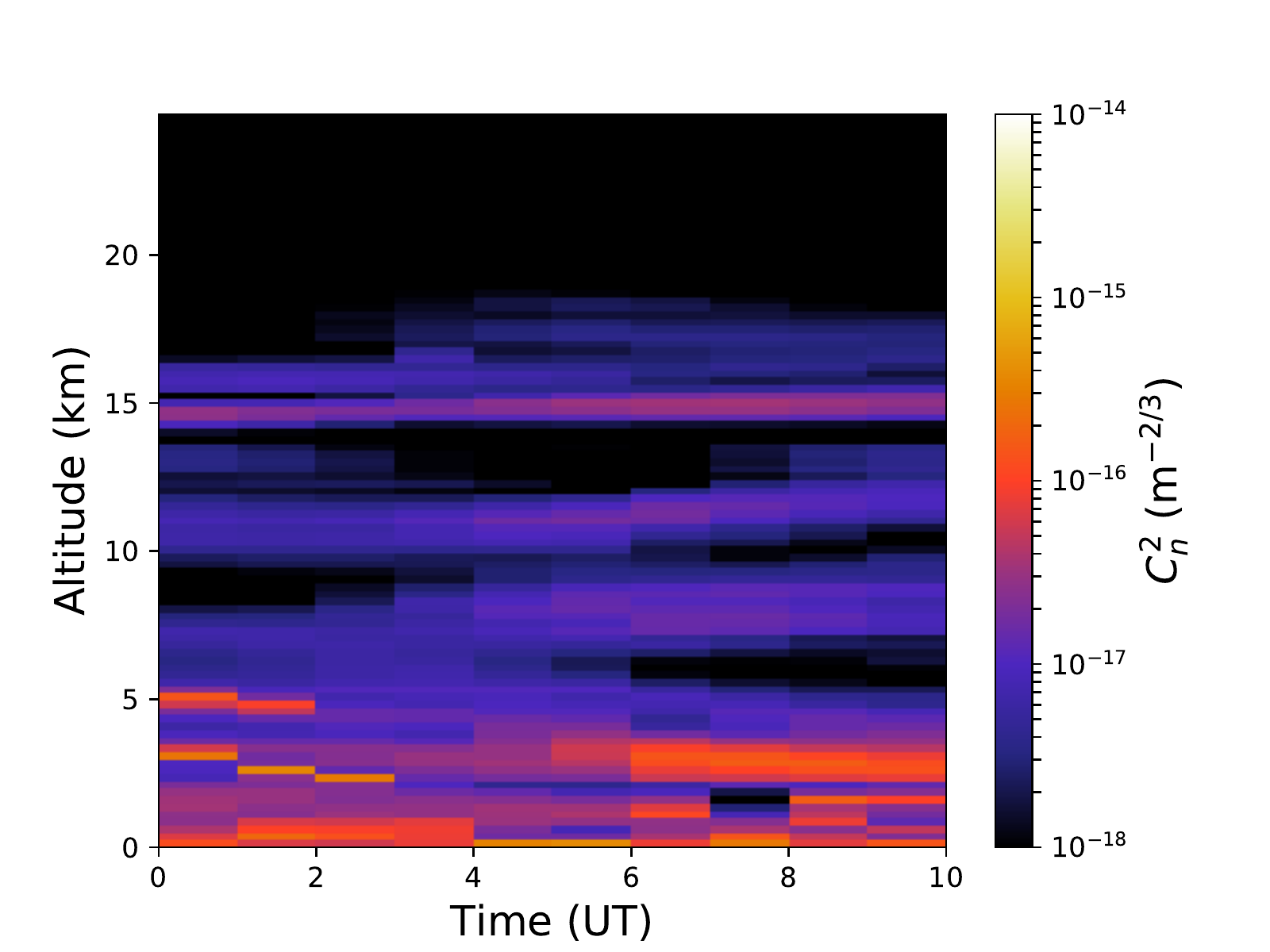} \\
    	\includegraphics[width=0.23\textwidth,trim={2cm 0 1cm 0}]{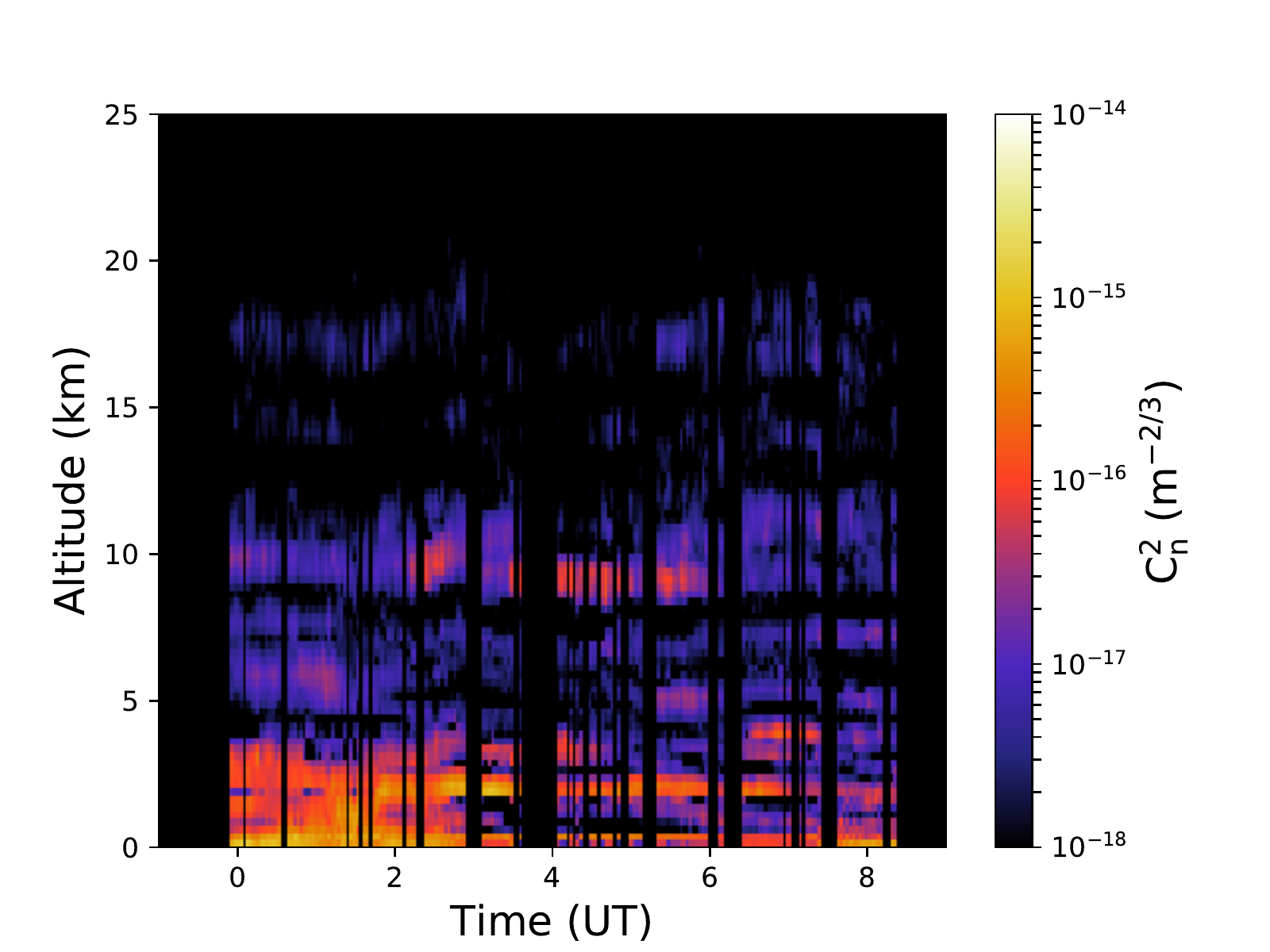} &
    	\includegraphics[width=0.23\textwidth,trim={2cm 0 1cm 0}]{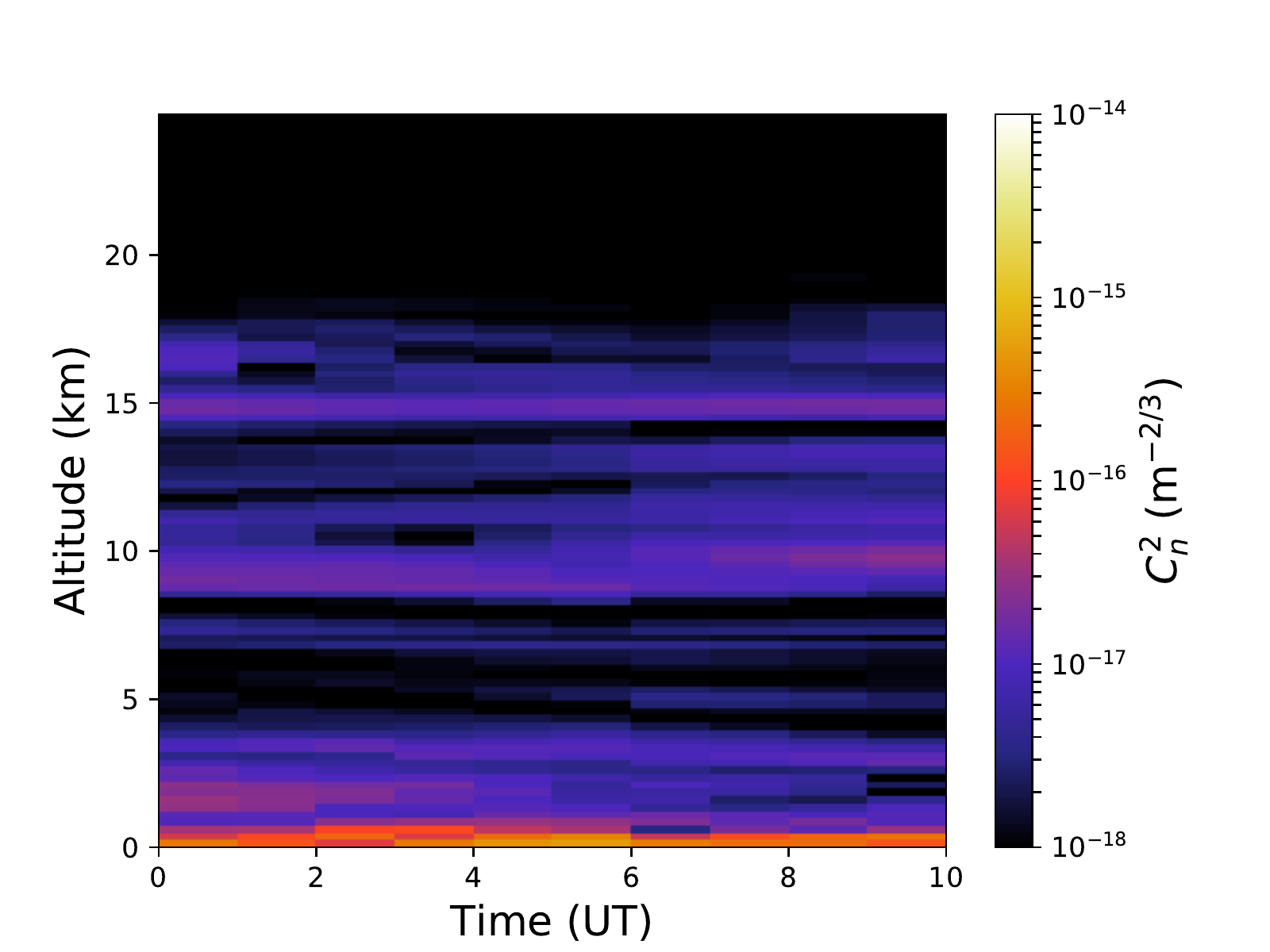} &

    	\includegraphics[width=0.23\textwidth,trim={2cm 0 1cm 0}]{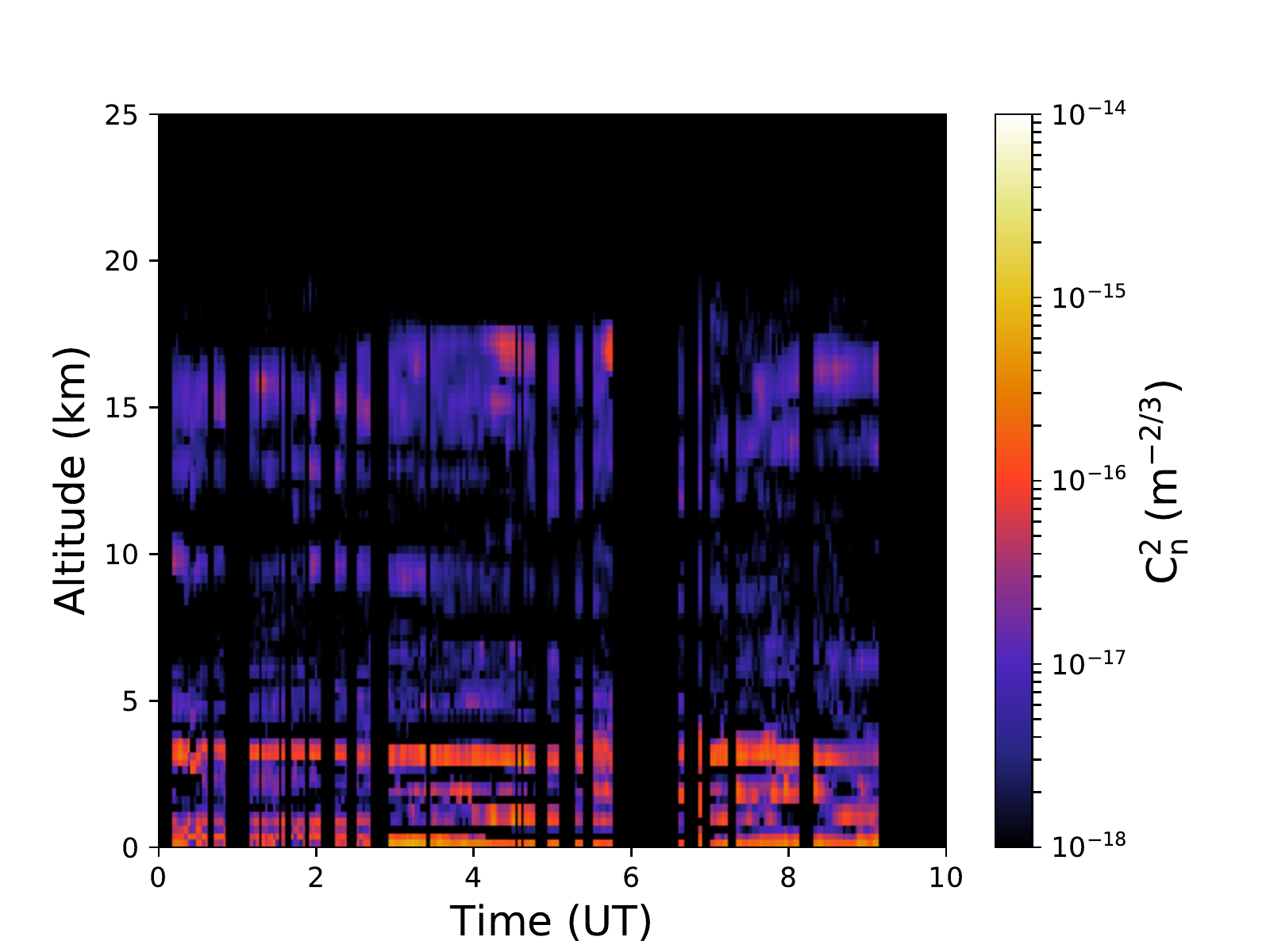} &
    	\includegraphics[width=0.23\textwidth,trim={2cm 0 1cm 0}]{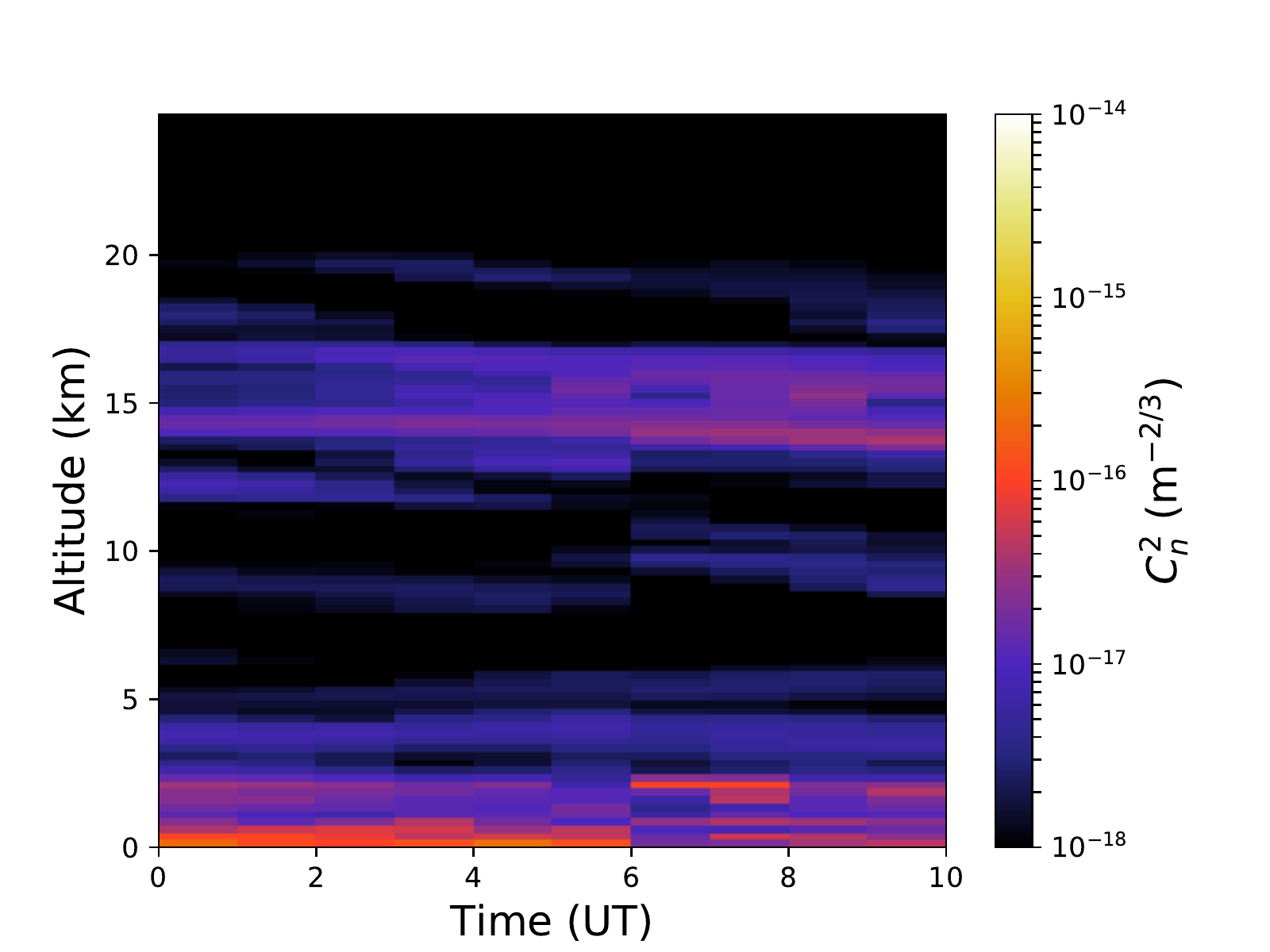} \\
	\includegraphics[width=0.23\textwidth,trim={2cm 0 1cm 0}]{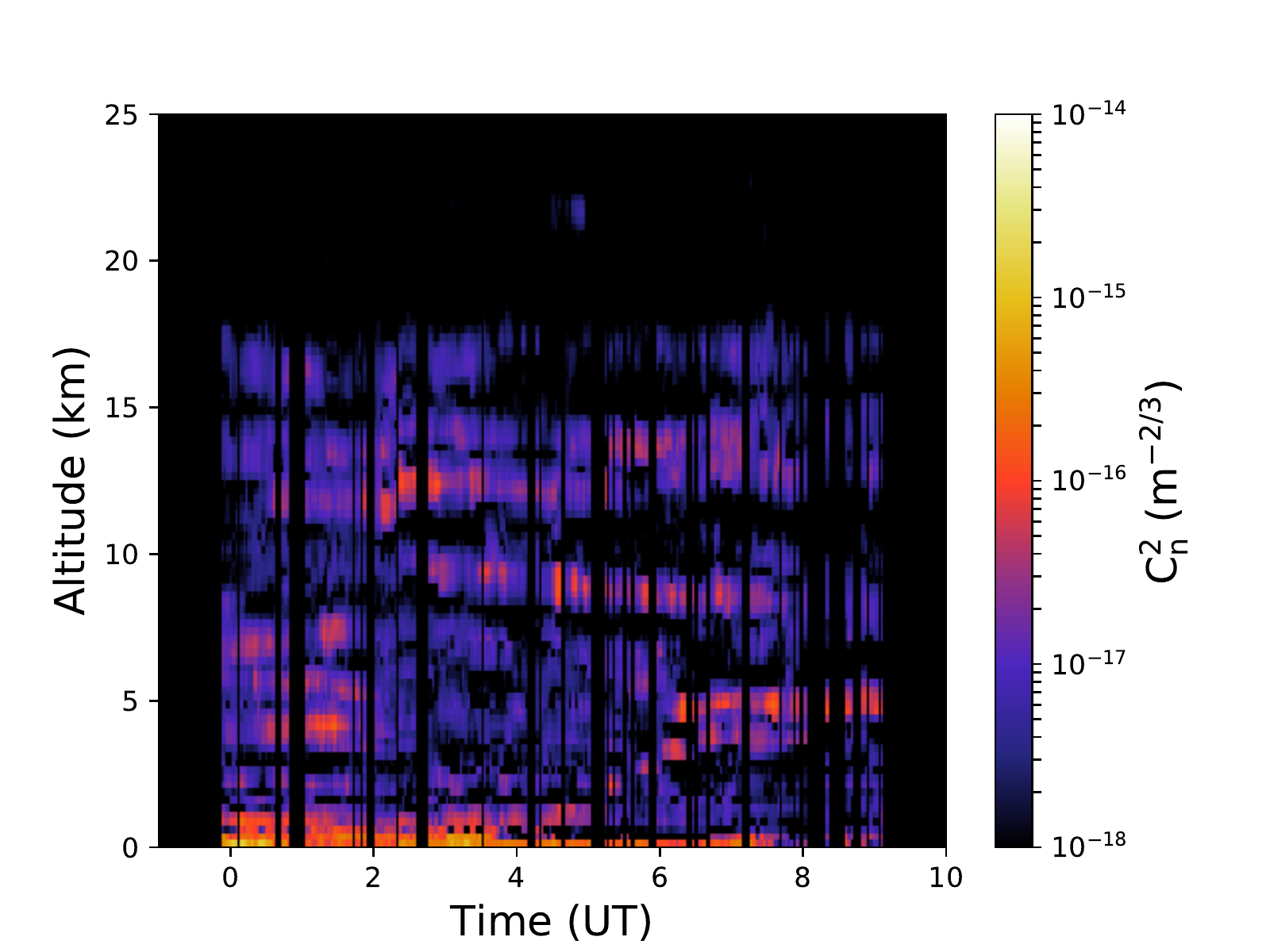} &
    	\includegraphics[width=0.23\textwidth,trim={2cm 0 1cm 0}]{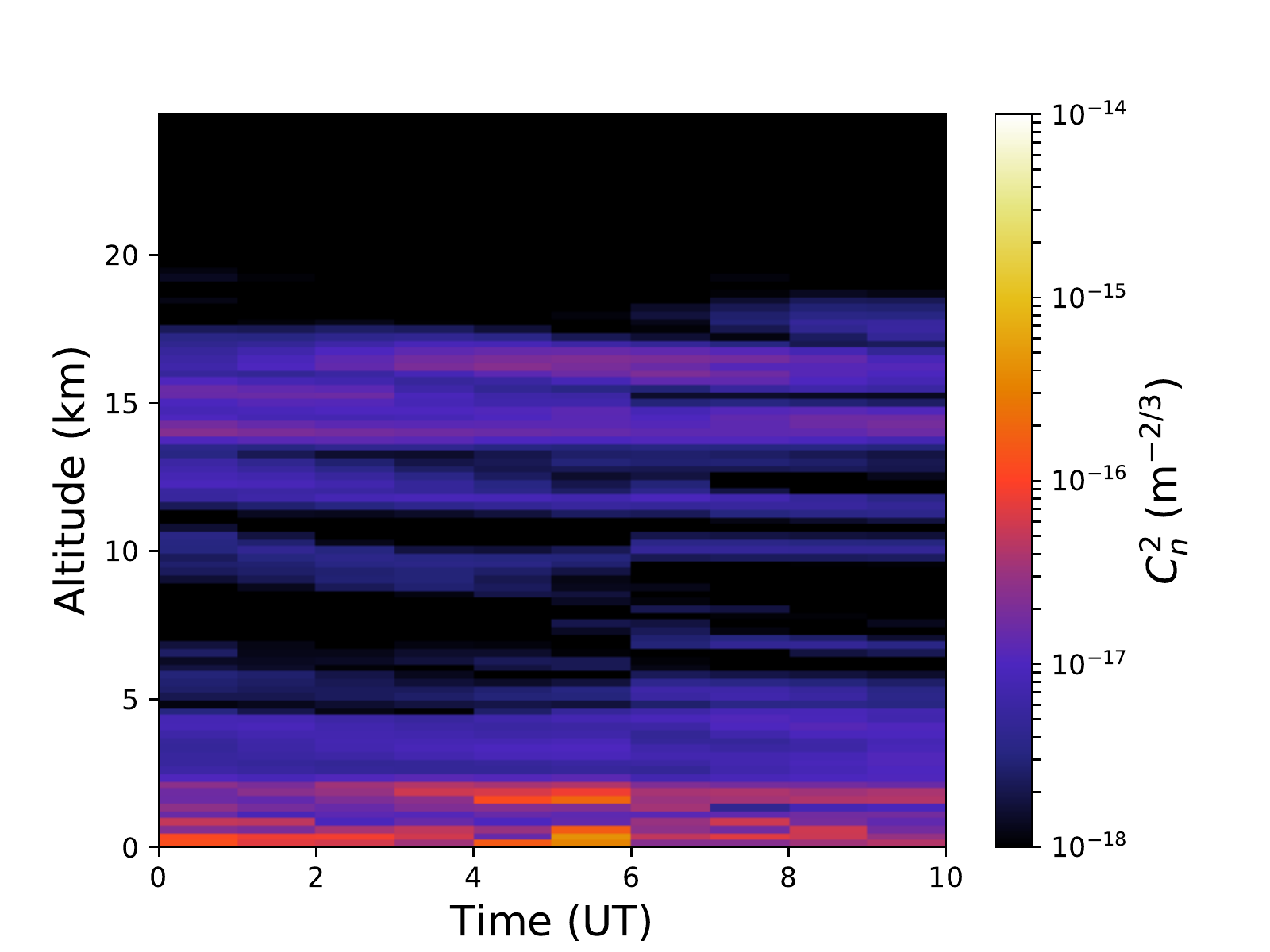} &	
    	\includegraphics[width=0.23\textwidth,trim={2cm 0 1cm 0}]{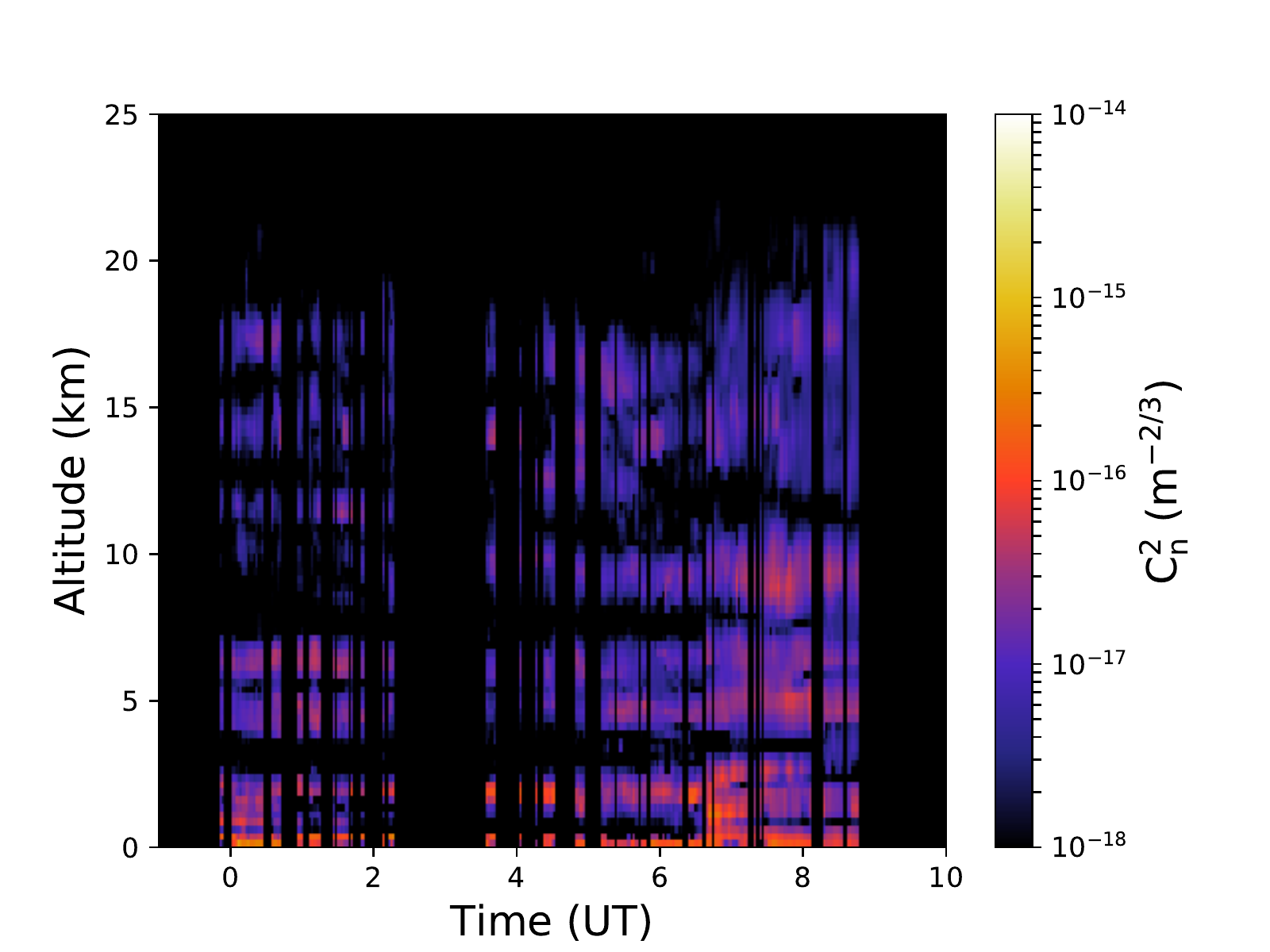} &
    	\includegraphics[width=0.23\textwidth,trim={2cm 0 1cm 0}]{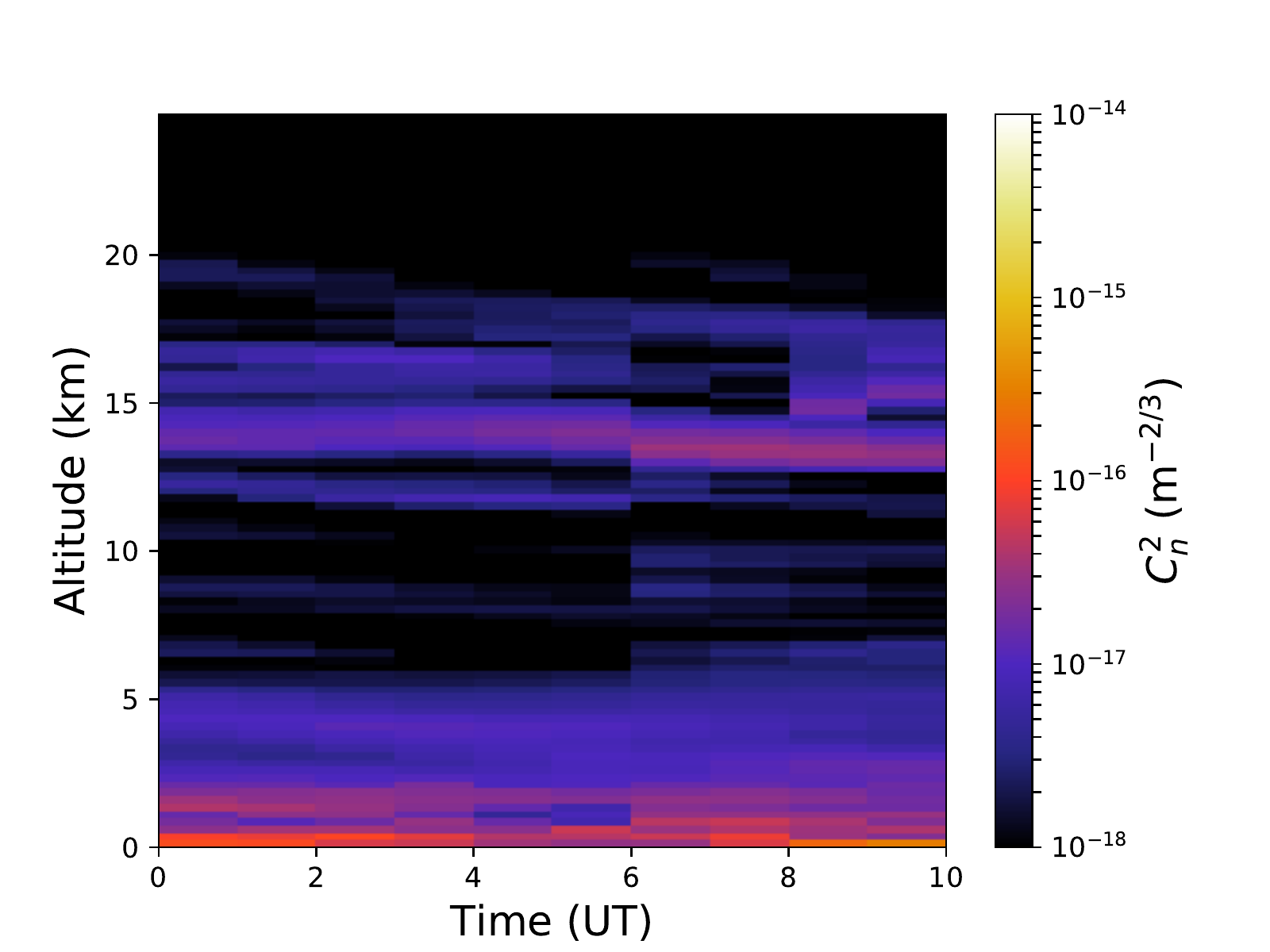}\\

	\includegraphics[width=0.23\textwidth,trim={2cm 0 1cm 0}]{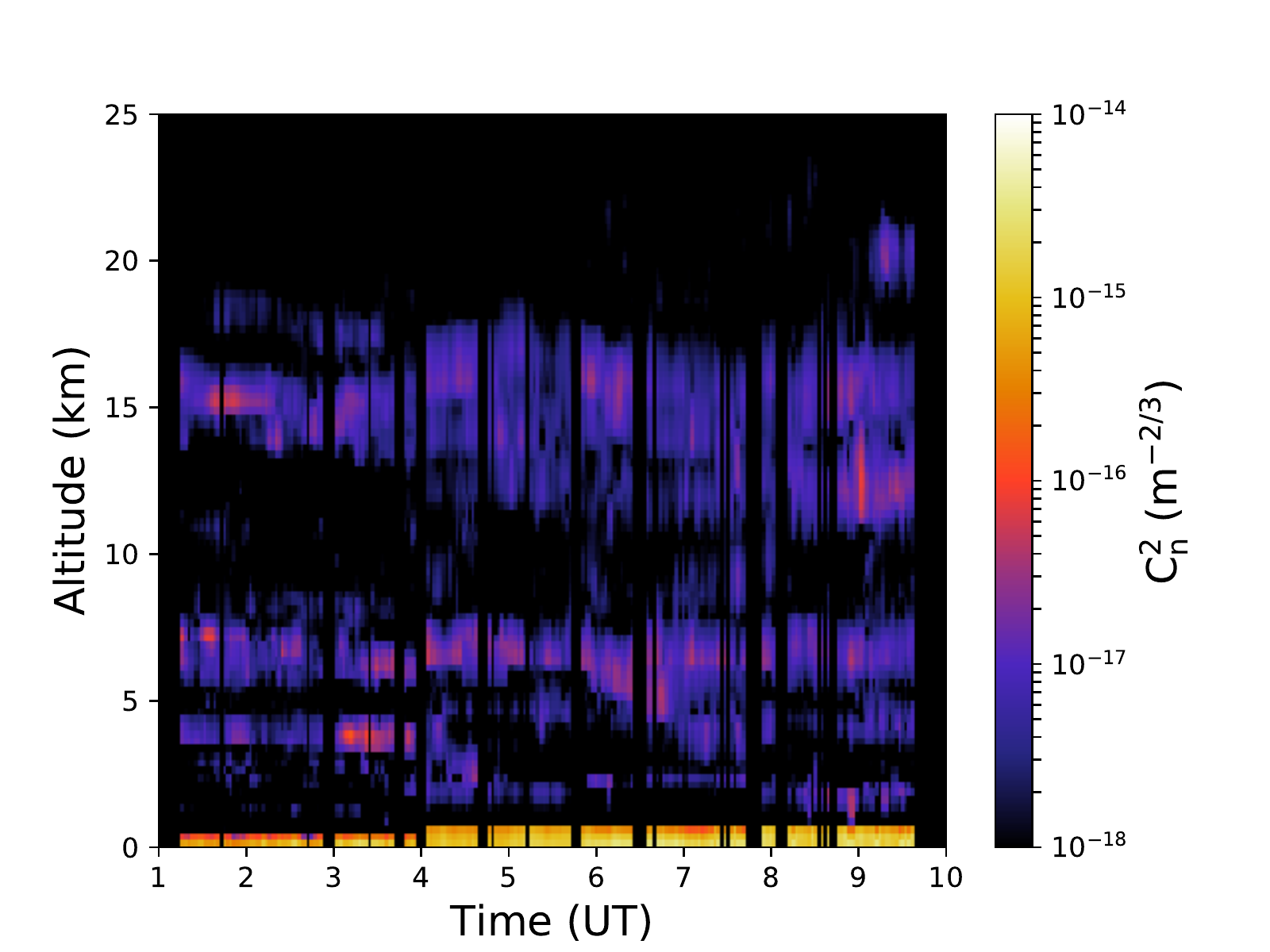} &
    	\includegraphics[width=0.23\textwidth,trim={2cm 0 1cm 0}]{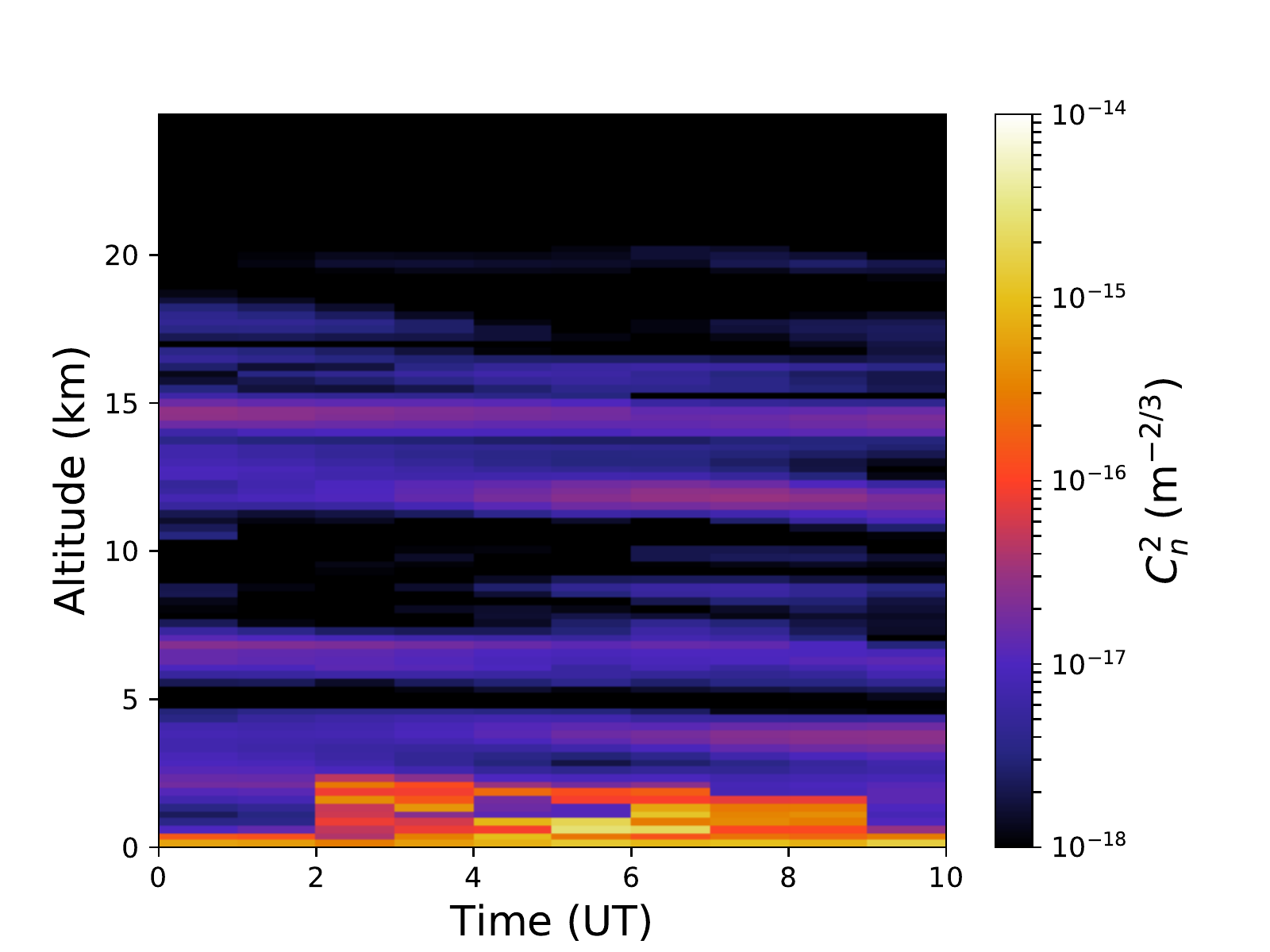} &
    	\includegraphics[width=0.23\textwidth,trim={2cm 0 1cm 0}]{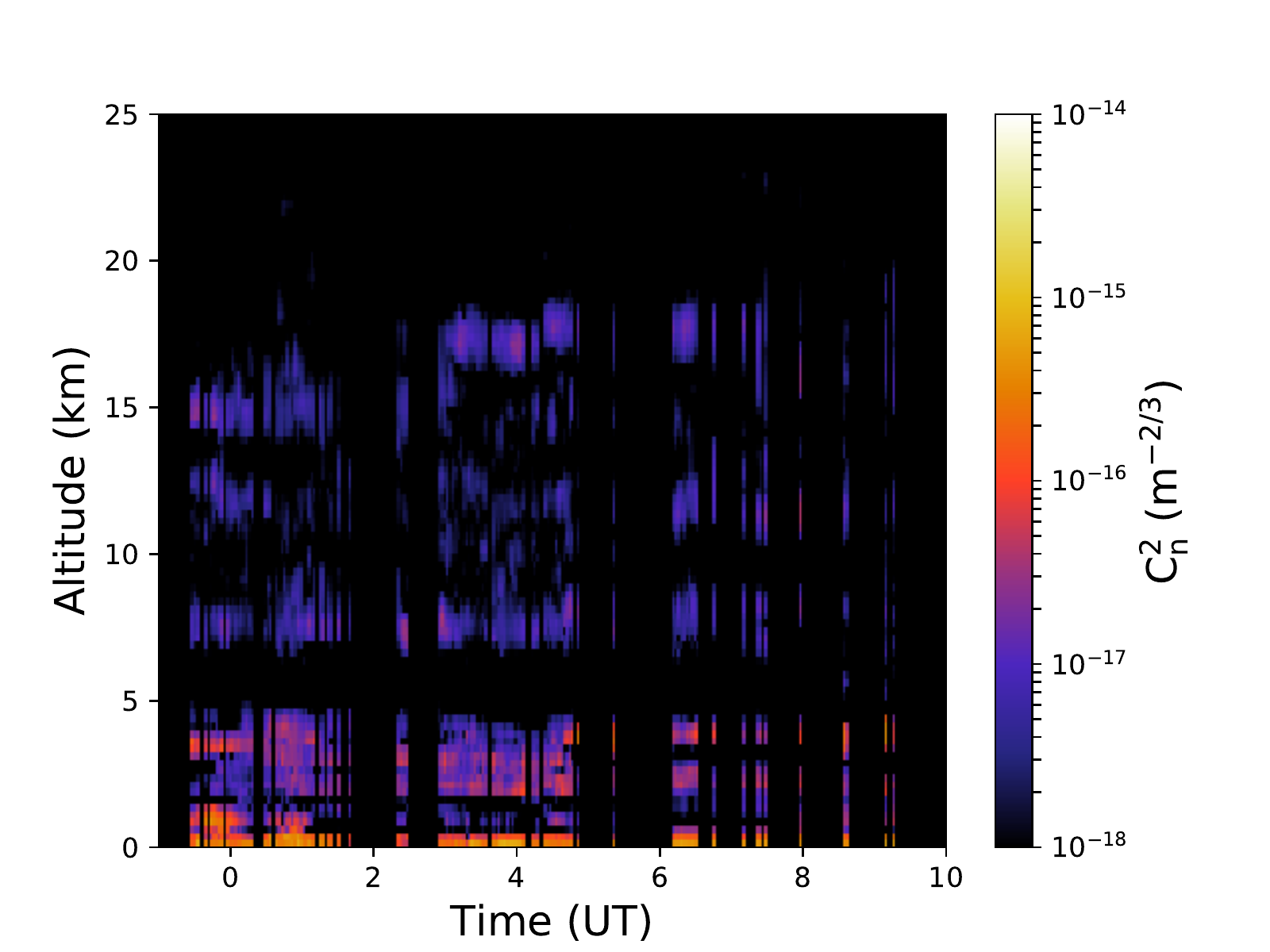} &
    	\includegraphics[width=0.23\textwidth,trim={2cm 0 1cm 0}]{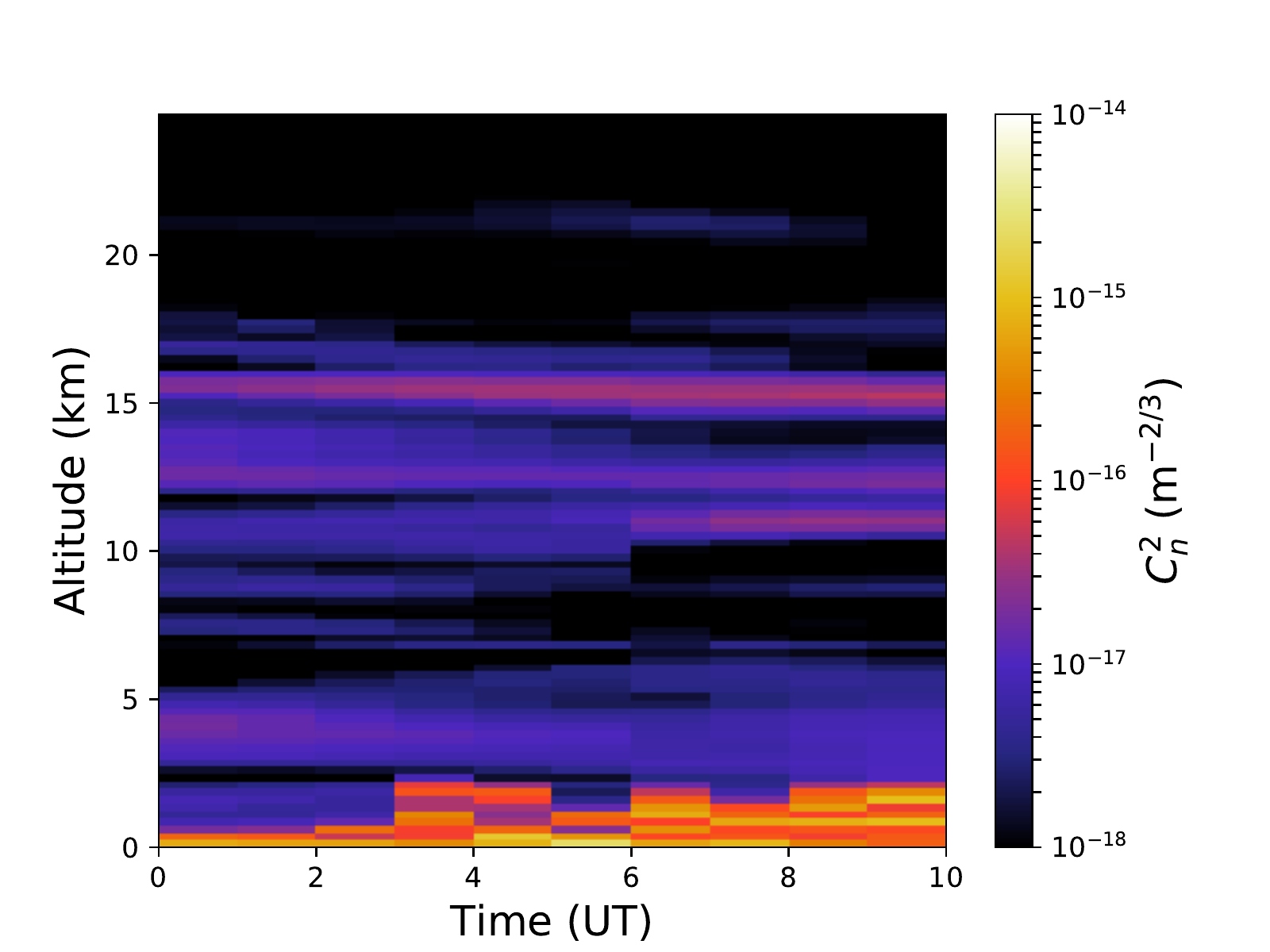} \\
	\includegraphics[width=0.23\textwidth,trim={2cm 0 1cm 0}]{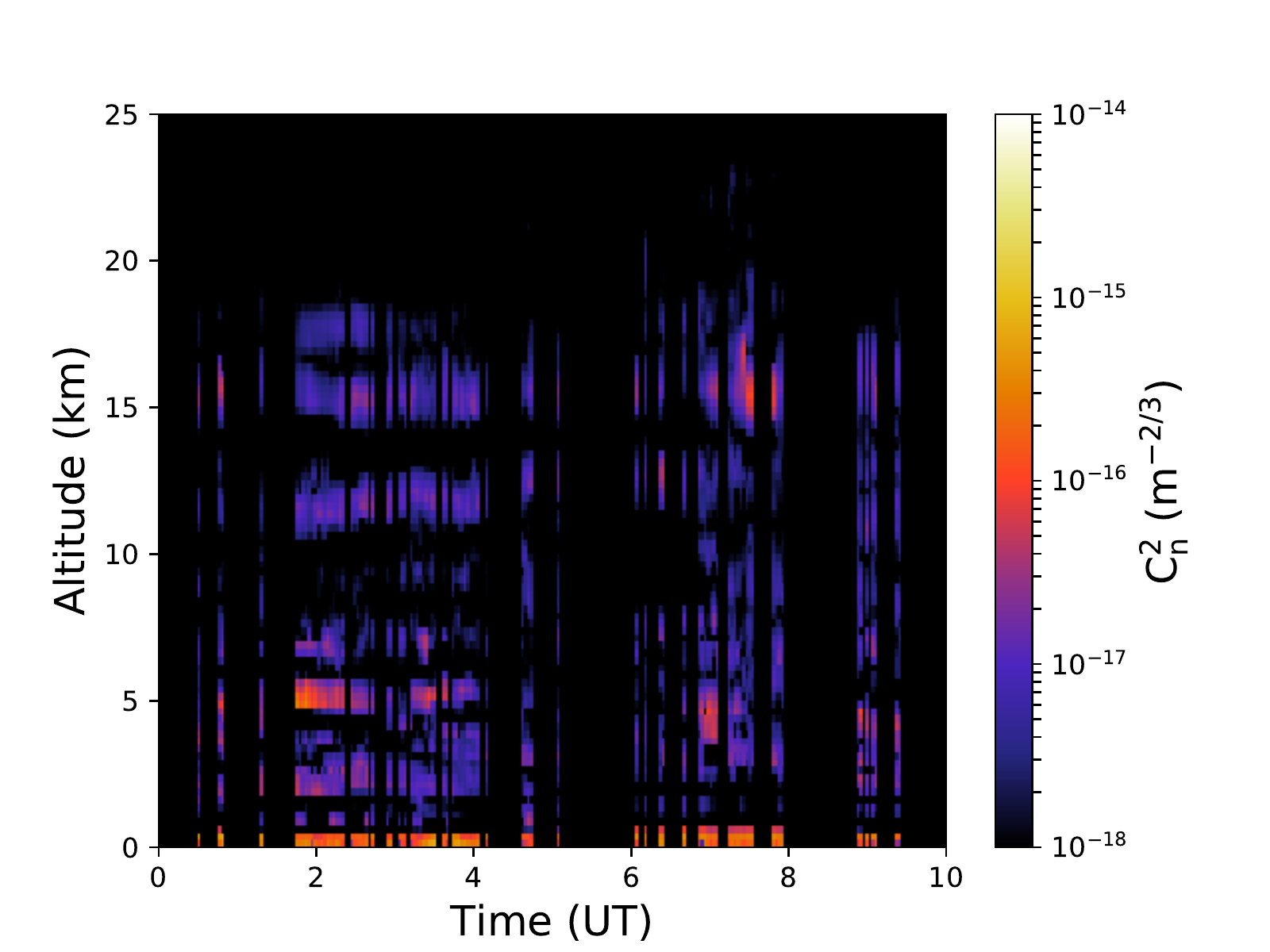} &
    	\includegraphics[width=0.23\textwidth,trim={2cm 0 1cm 0}]{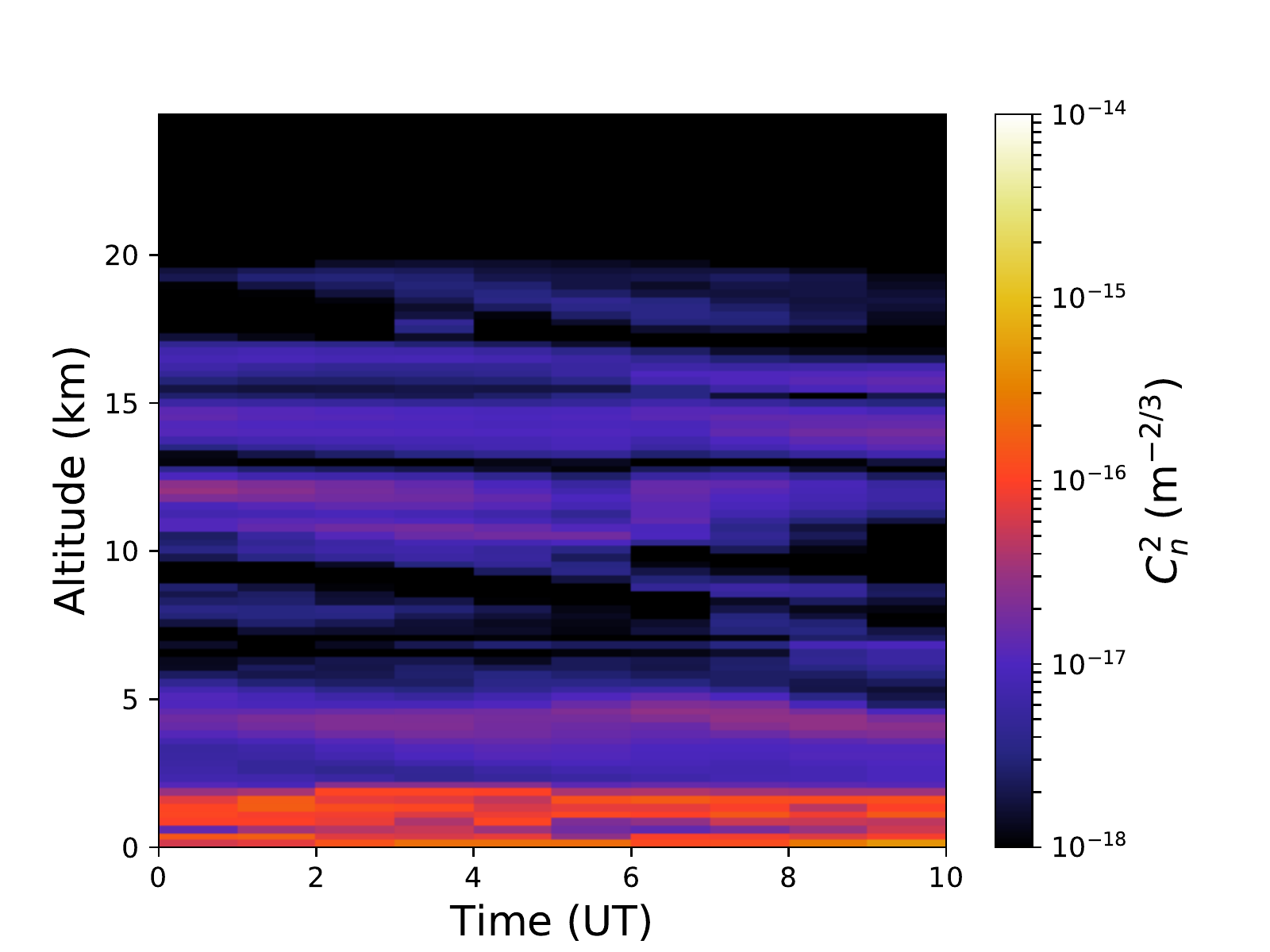} &
    	\includegraphics[width=0.23\textwidth,trim={2cm 0 1cm 0}]{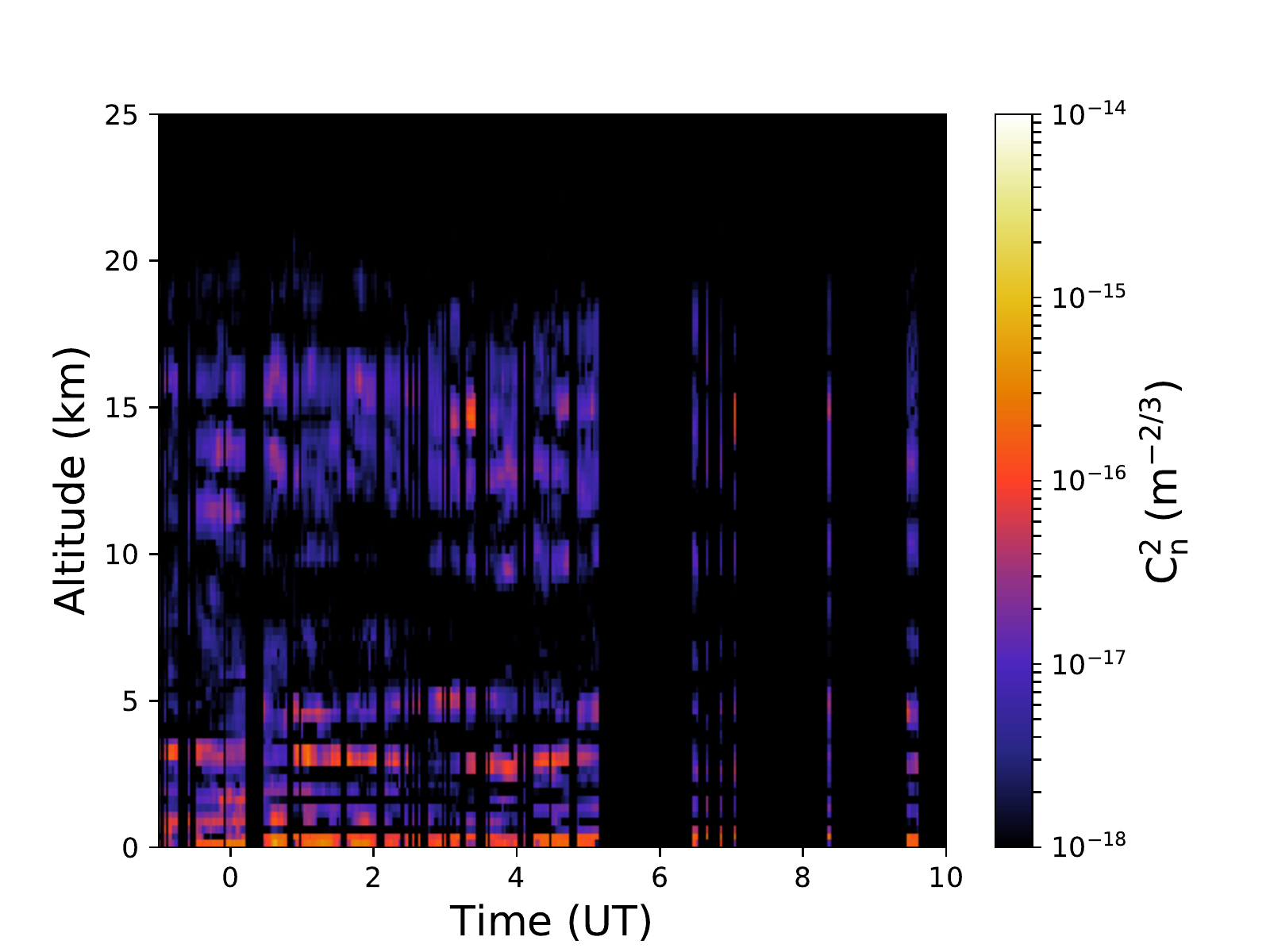} &
    	\includegraphics[width=0.23\textwidth,trim={2cm 0 1cm 0}]{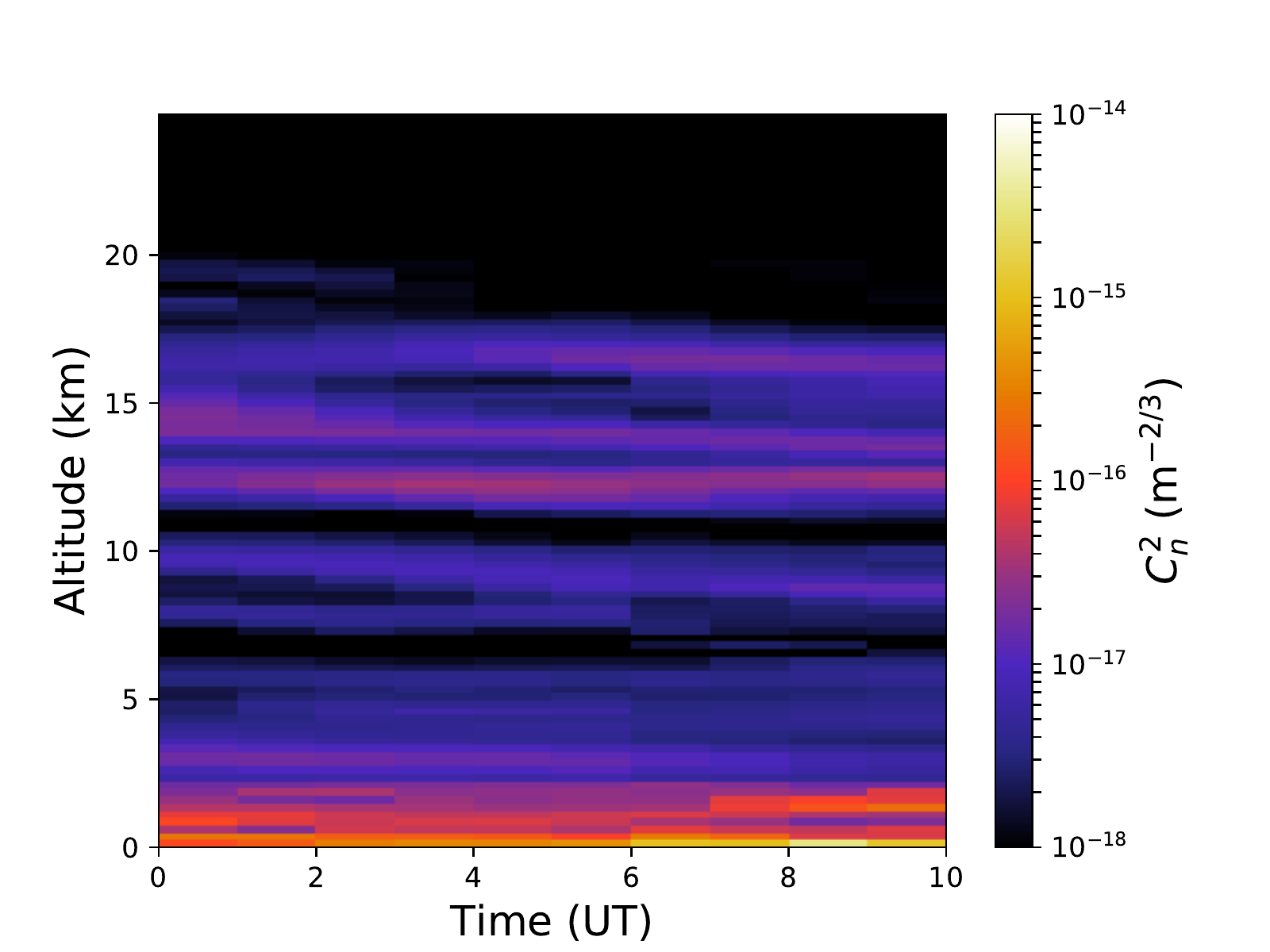} \\
	
	\includegraphics[width=0.23\textwidth,trim={2cm 0 1cm 0}]{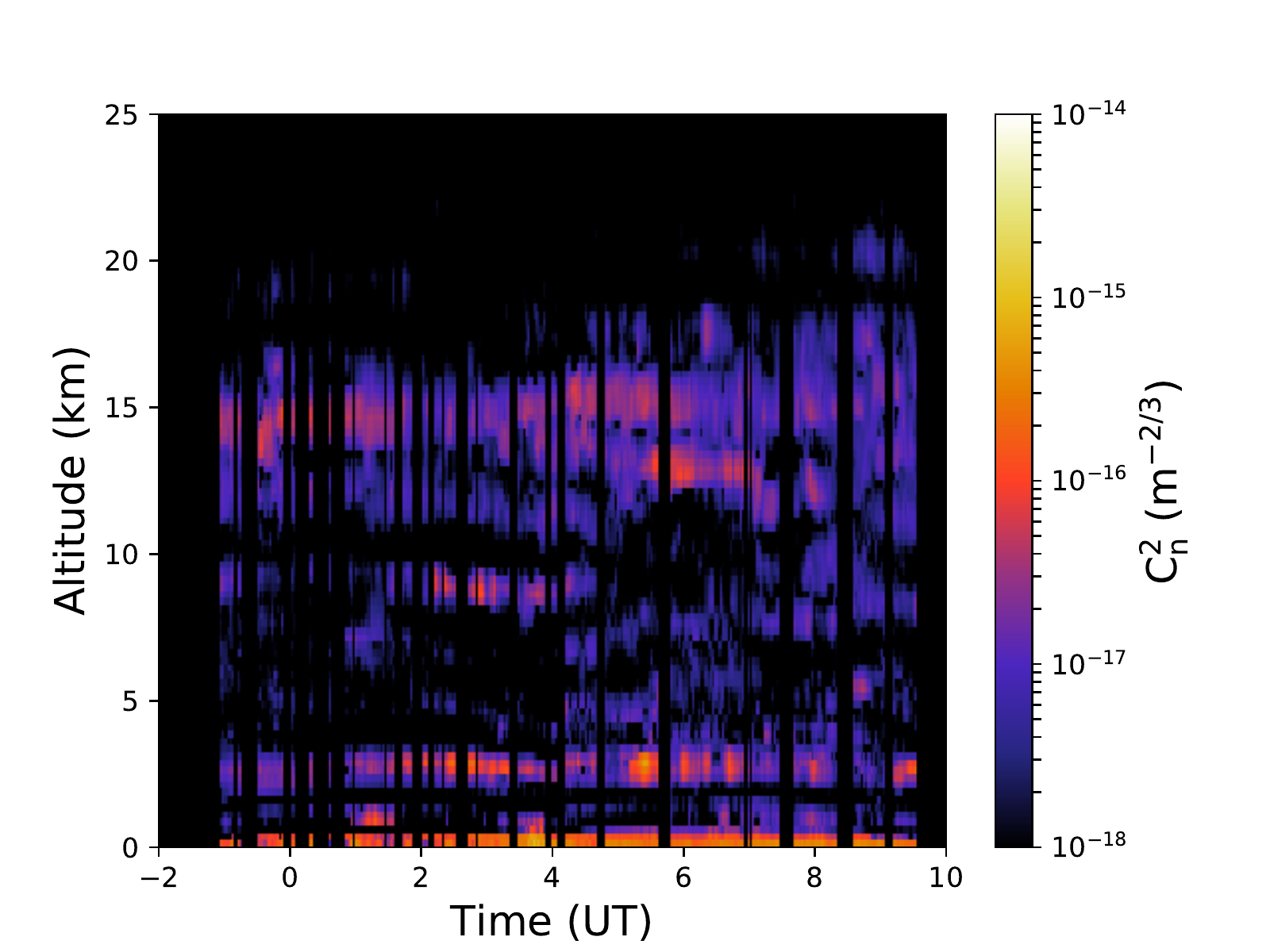} &
    	\includegraphics[width=0.23\textwidth,trim={2cm 0 1cm 0}]{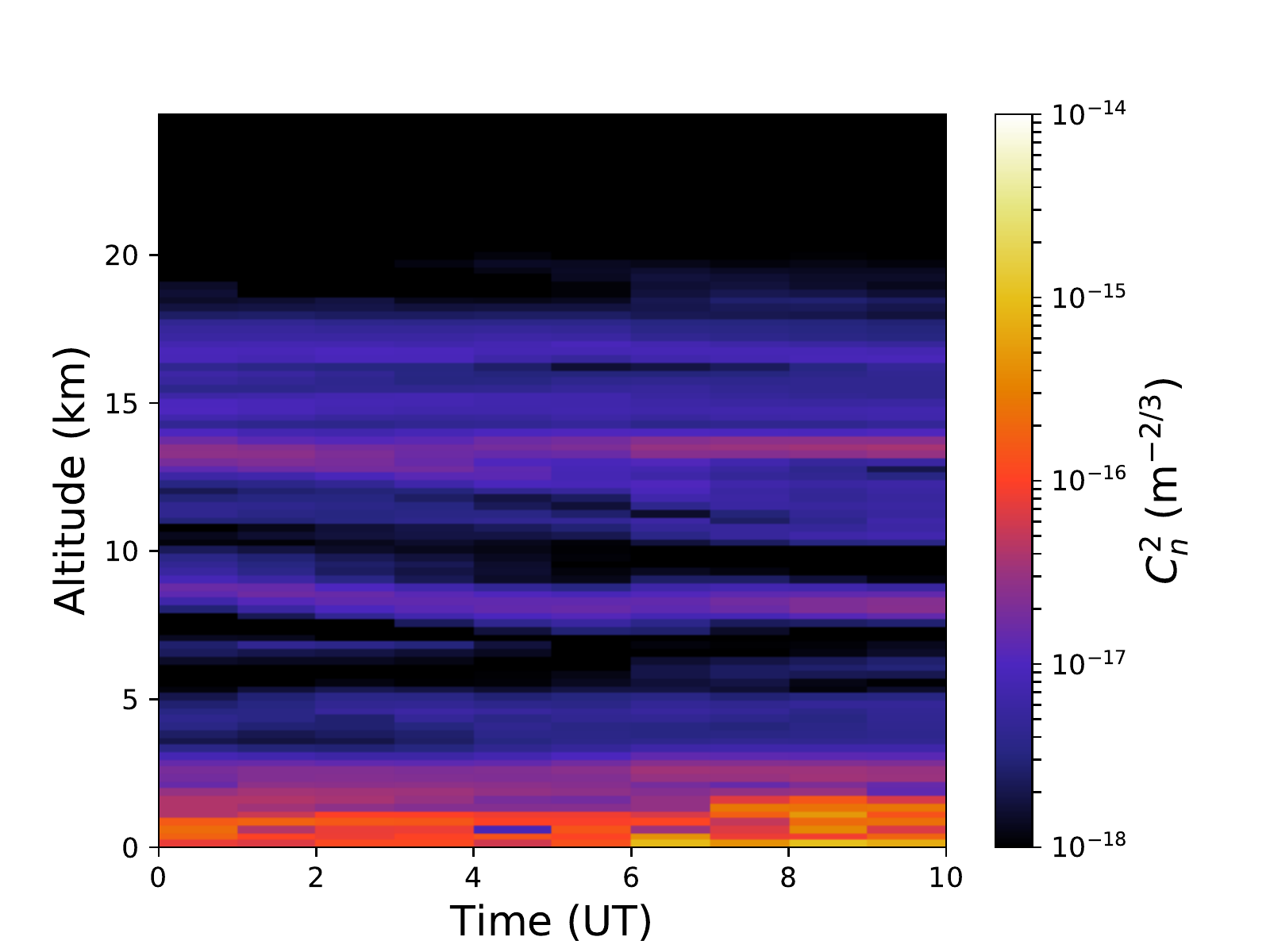} &
    	\includegraphics[width=0.23\textwidth,trim={2cm 0 1cm 0}]{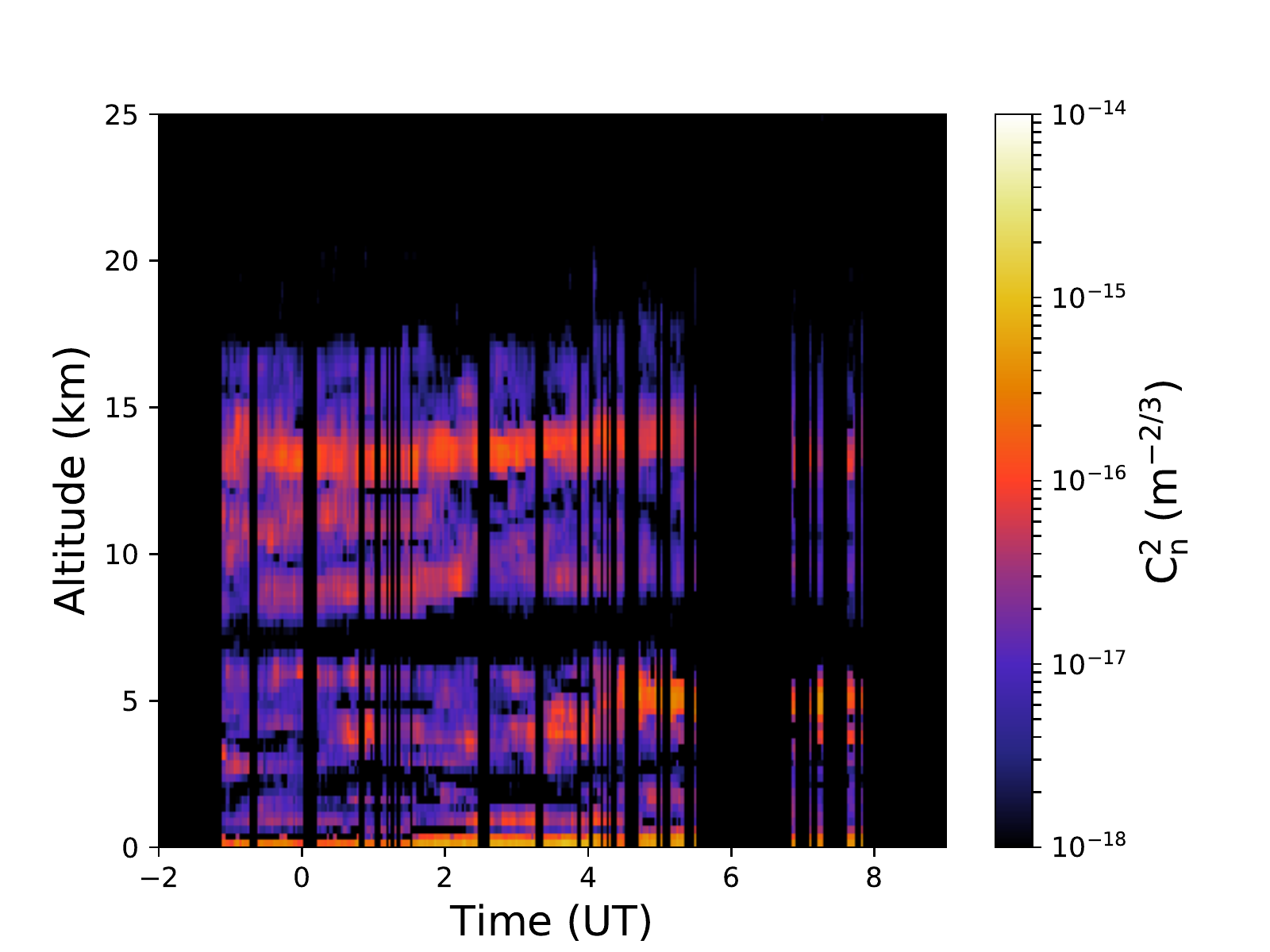} &
    	\includegraphics[width=0.23\textwidth,trim={2cm 0 1cm 0}]{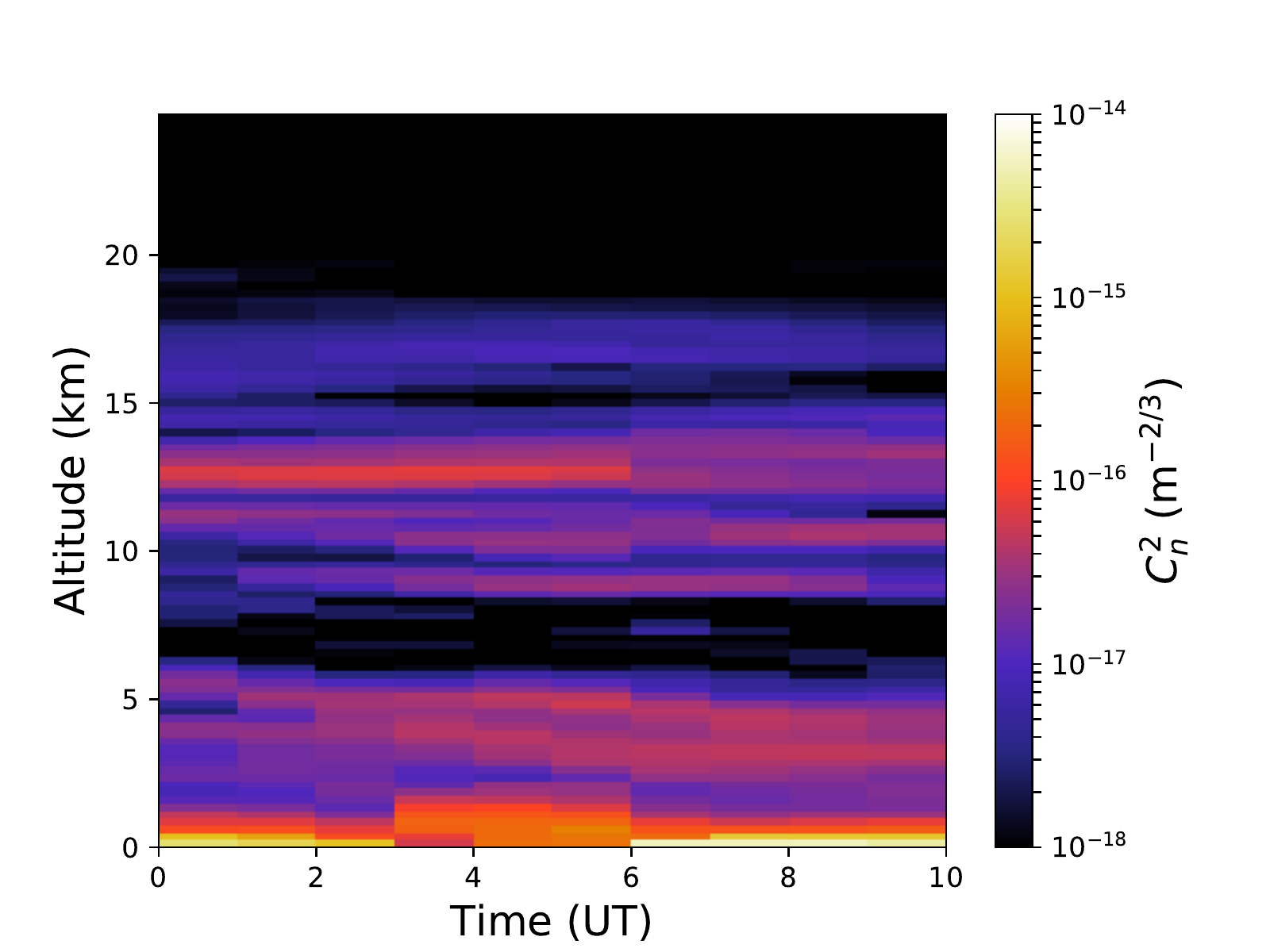} \\

\end{array}$
\caption{Example vertical profiles as measured by the stereo-SCIDAR (green) and estimated by the ECMWF GCM model (red). The profiles shown are the median for an individual night of observation. The coloured region shows the interquartile range. These profiles are from the nights beginning10th -12th December 2016, 7th - 9th March and 12th - 17th April 2017.}
\label{fig:seqProfiles2}
\end{figure*}

\begin{figure*}
\centering
$\begin{array}{cccc}

	\includegraphics[width=0.23\textwidth,trim={2cm 0 1cm 0}]{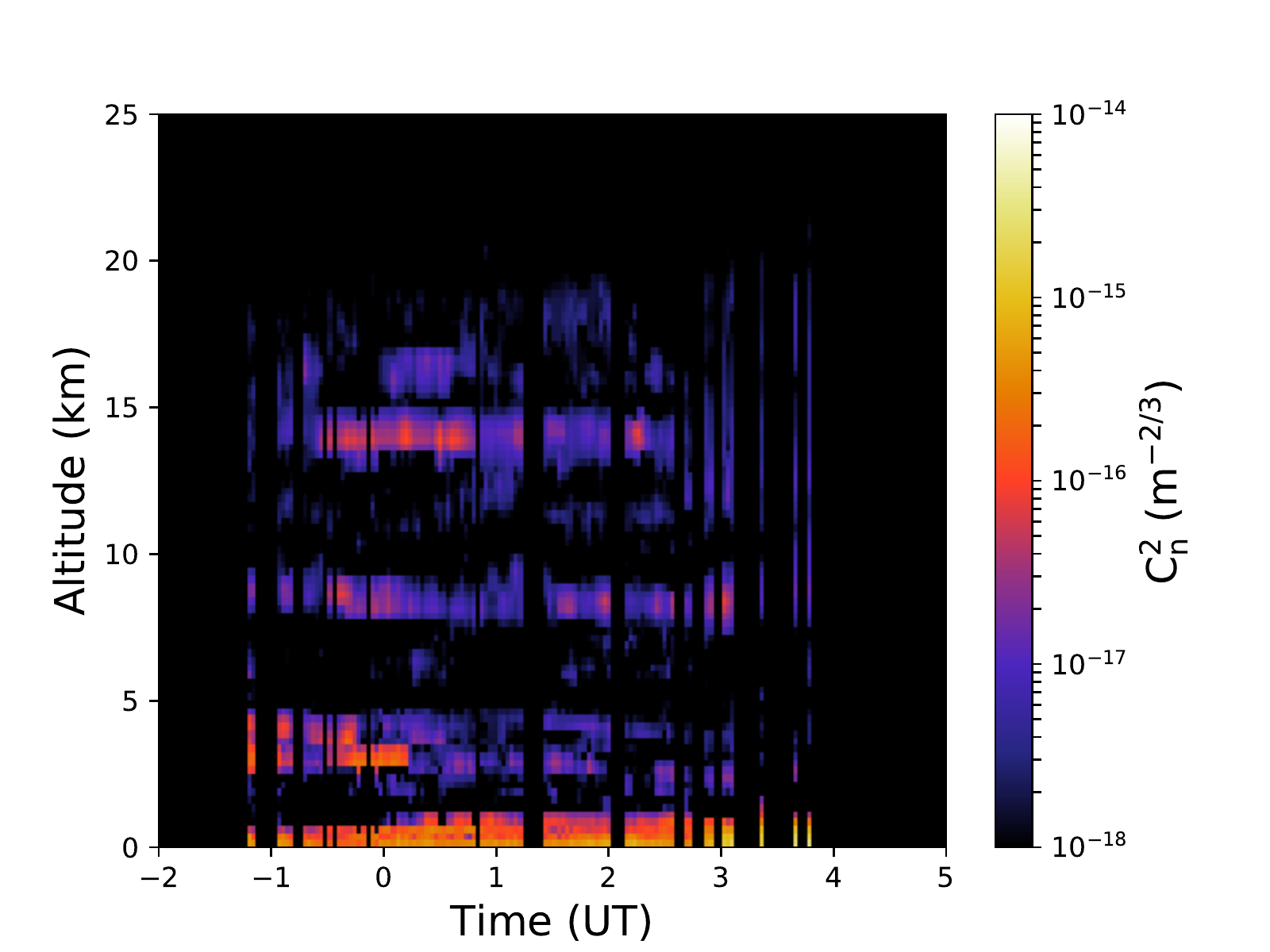} &
    	\includegraphics[width=0.23\textwidth,trim={2cm 0 1cm 0}]{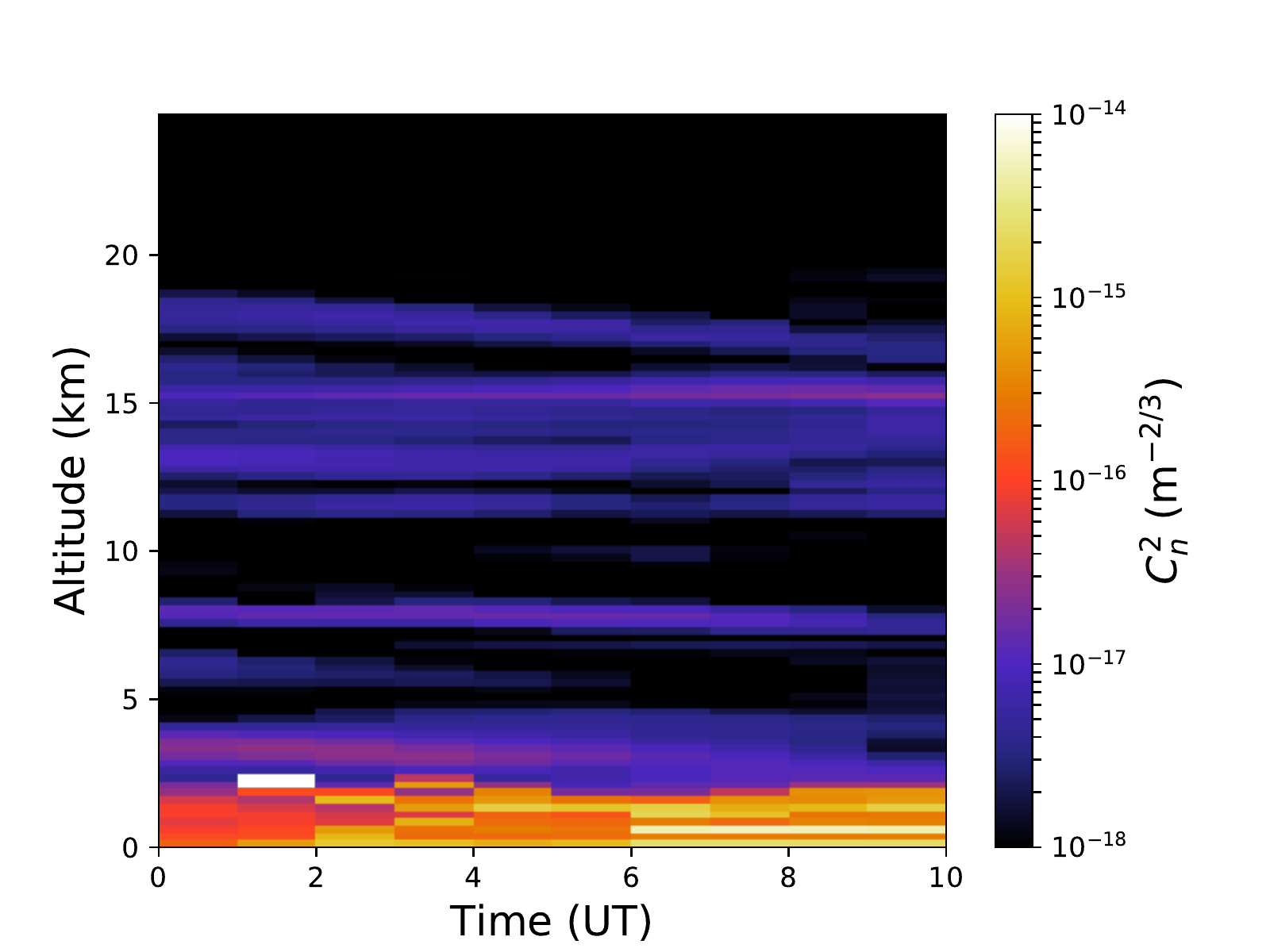} &
    	\includegraphics[width=0.23\textwidth,trim={2cm 0 1cm 0}]{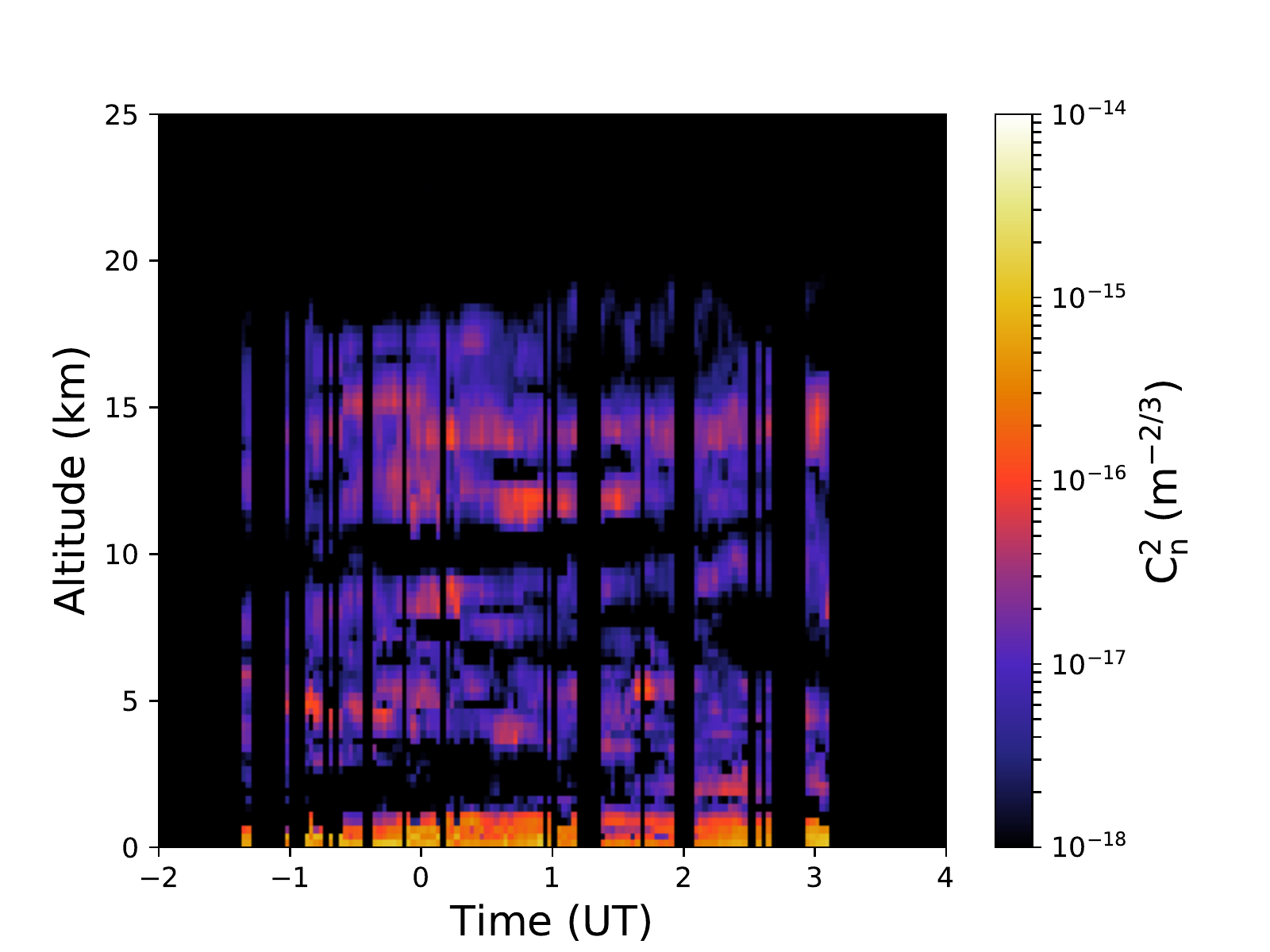} &
    	\includegraphics[width=0.23\textwidth,trim={2cm 0 1cm 0}]{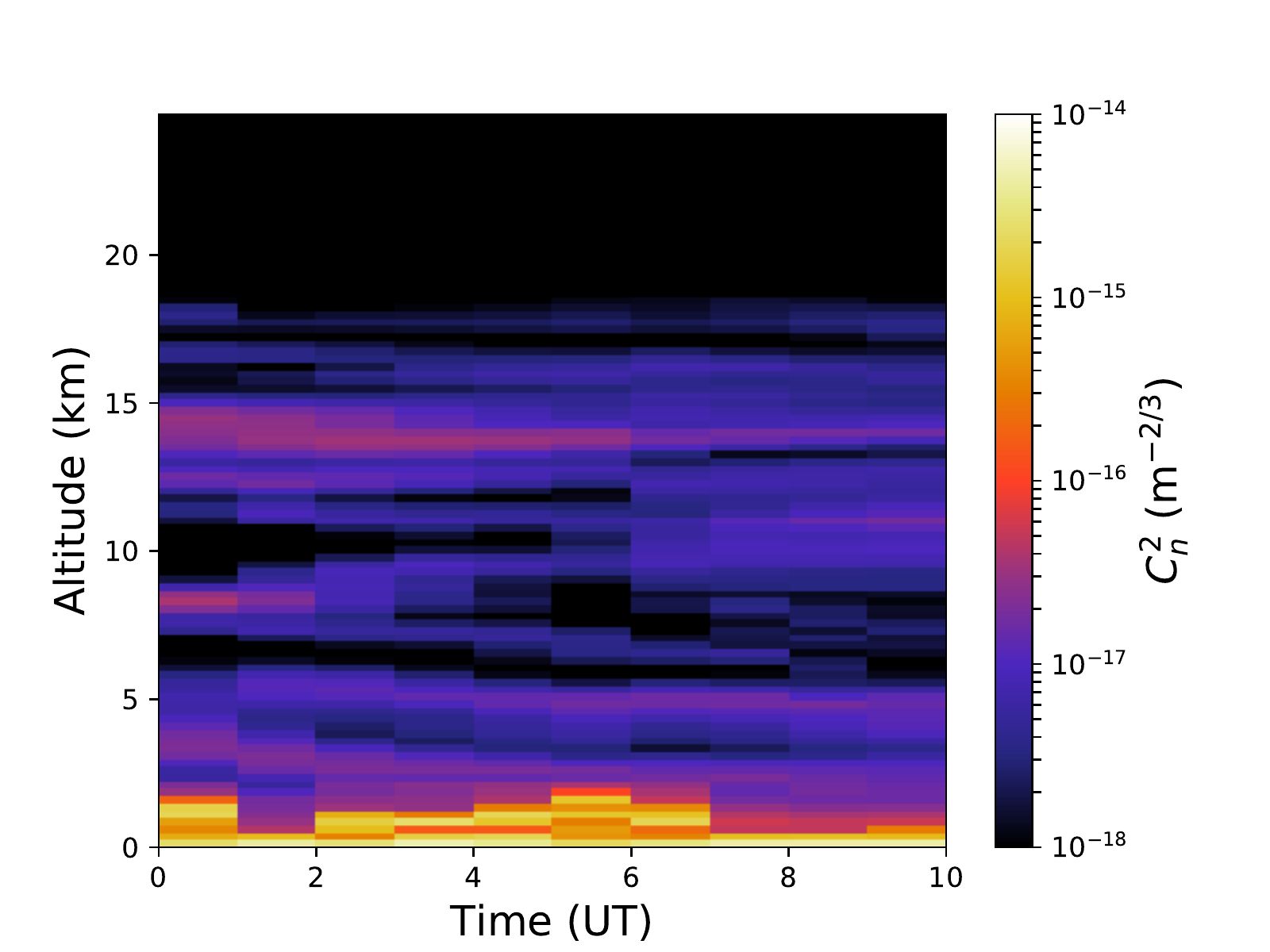} \\
	\includegraphics[width=0.23\textwidth,trim={2cm 0 1cm 0}]{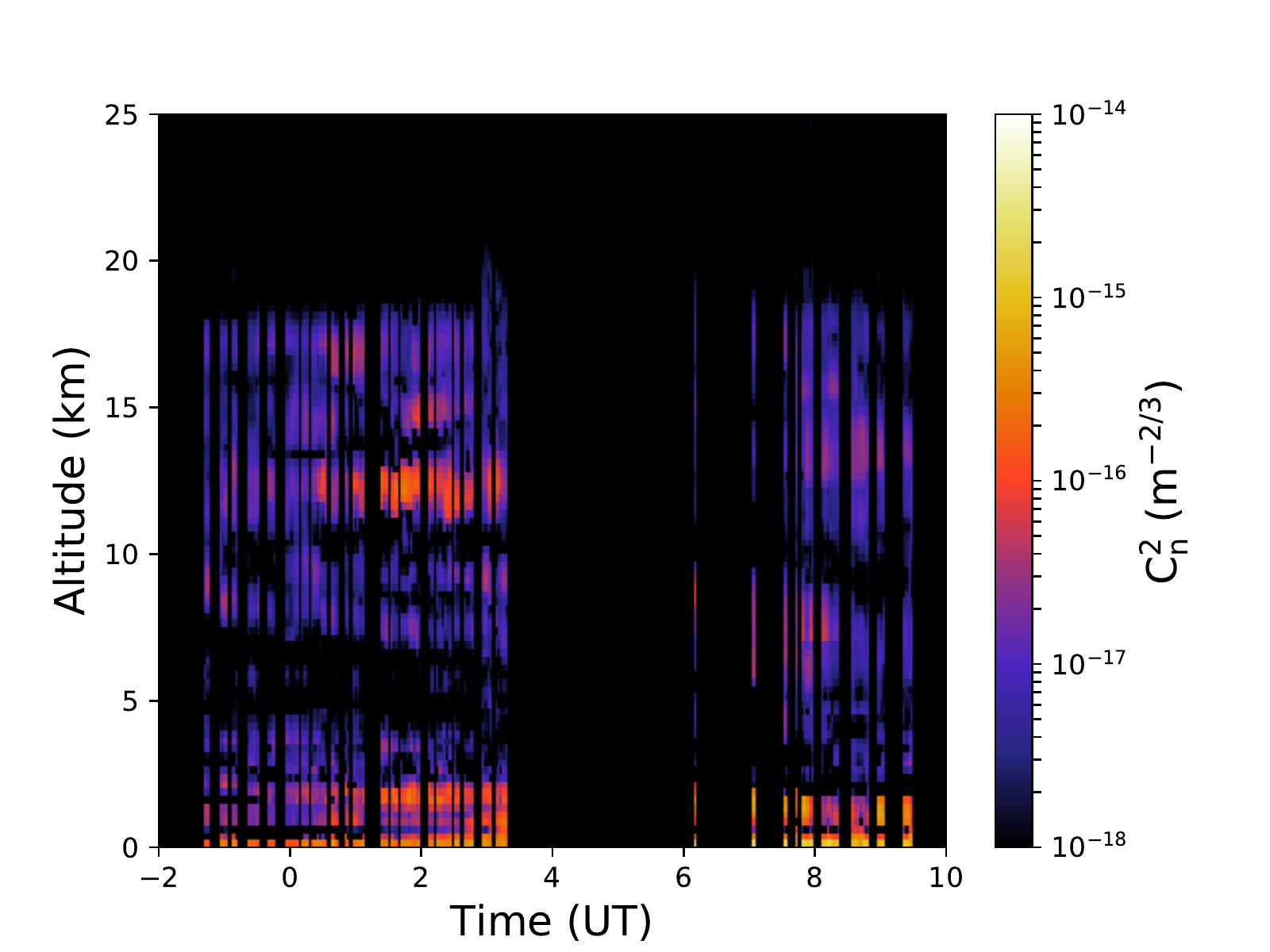} &
    	\includegraphics[width=0.23\textwidth,trim={2cm 0 1cm 0}]{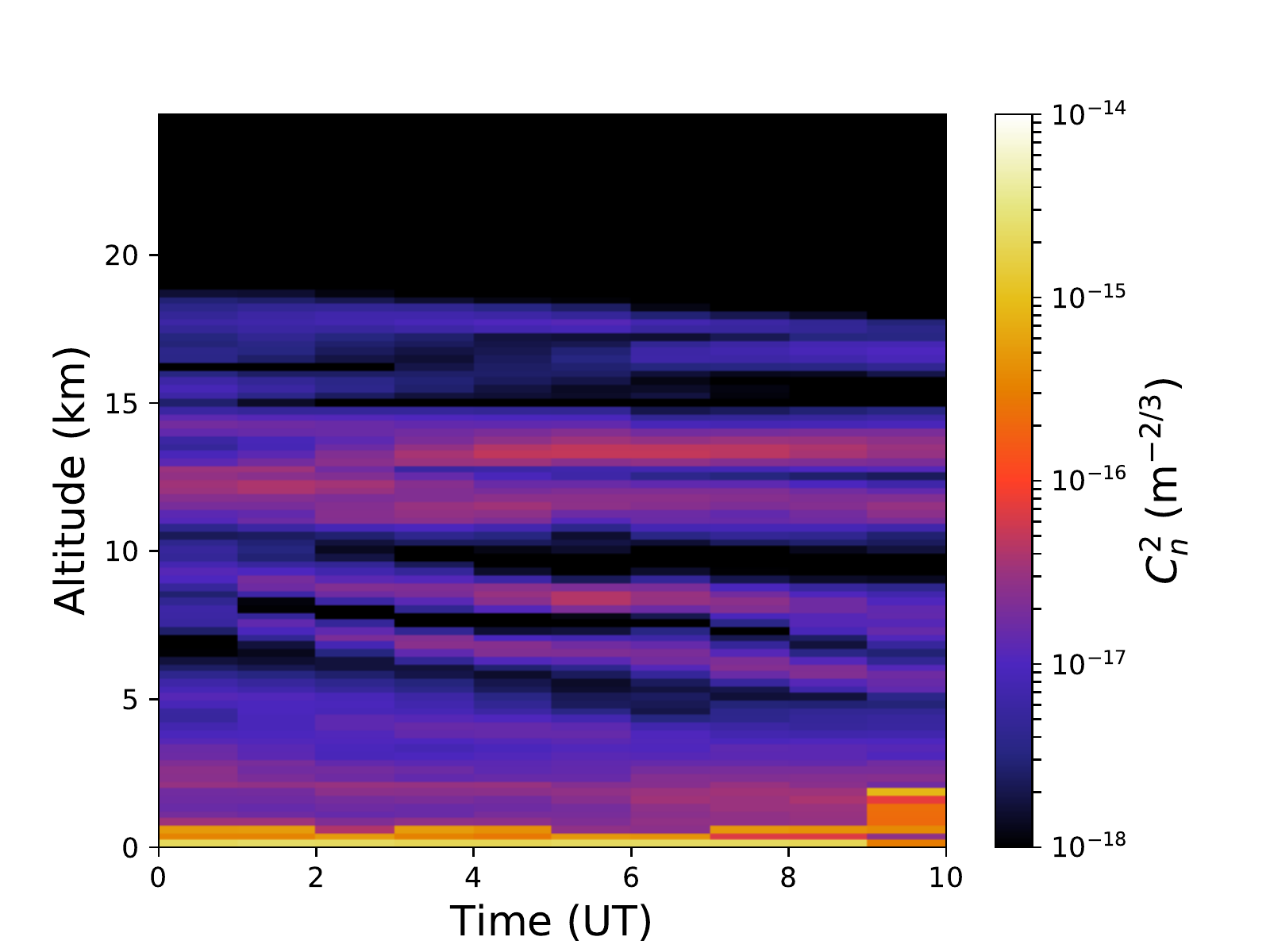} &
    	\includegraphics[width=0.23\textwidth,trim={2cm 0 1cm 0}]{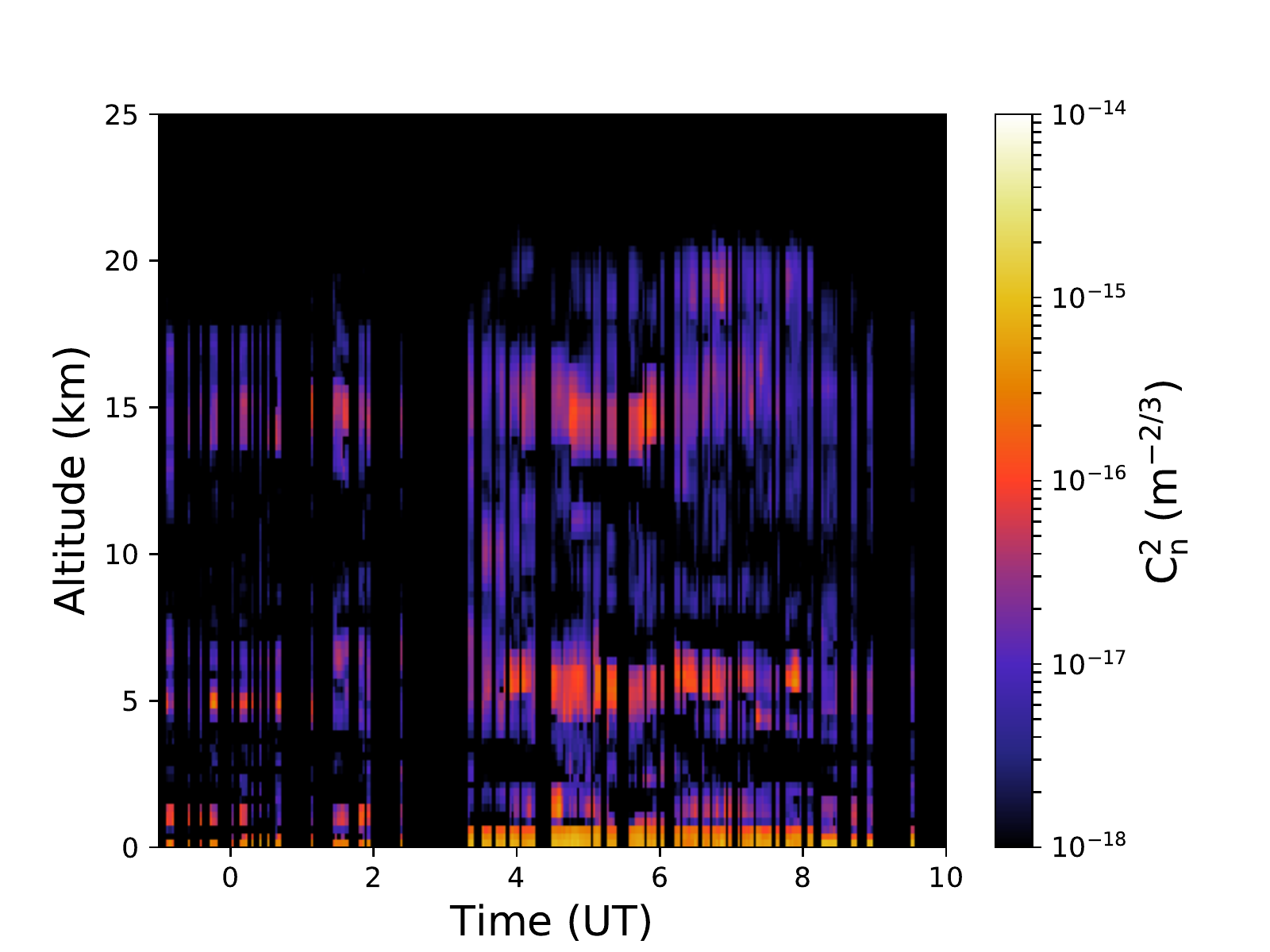} &
    	\includegraphics[width=0.23\textwidth,trim={2cm 0 1cm 0}]{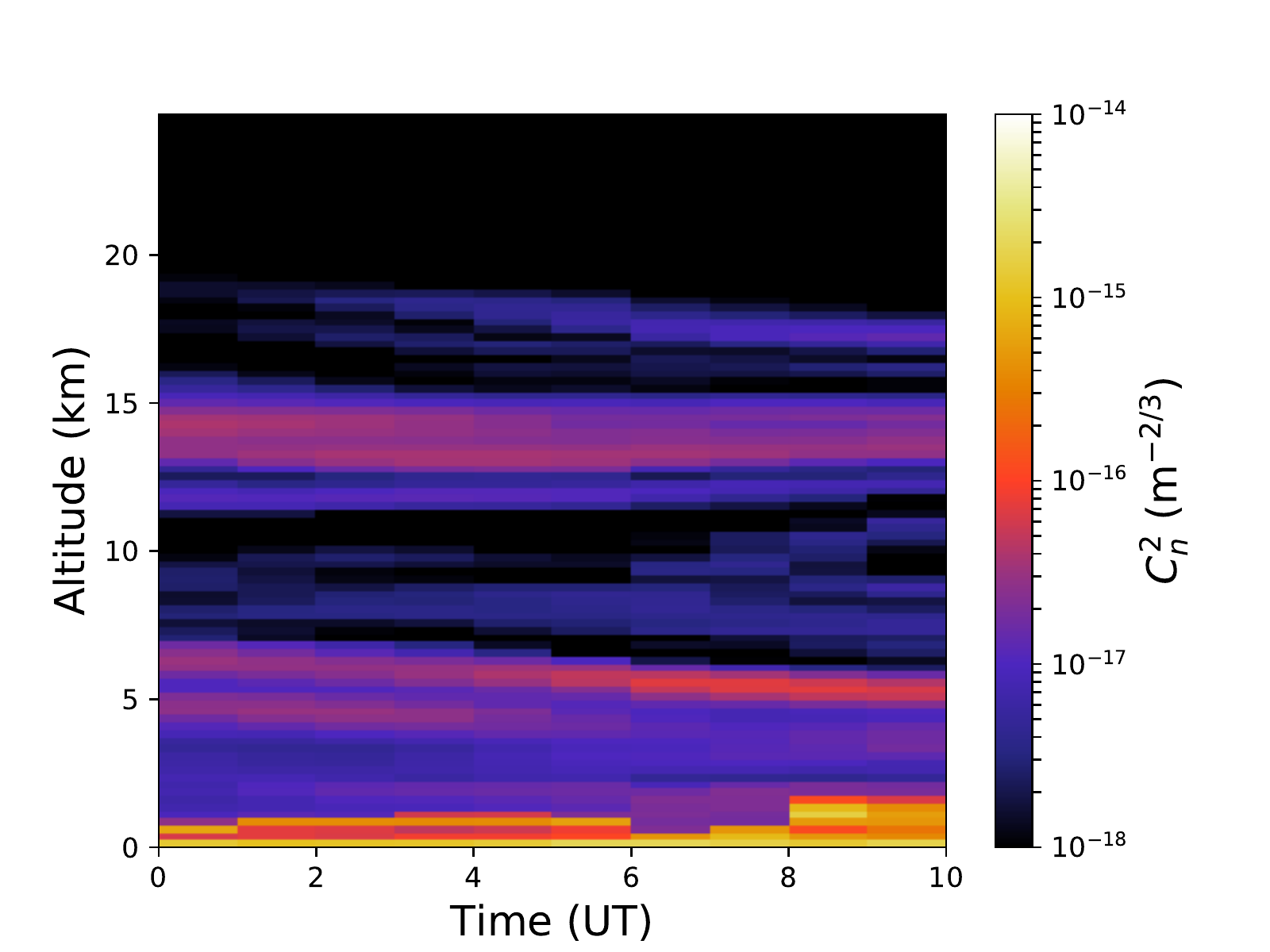} \\
	\includegraphics[width=0.23\textwidth,trim={2cm 0 1cm 0}]{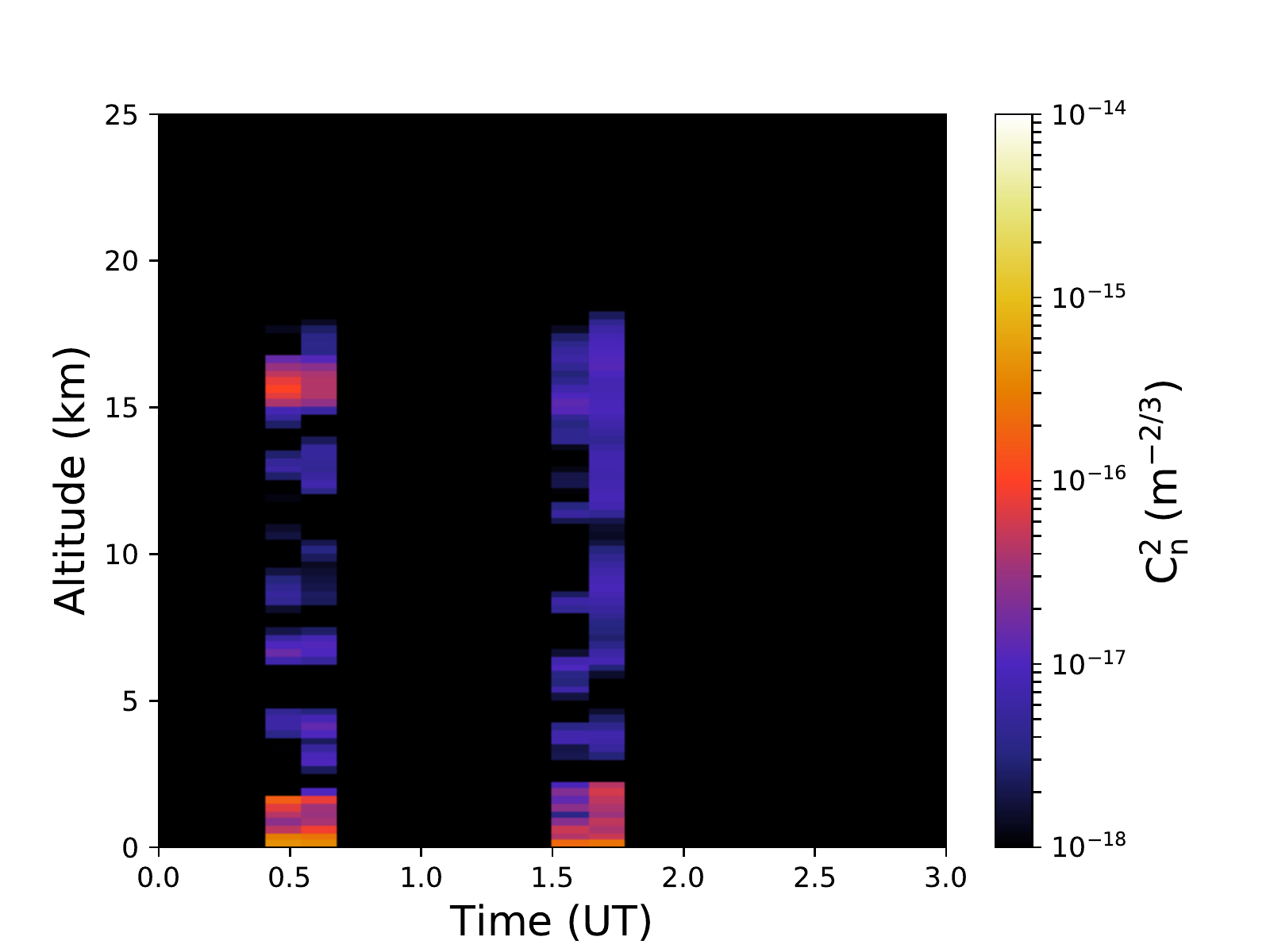} &
    	\includegraphics[width=0.23\textwidth,trim={2cm 0 1cm 0}]{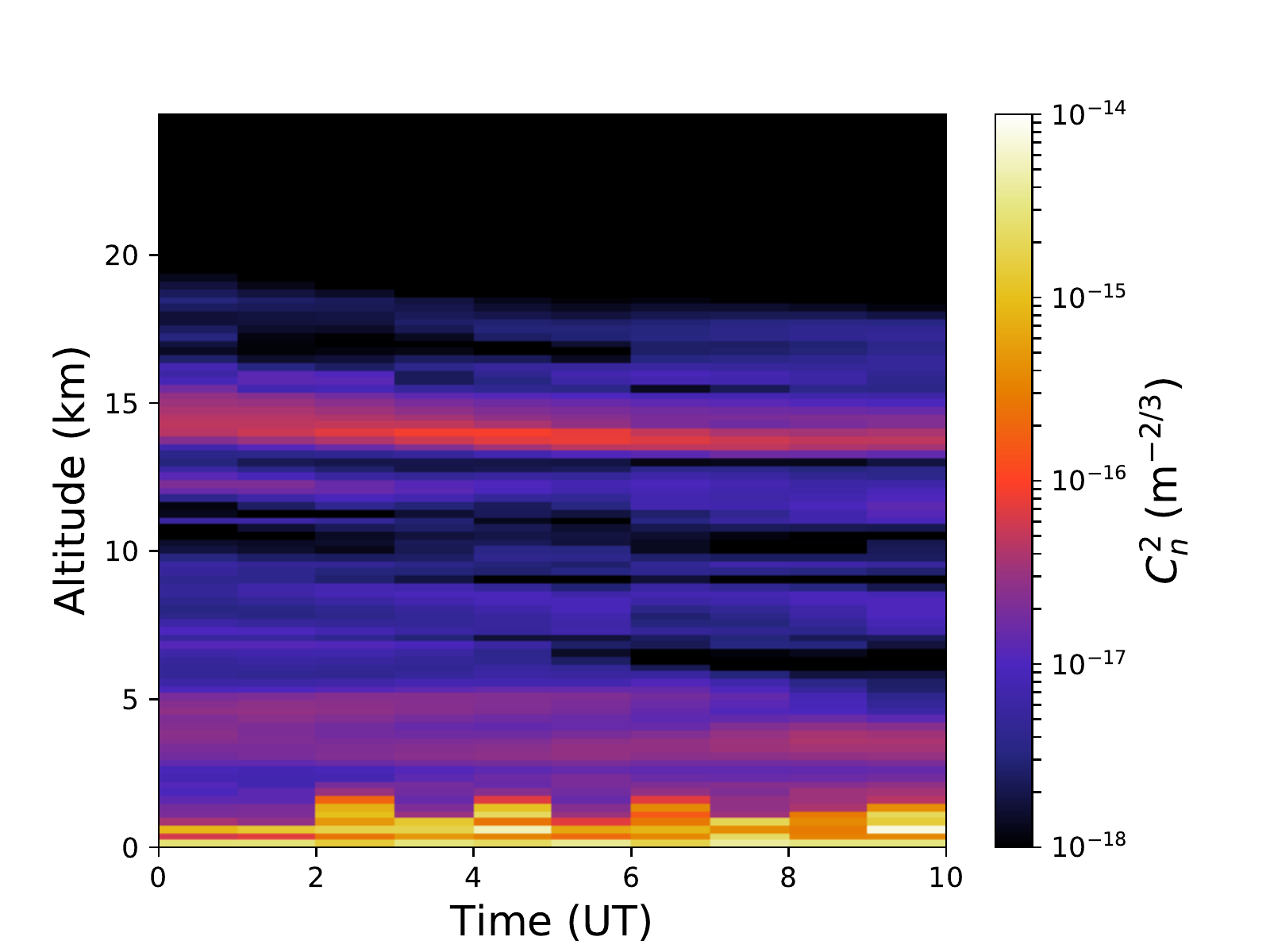} &

    	\includegraphics[width=0.23\textwidth,trim={2cm 0 1cm 0}]{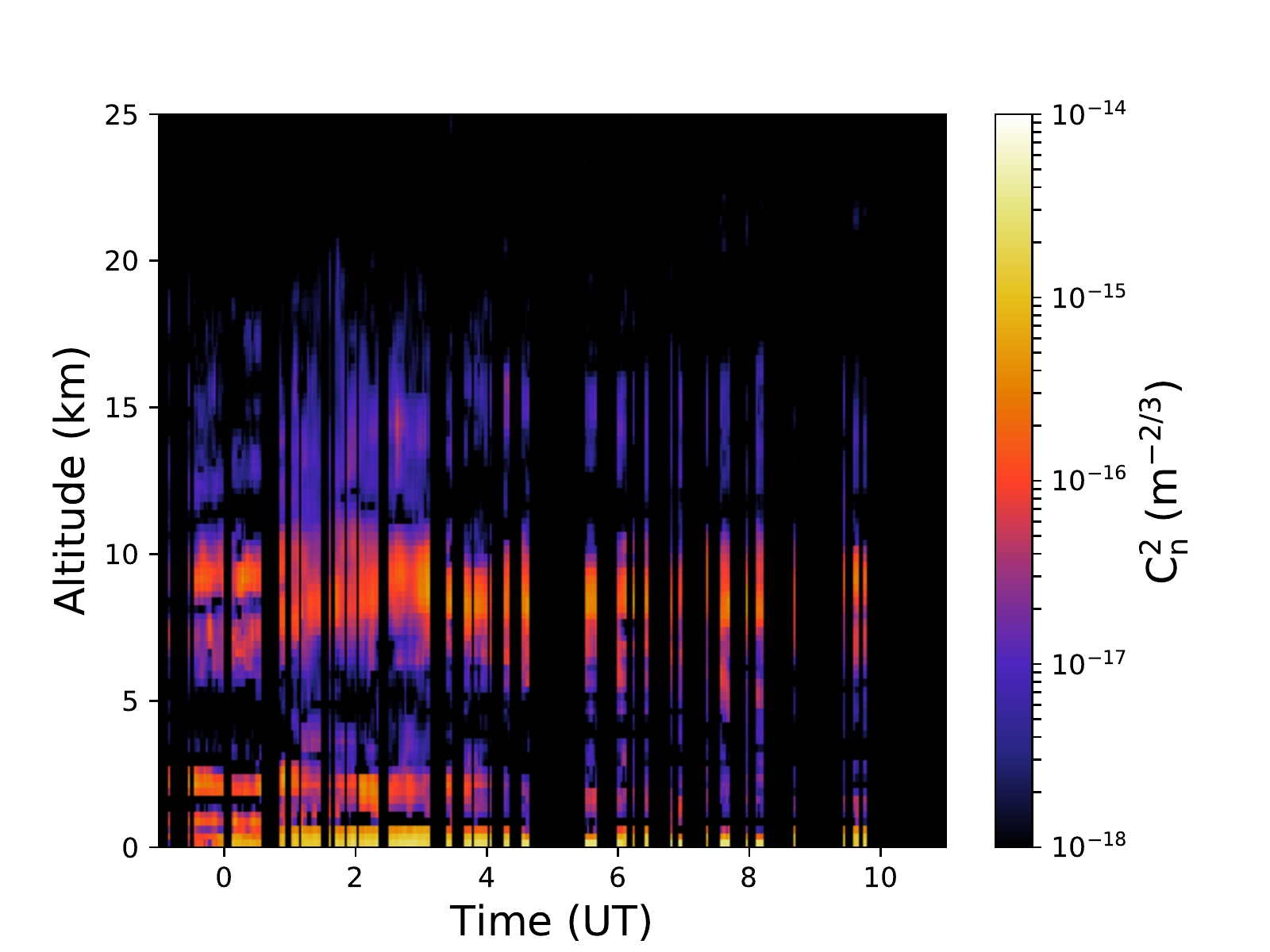} &
    	\includegraphics[width=0.23\textwidth,trim={2cm 0 1cm 0}]{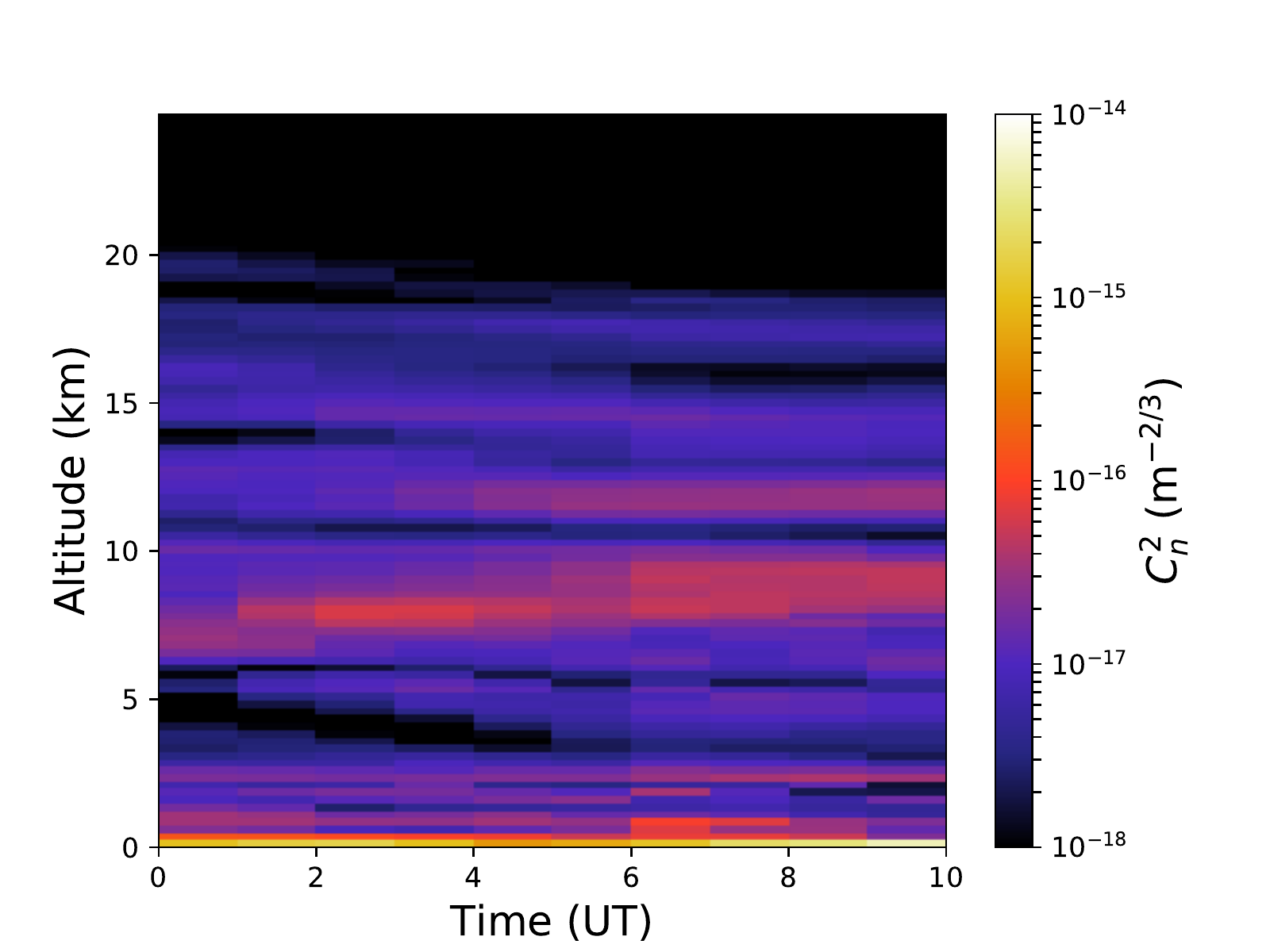} \\
	\includegraphics[width=0.23\textwidth,trim={2cm 0 1cm 0}]{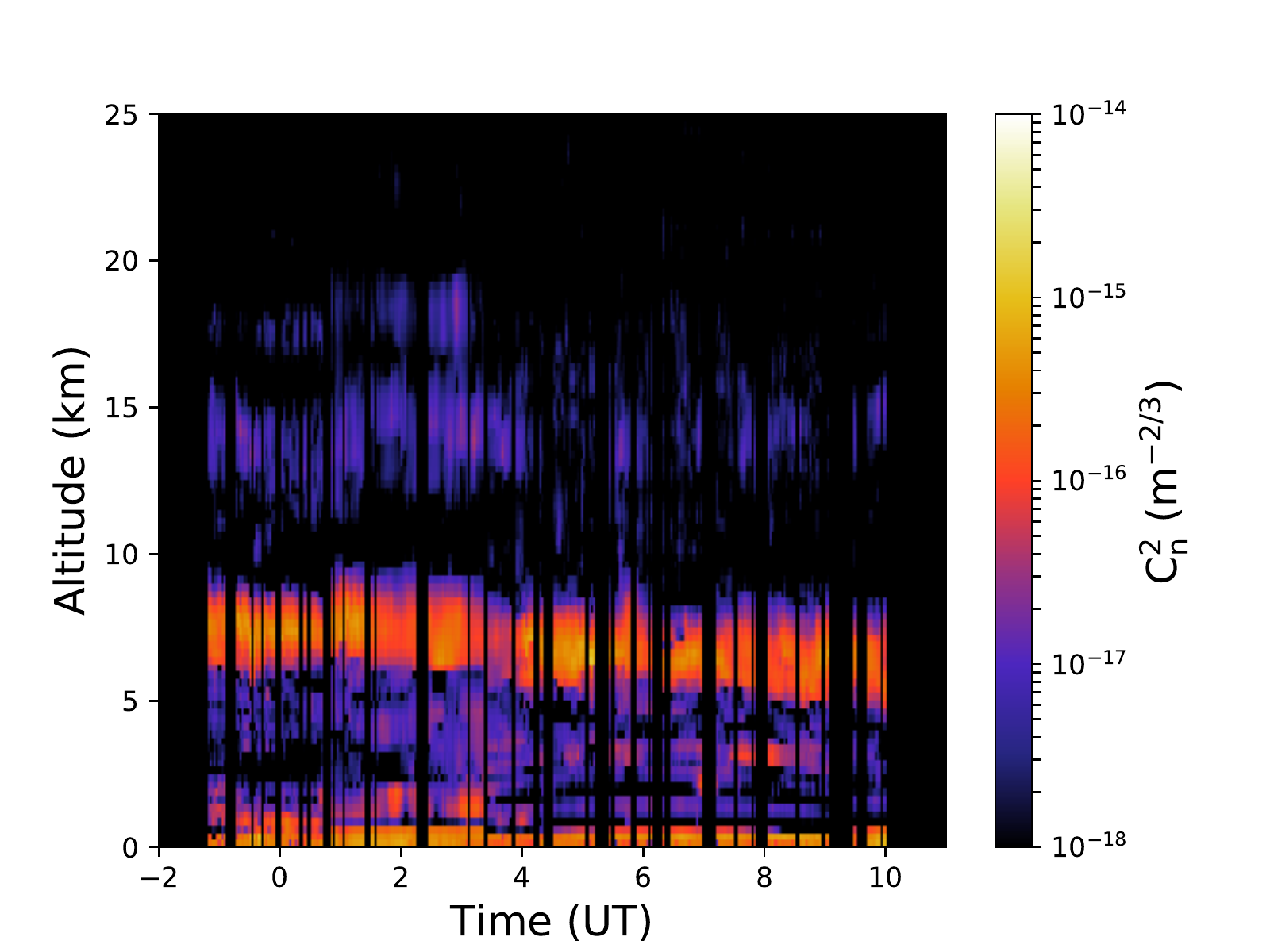} &
    	\includegraphics[width=0.23\textwidth,trim={2cm 0 1cm 0}]{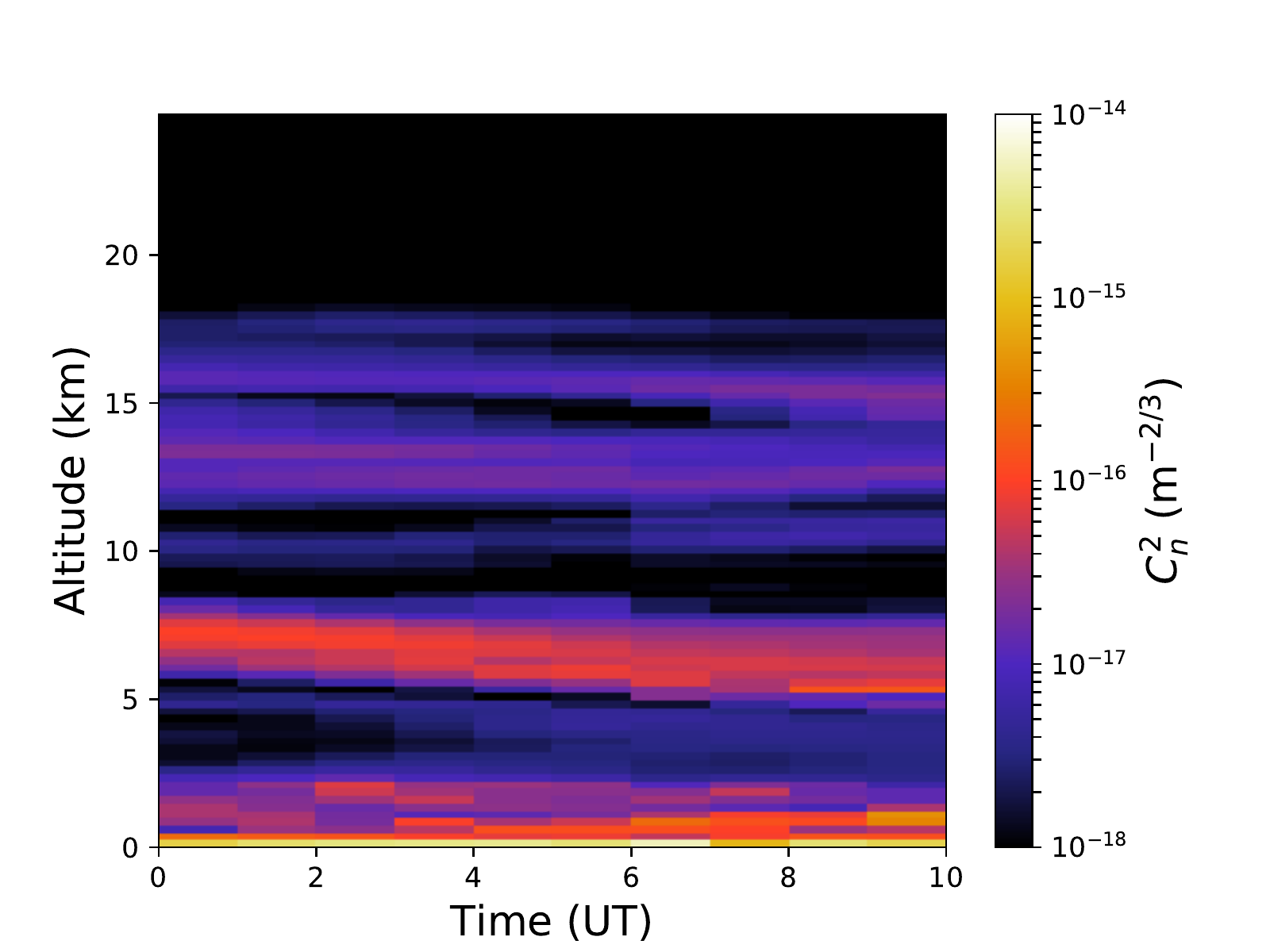} &
	\includegraphics[width=0.23\textwidth,trim={2cm 0 1cm 0}]{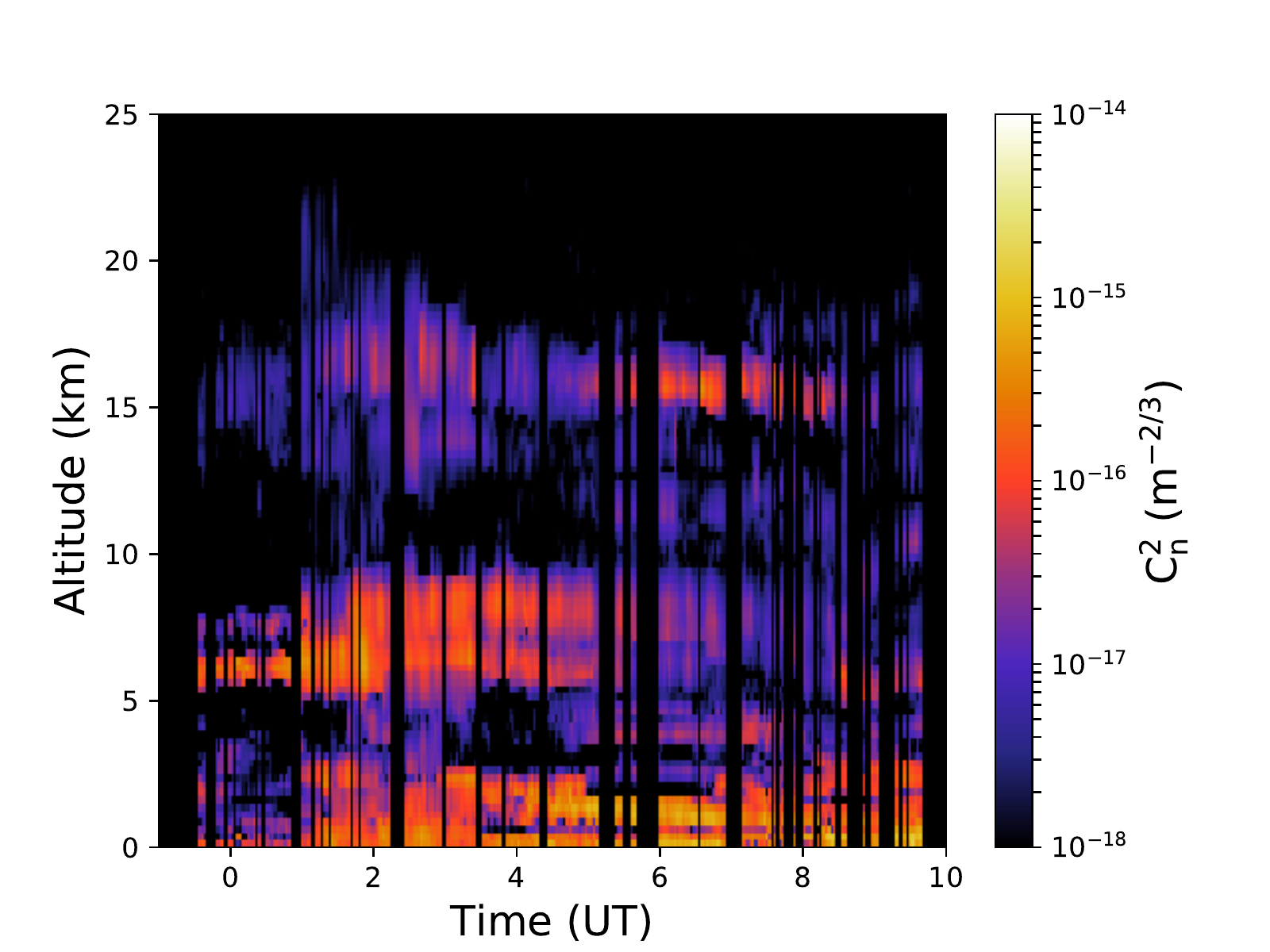} &
    	\includegraphics[width=0.23\textwidth,trim={2cm 0 1cm 0}]{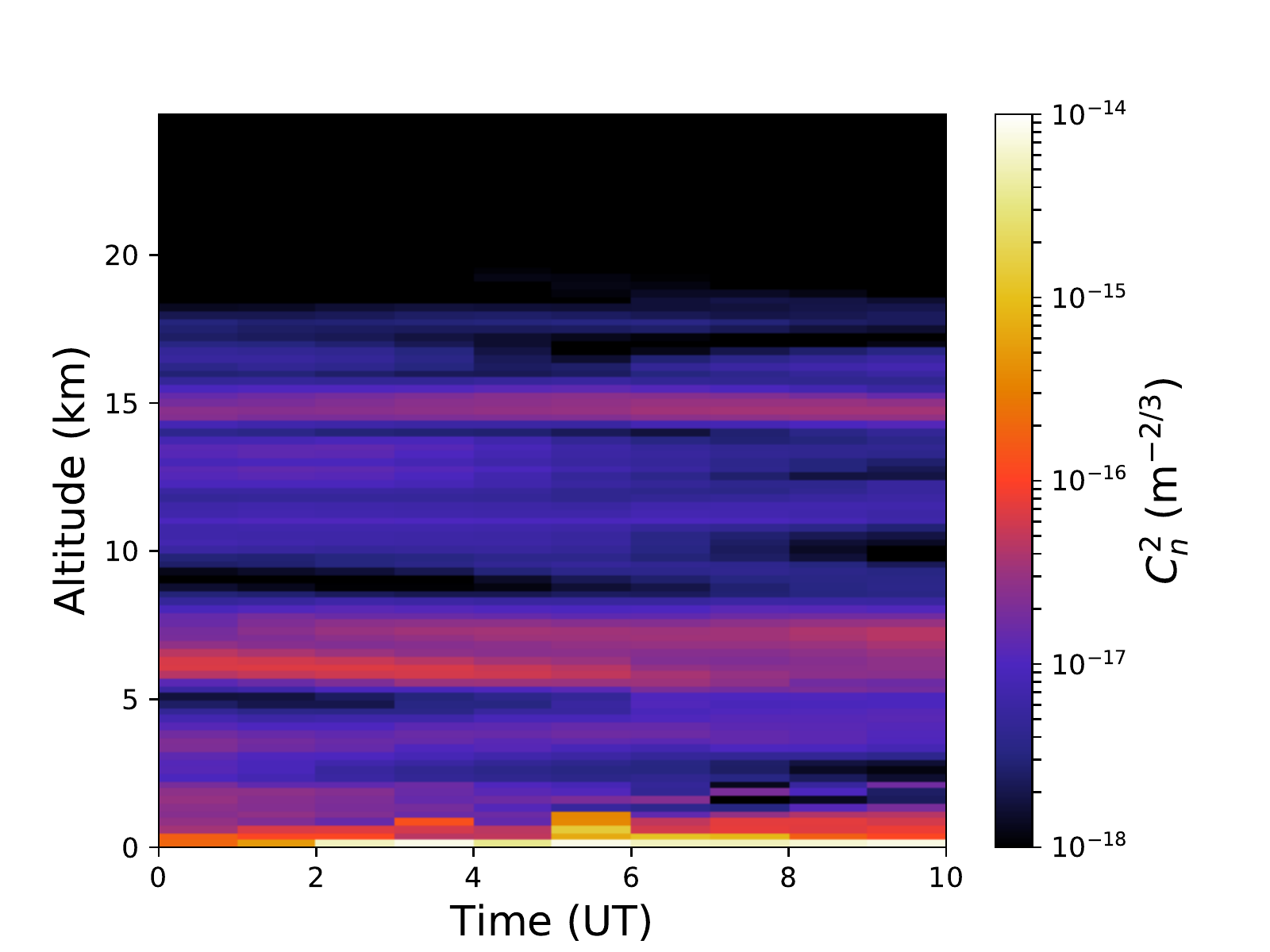} \\
	
	\includegraphics[width=0.23\textwidth,trim={2cm 0 1cm 0}]{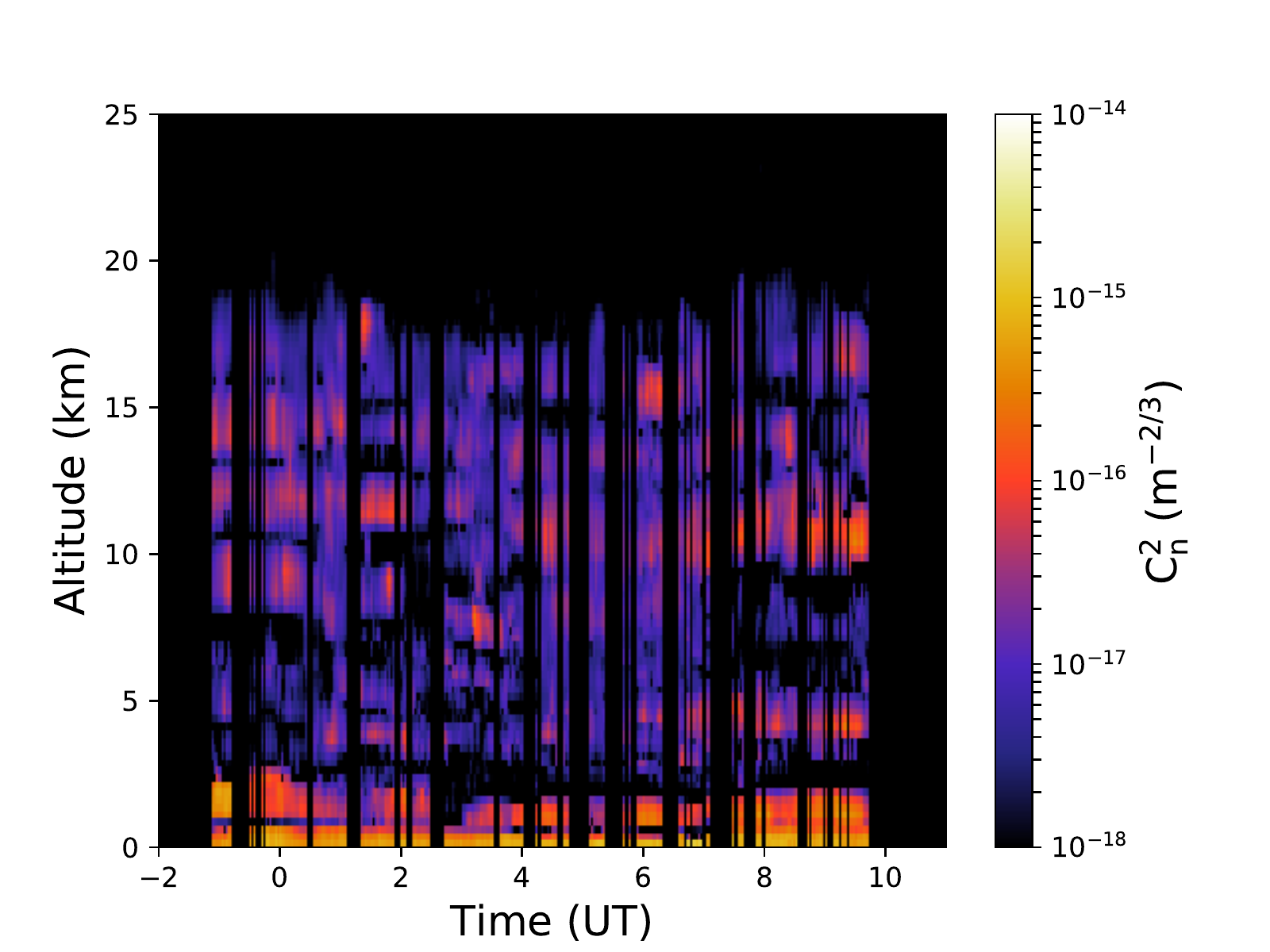} &
	\includegraphics[width=0.23\textwidth,trim={2cm 0 1cm 0}]{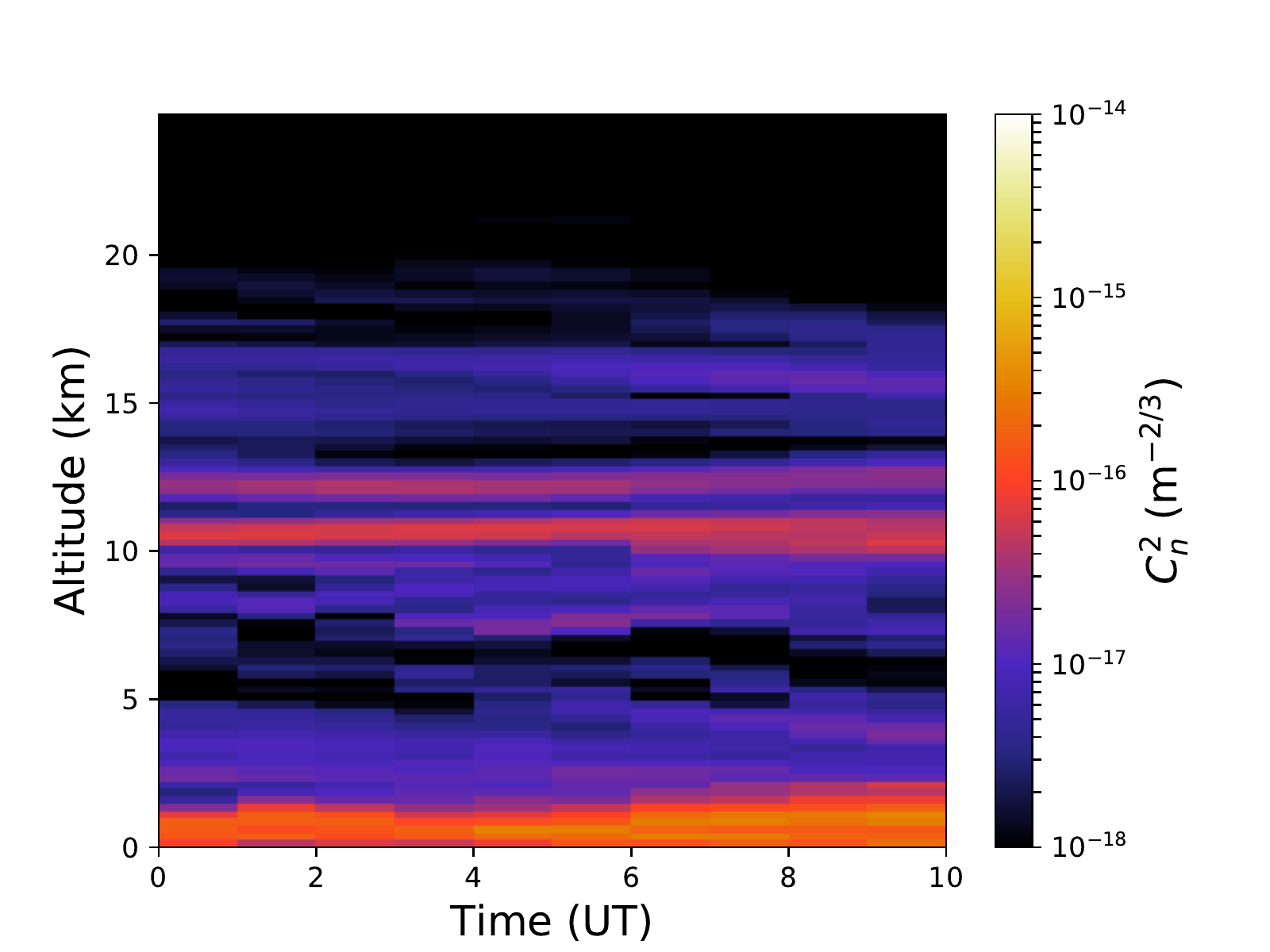} &
	\includegraphics[width=0.23\textwidth,trim={2cm 0 1cm 0}]{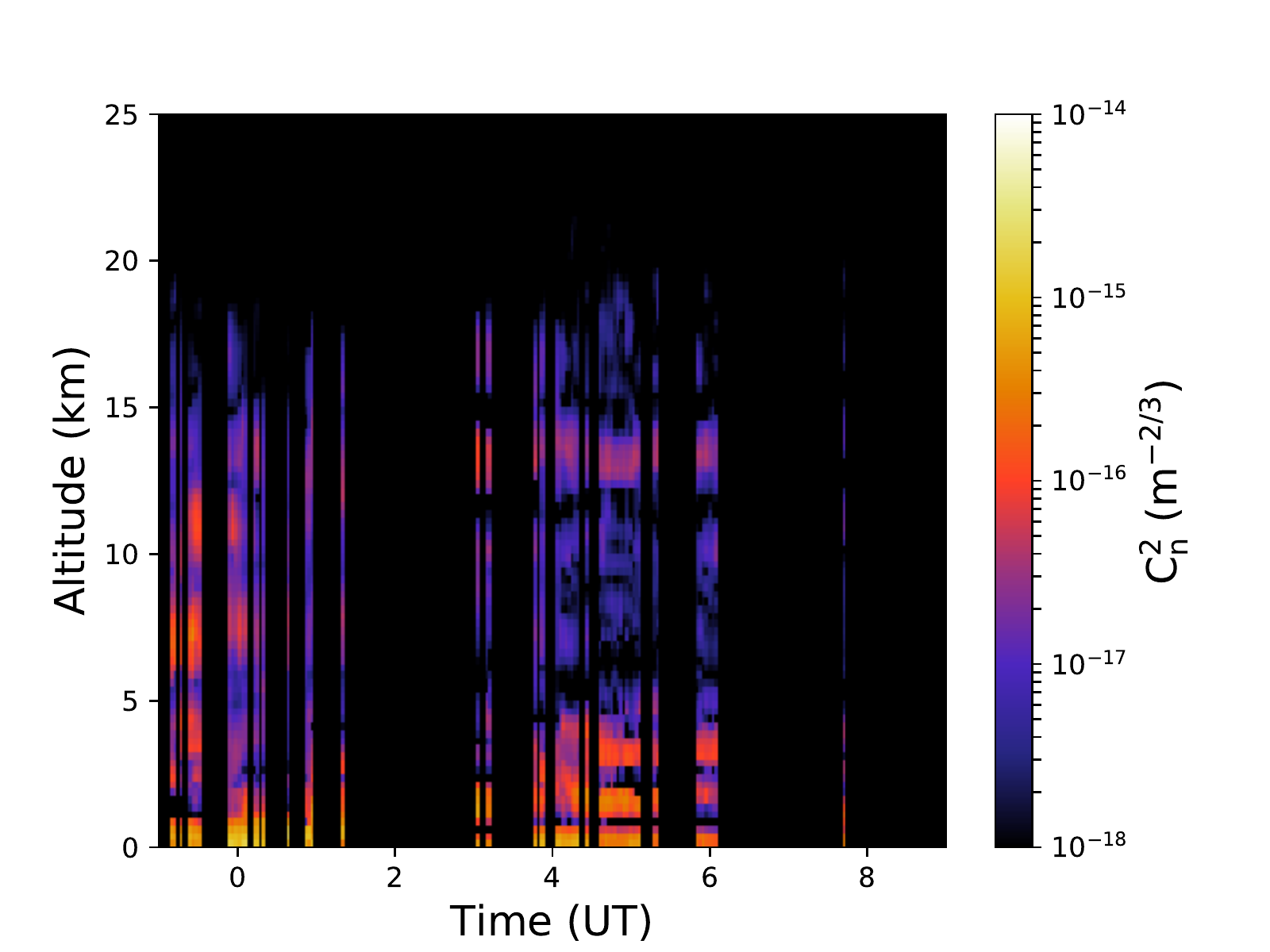} &
    	\includegraphics[width=0.23\textwidth,trim={2cm 0 1cm 0}]{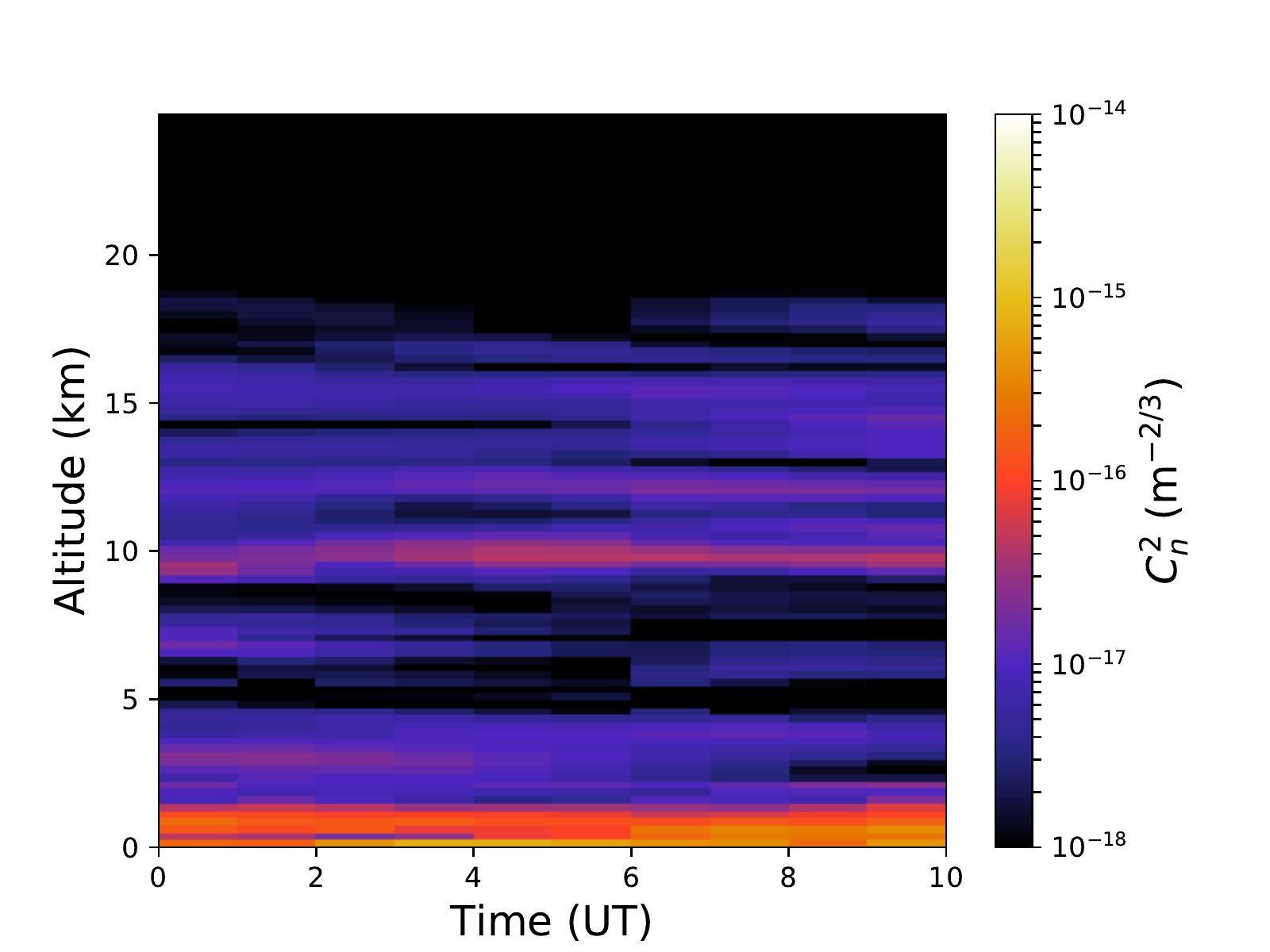} \\
	\includegraphics[width=0.23\textwidth,trim={2cm 0 1cm 0}]{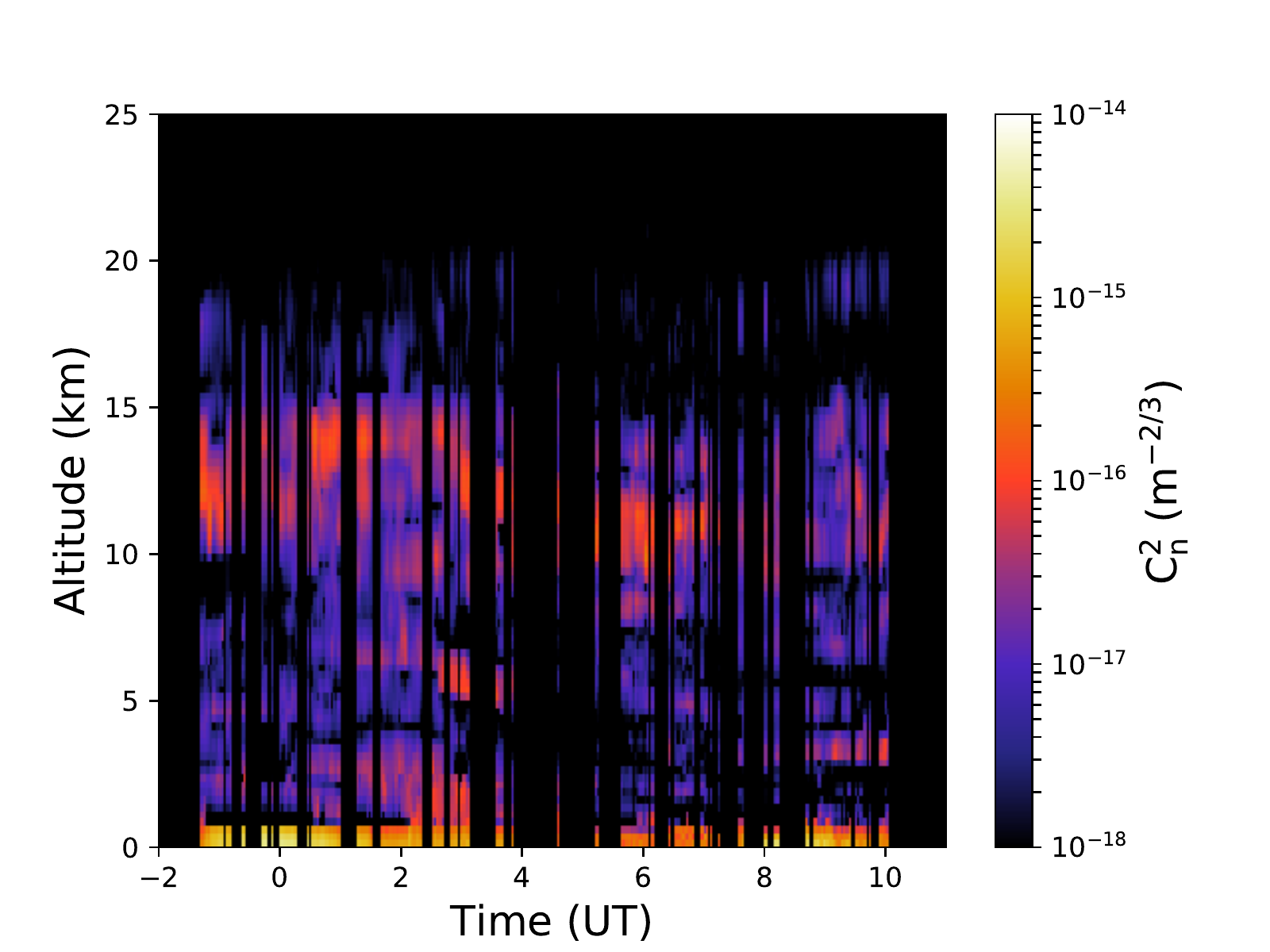} &
    	\includegraphics[width=0.23\textwidth,trim={2cm 0 1cm 0}]{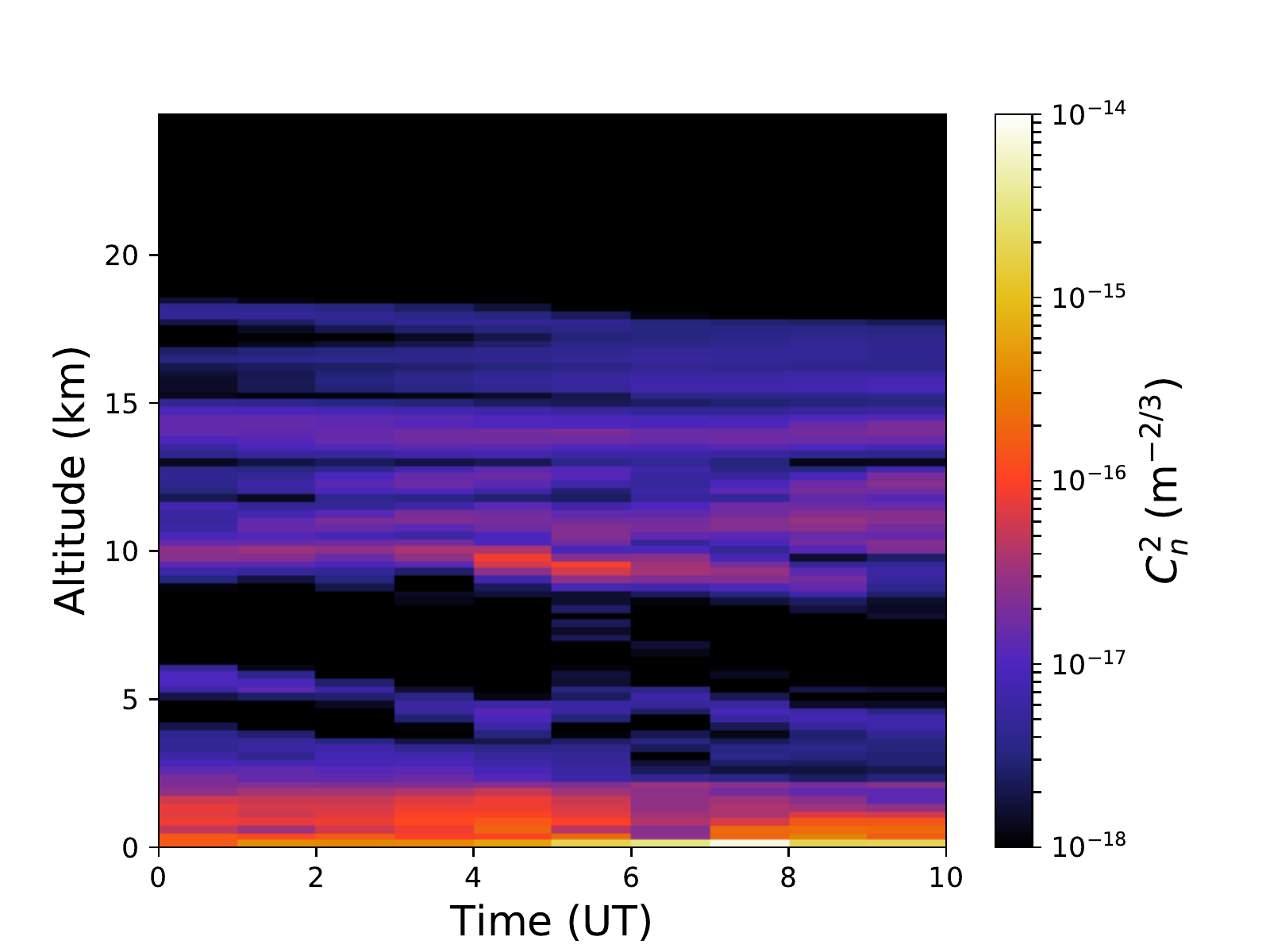} &
	\includegraphics[width=0.23\textwidth,trim={2cm 0 1cm 0}]{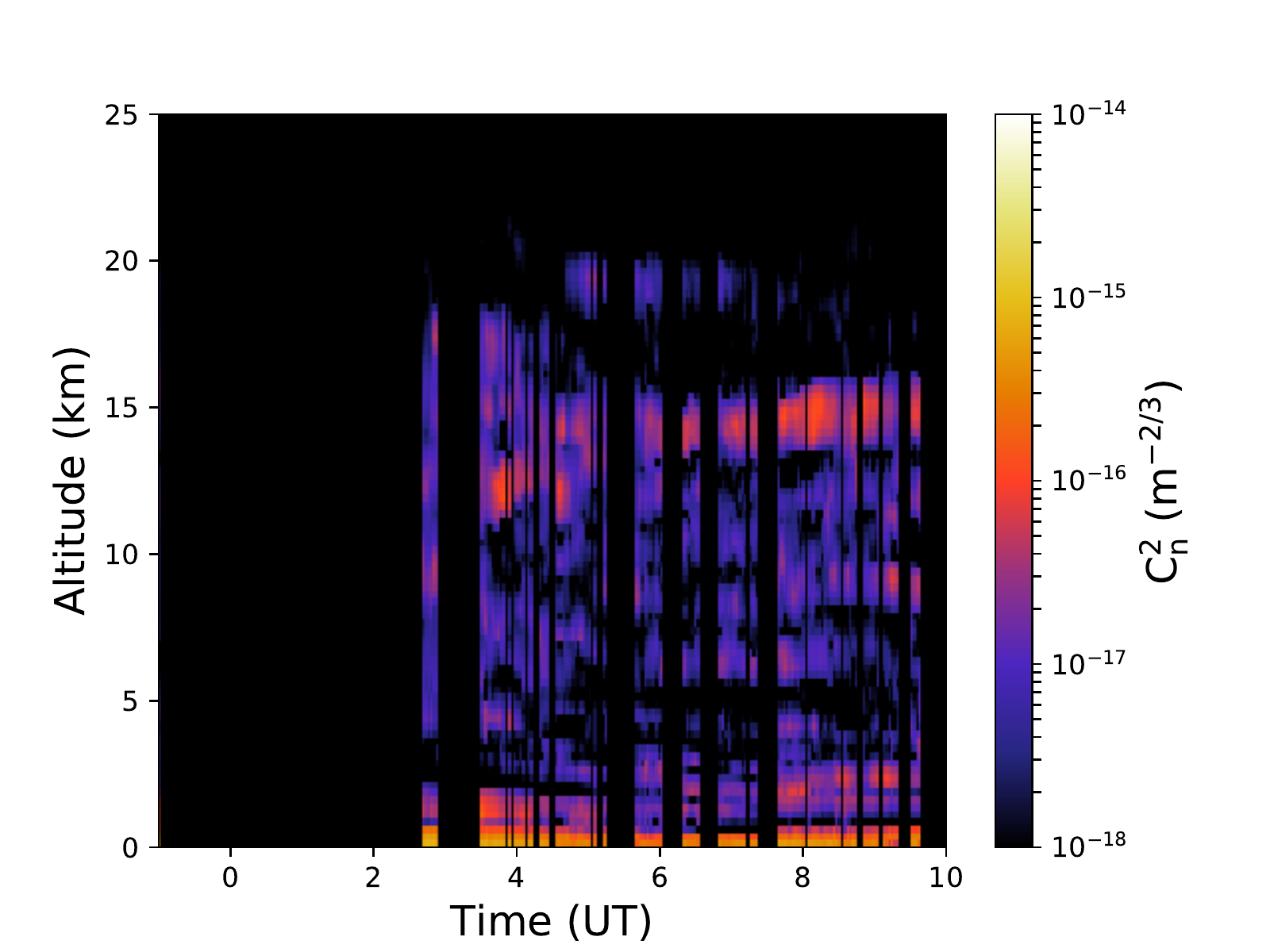} &
    	\includegraphics[width=0.23\textwidth,trim={2cm 0 1cm 0}]{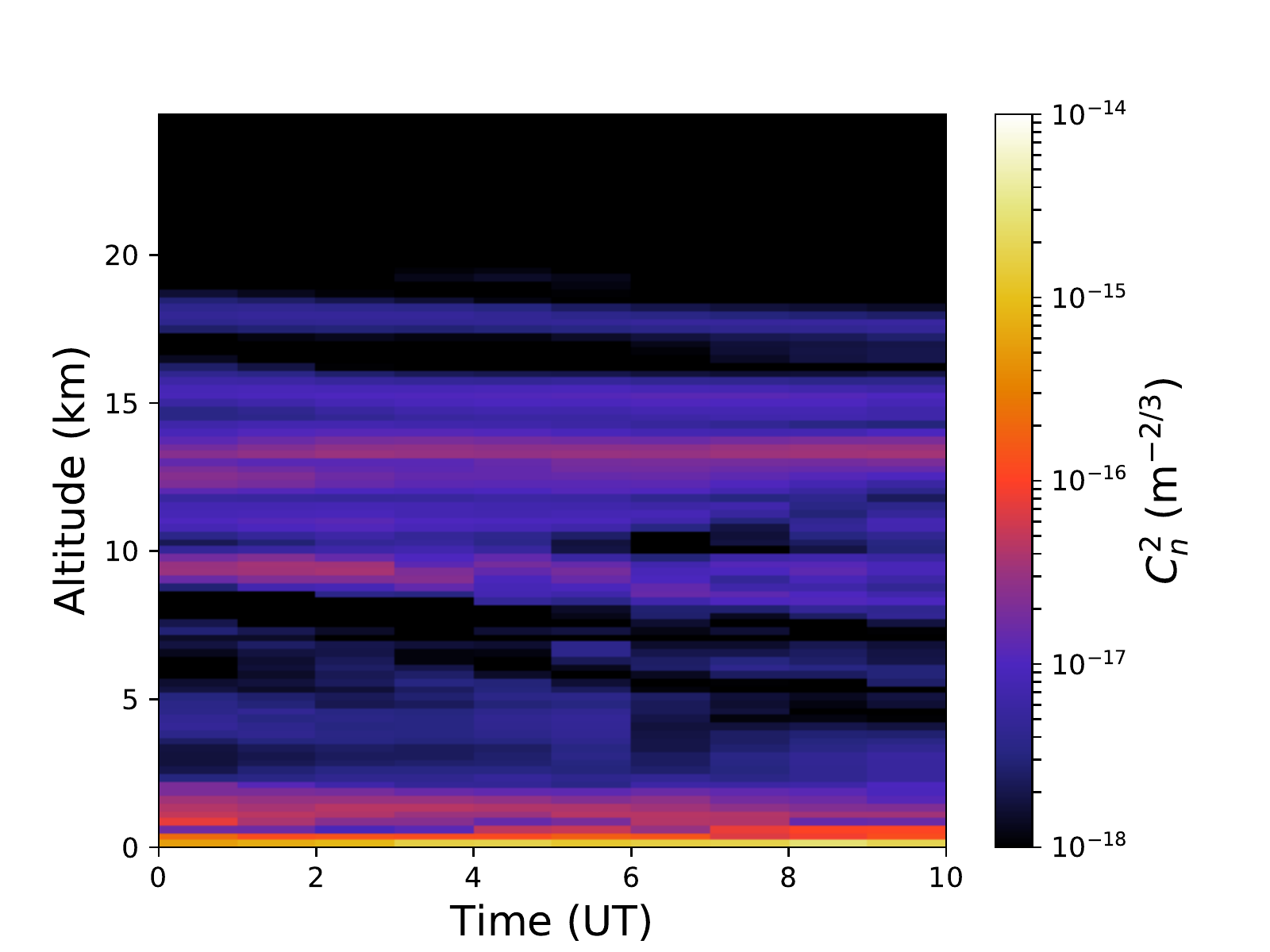}

\end{array}$
\caption{Example vertical profiles as measured by the stereo-SCIDAR (green) and estimated by the ECMWF GCM model (red). The profiles shown are the median for an individual night of observation. The coloured region shows the interquartile range. These profiles are from the nights beginning 5th - 9th May, 8th - 10th June and 3rd - 6th July 2017.}
\label{fig:seqProfiles3}
\end{figure*}

\begin{figure*}
\centering
$\begin{array}{cccc}

	\includegraphics[width=0.23\textwidth,trim={2cm 0 1cm 0}]{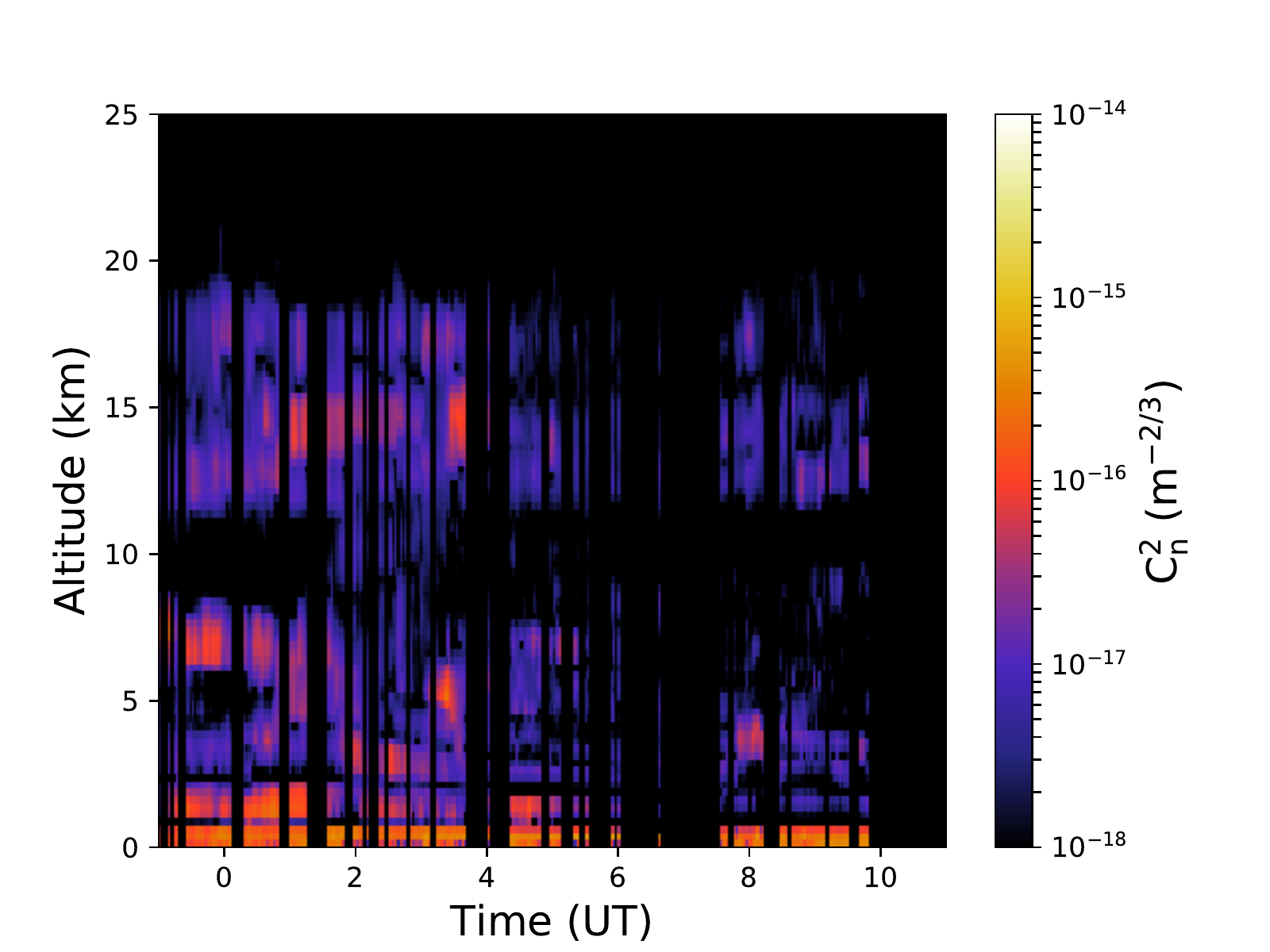} &
    	\includegraphics[width=0.23\textwidth,trim={2cm 0 1cm 0}]{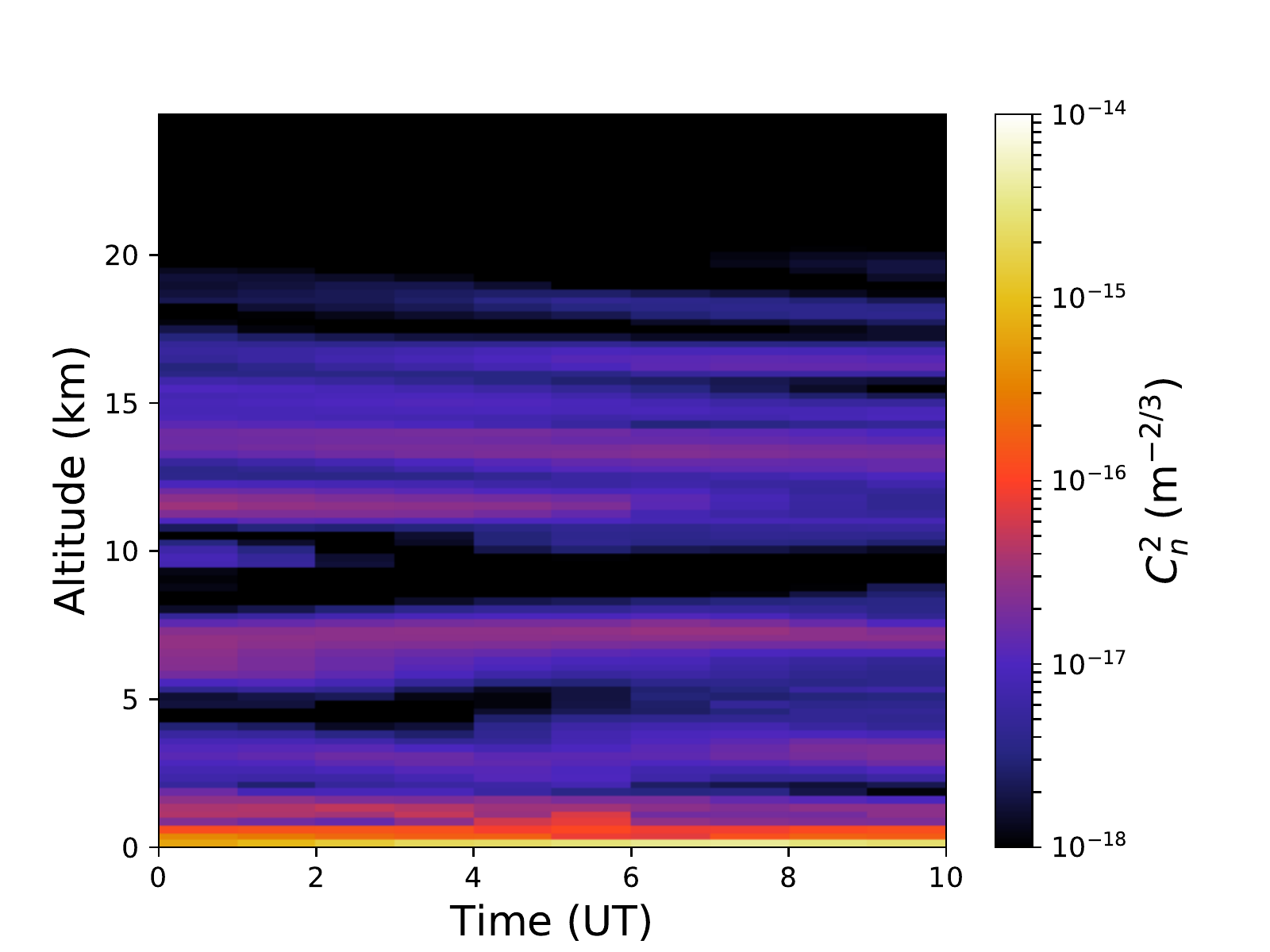} &
	
    	\includegraphics[width=0.23\textwidth,trim={2cm 0 1cm 0}]{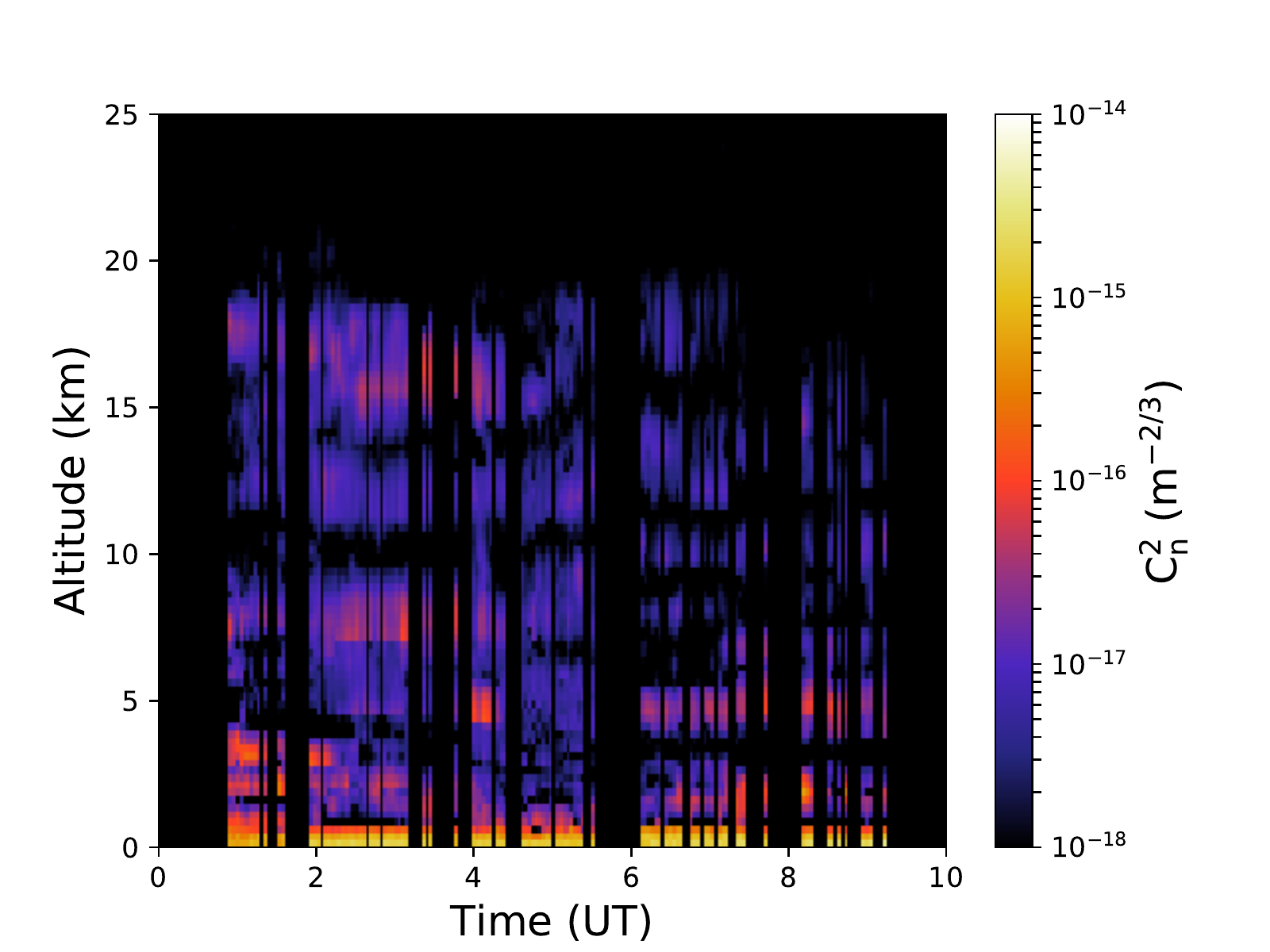} &
    	\includegraphics[width=0.23\textwidth,trim={2cm 0 1cm 0}]{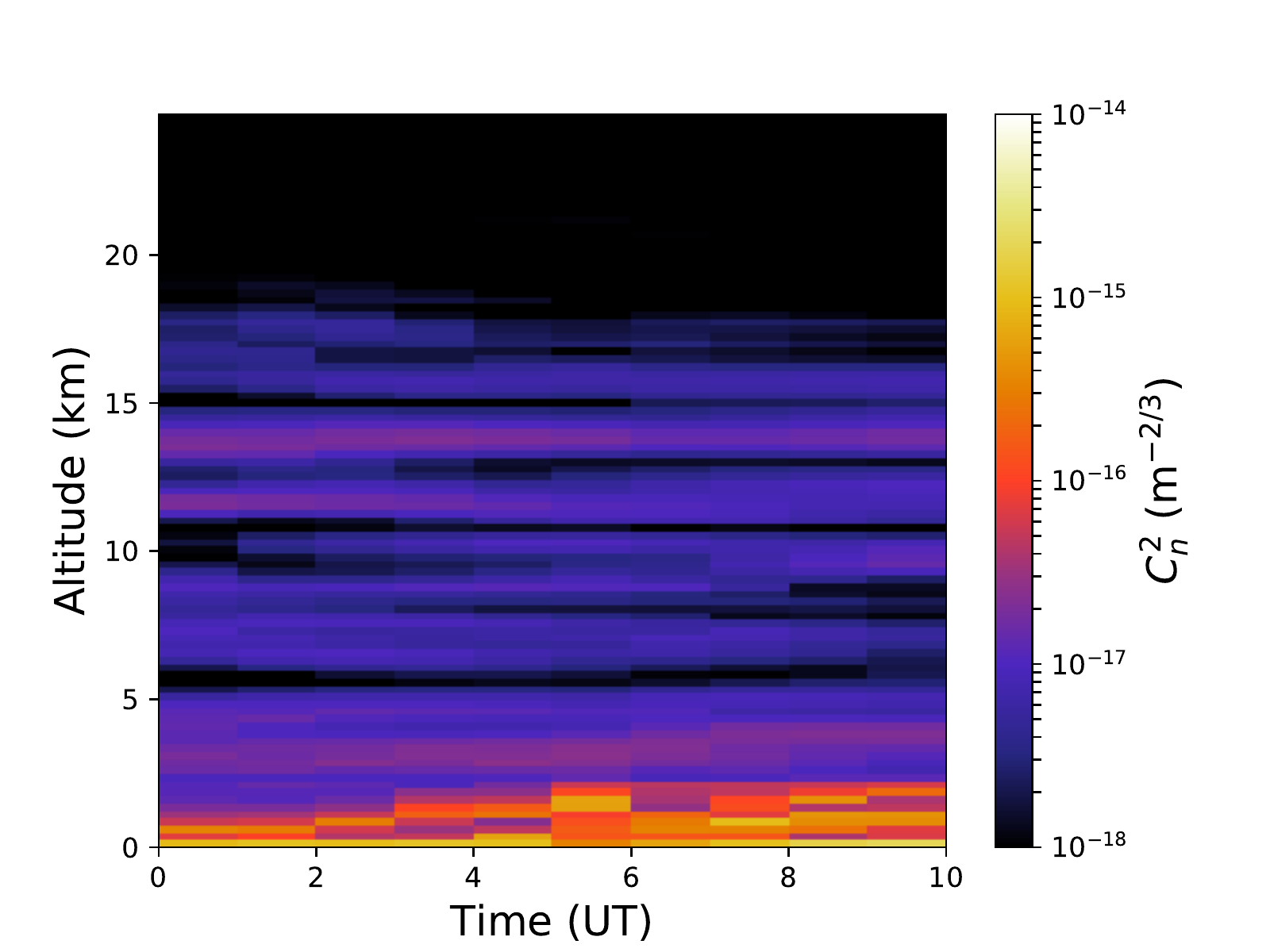} \\
	\includegraphics[width=0.23\textwidth,trim={2cm 0 1cm 0}]{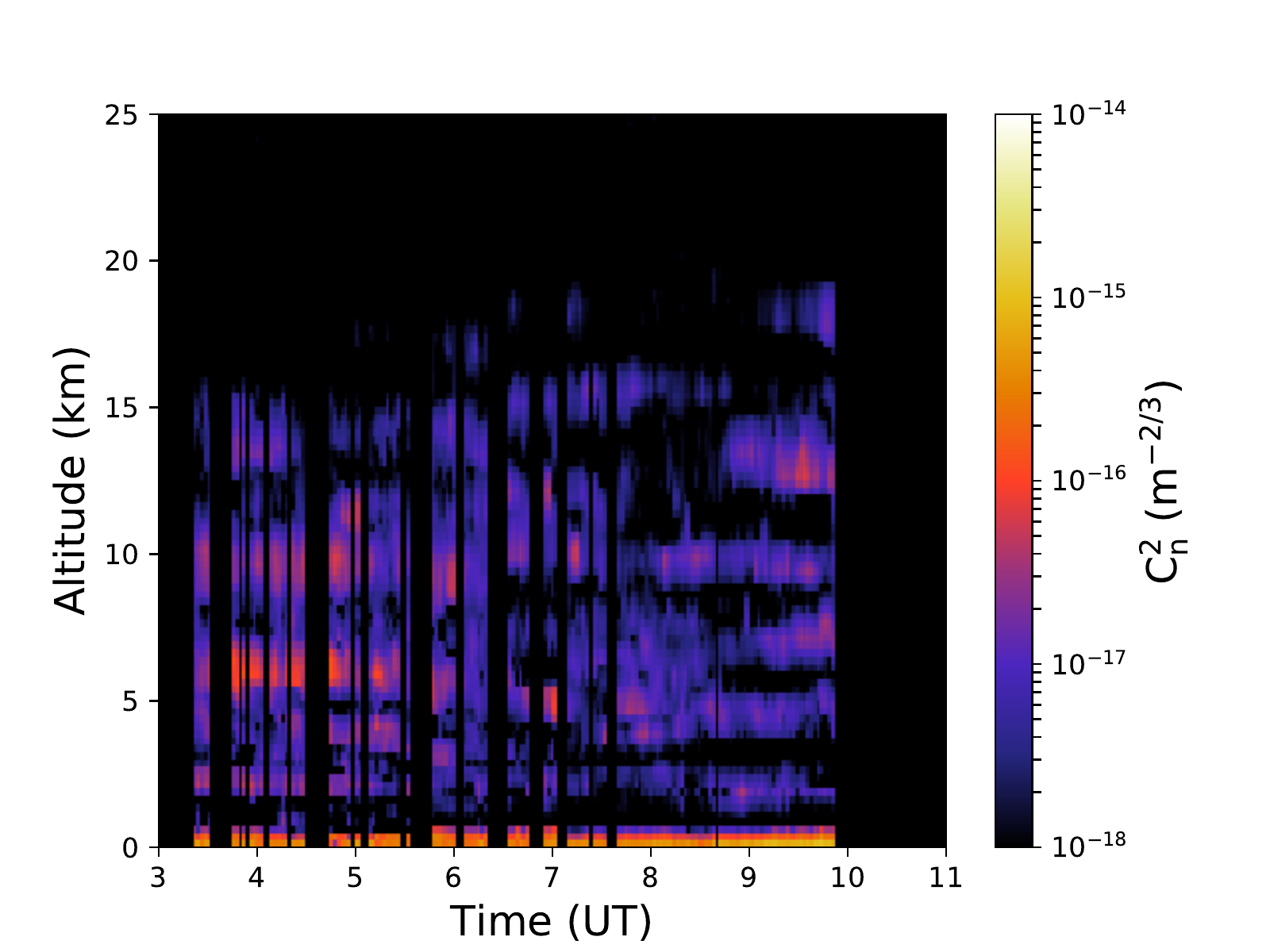} &
    	\includegraphics[width=0.23\textwidth,trim={2cm 0 1cm 0}]{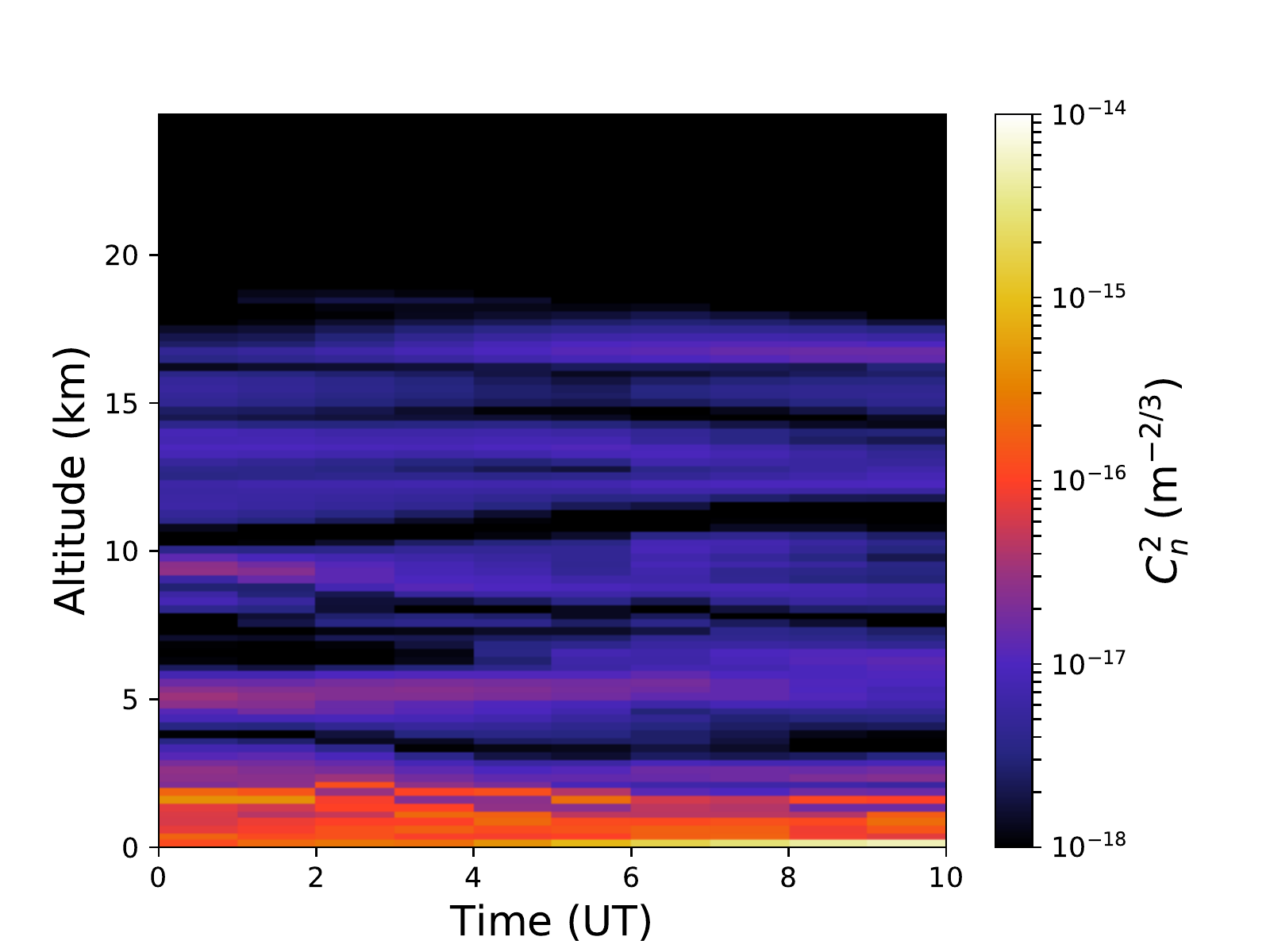} &
    	\includegraphics[width=0.23\textwidth,trim={2cm 0 1cm 0}]{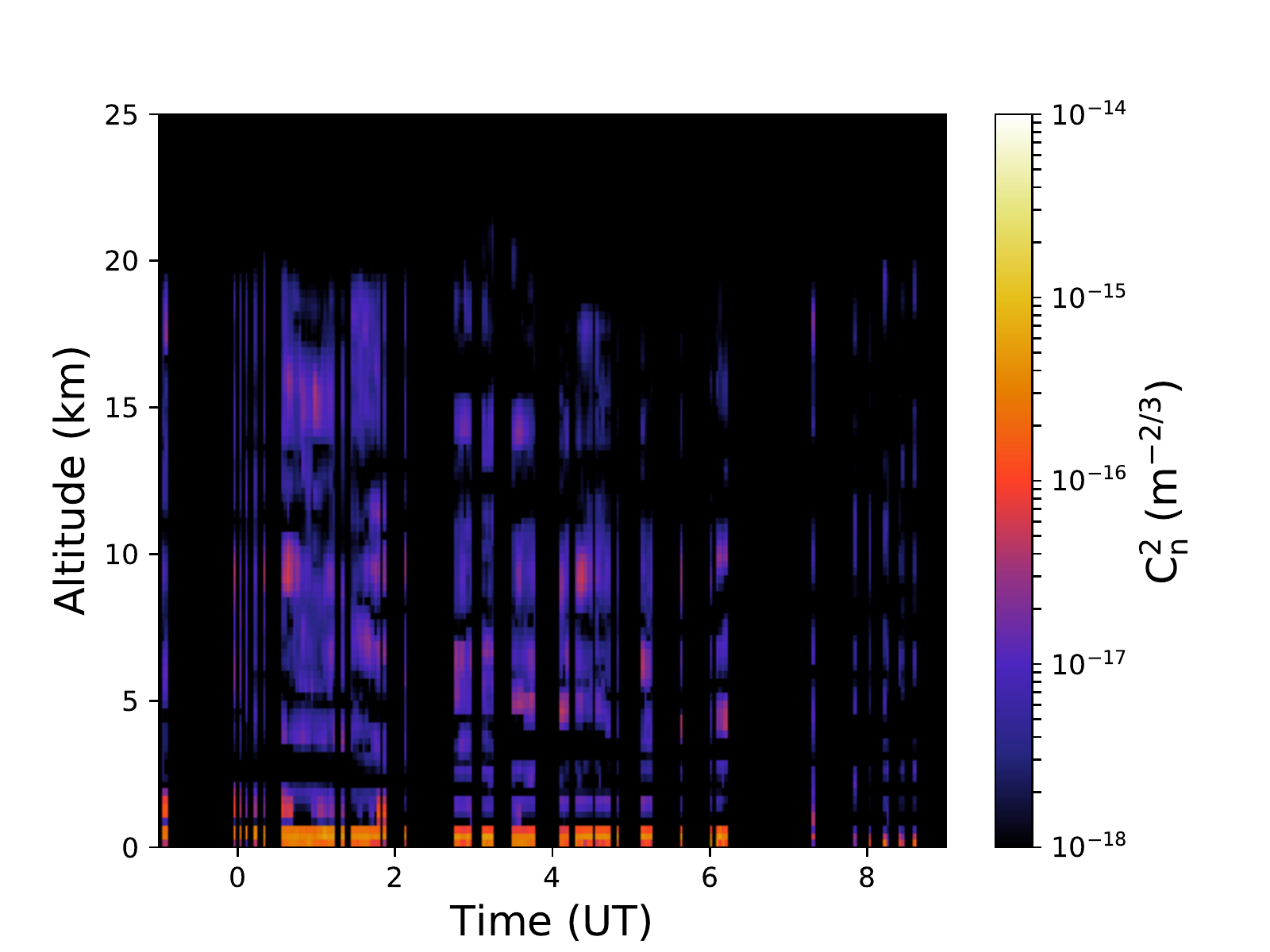} &
    	\includegraphics[width=0.23\textwidth,trim={2cm 0 1cm 0}]{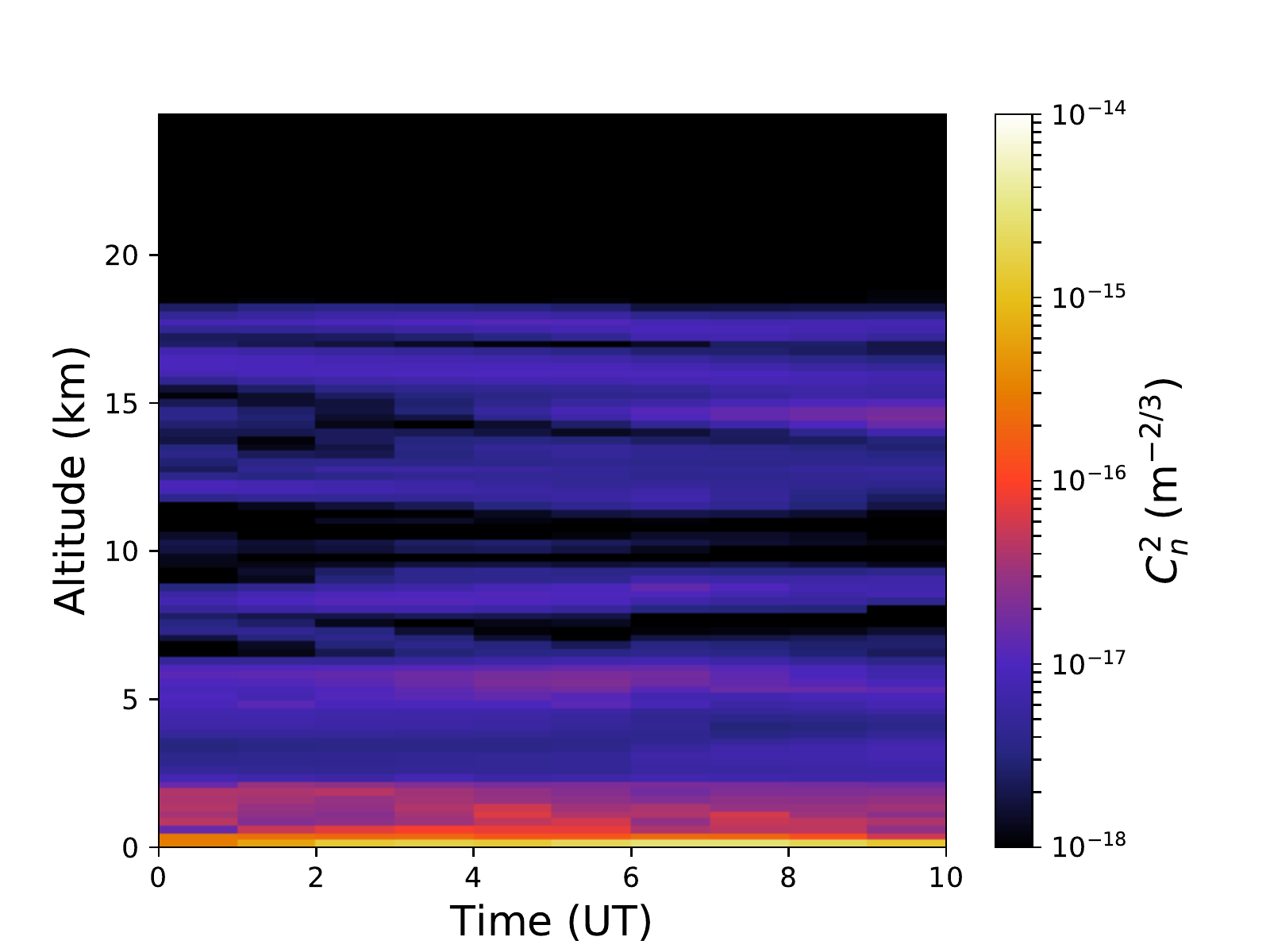} \\
	\includegraphics[width=0.23\textwidth,trim={2cm 0 1cm 0}]{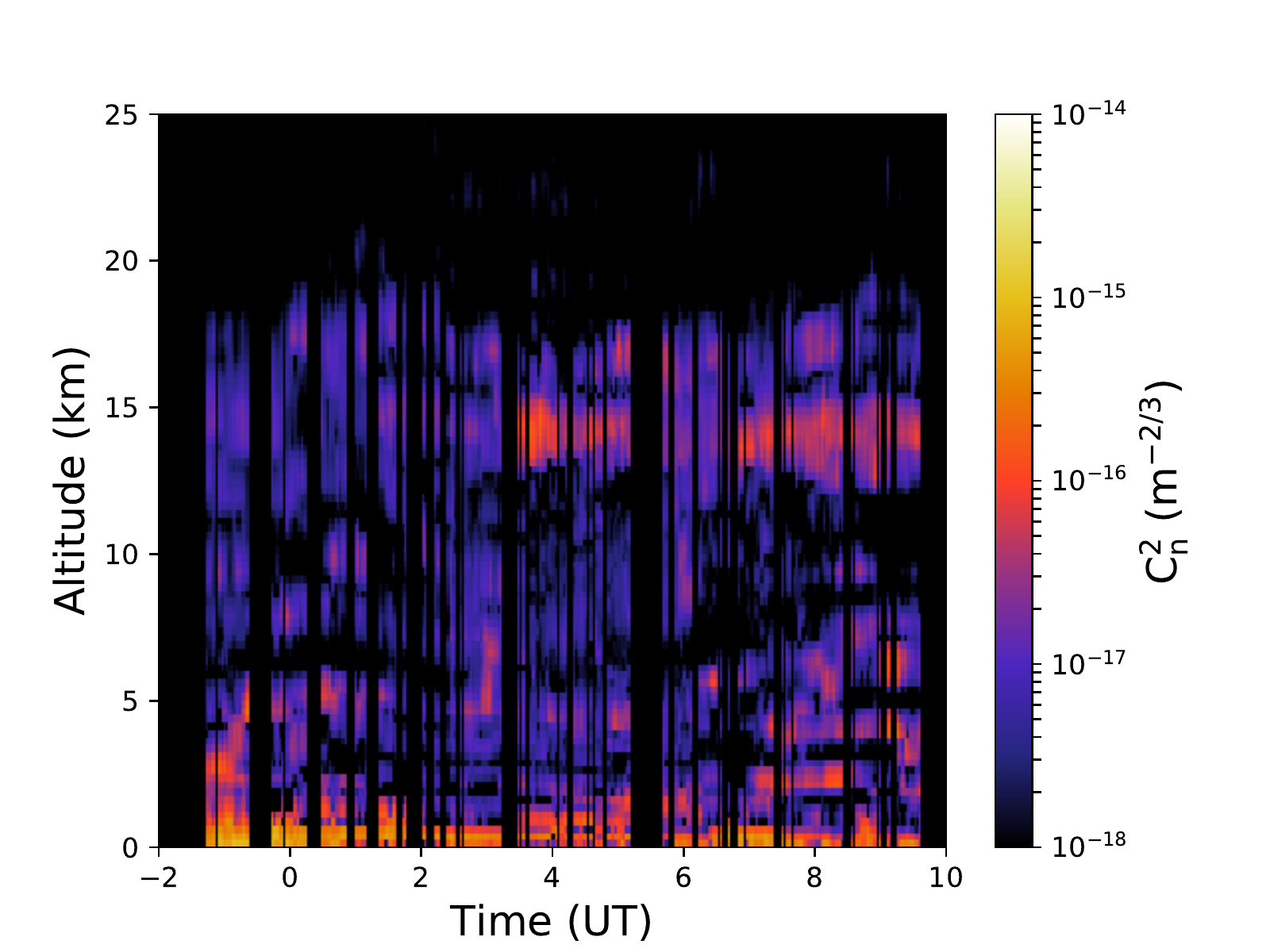} &
    	\includegraphics[width=0.23\textwidth,trim={2cm 0 1cm 0}]{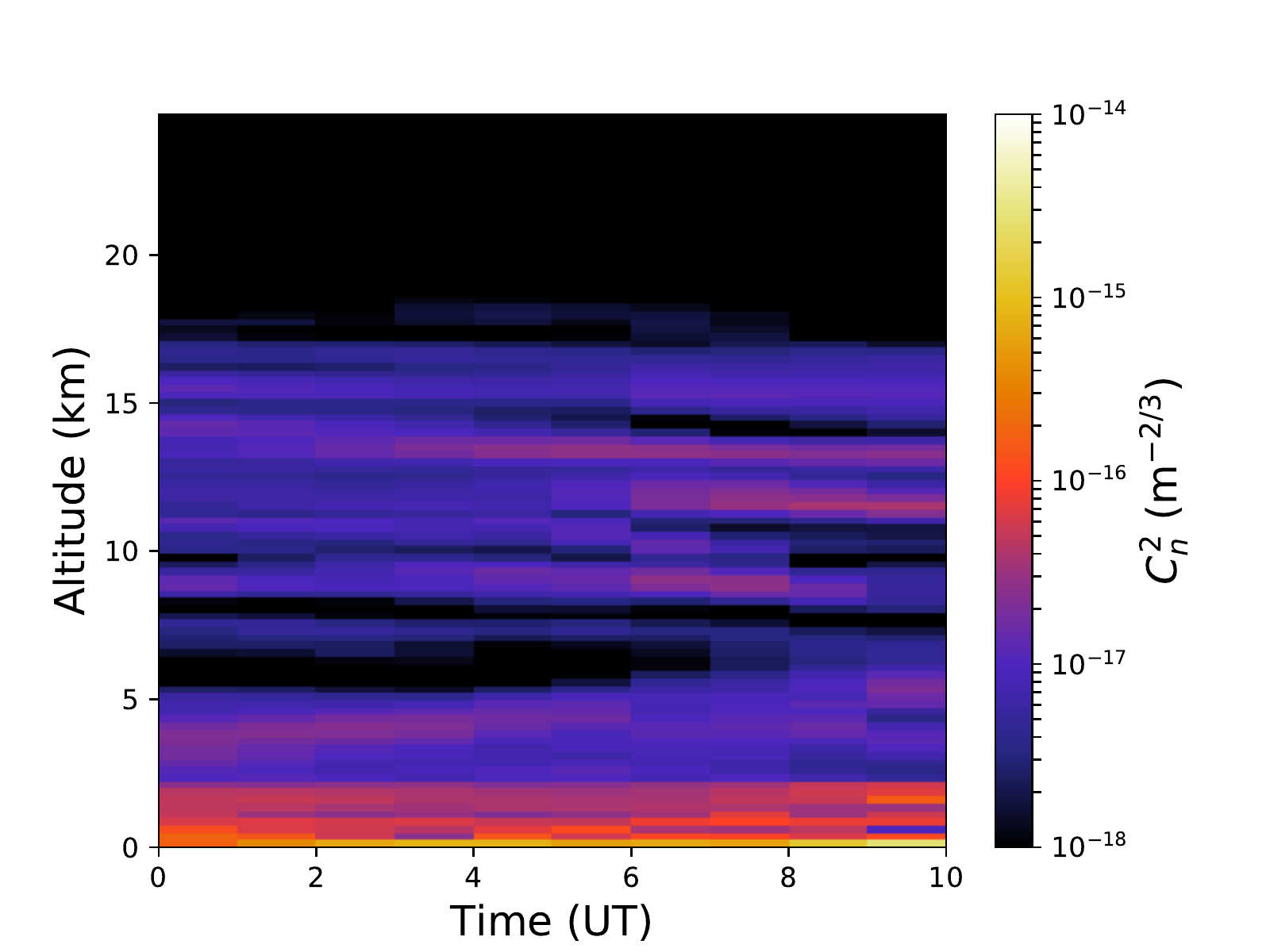} &
    	\includegraphics[width=0.23\textwidth,trim={2cm 0 1cm 0}]{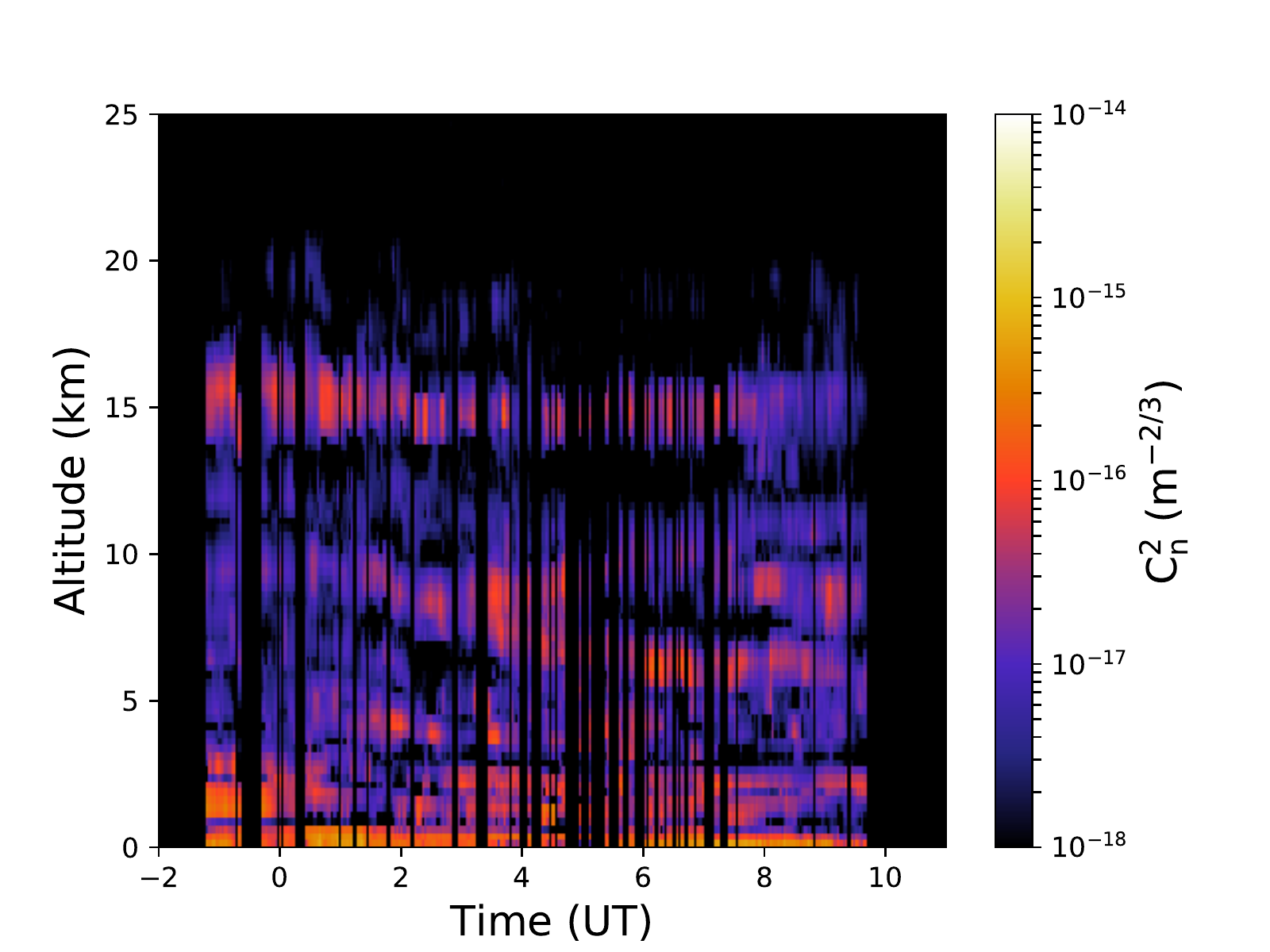} &
    	\includegraphics[width=0.23\textwidth,trim={2cm 0 1cm 0}]{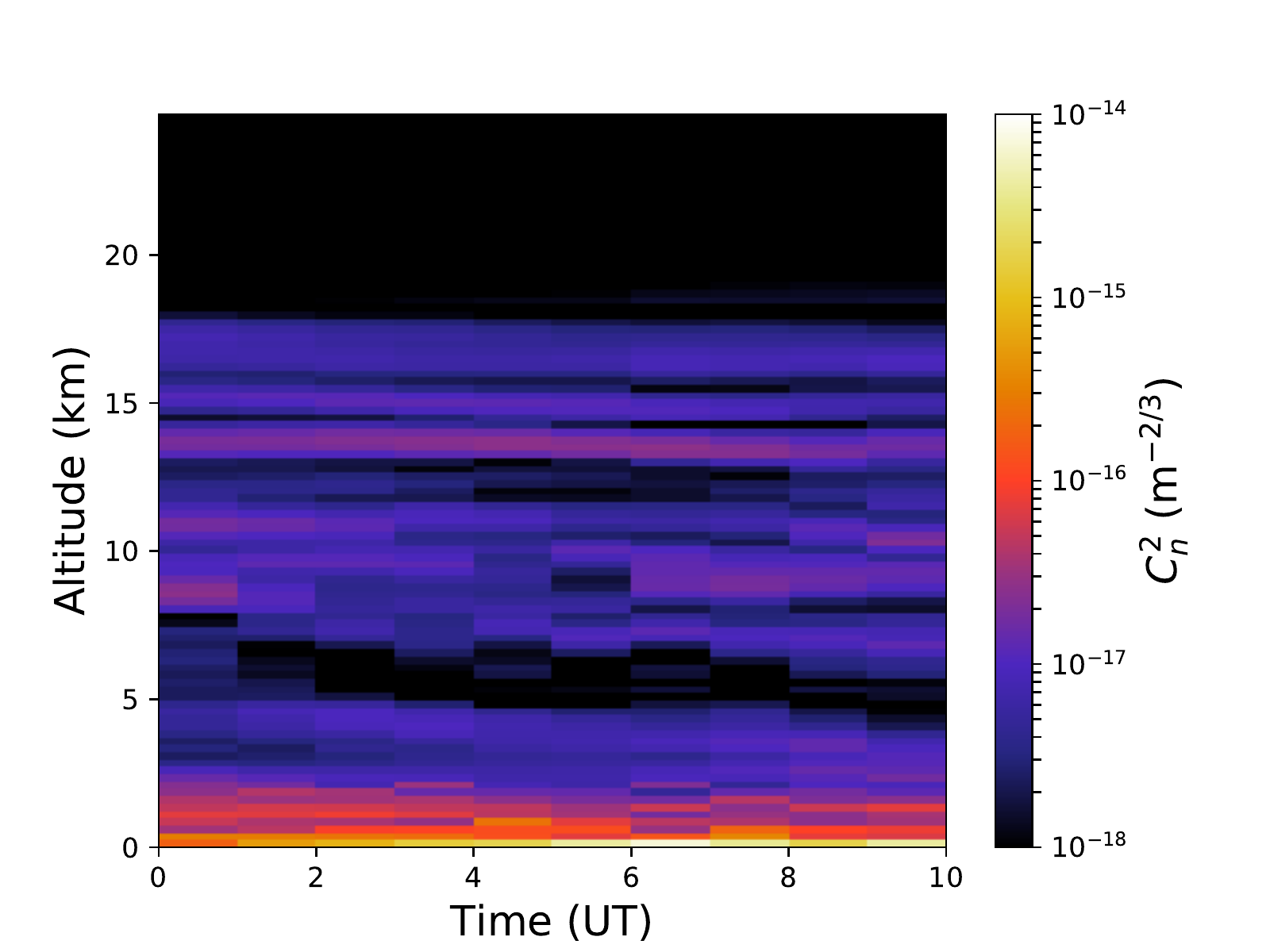} \\
	\includegraphics[width=0.23\textwidth,trim={2cm 0 1cm 0}]{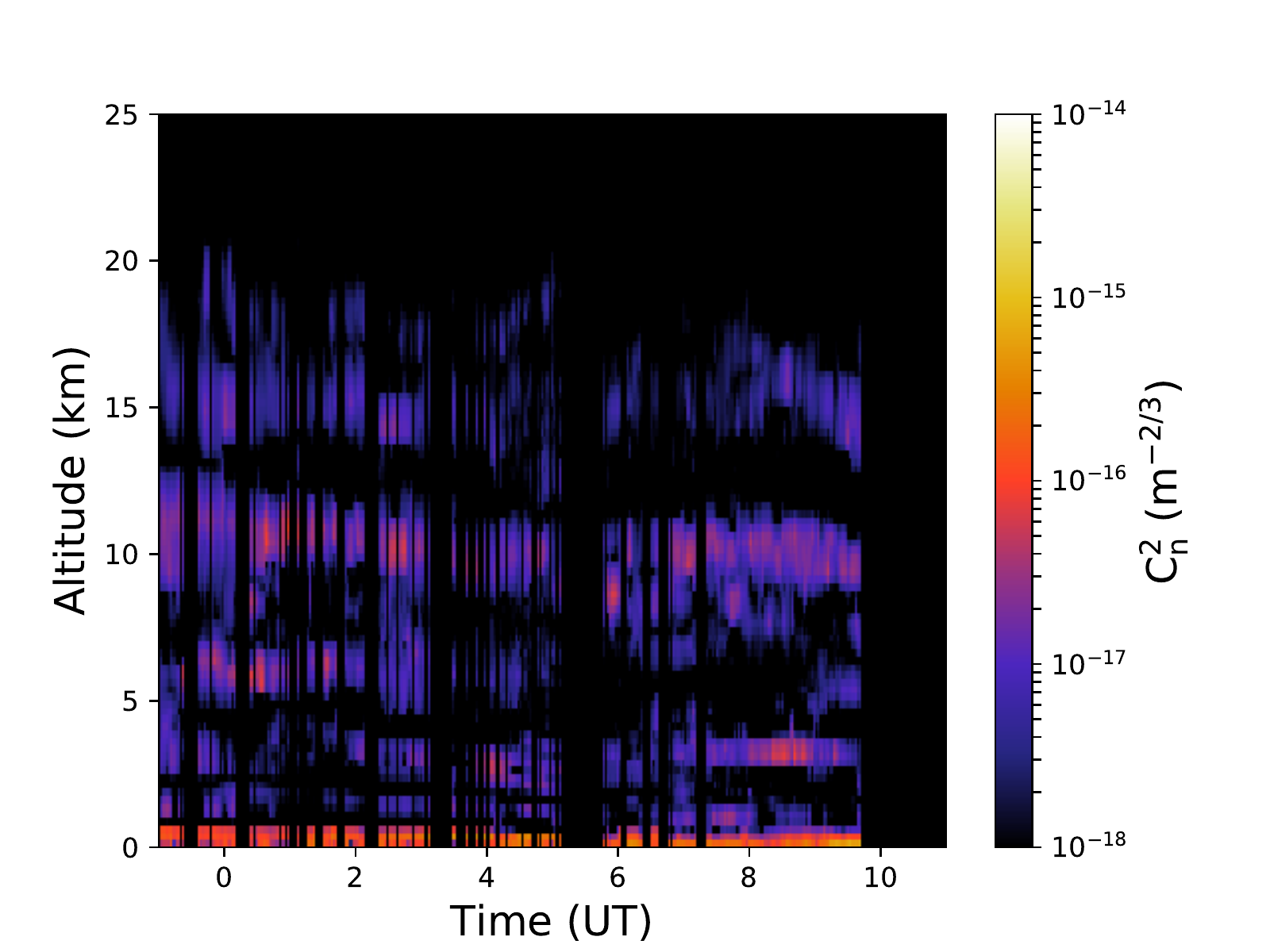} &
    	\includegraphics[width=0.23\textwidth,trim={2cm 0 1cm 0}]{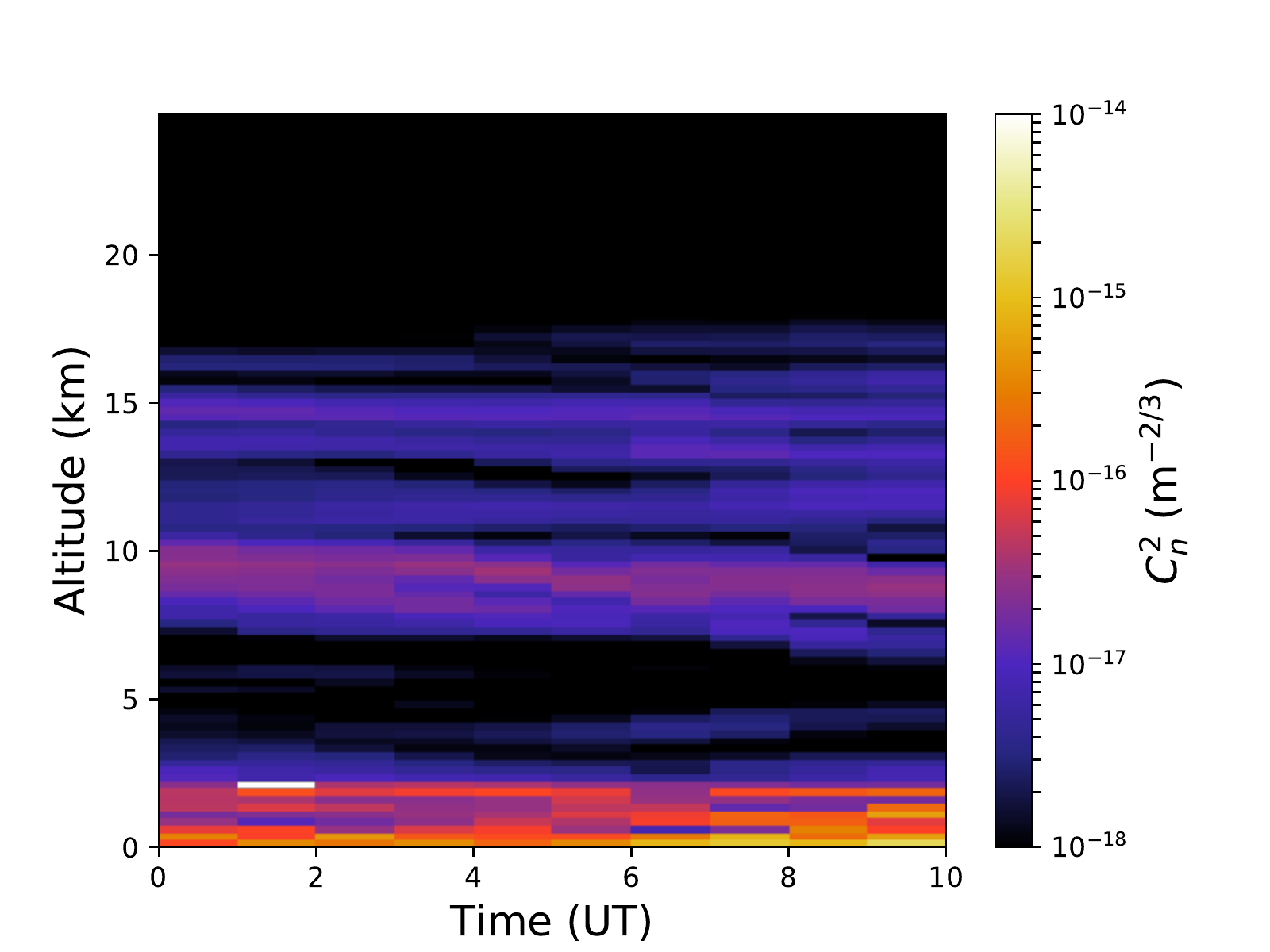} &
	
	\includegraphics[width=0.23\textwidth,trim={2cm 0 1cm 0}]{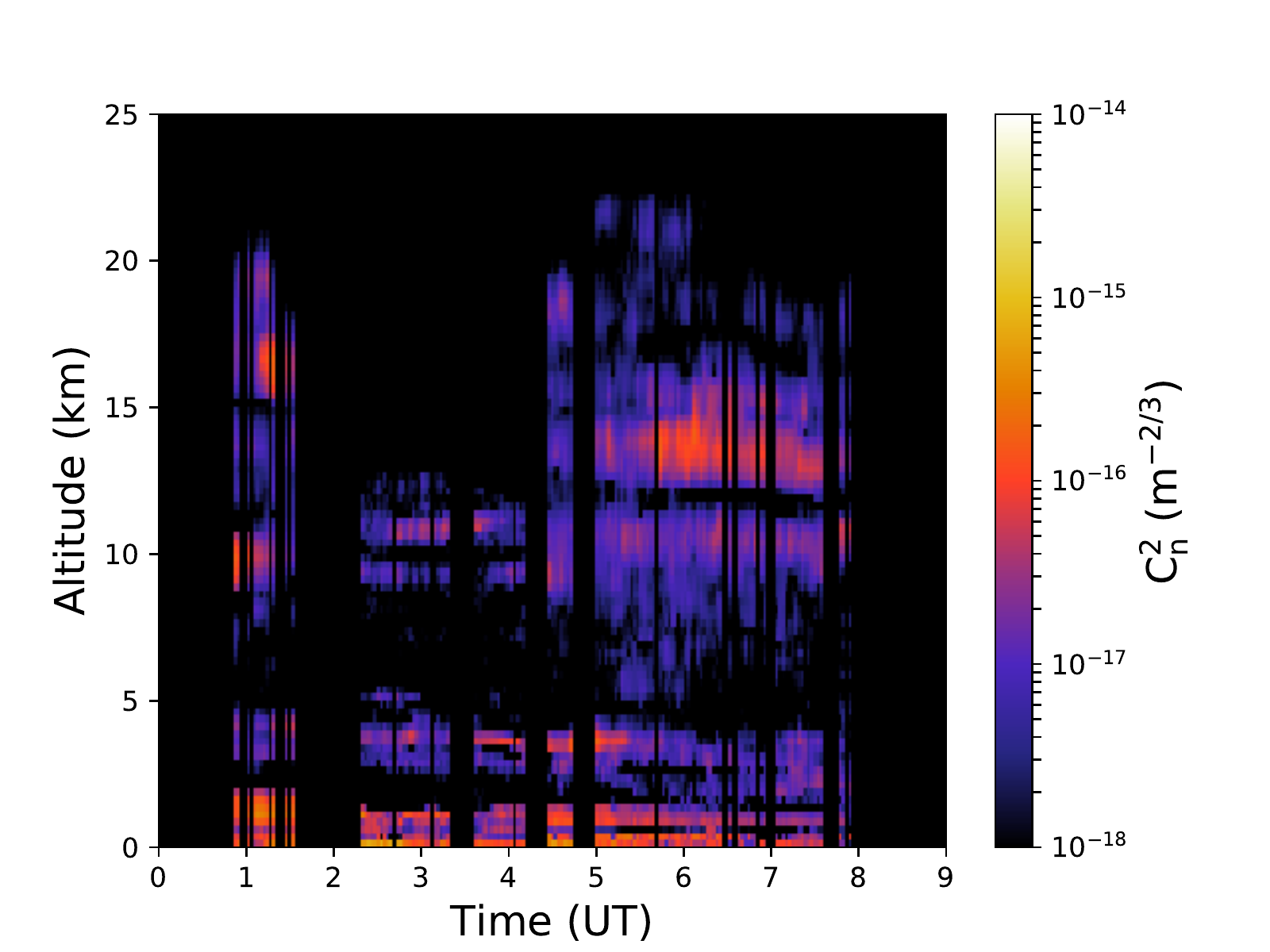} &
    	\includegraphics[width=0.23\textwidth,trim={2cm 0 1cm 0}]{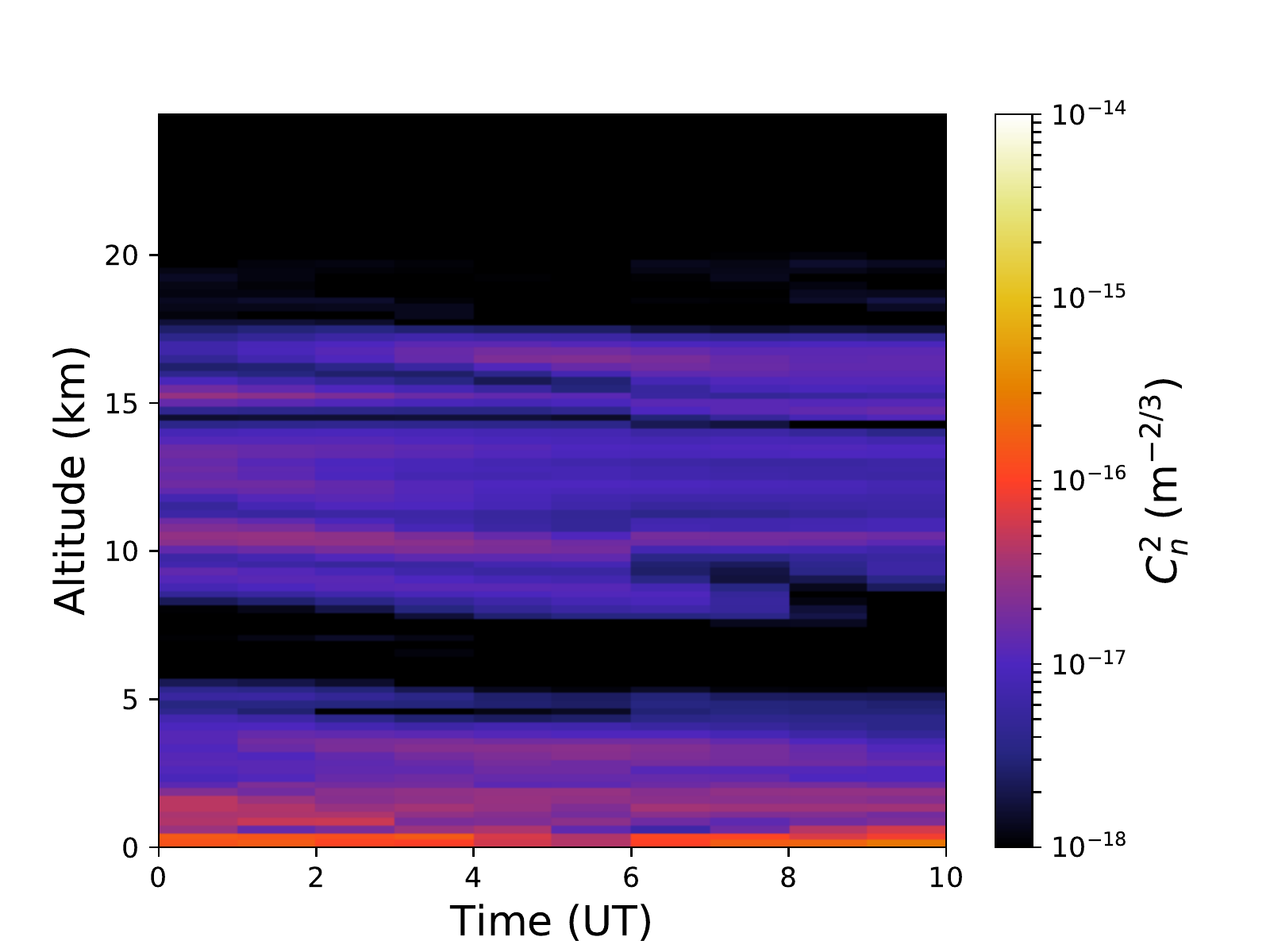} \\
	\includegraphics[width=0.23\textwidth,trim={2cm 0 1cm 0}]{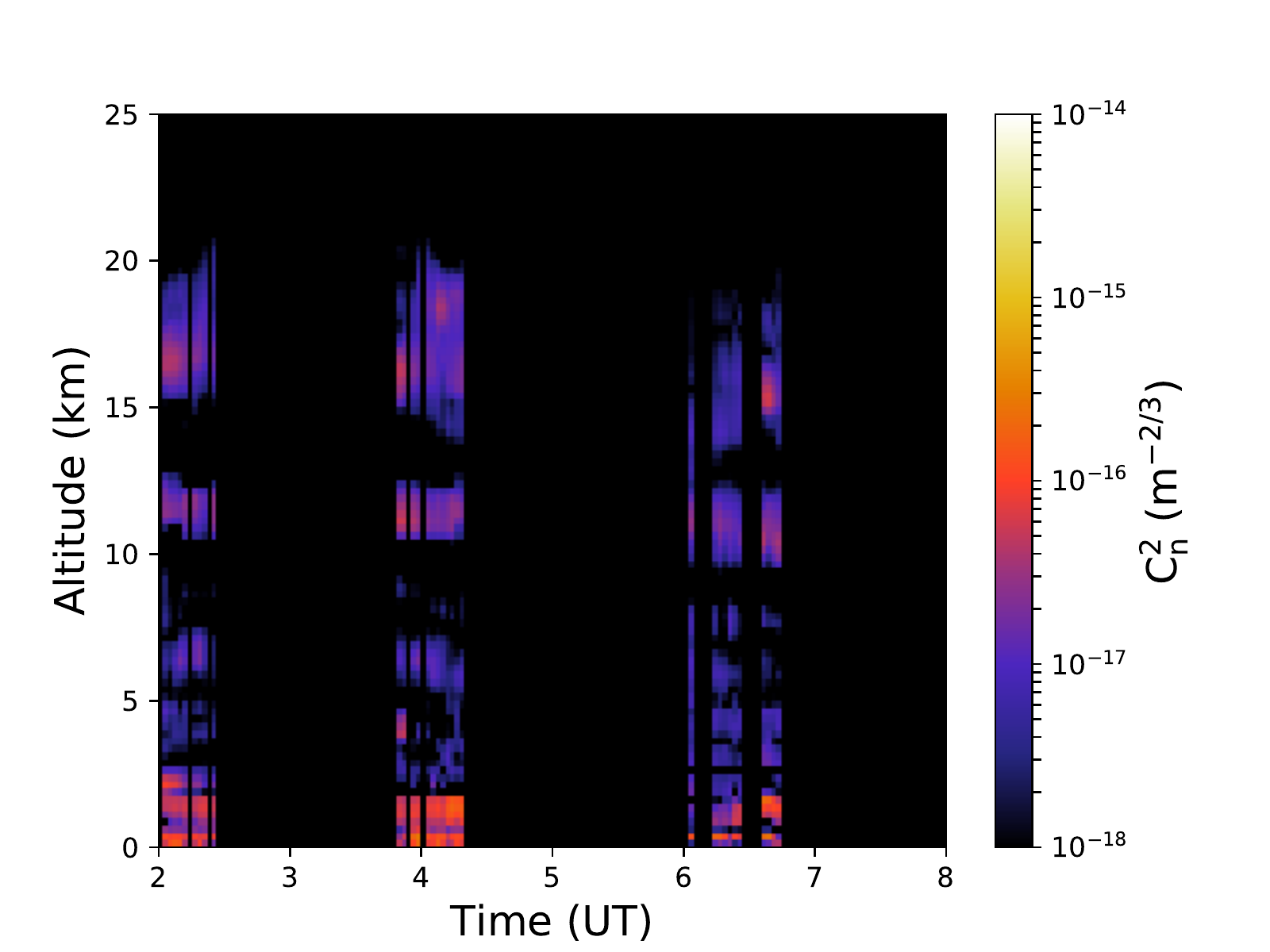} &
    	\includegraphics[width=0.23\textwidth,trim={2cm 0 1cm 0}]{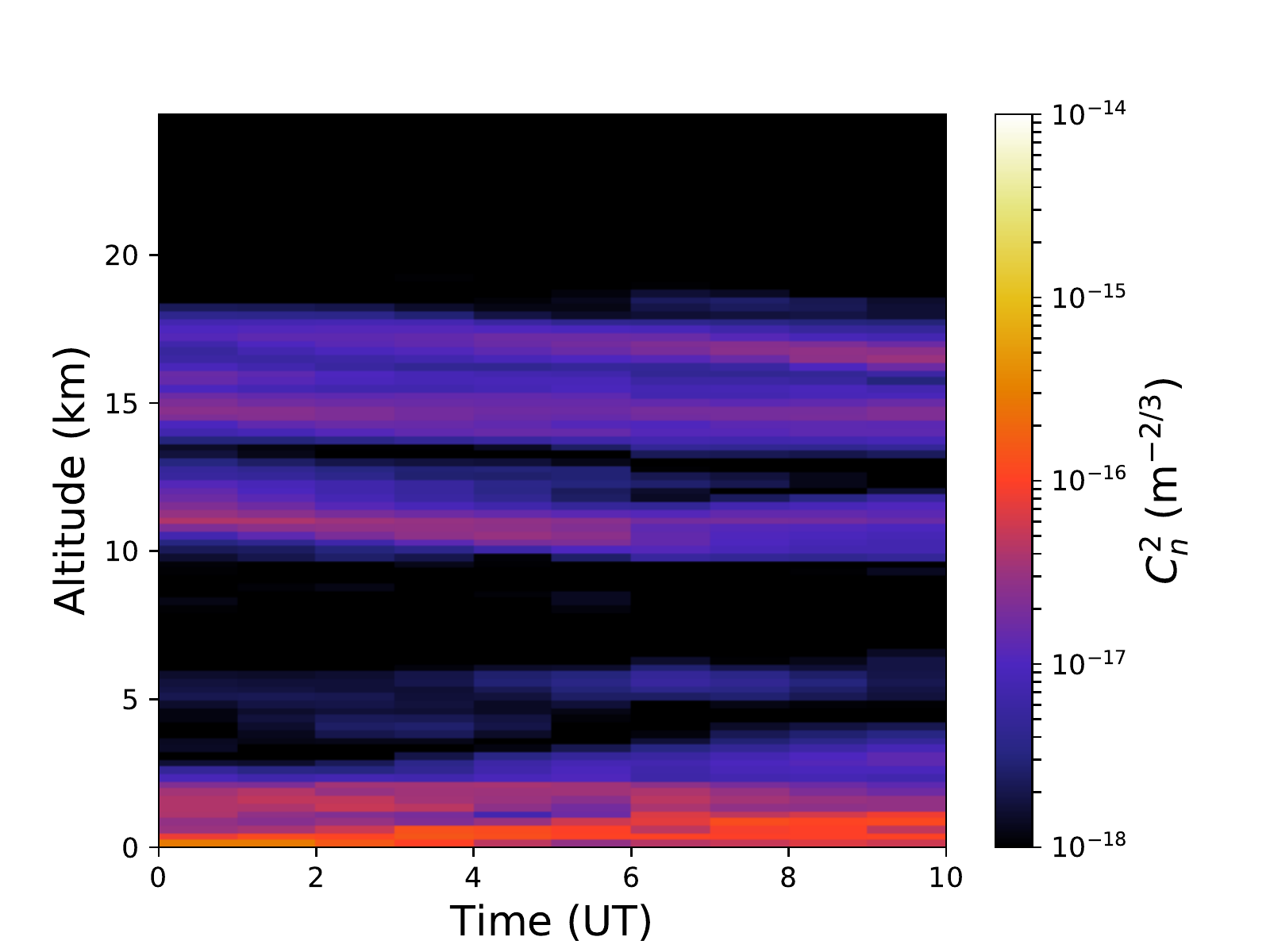} &	
	\includegraphics[width=0.23\textwidth,trim={2cm 0 1cm 0}]{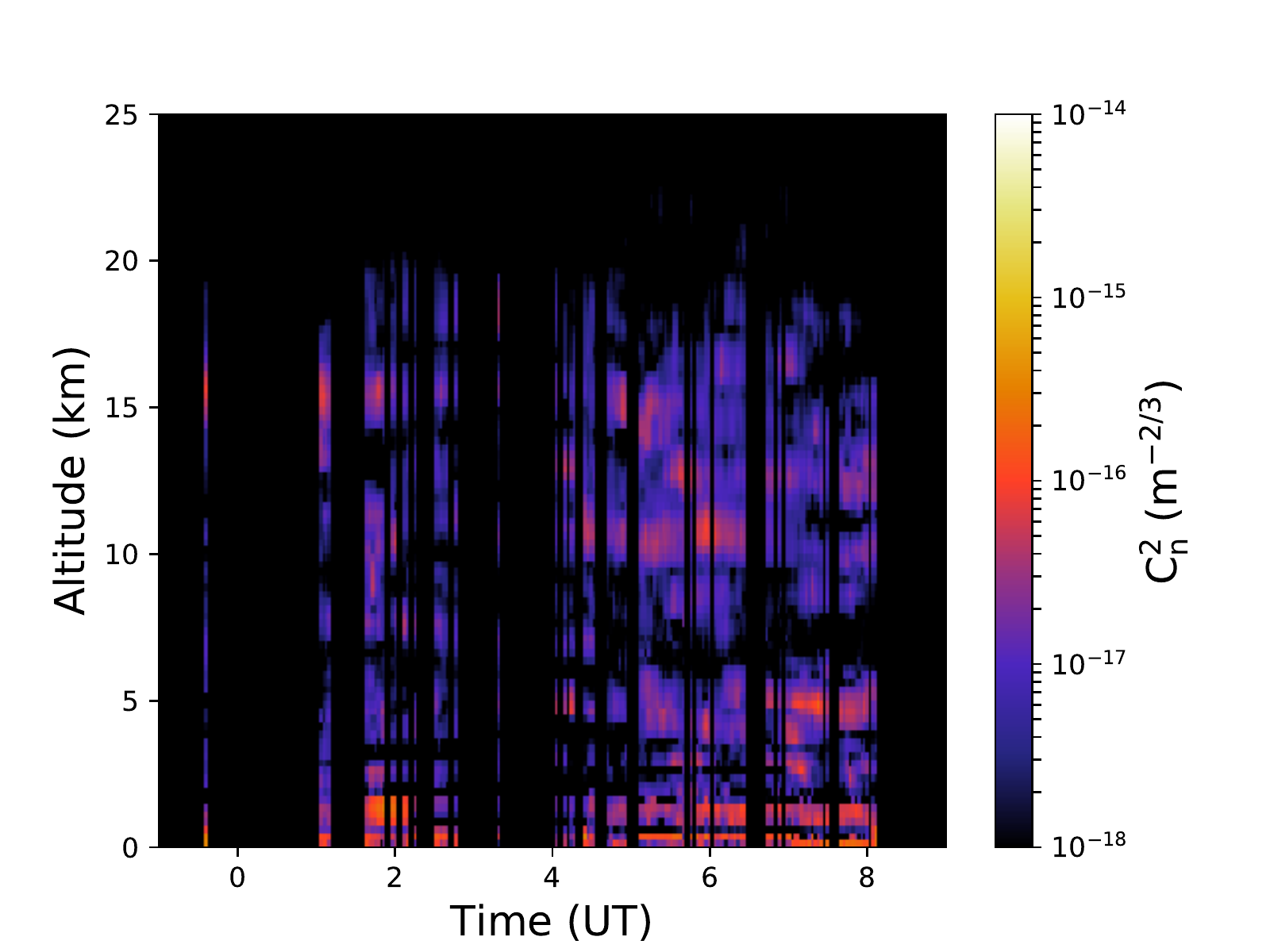} &
    	\includegraphics[width=0.23\textwidth,trim={2cm 0 1cm 0}]{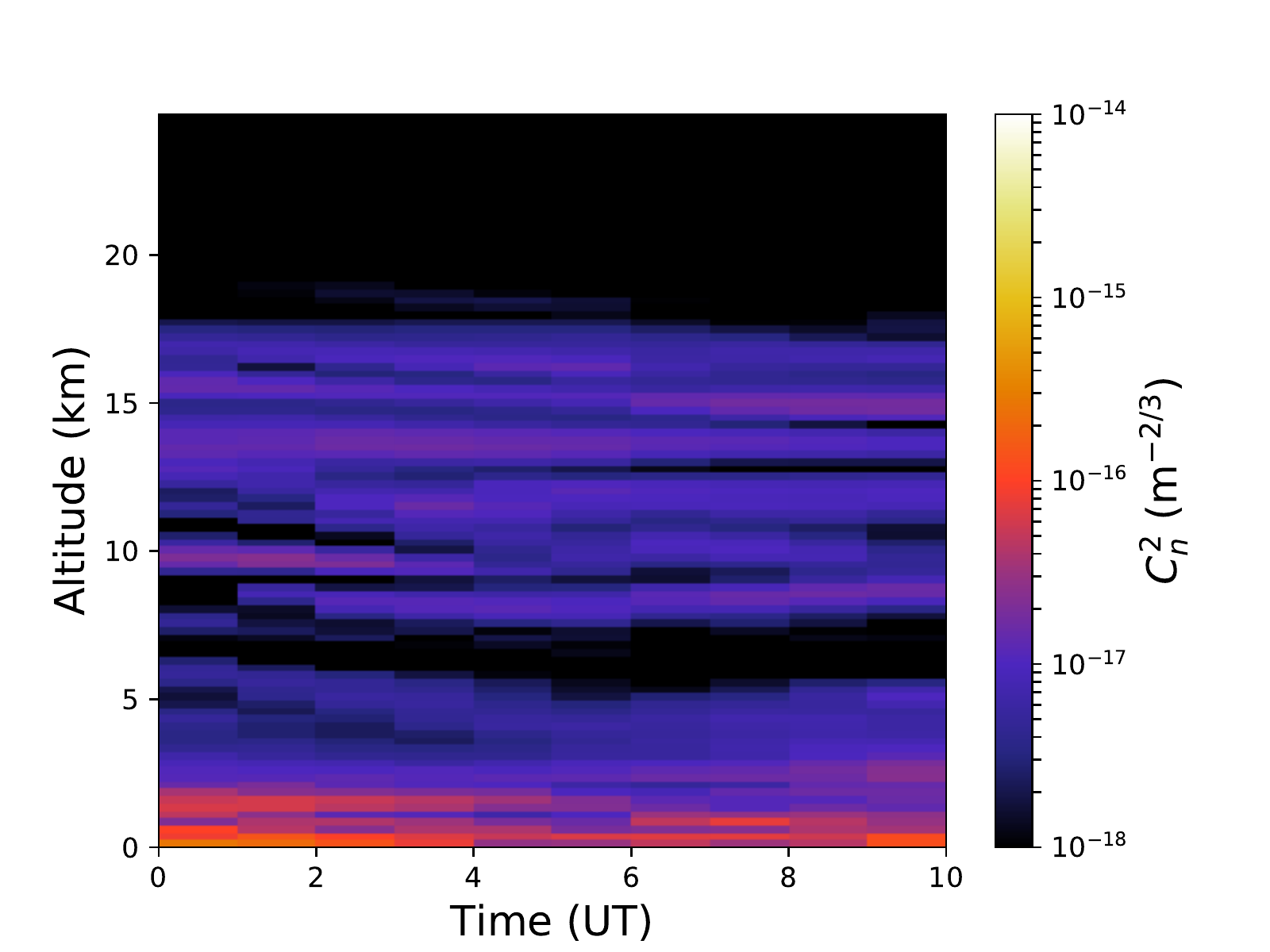} \\
	\includegraphics[width=0.23\textwidth,trim={2cm 0 1cm 0}]{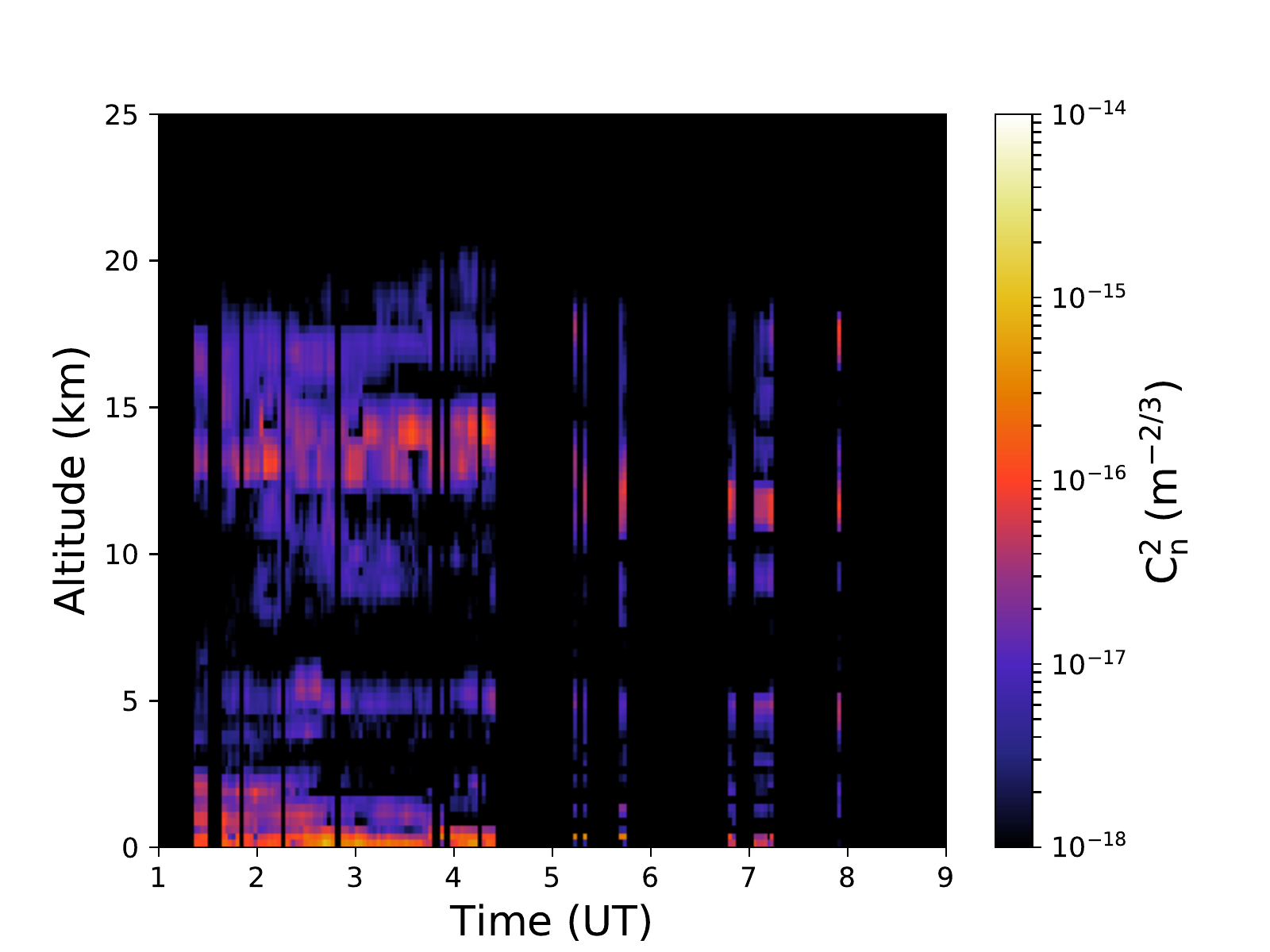} &
    	\includegraphics[width=0.23\textwidth,trim={2cm 0 1cm 0}]{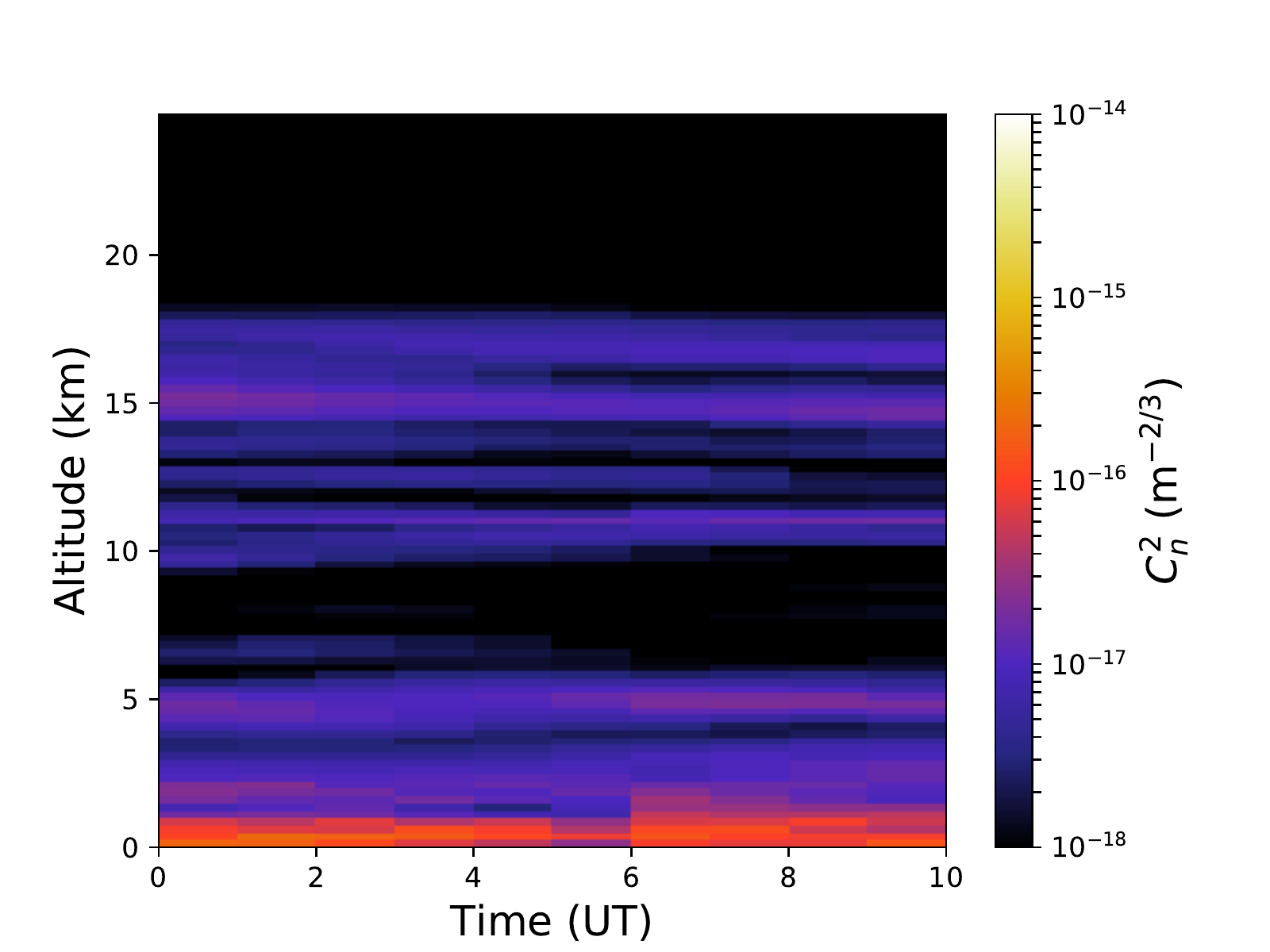} &
	\includegraphics[width=0.23\textwidth,trim={2cm 0 1cm 0}]{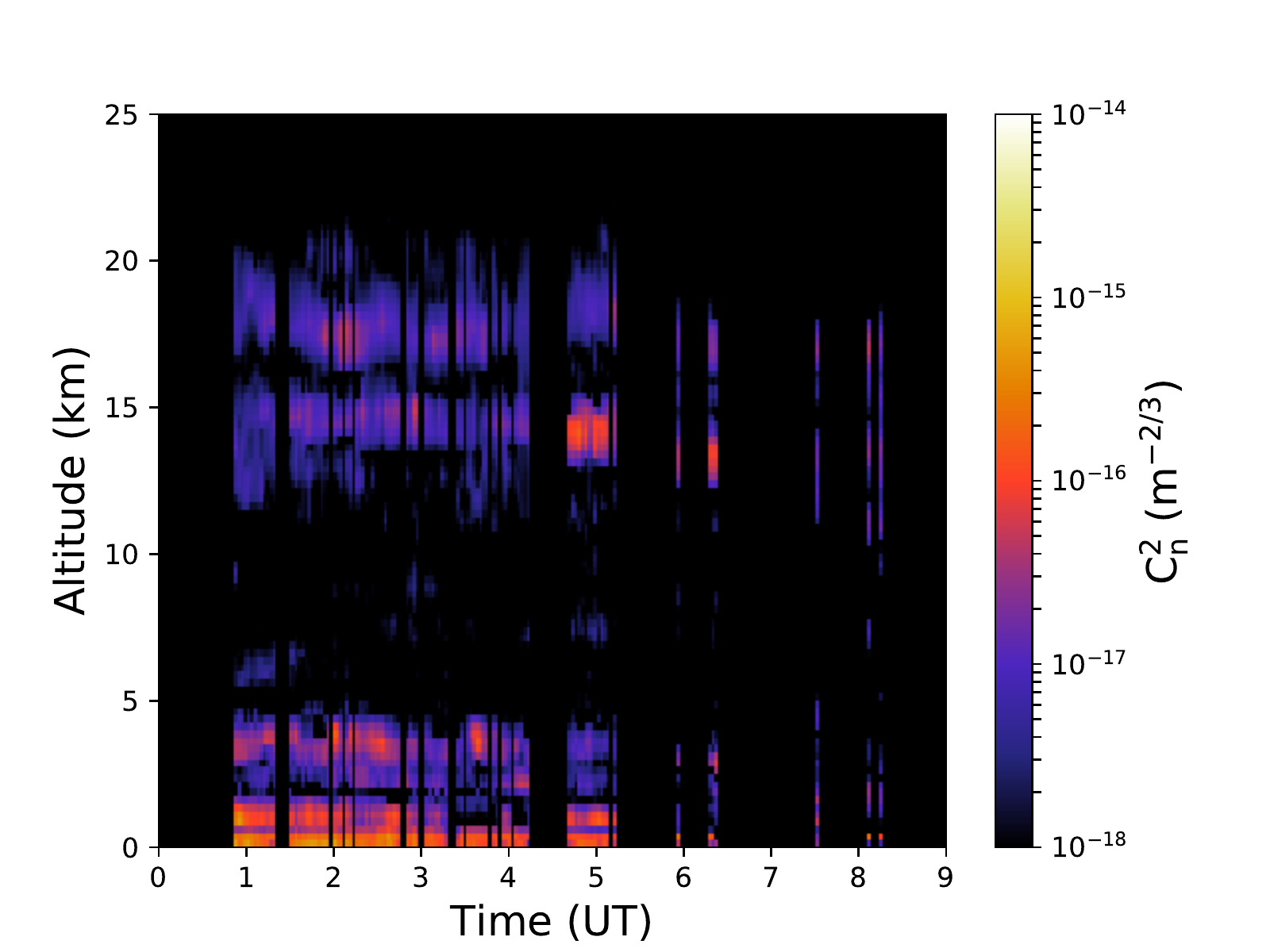} &
    	\includegraphics[width=0.23\textwidth,trim={2cm 0 1cm 0}]{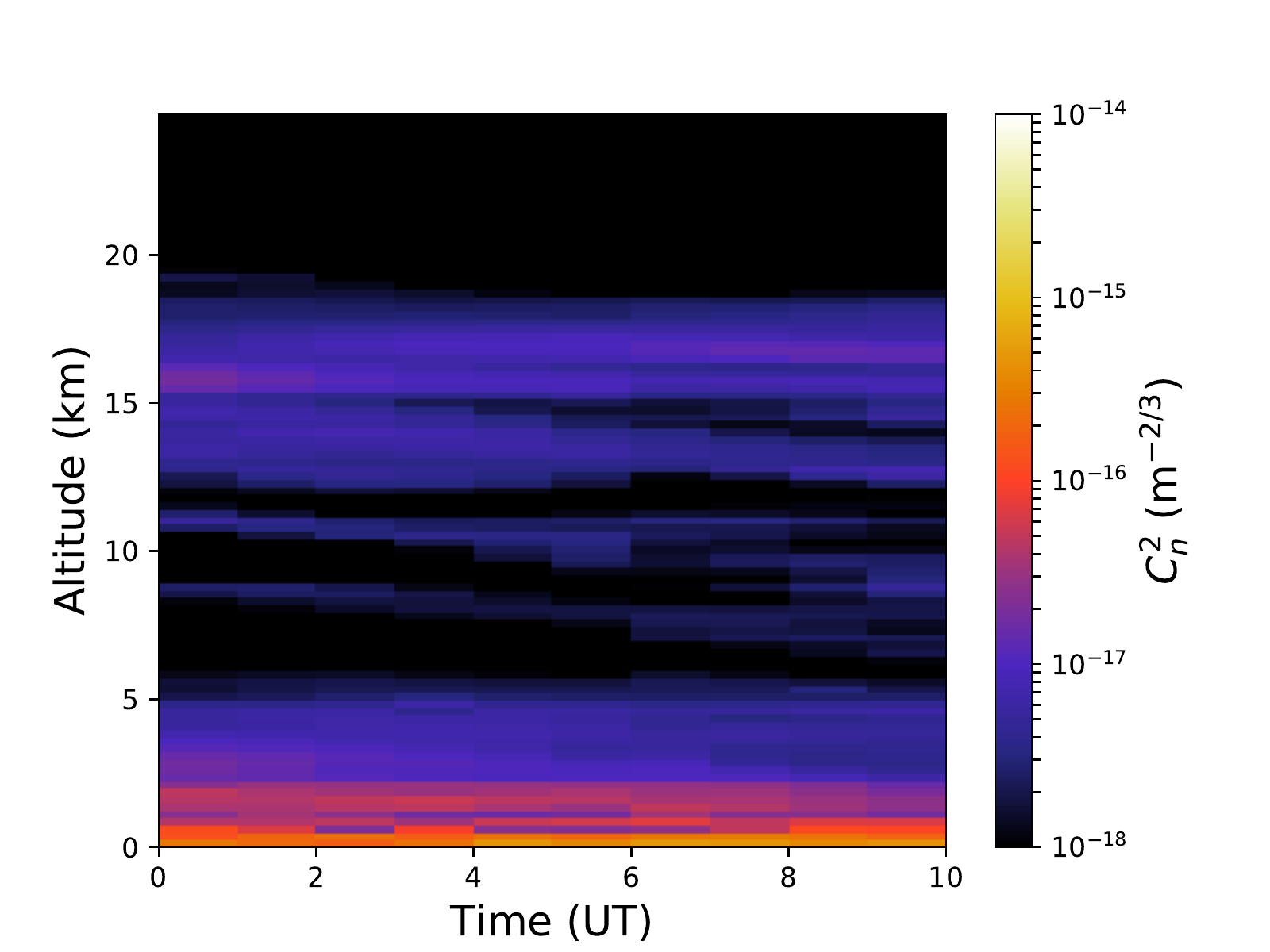} \\
	
\end{array}$
\caption{Example vertical profiles as measured by the stereo-SCIDAR (green) and estimated by the ECMWF GCM model (red). The profiles shown are the median for an individual night of observation. The coloured region shows the interquartile range. These profiles are from the nights beginning 8th July, 3rd - 8th August and 4th - 8th November 2017.}
\label{fig:seqProfiles4}
\end{figure*}

Figures~\ref{fig:seqProfiles1}, \ref{fig:seqProfiles2}, \ref{fig:seqProfiles3} and \ref{fig:seqProfiles4} show the nightly median for all the stereo-SCIDAR nights at Cerro Paranal in 2016. The ECMWF turbulence forecast is also shown.

\begin{figure*}
\centering
$\begin{array}{cccc}
	
    	\includegraphics[width=0.23\textwidth,trim={2cm 0 1cm 0}]{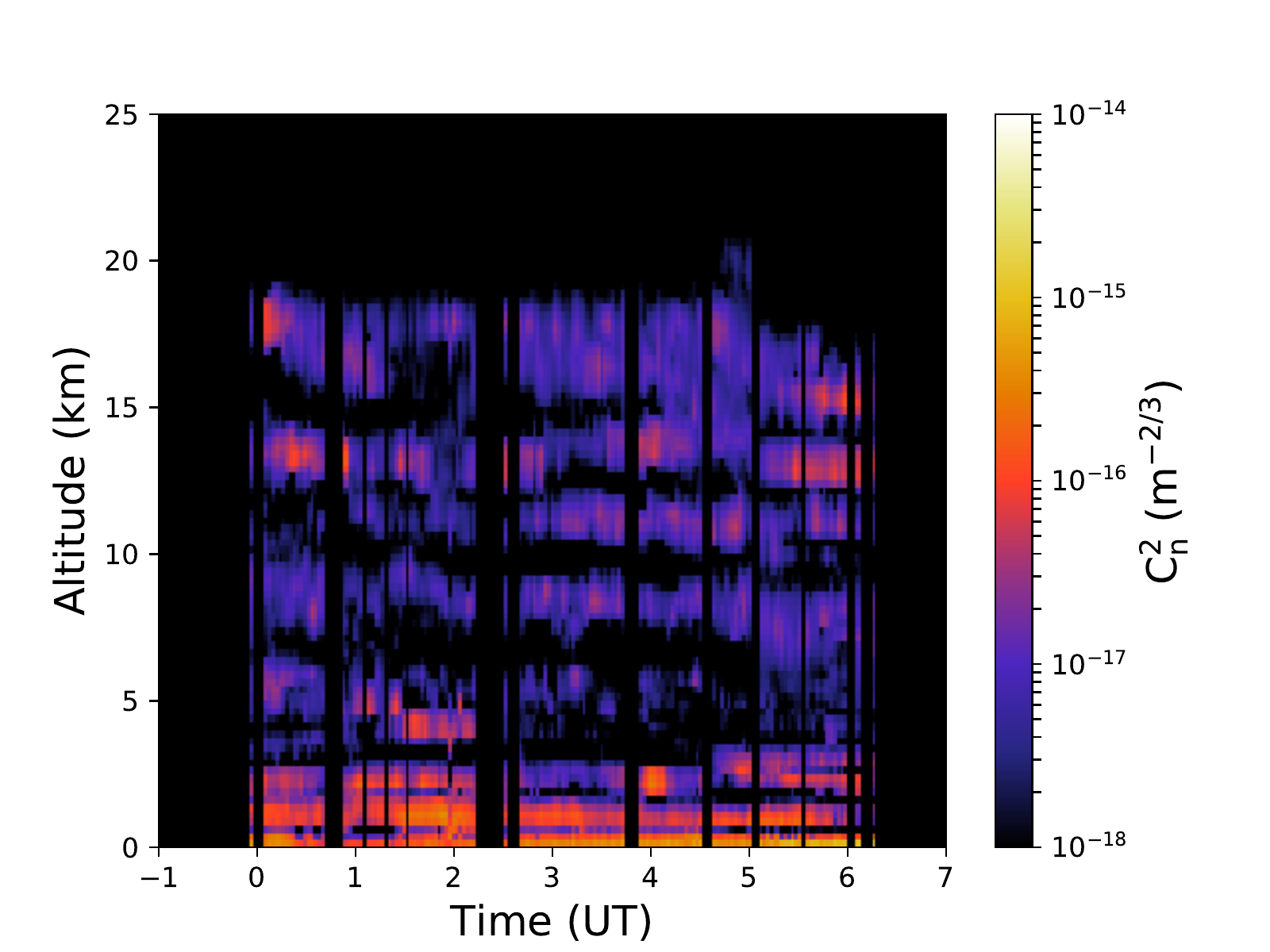} &
    	\includegraphics[width=0.23\textwidth,trim={2cm 0 1cm 0}]{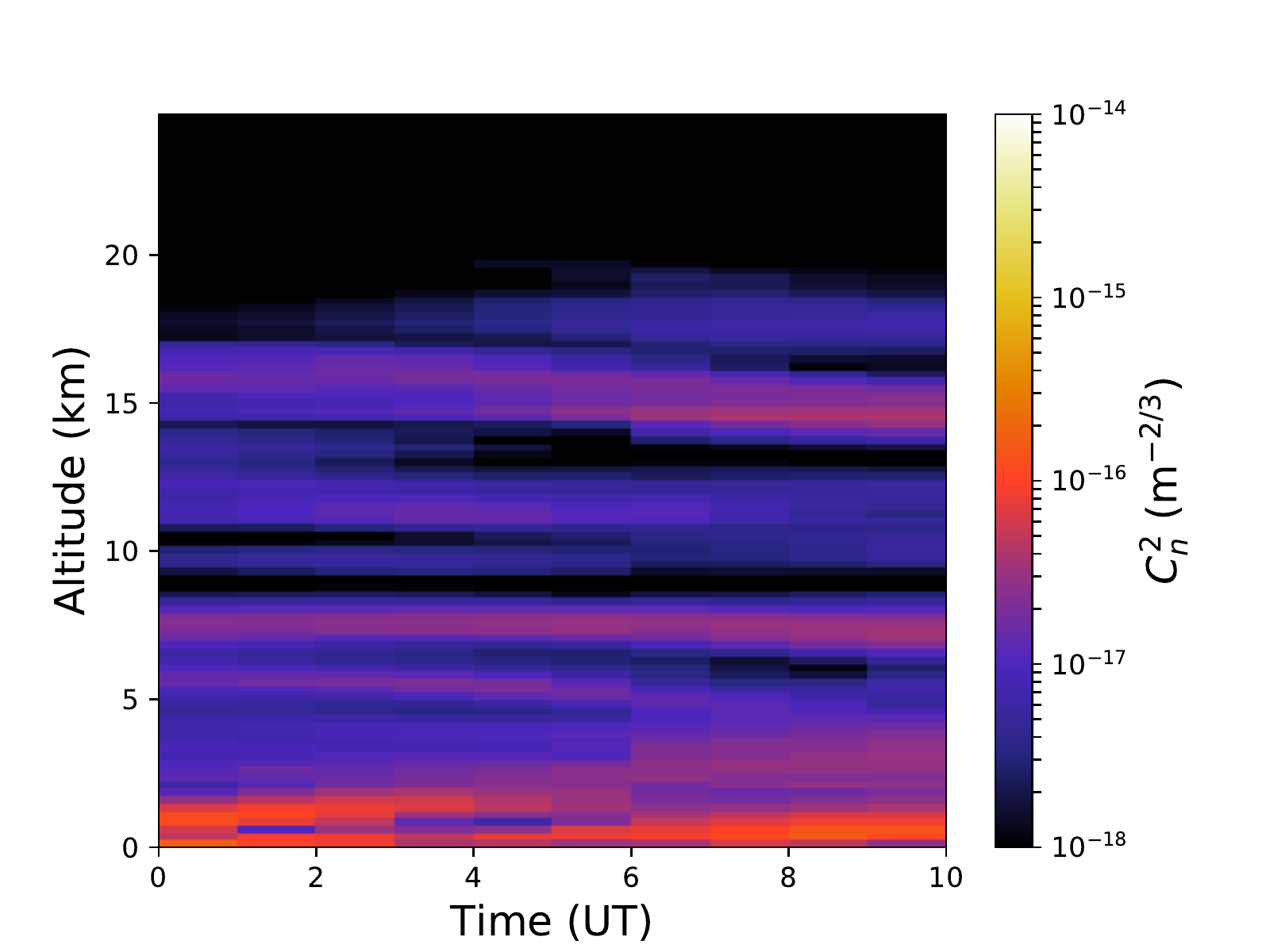} &
	\includegraphics[width=0.23\textwidth,trim={2cm 0 1cm 0}]{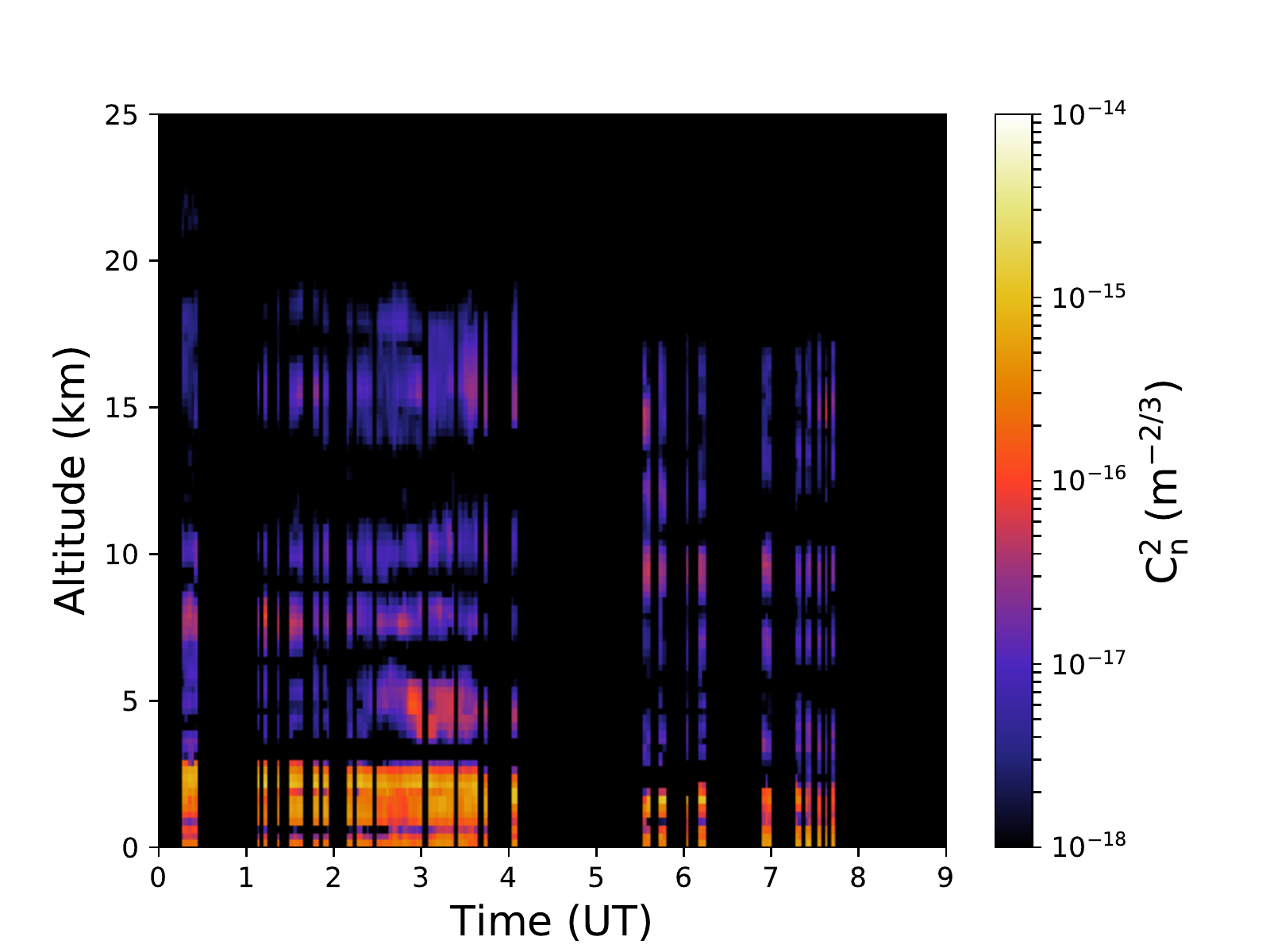} &
    	\includegraphics[width=0.23\textwidth,trim={2cm 0 1cm 0}]{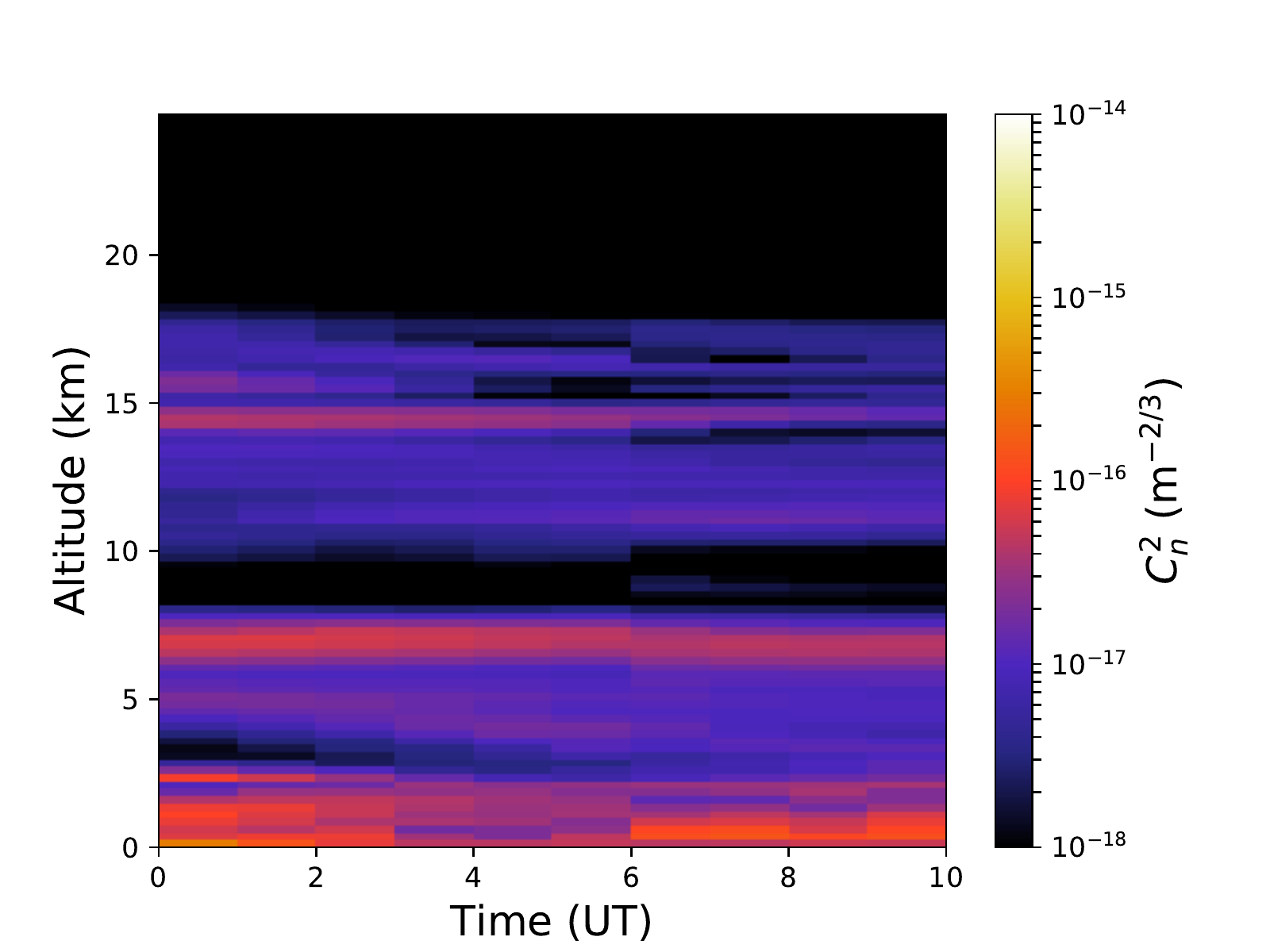} \\
	
    	\includegraphics[width=0.23\textwidth,trim={2cm 0 1cm 0}]{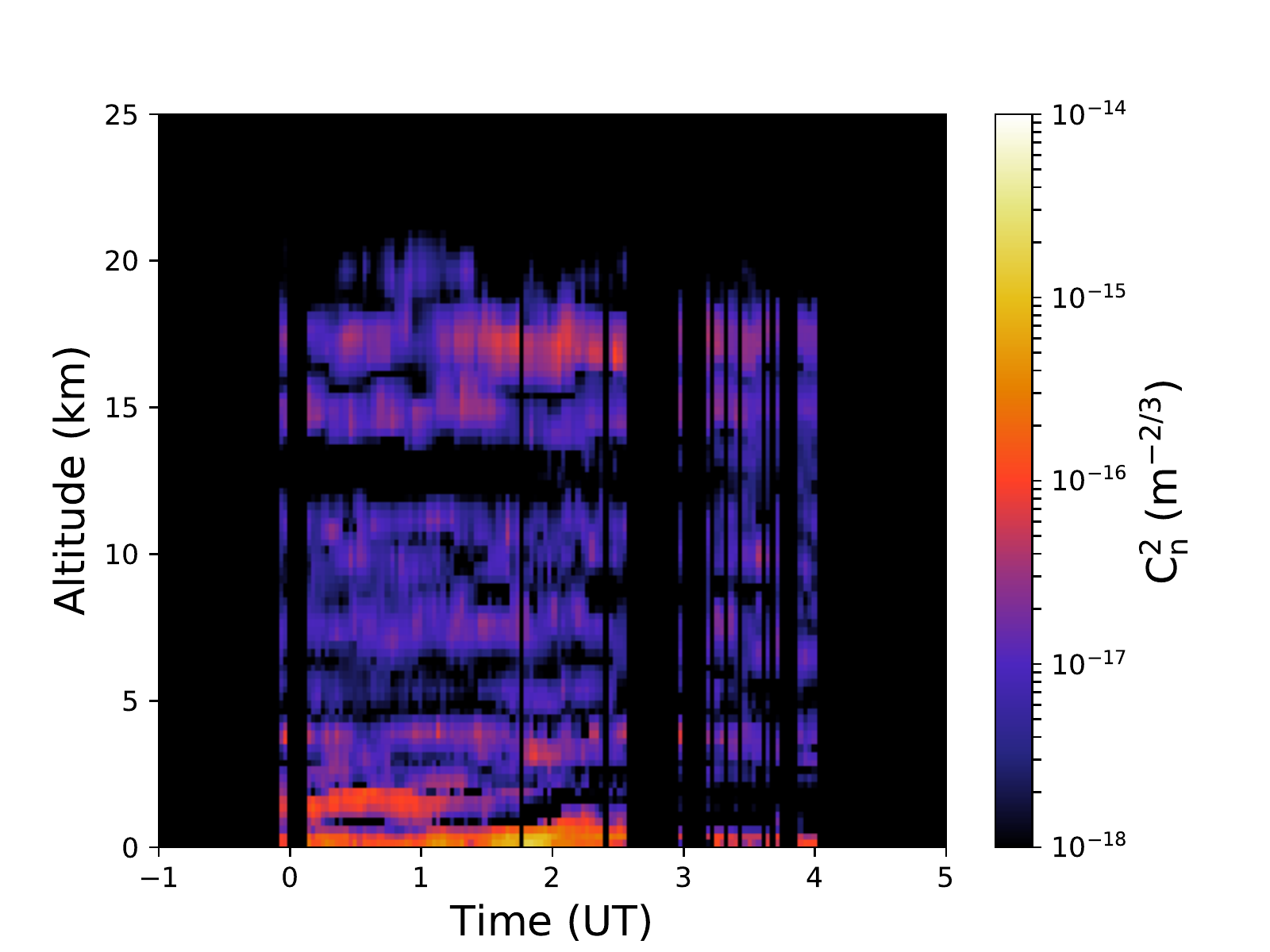} &
    	\includegraphics[width=0.23\textwidth,trim={2cm 0 1cm 0}]{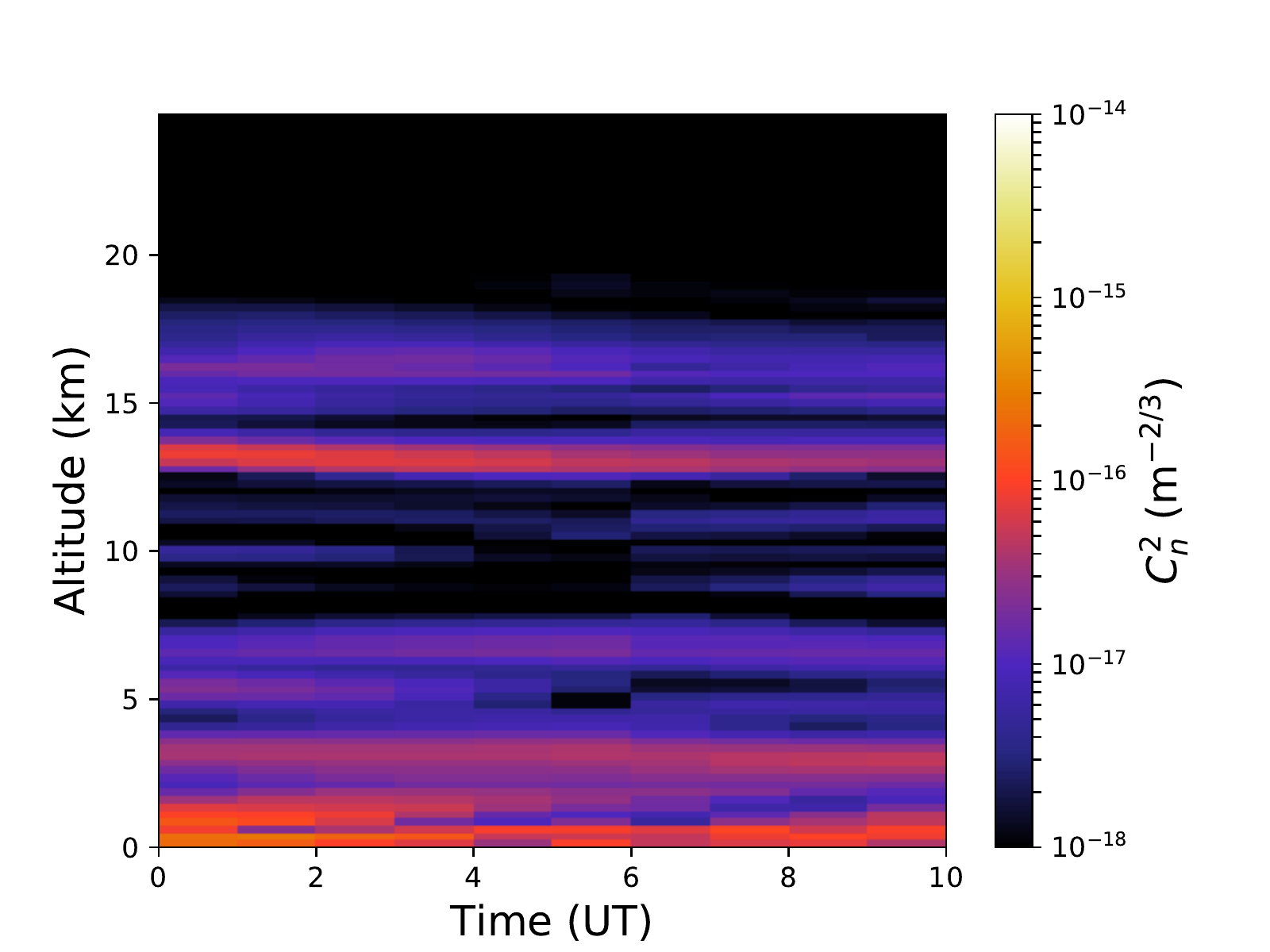} &		
	\includegraphics[width=0.23\textwidth,trim={2cm 0 1cm 0}]{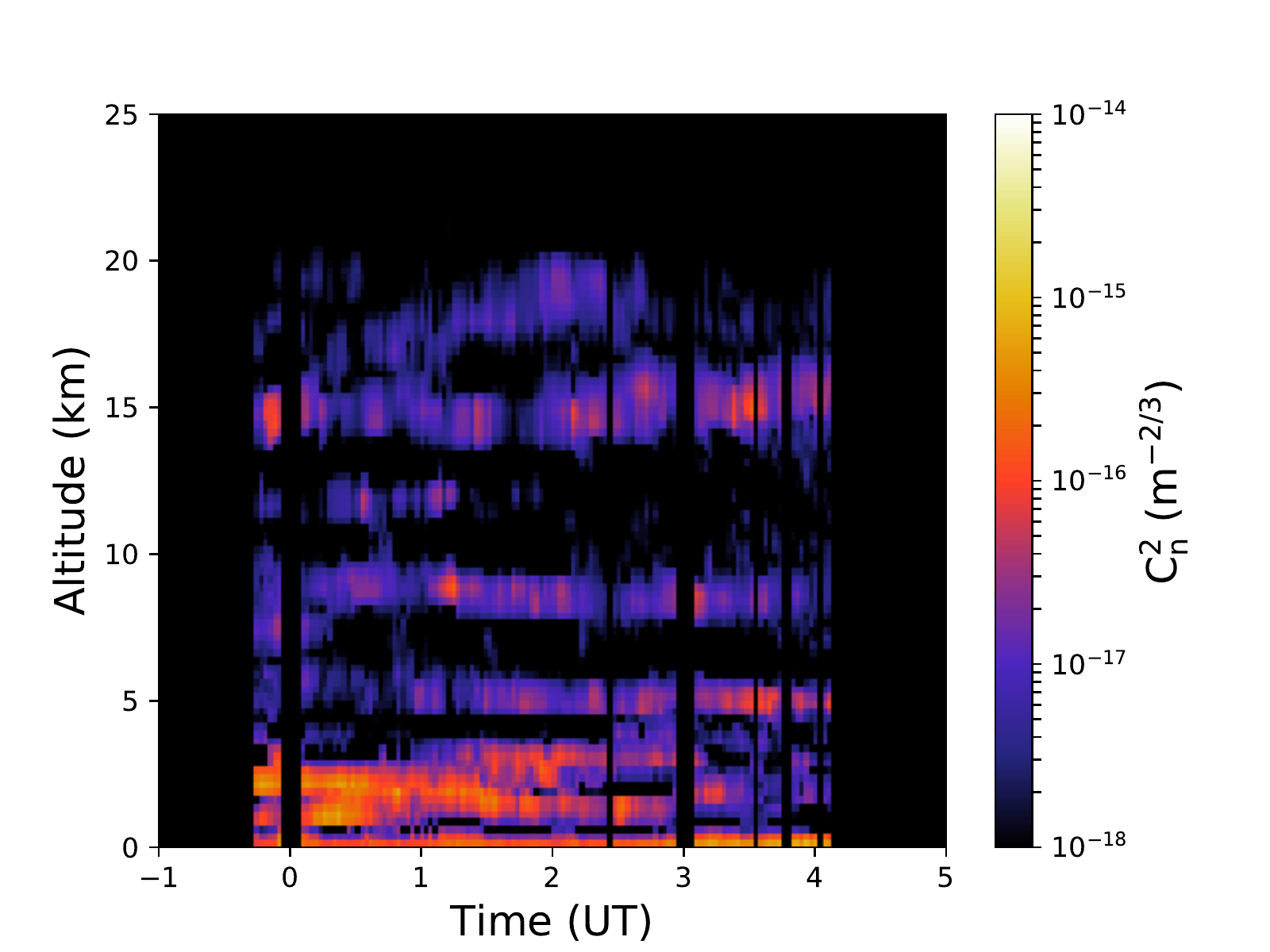} &
    	\includegraphics[width=0.23\textwidth,trim={2cm 0 1cm 0}]{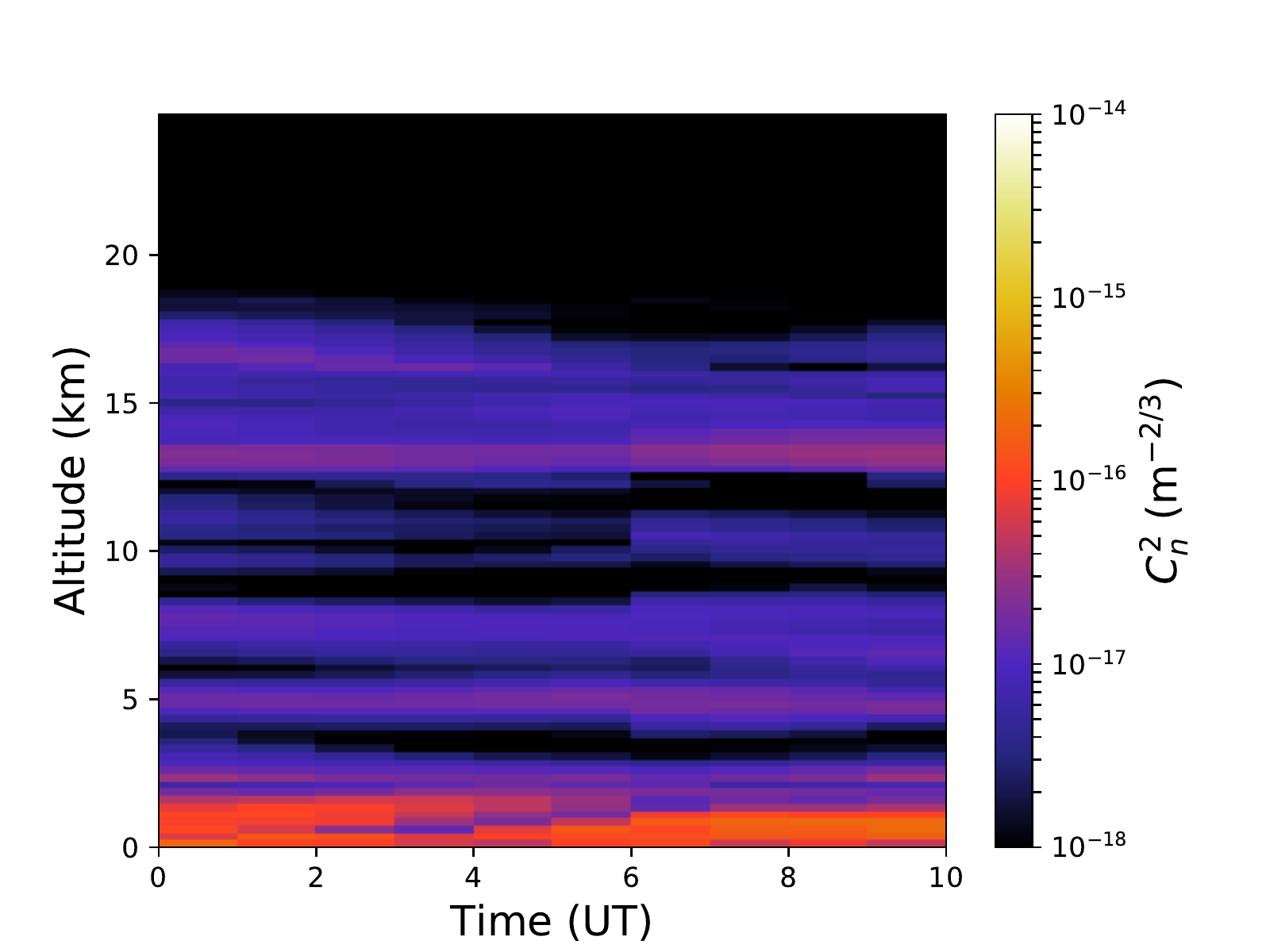} \\
    	\includegraphics[width=0.23\textwidth,trim={2cm 0 1cm 0}]{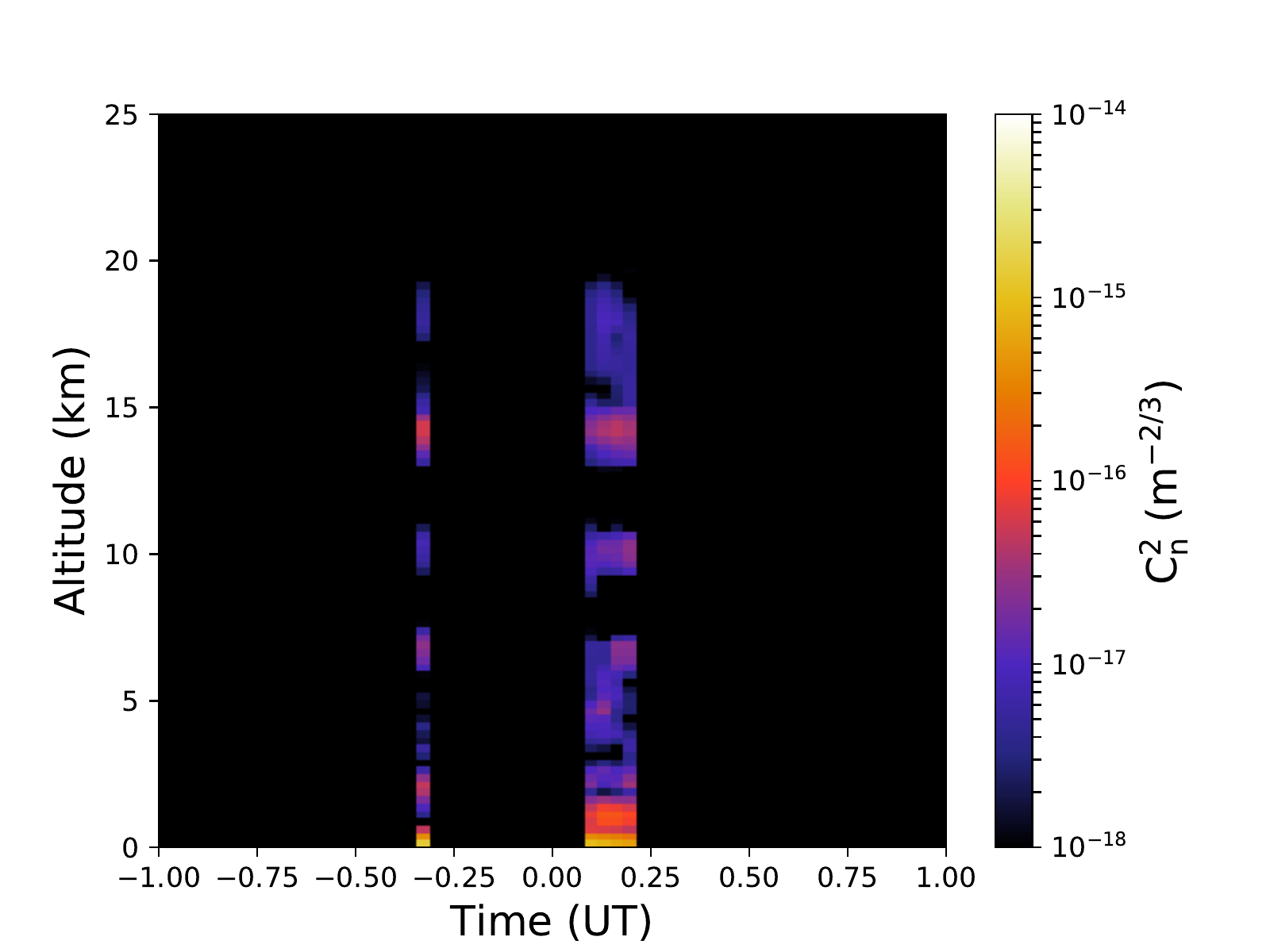} &
    	\includegraphics[width=0.23\textwidth,trim={2cm 0 1cm 0}]{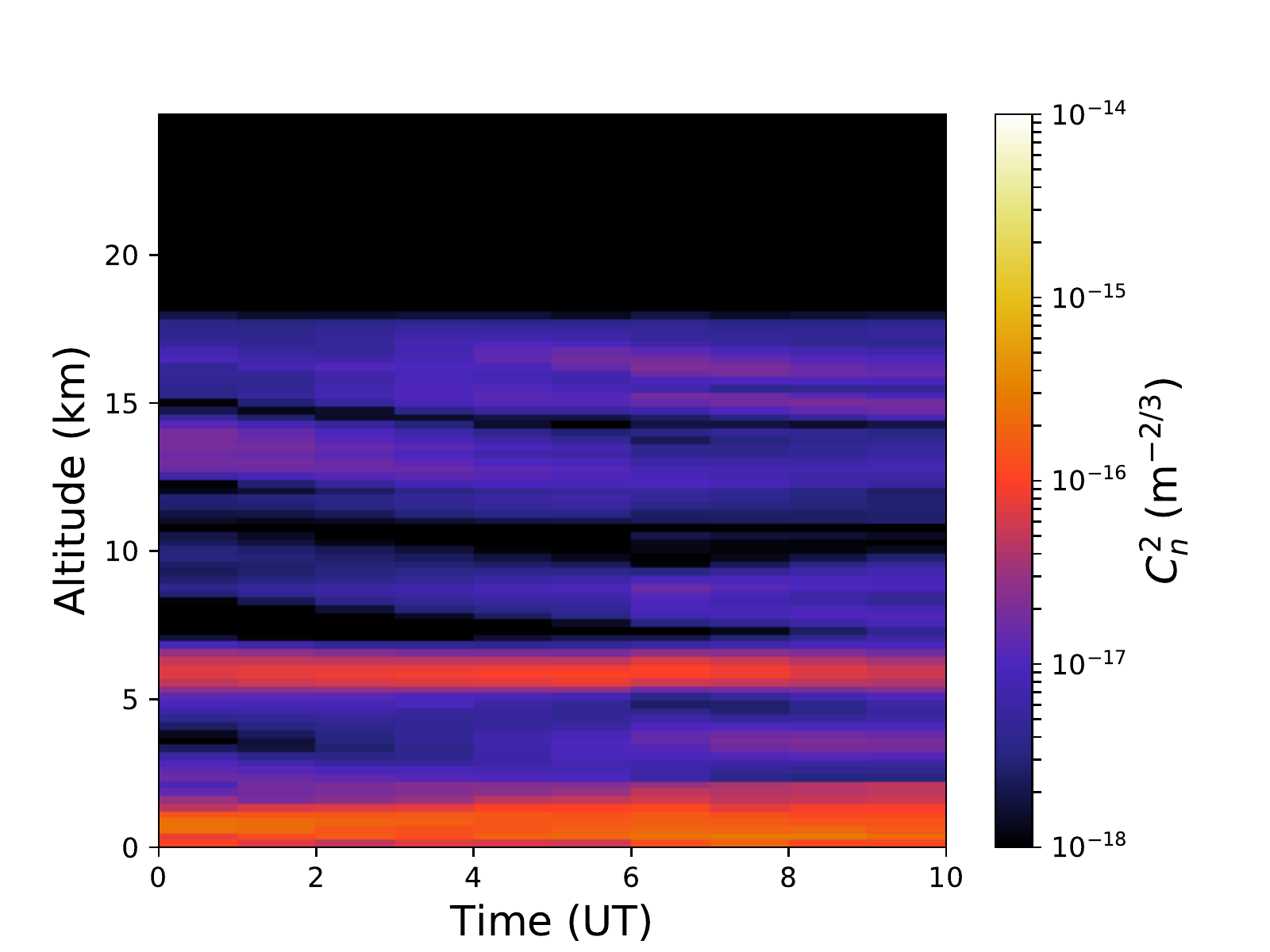} &	
	
	\includegraphics[width=0.23\textwidth,trim={2cm 0 1cm 0}]{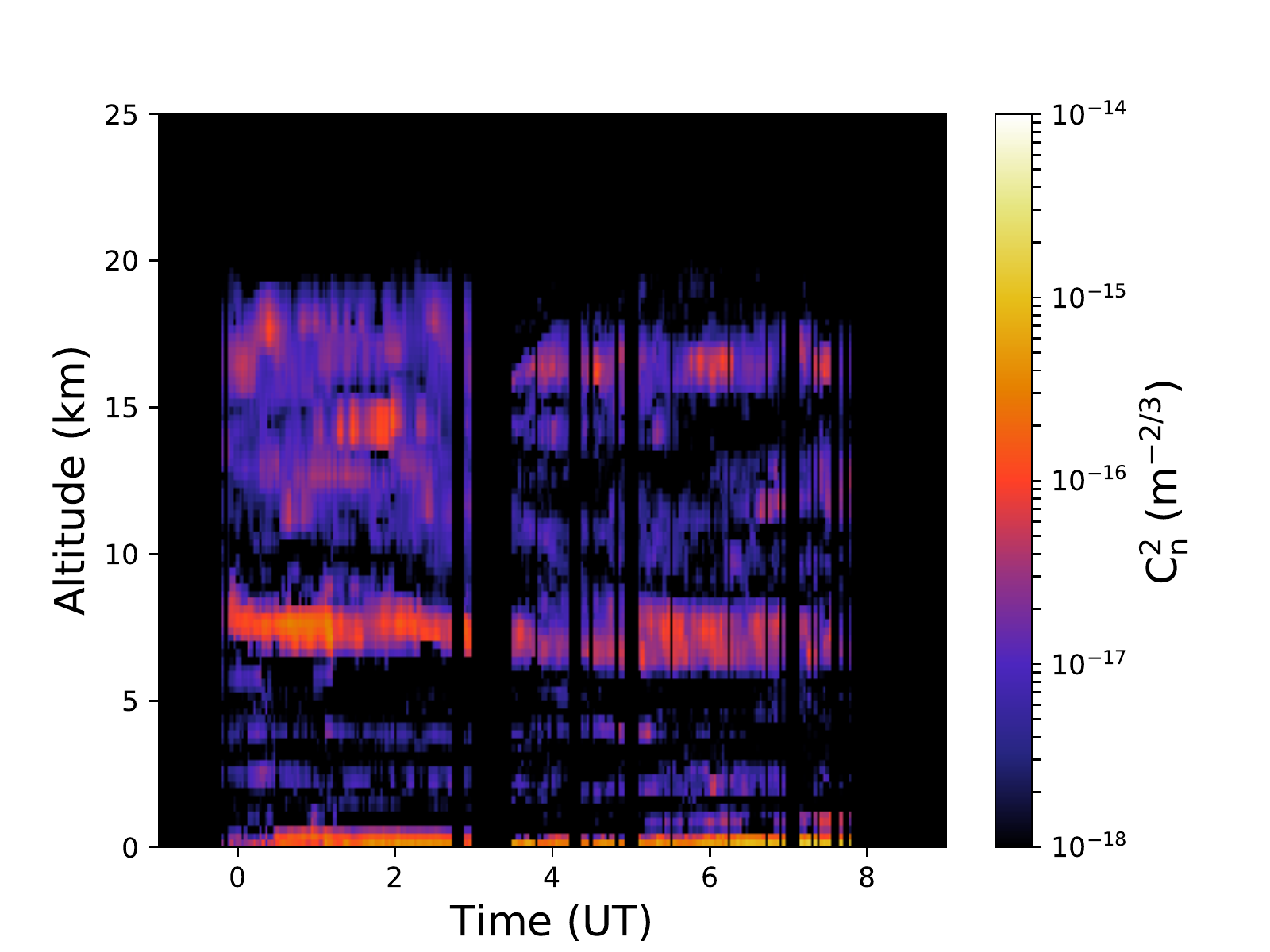} &
    	\includegraphics[width=0.23\textwidth,trim={2cm 0 1cm 0}]{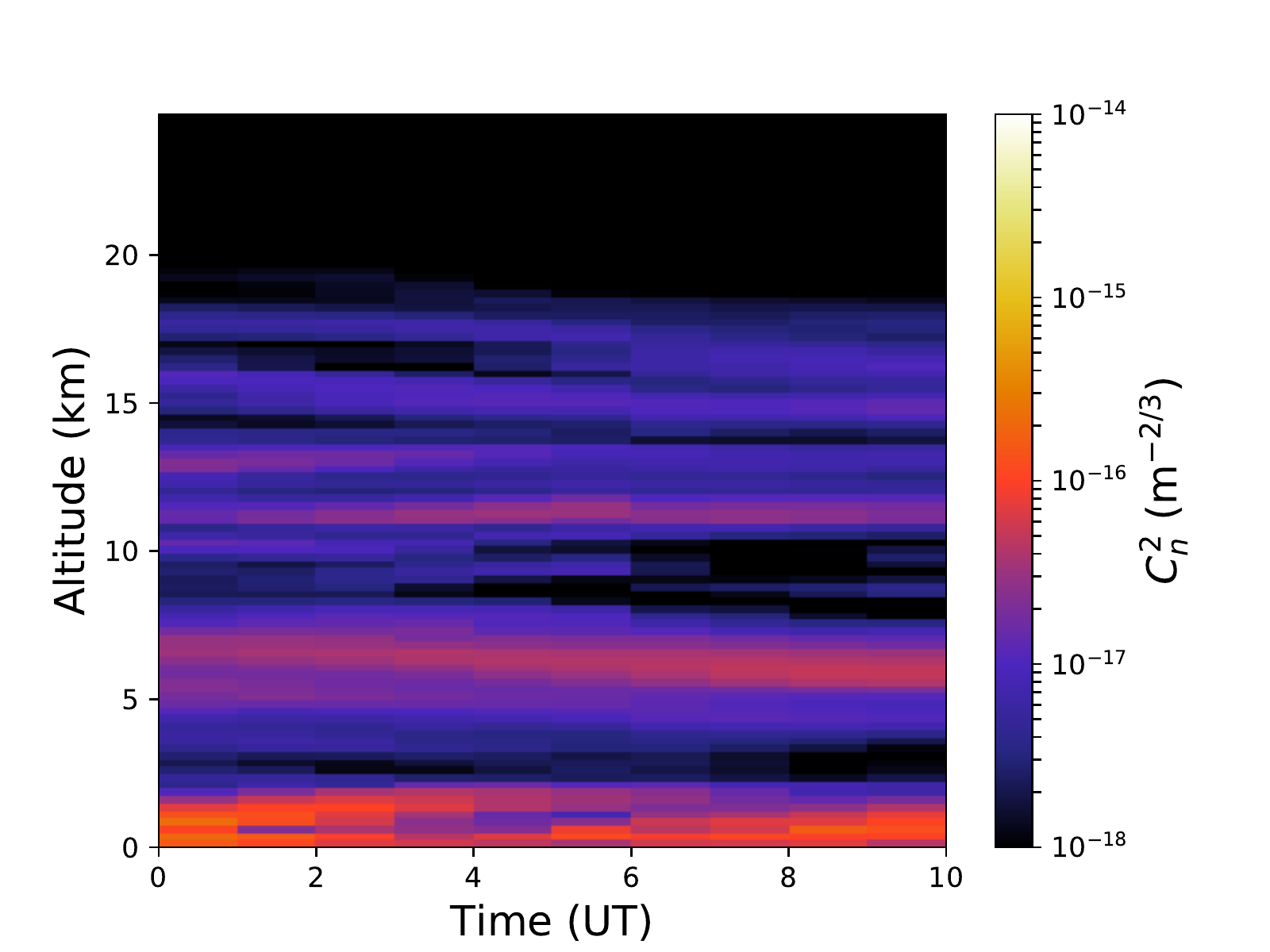} \\
	
    	\includegraphics[width=0.23\textwidth,trim={2cm 0 1cm 0}]{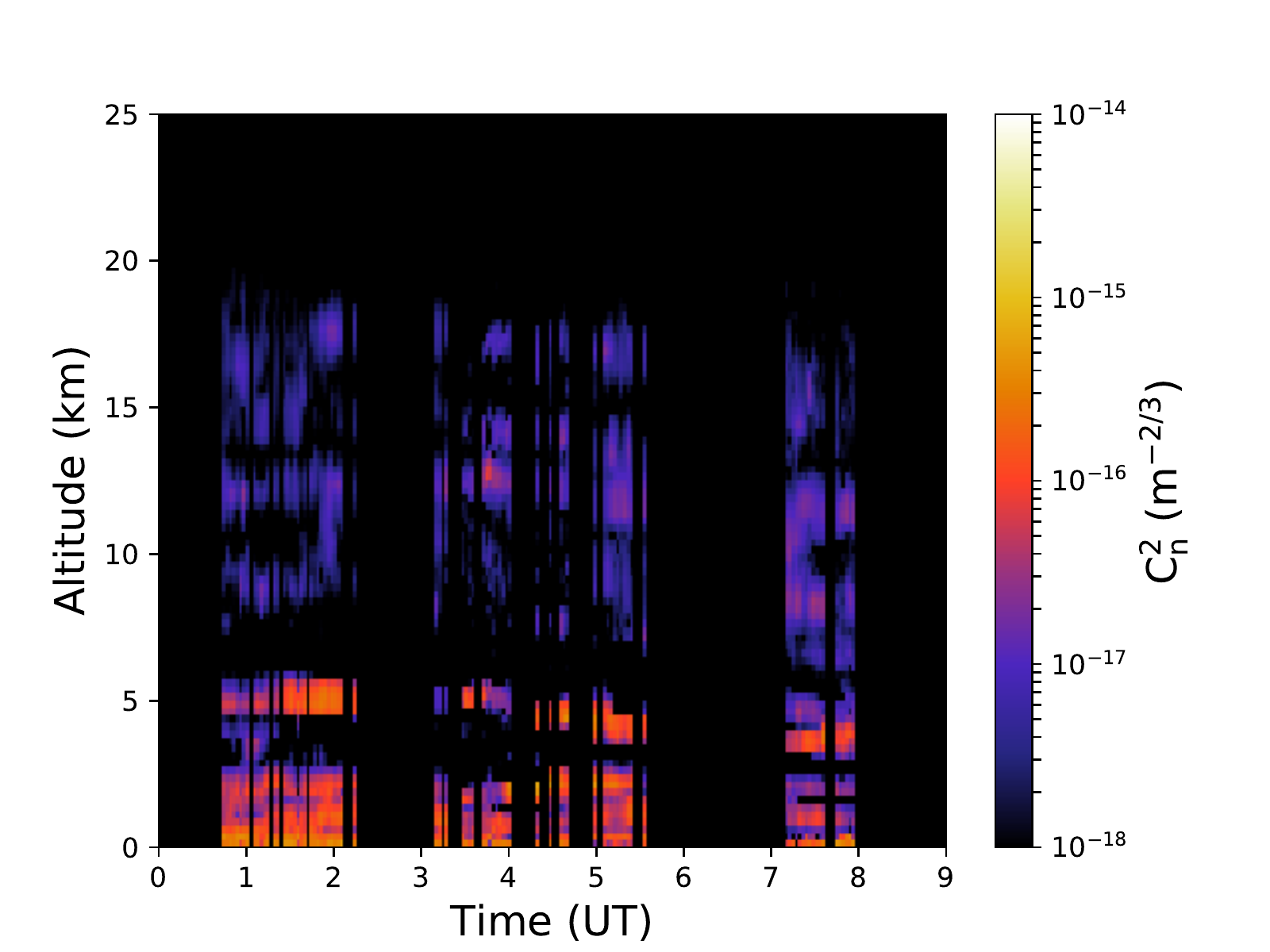} &
    	\includegraphics[width=0.23\textwidth,trim={2cm 0 1cm 0}]{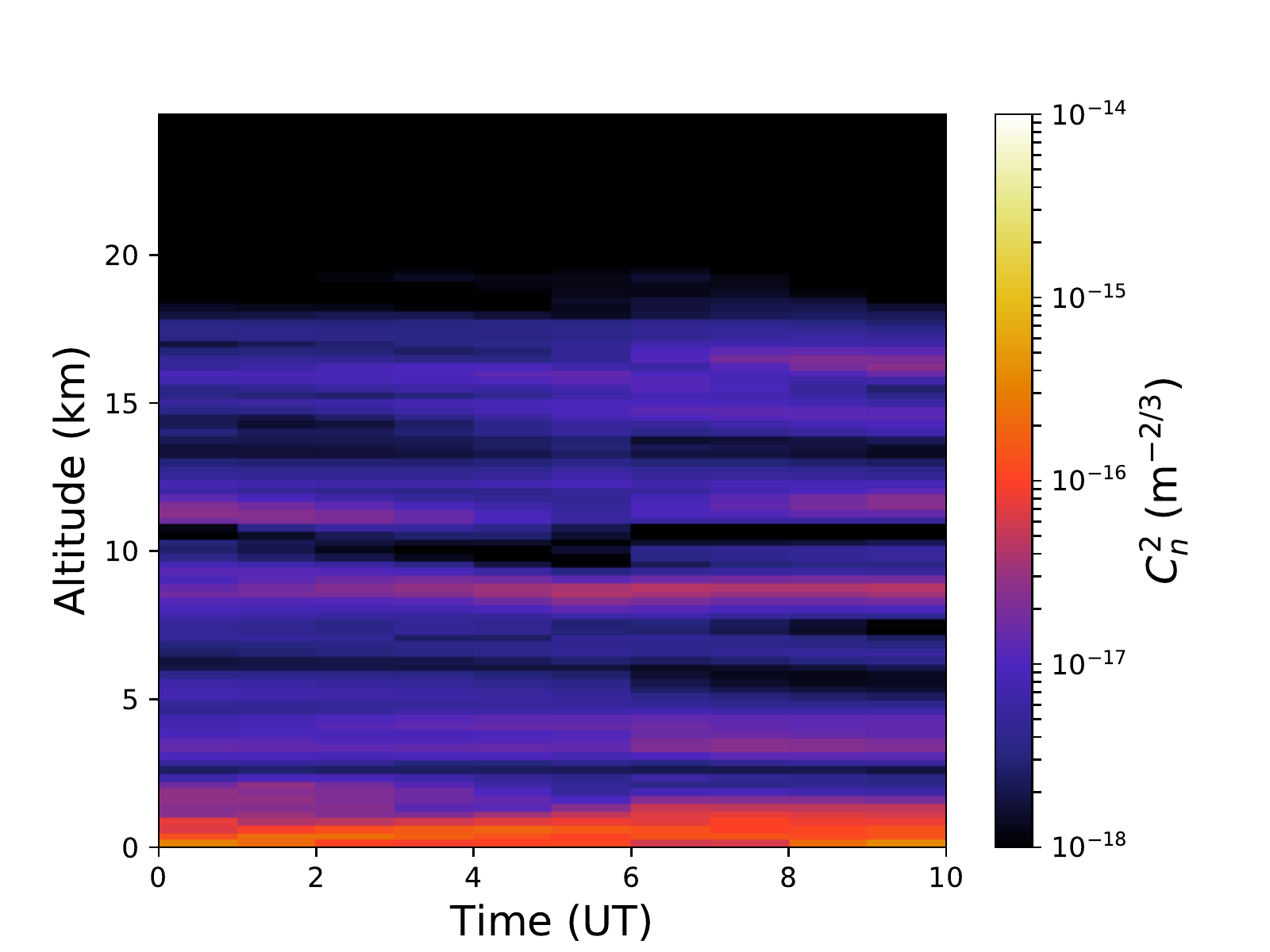} &	
	\includegraphics[width=0.23\textwidth,trim={2cm 0 1cm 0}]{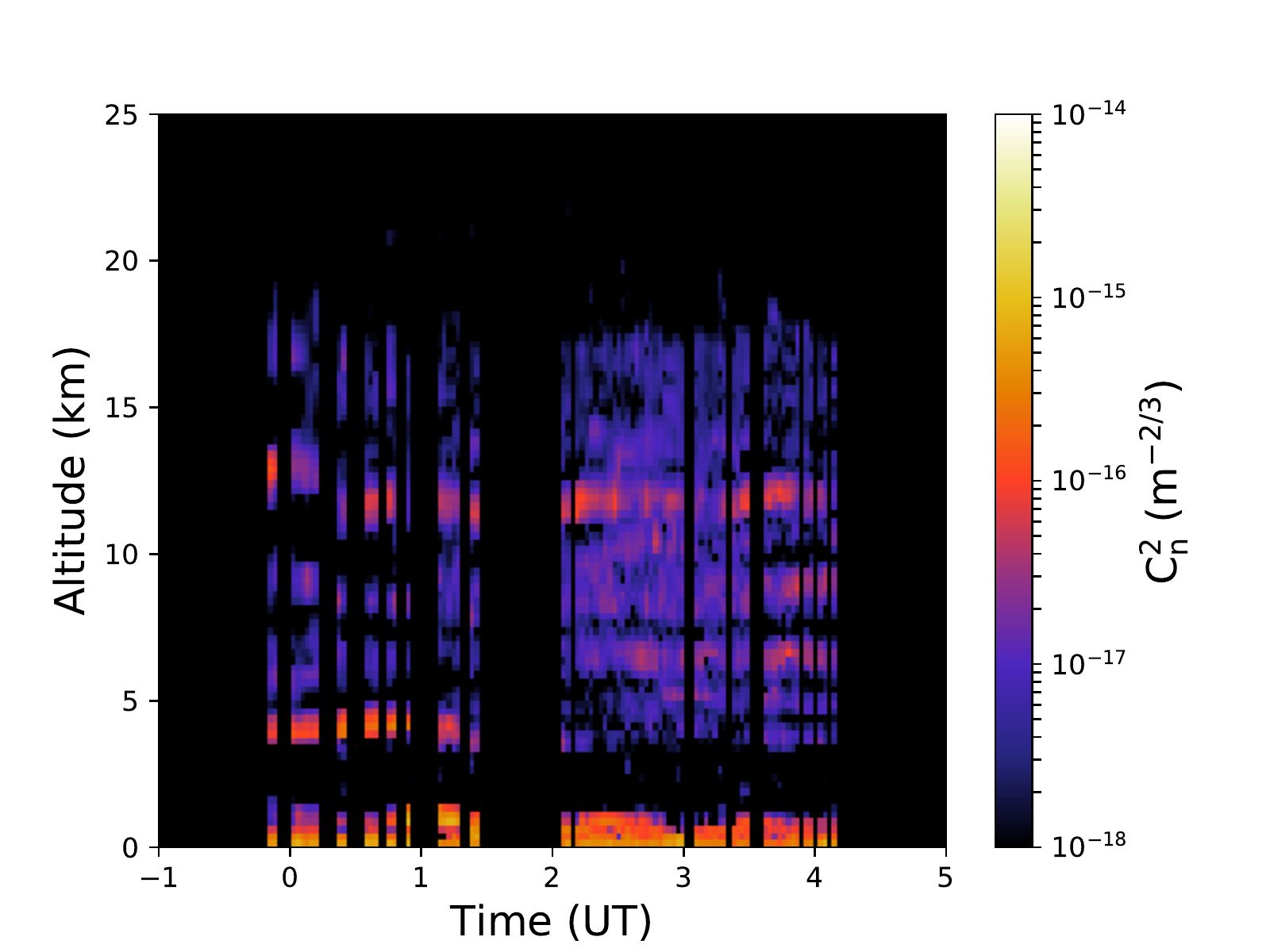} &
    	\includegraphics[width=0.23\textwidth,trim={2cm 0 1cm 0}]{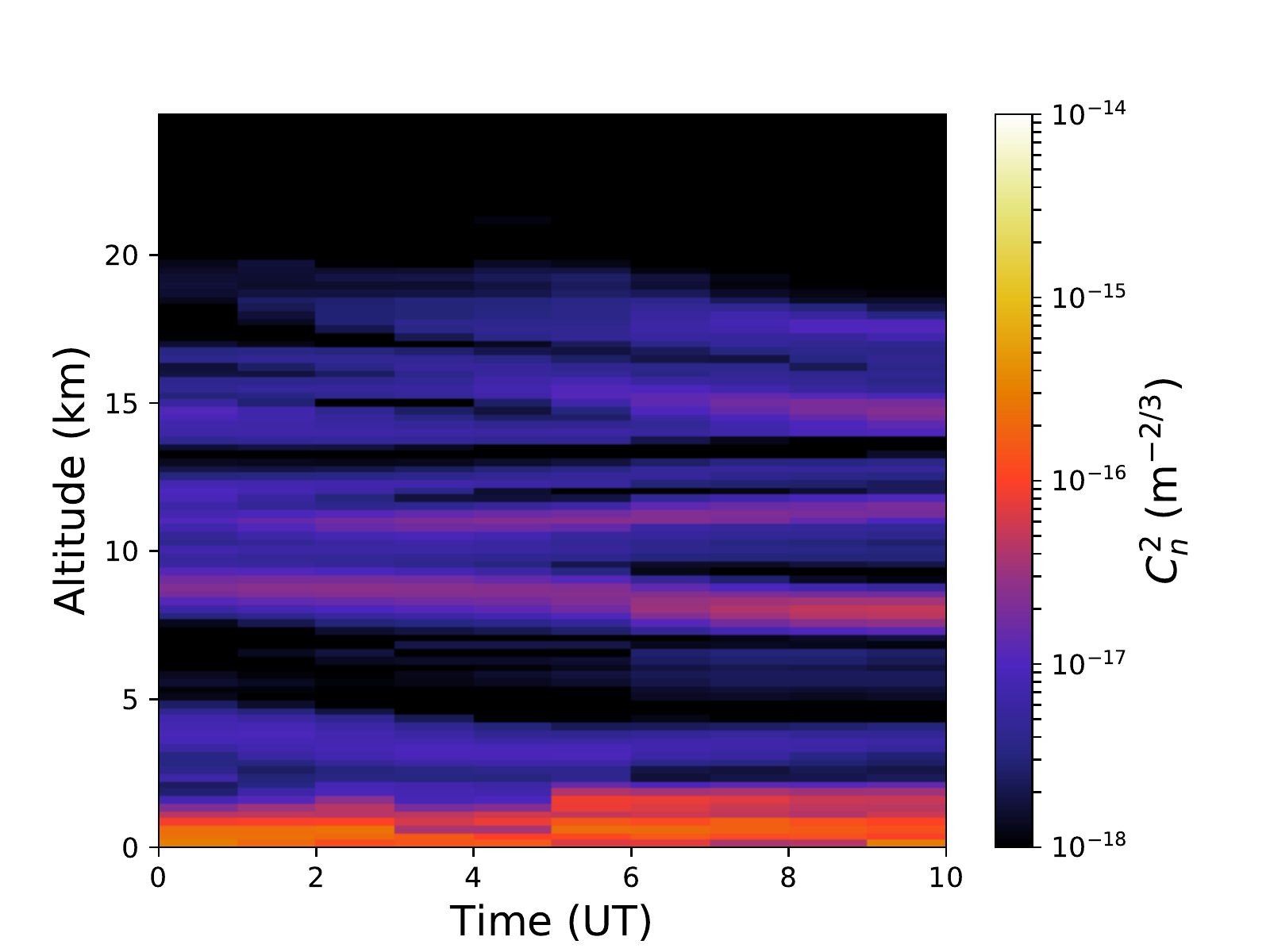} \\
    	\includegraphics[width=0.23\textwidth,trim={2cm 0 1cm 0}]{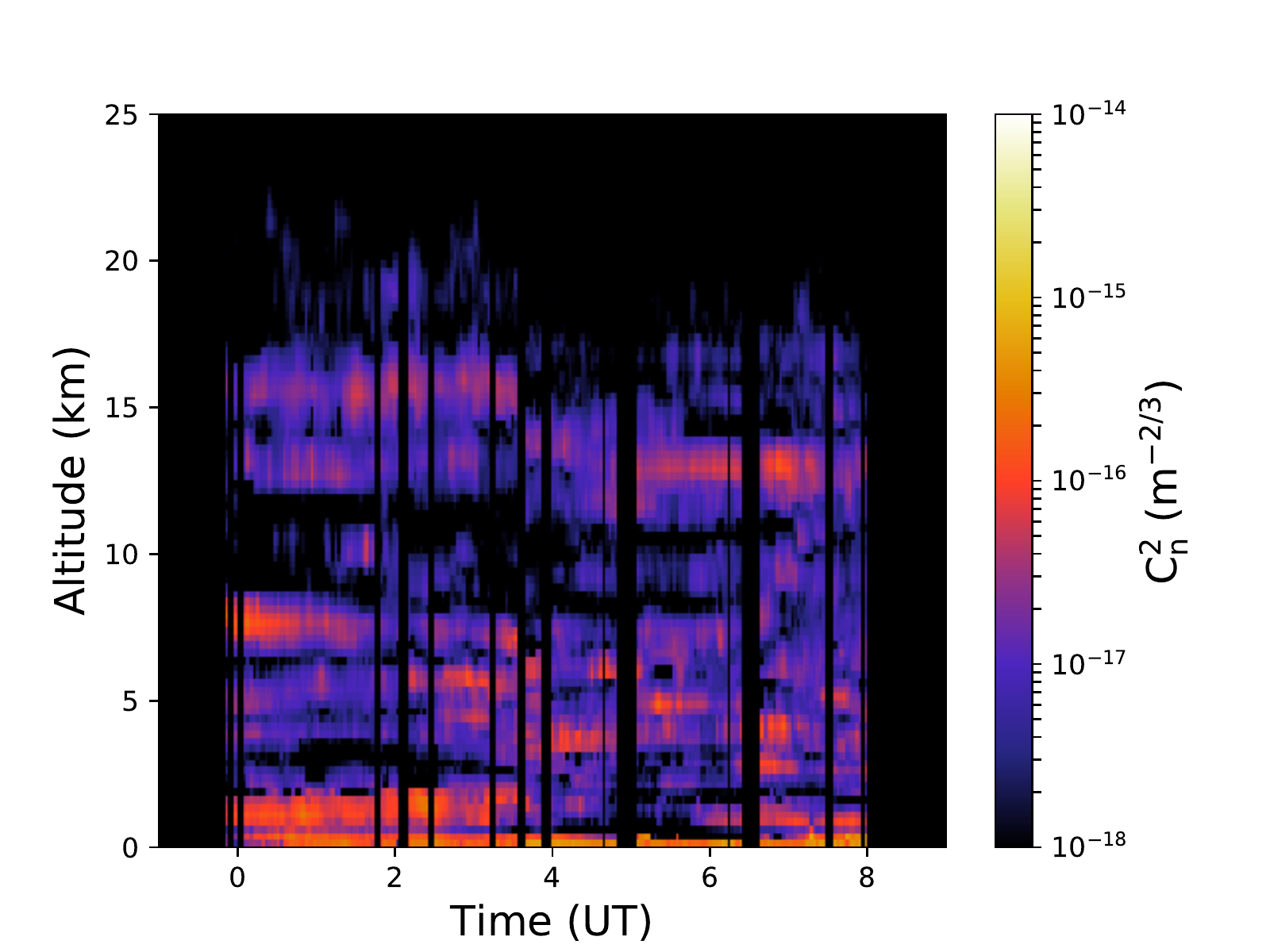} &
    	\includegraphics[width=0.23\textwidth,trim={2cm 0 1cm 0}]{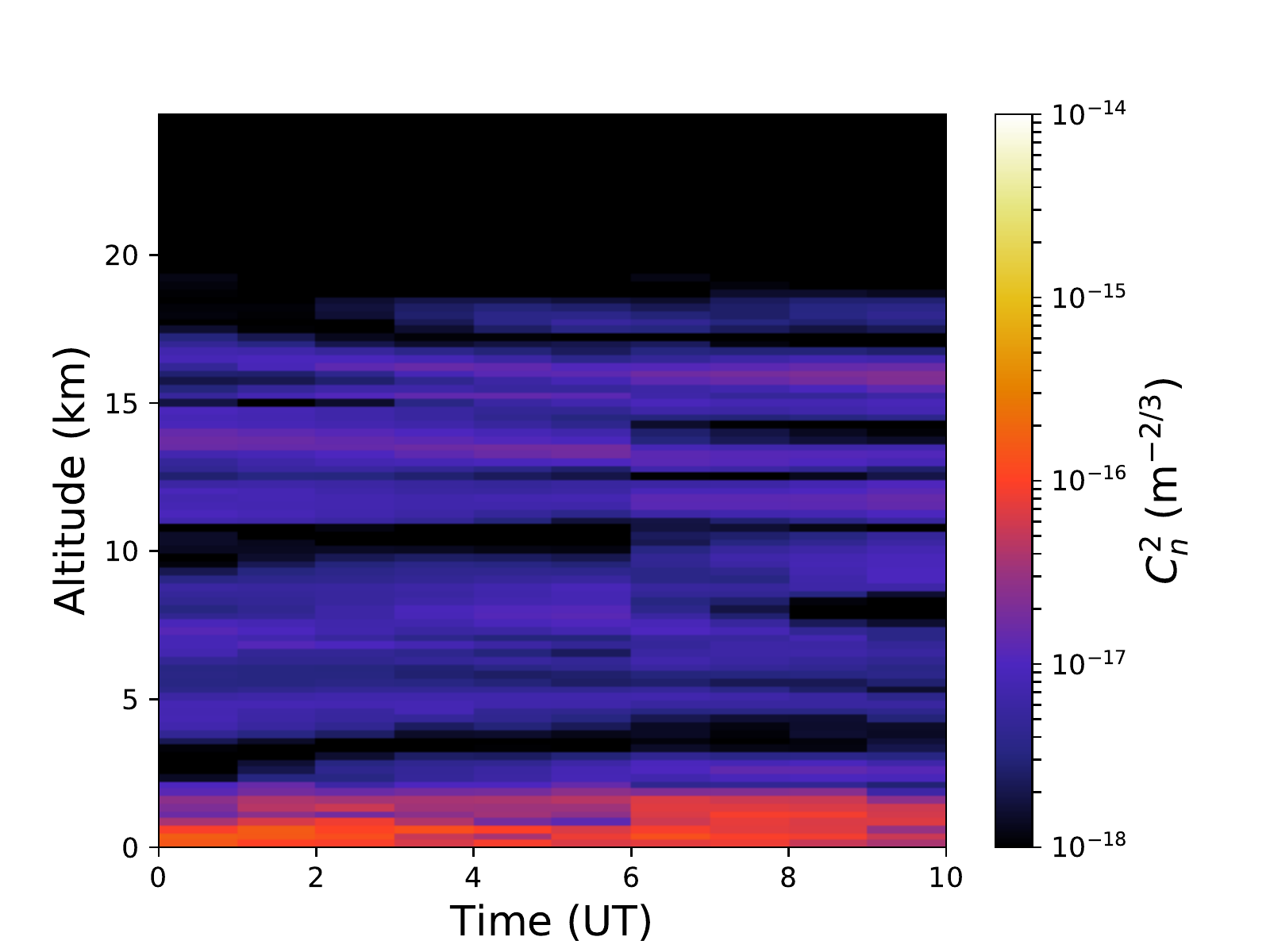} &	
	\includegraphics[width=0.23\textwidth,trim={2cm 0 1cm 0}]{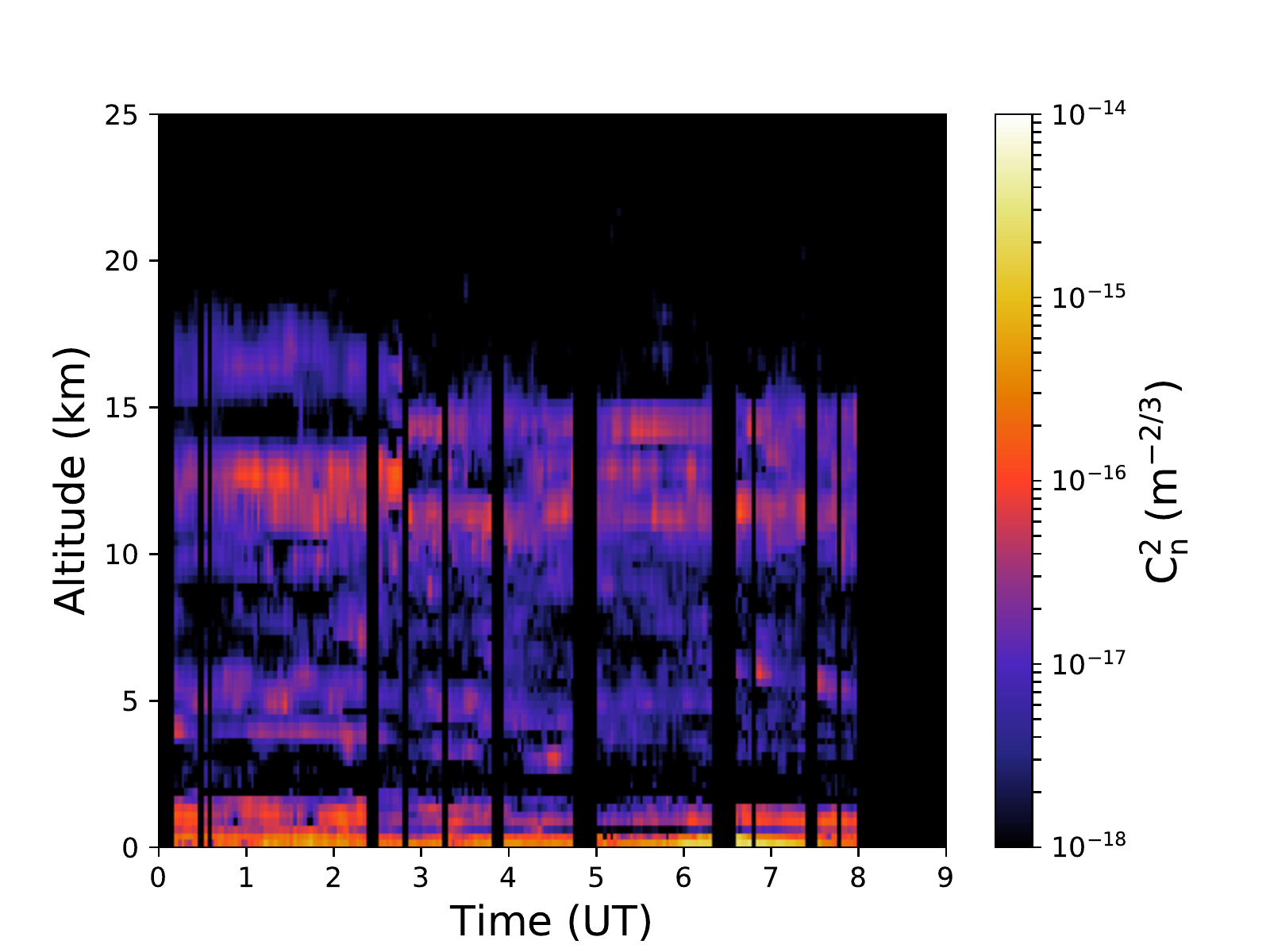} &
    	\includegraphics[width=0.23\textwidth,trim={2cm 0 1cm 0}]{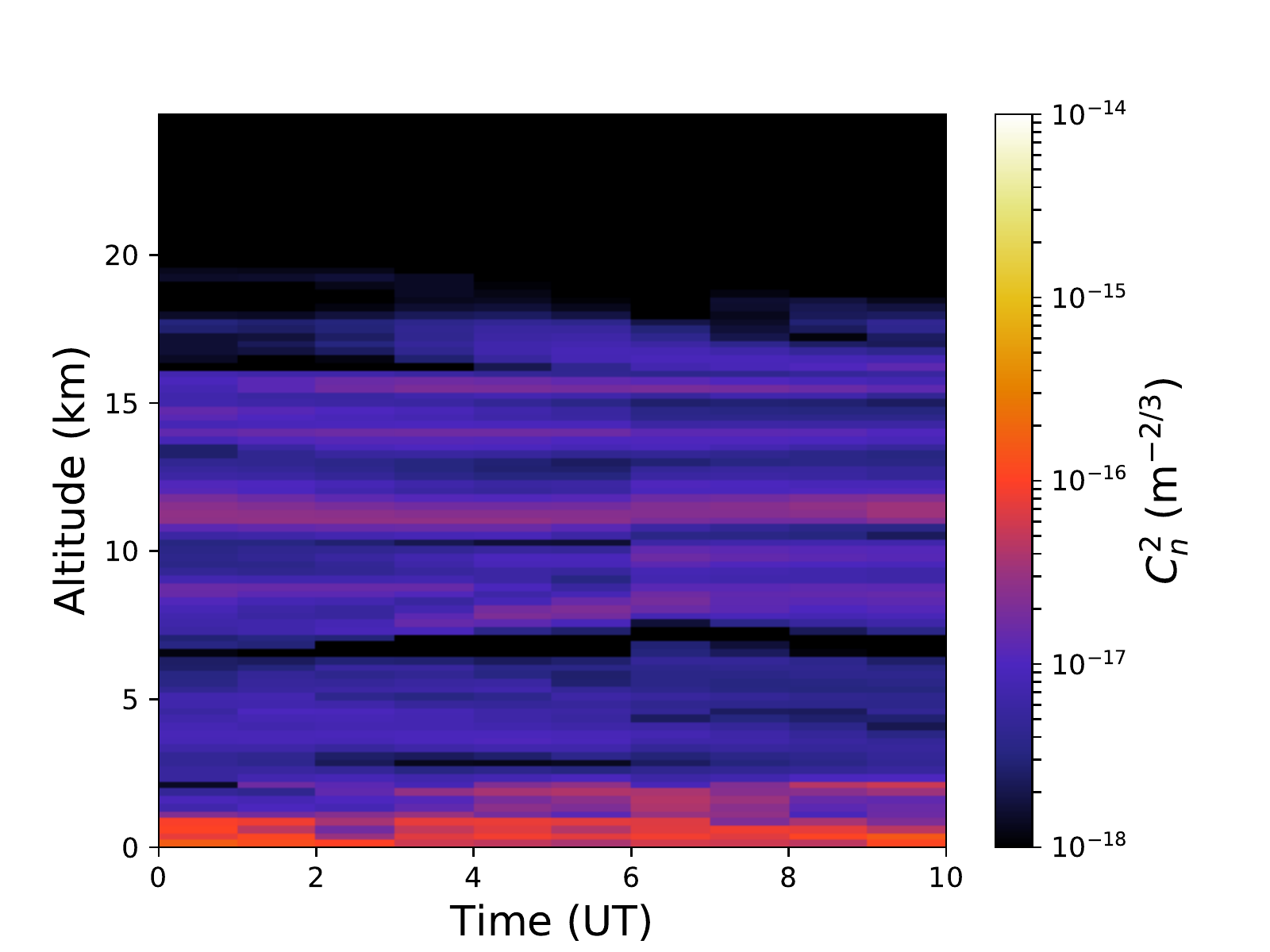} \\
    	\includegraphics[width=0.23\textwidth,trim={2cm 0 1cm 0}]{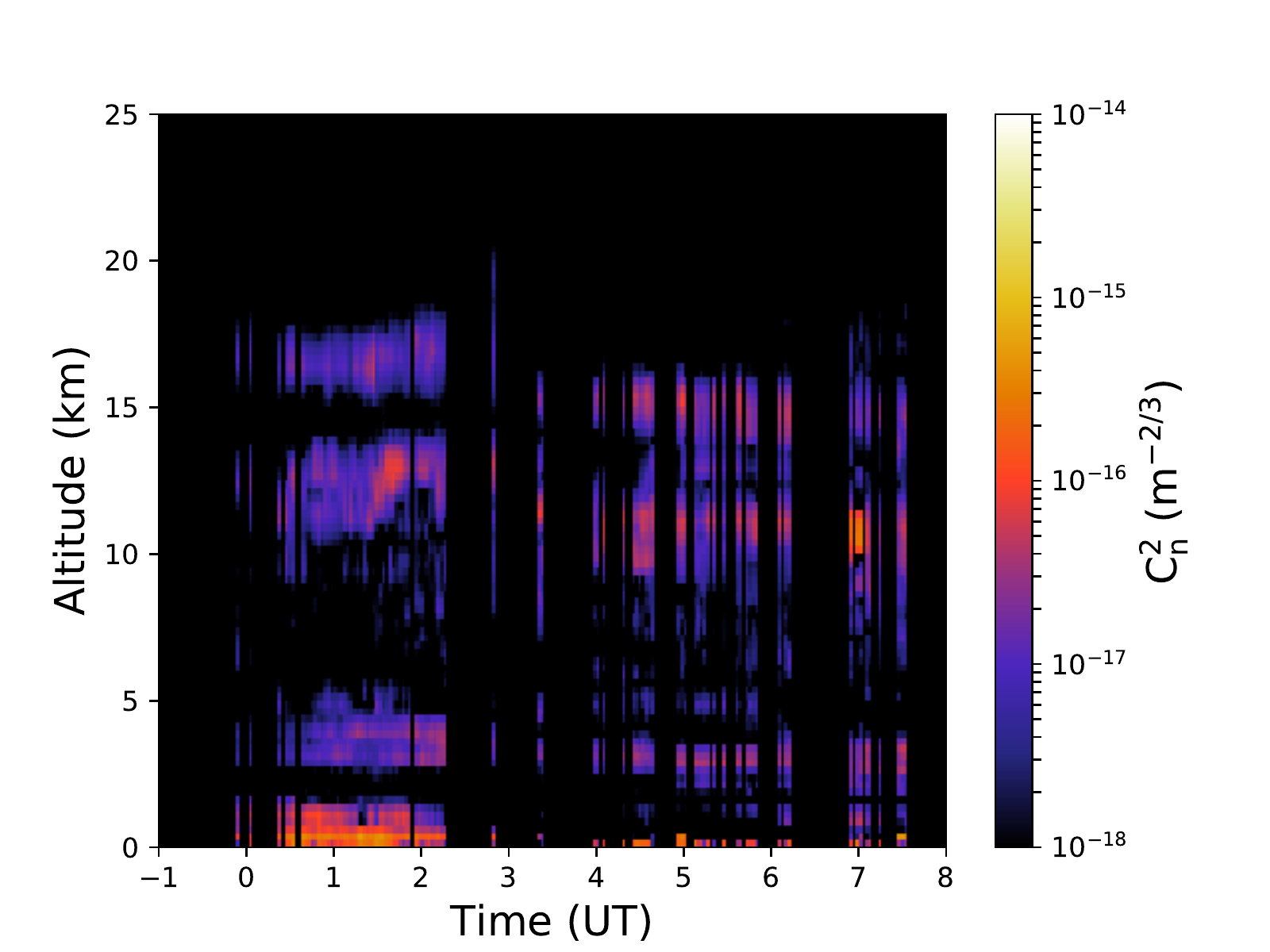} &
    	\includegraphics[width=0.23\textwidth,trim={2cm 0 1cm 0}]{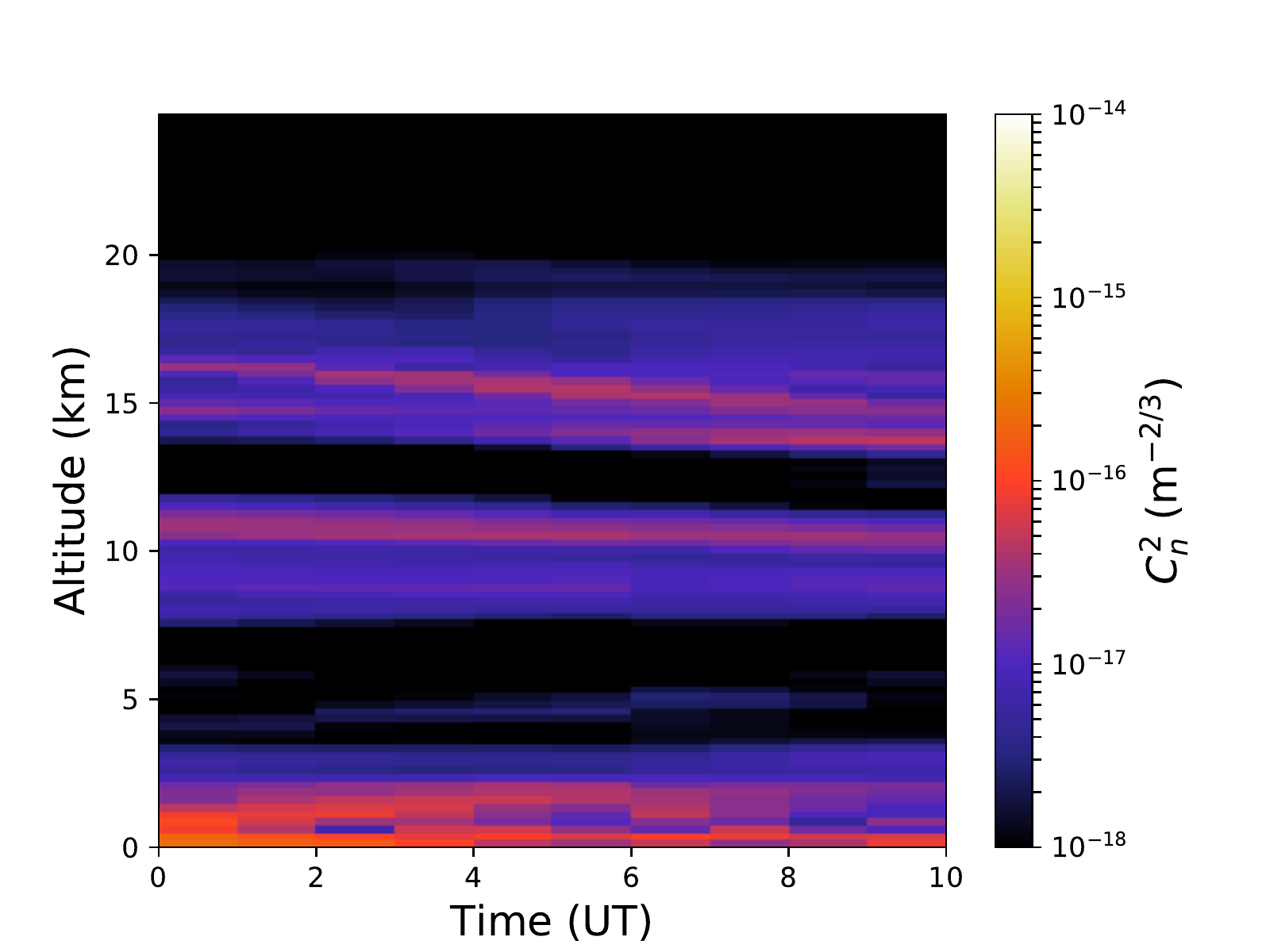} &
	\includegraphics[width=0.23\textwidth,trim={2cm 0 1cm 0}]{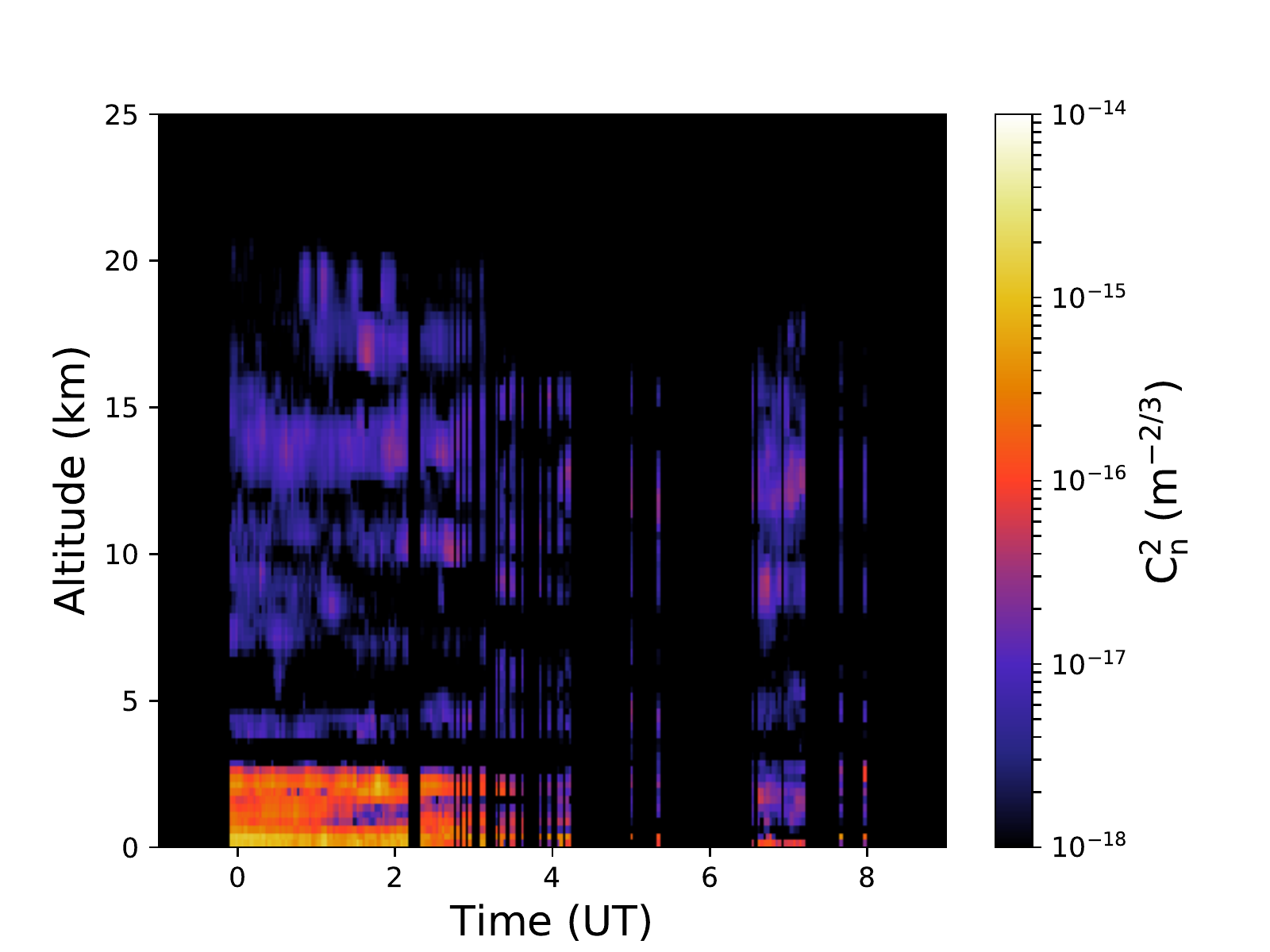} &
    	\includegraphics[width=0.23\textwidth,trim={2cm 0 1cm 0}]{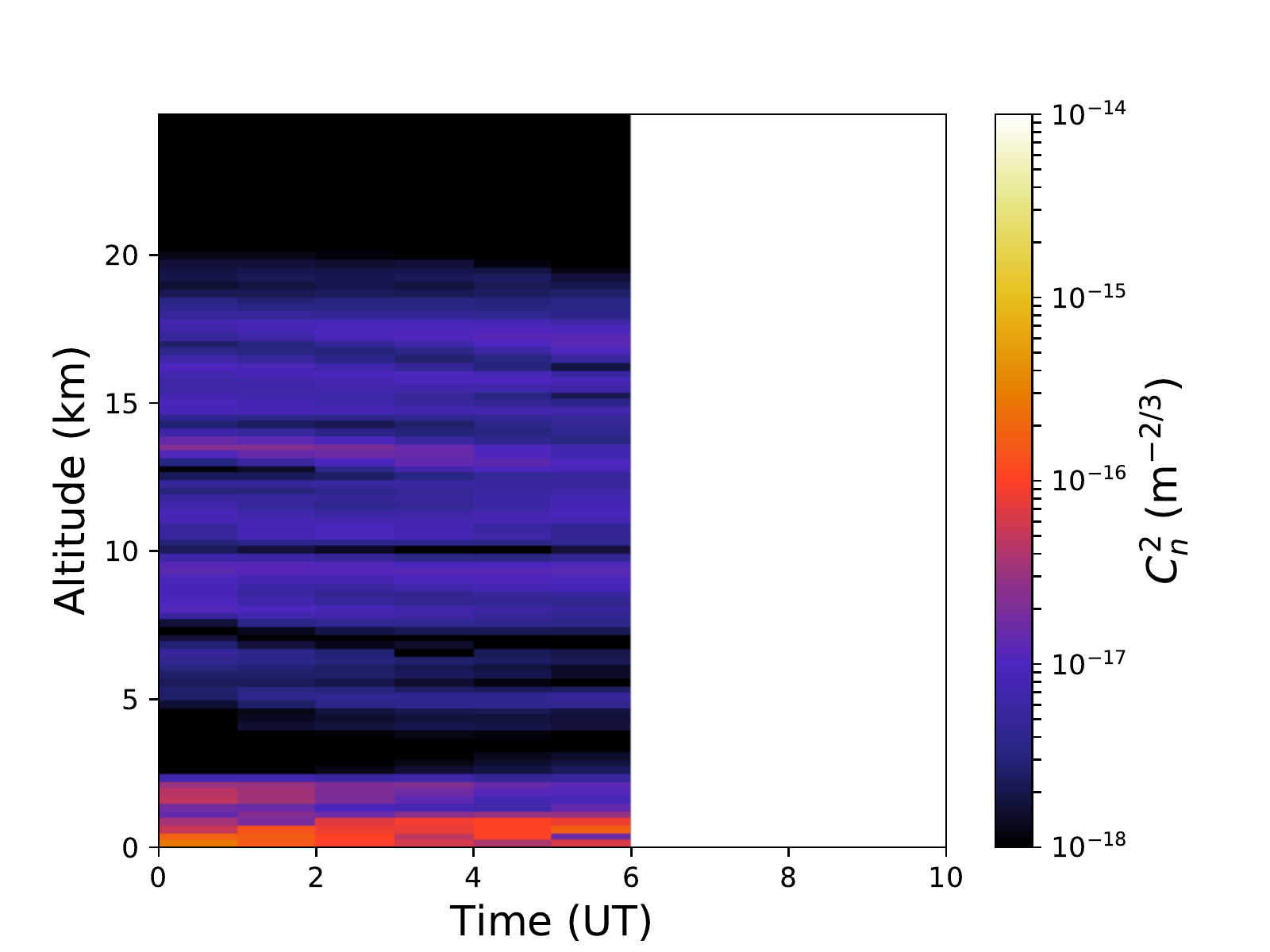} \\

\end{array}$
\caption{Example vertical profiles as measured by the stereo-SCIDAR (green) and estimated by the ECMWF GCM model (red). The profiles shown are the median for an individual night of observation. The coloured region shows the interquartile range. These profiles are from the nights beginning 18th - 19th, 29th - 30th November 2017, 1st, 5th, 8th-13th December.}
\label{fig:seqProfiles5}
\end{figure*}
	
\begin{figure*}
\centering
$\begin{array}{cccc}	
	\includegraphics[width=0.23\textwidth,trim={2cm 0 1cm 0}]{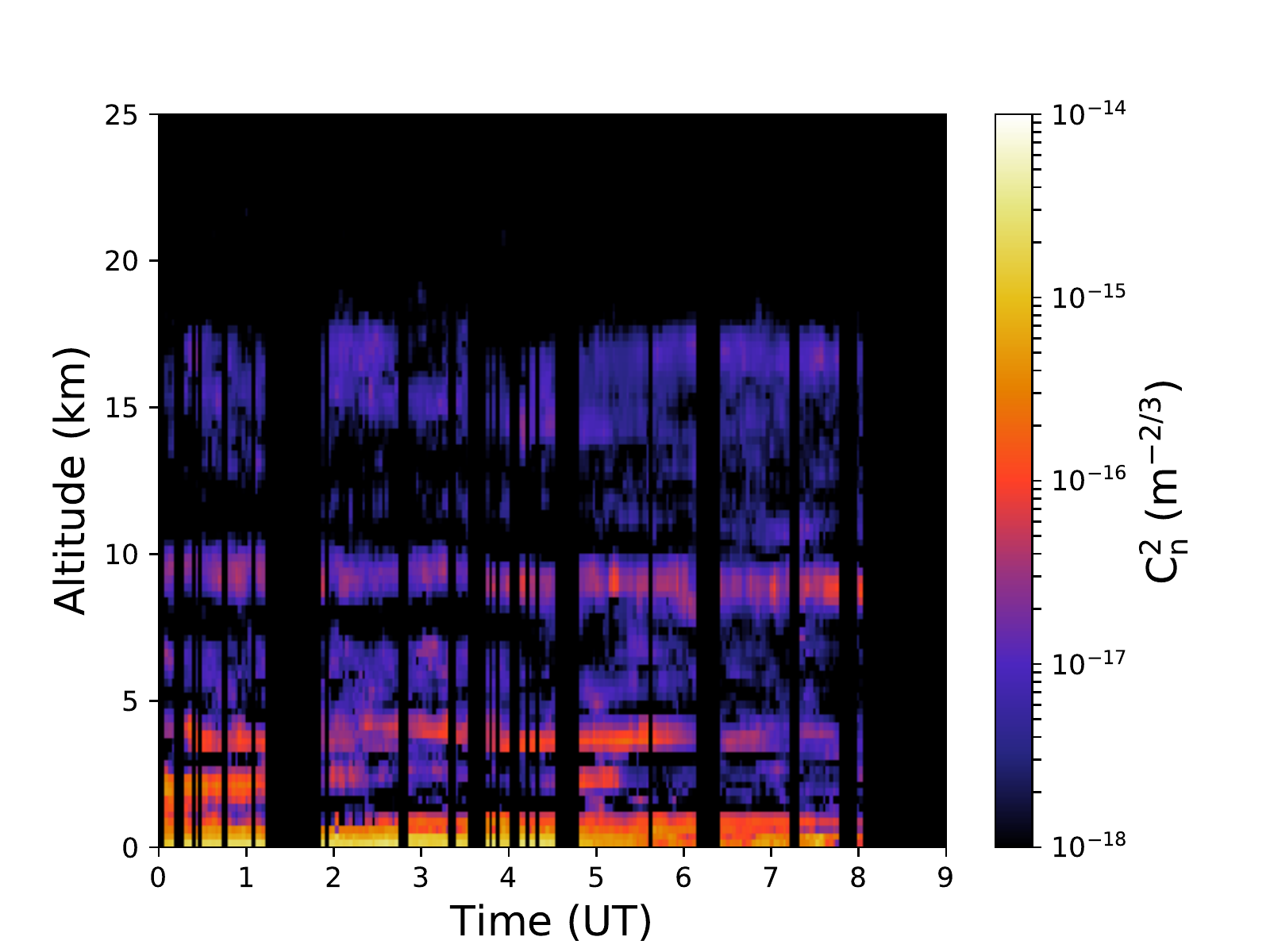} &
    	\includegraphics[width=0.23\textwidth,trim={2cm 0 1cm 0}]{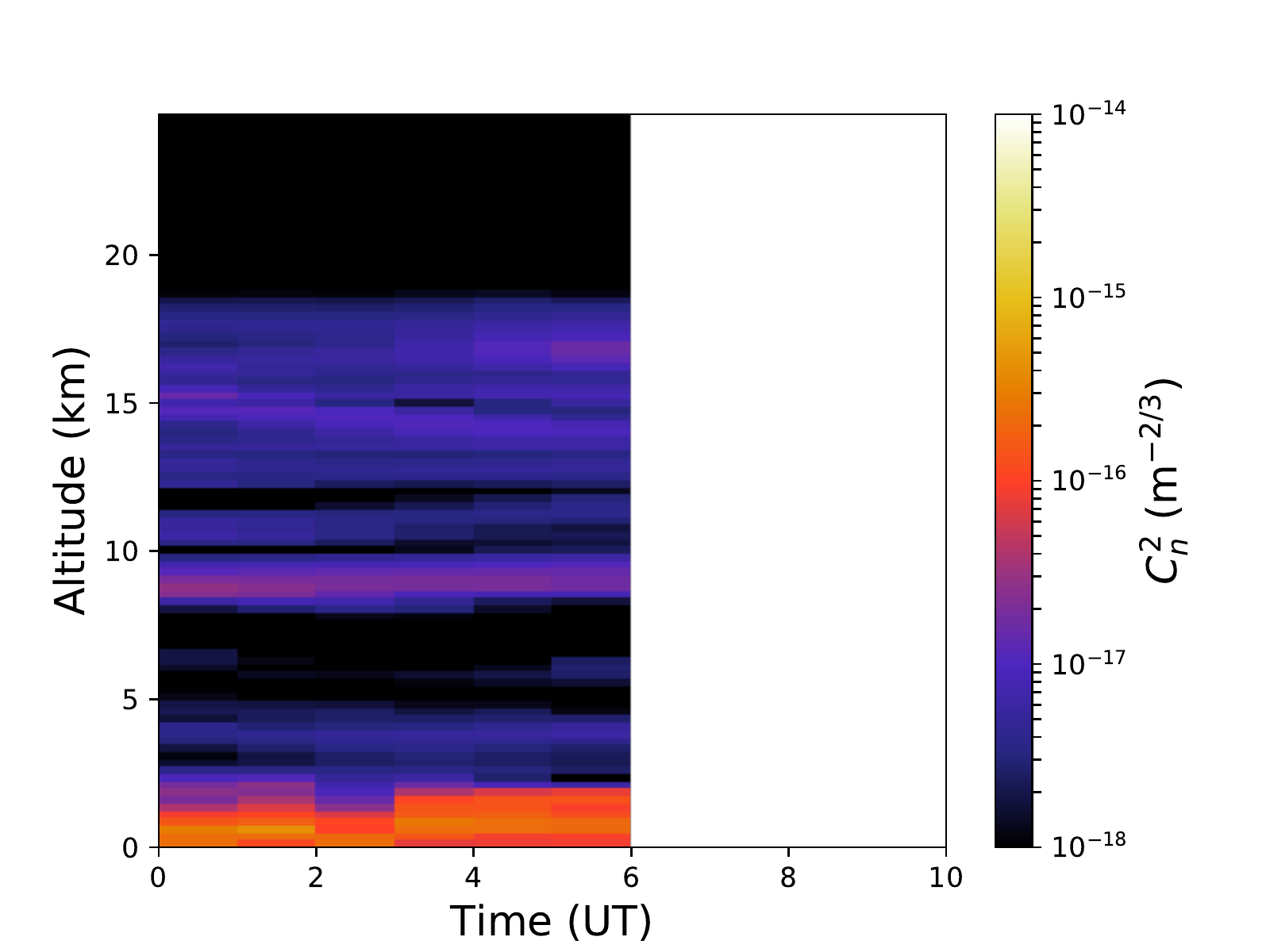} &	
	\includegraphics[width=0.23\textwidth,trim={2cm 0 1cm 0}]{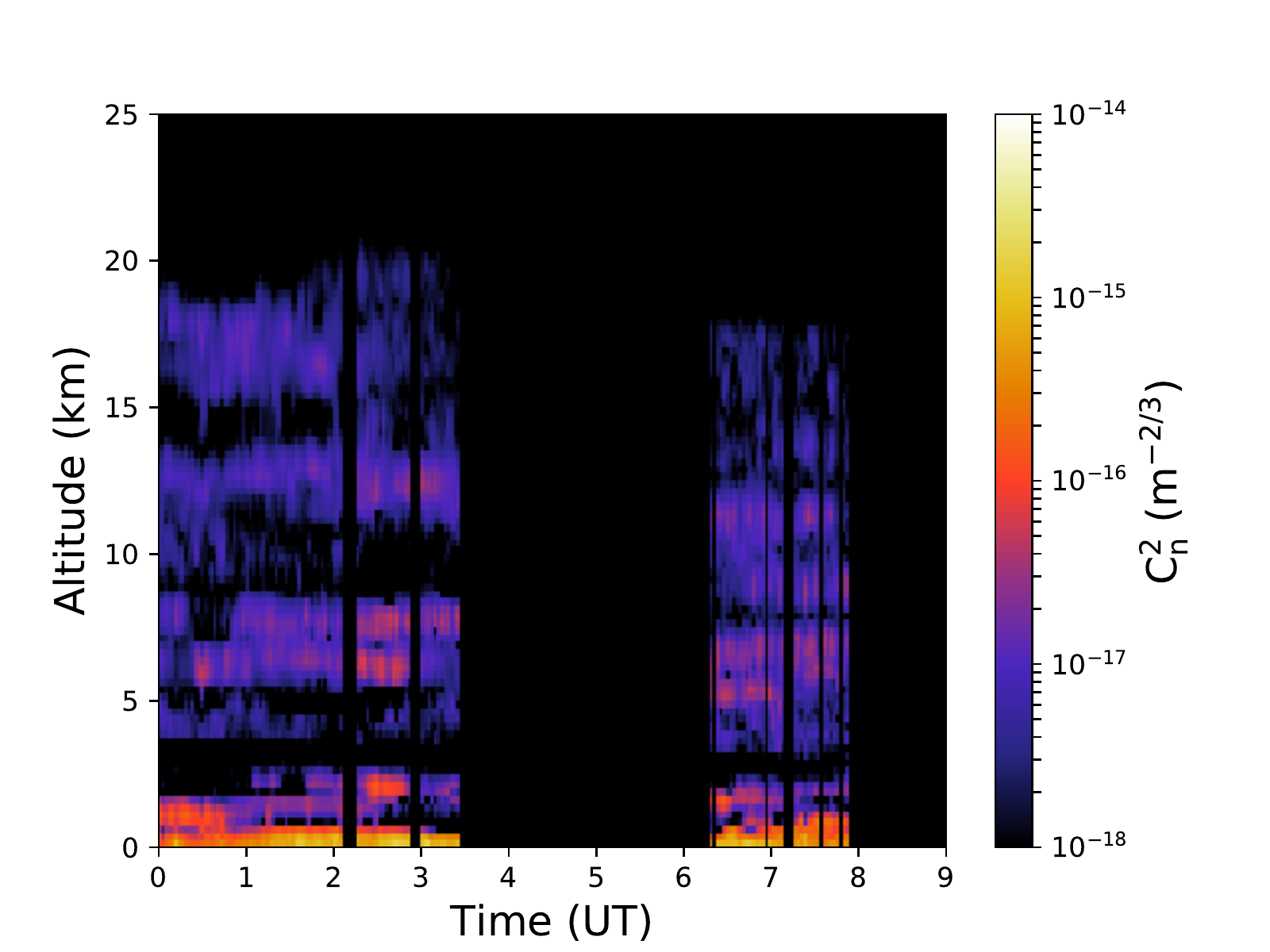} &
    	\includegraphics[width=0.23\textwidth,trim={2cm 0 1cm 0}]{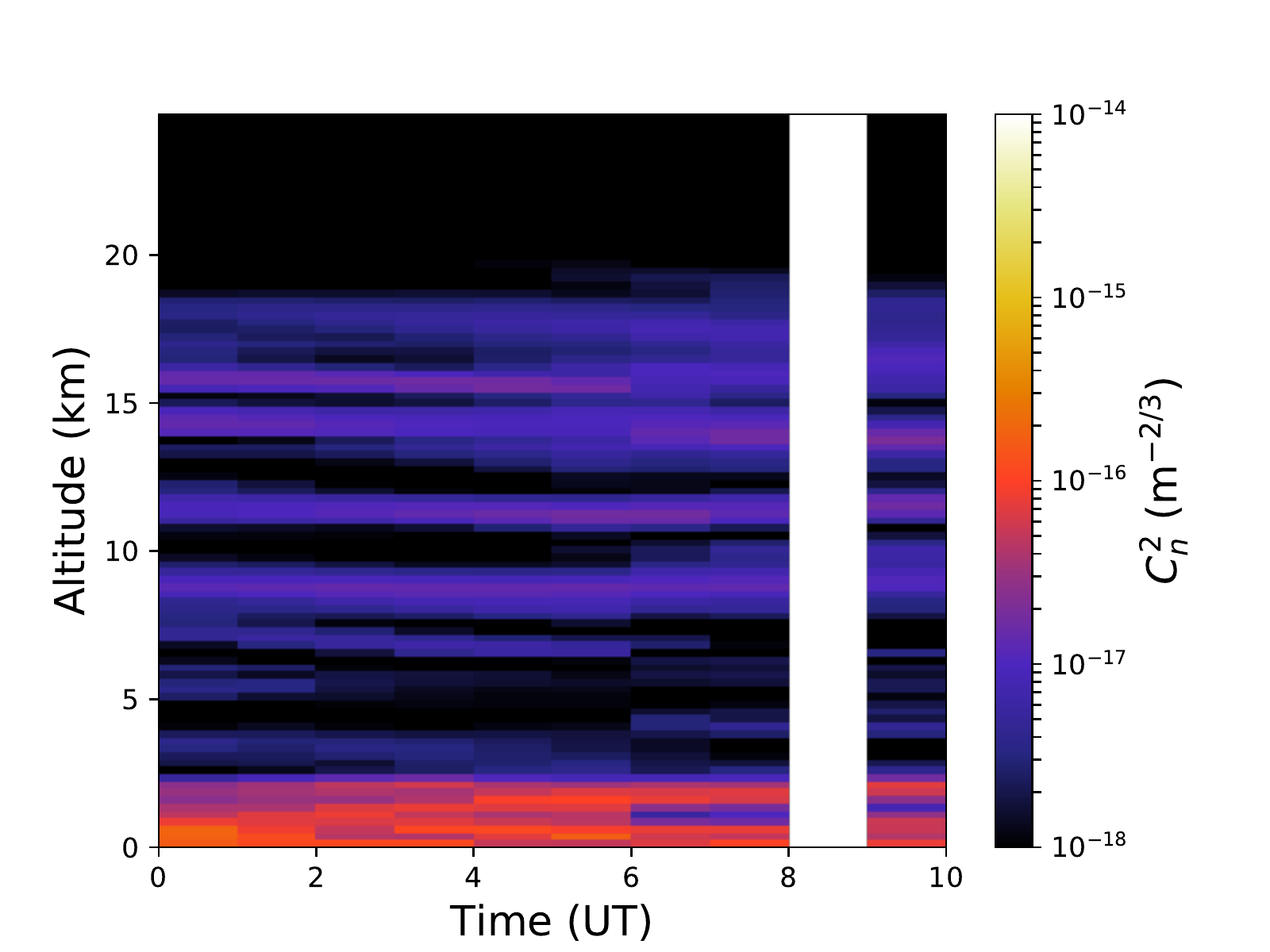} \\
    	\includegraphics[width=0.23\textwidth,trim={2cm 0 1cm 0}]{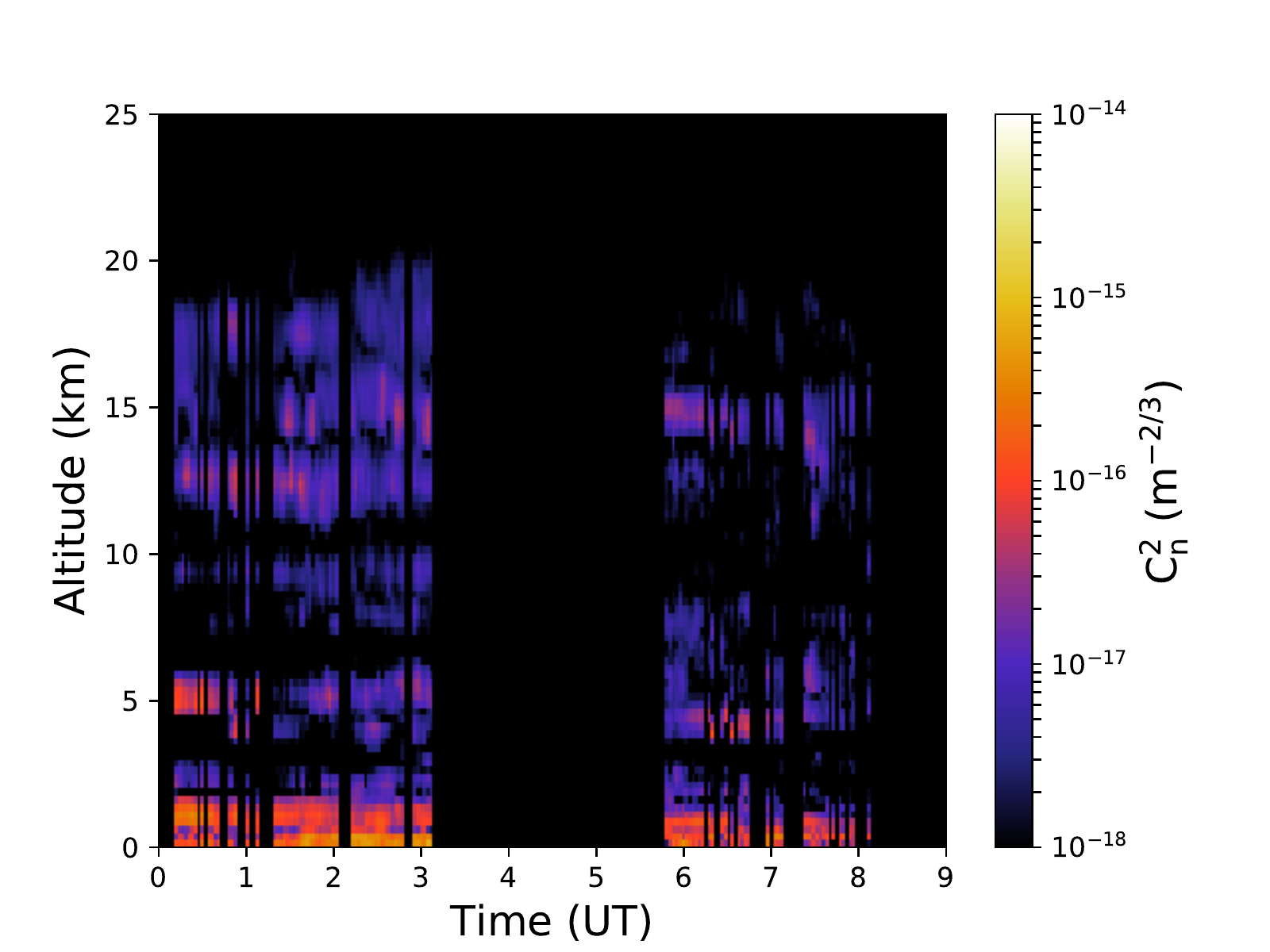} &
    	\includegraphics[width=0.23\textwidth,trim={2cm 0 1cm 0}]{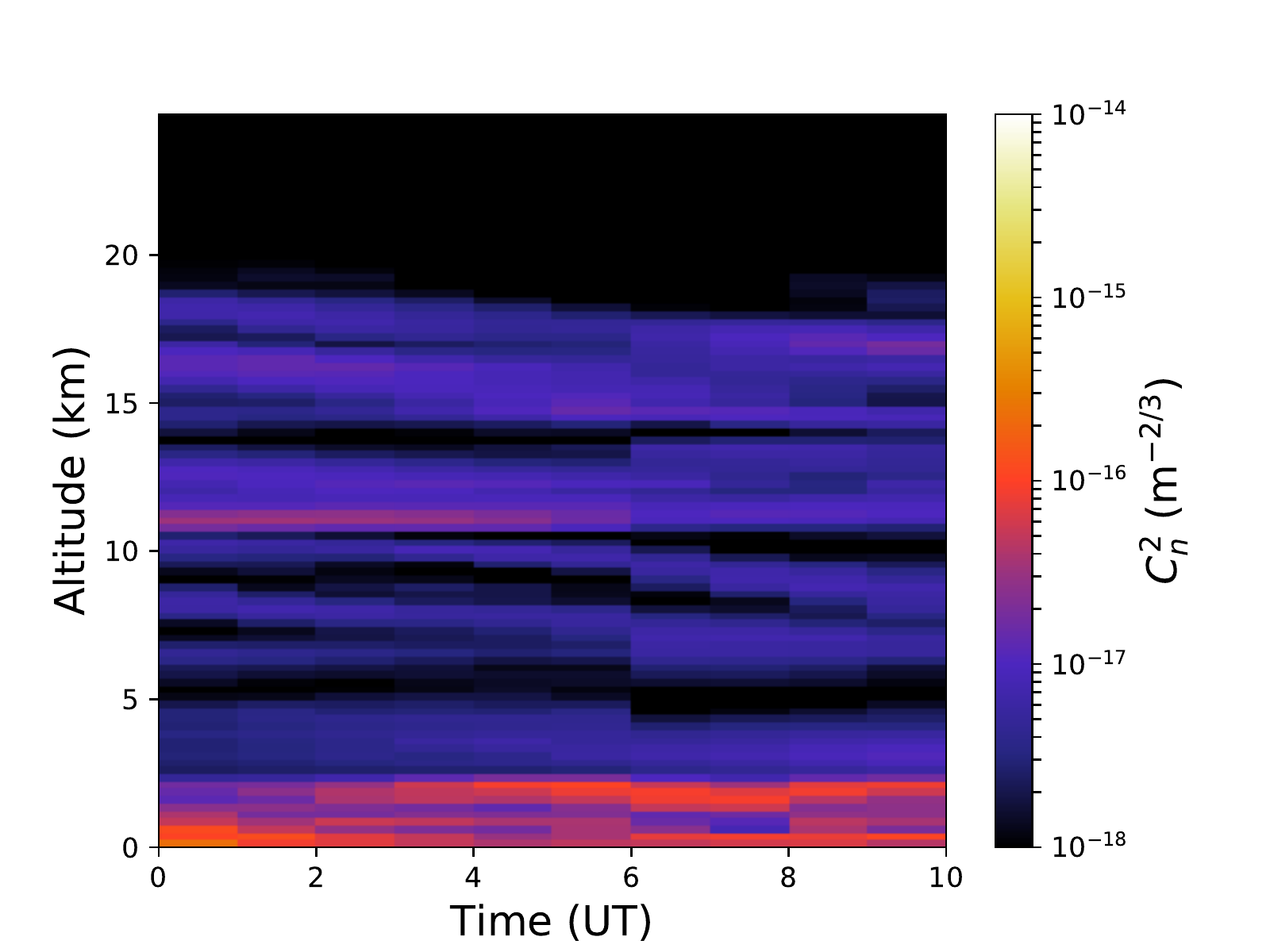} &	
	\includegraphics[width=0.23\textwidth,trim={2cm 0 1cm 0}]{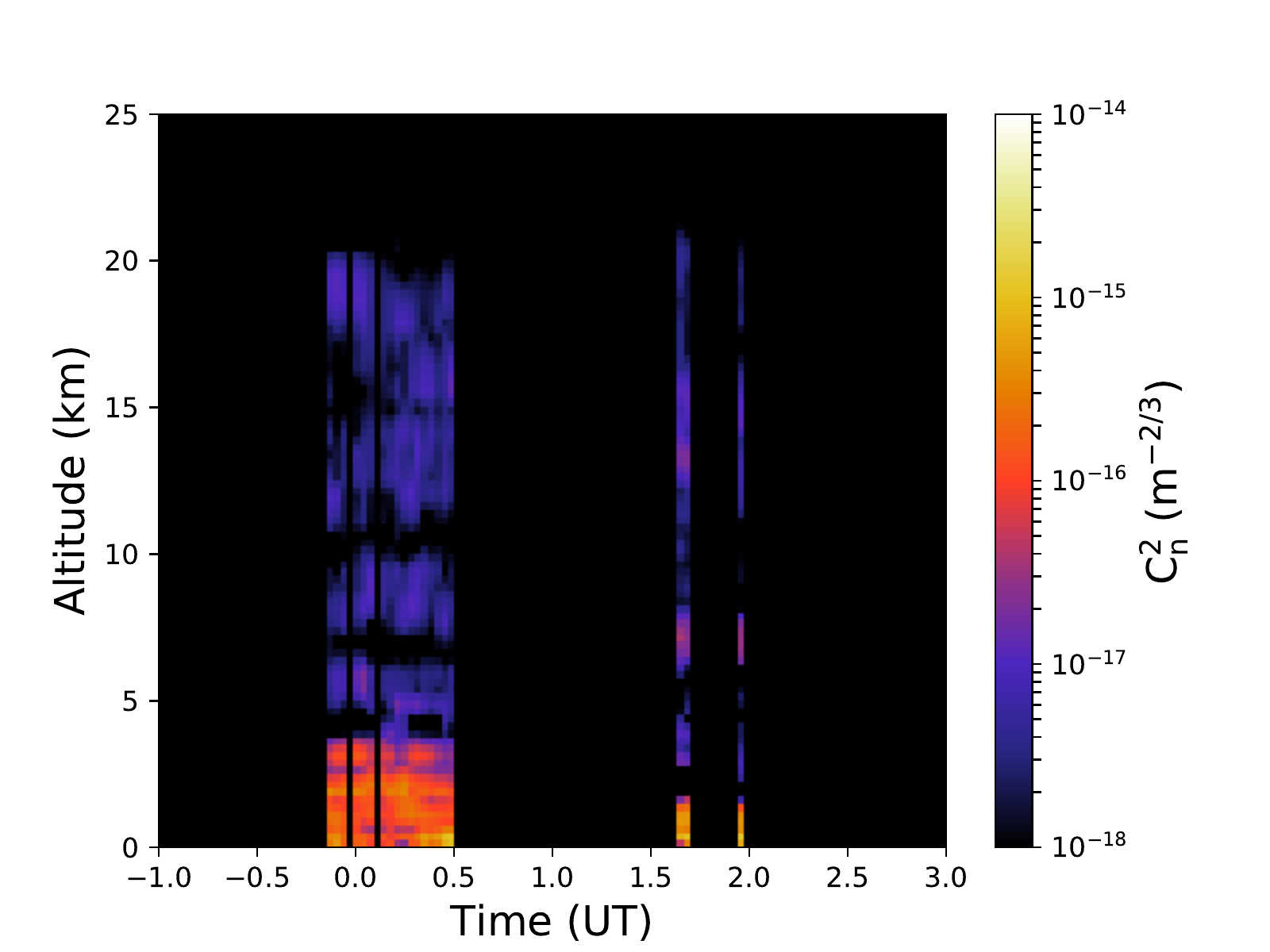} &
    	\includegraphics[width=0.23\textwidth,trim={2cm 0 1cm 0}]{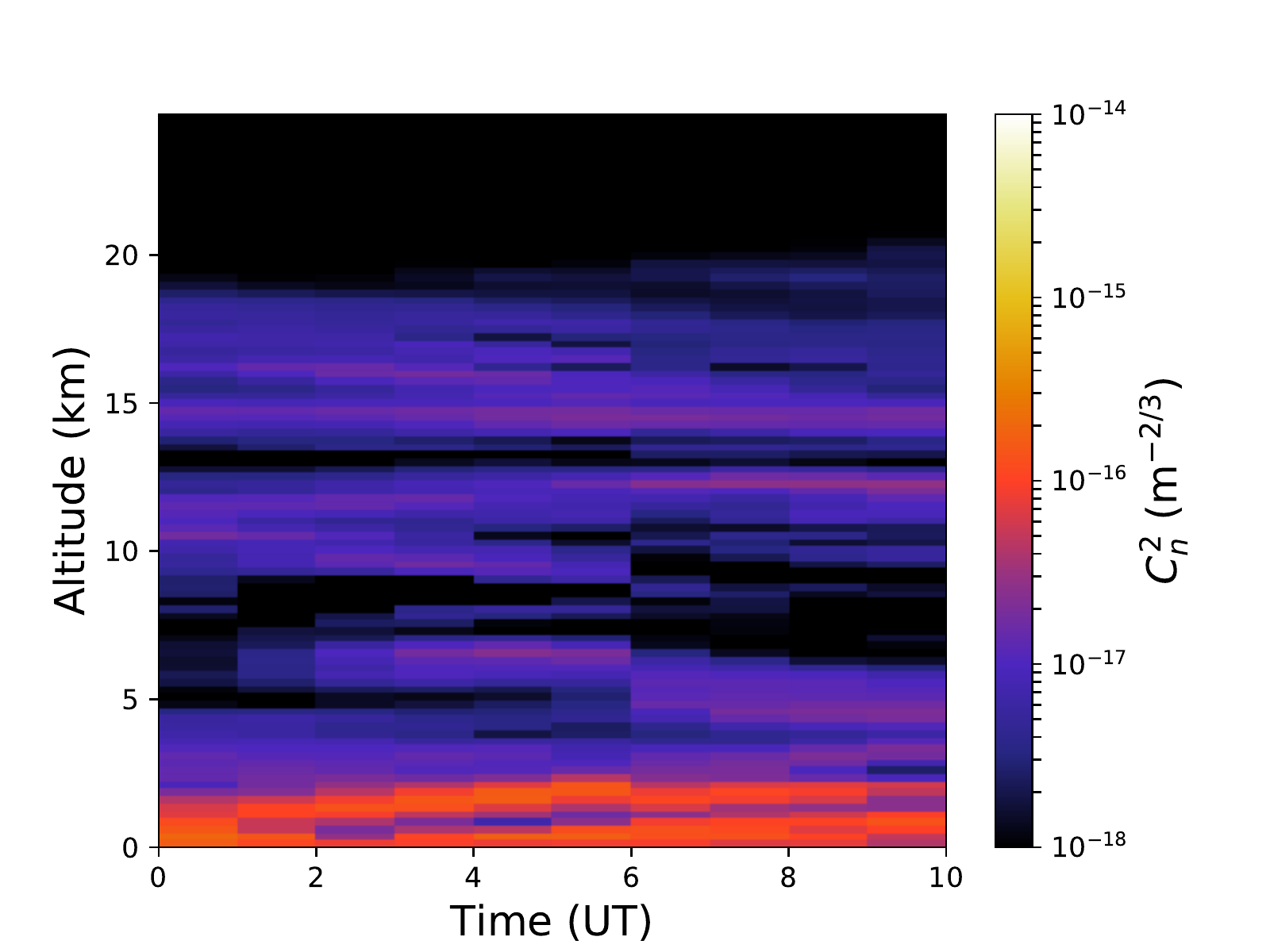} \\
	
	\includegraphics[width=0.23\textwidth,trim={2cm 0 1cm 0}]{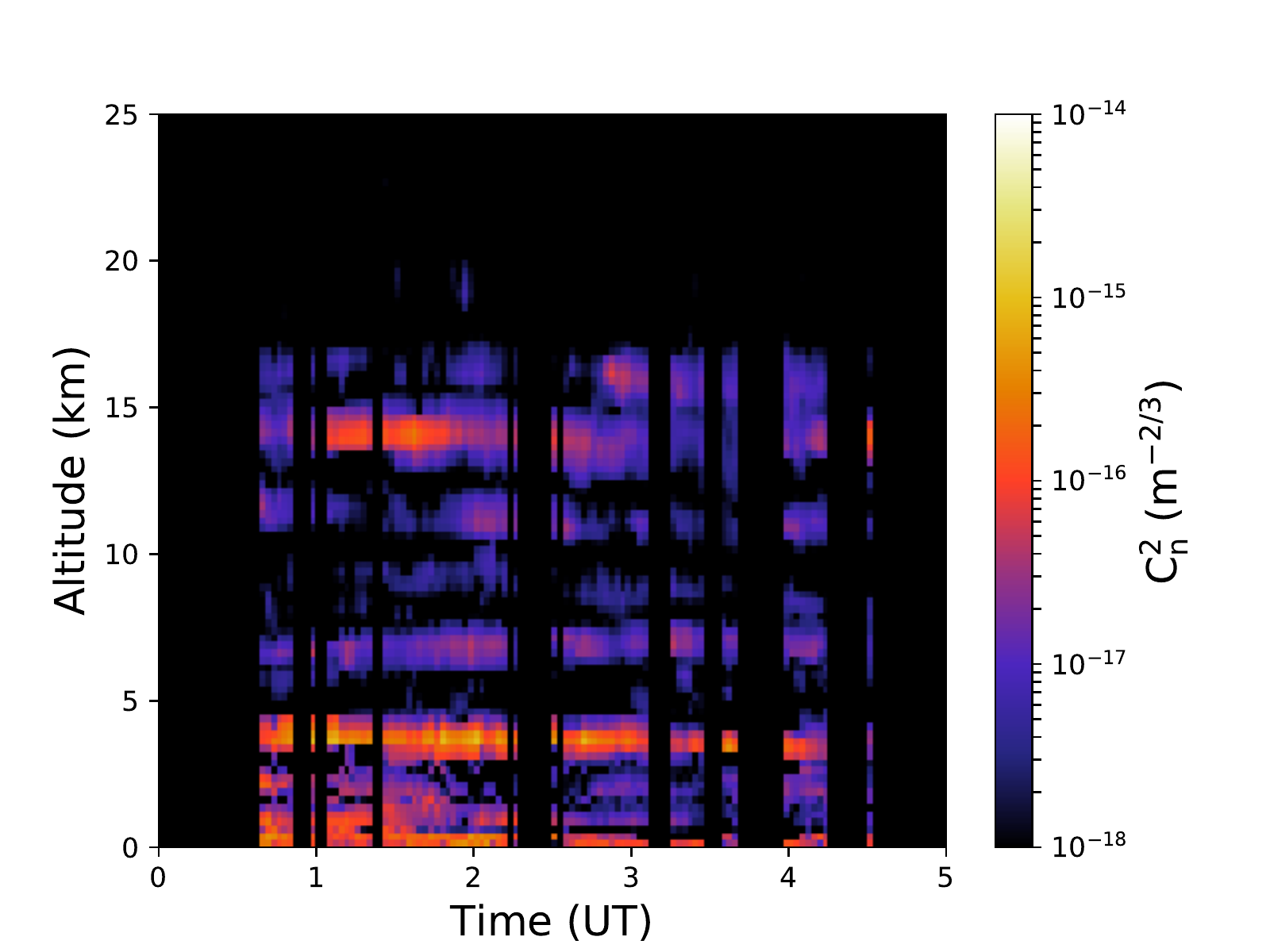} &
    	\includegraphics[width=0.23\textwidth,trim={2cm 0 1cm 0}]{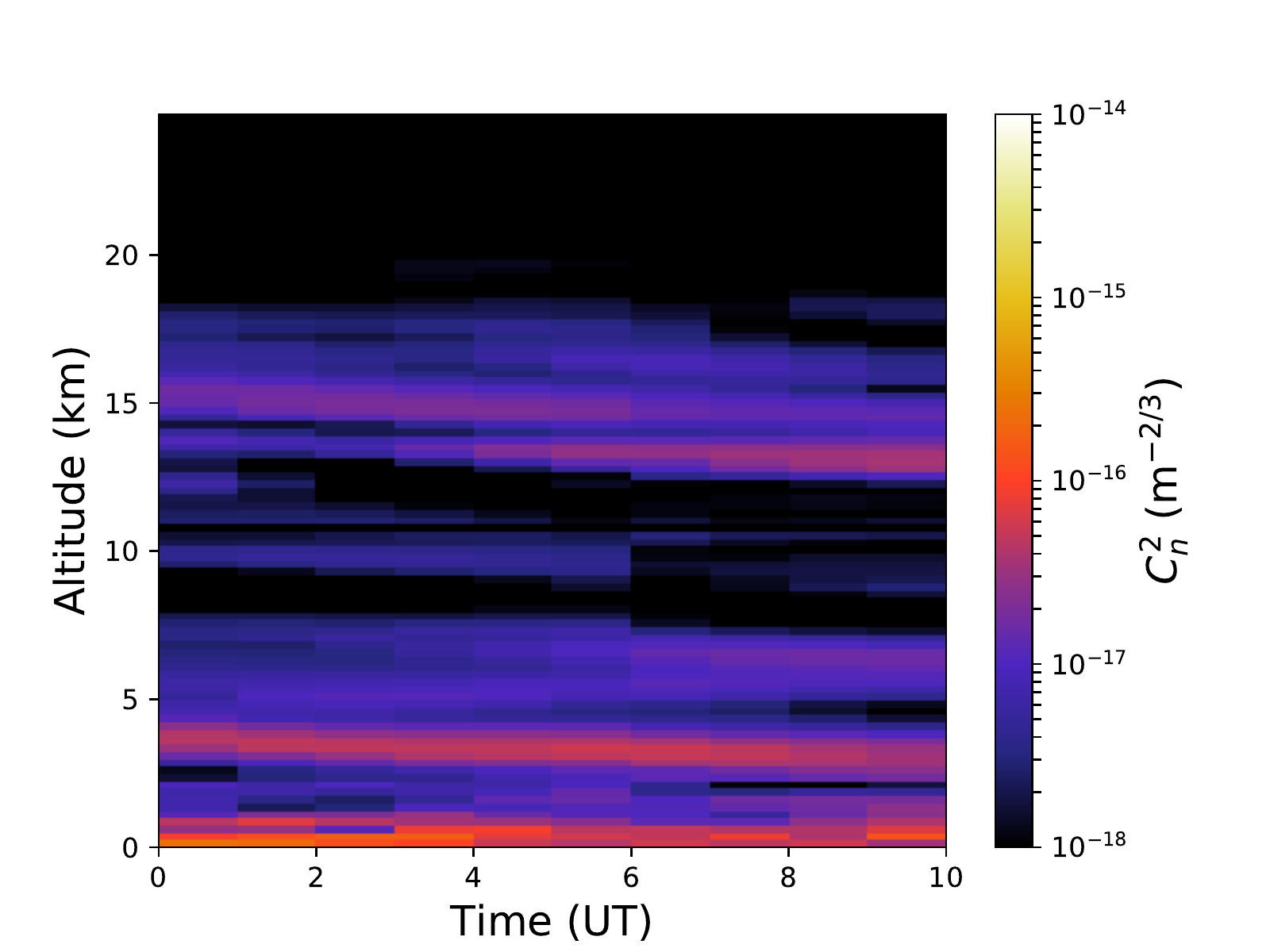} &
    	\includegraphics[width=0.23\textwidth,trim={2cm 0 1cm 0}]{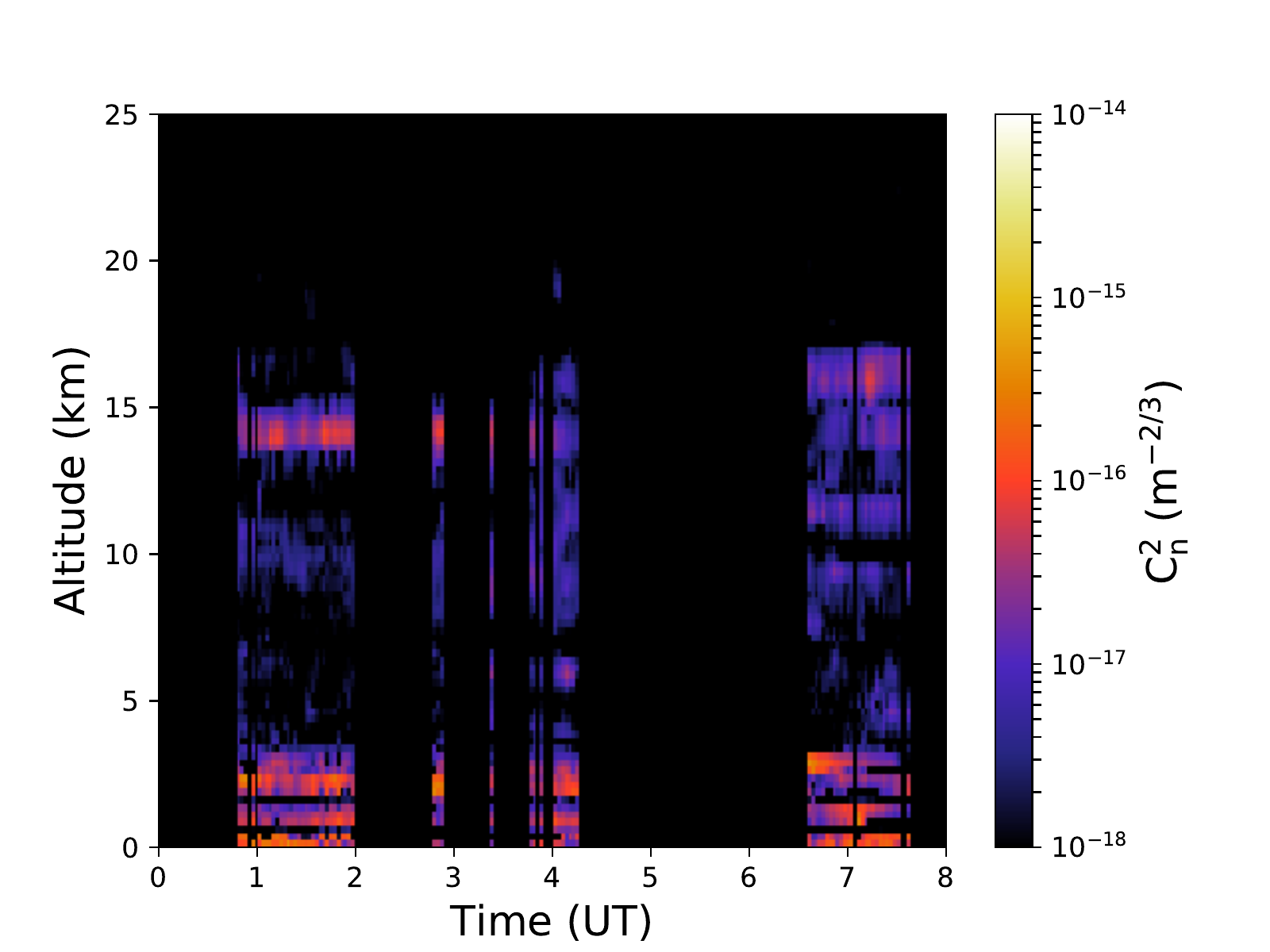} &
    	\includegraphics[width=0.23\textwidth,trim={2cm 0 1cm 0}]{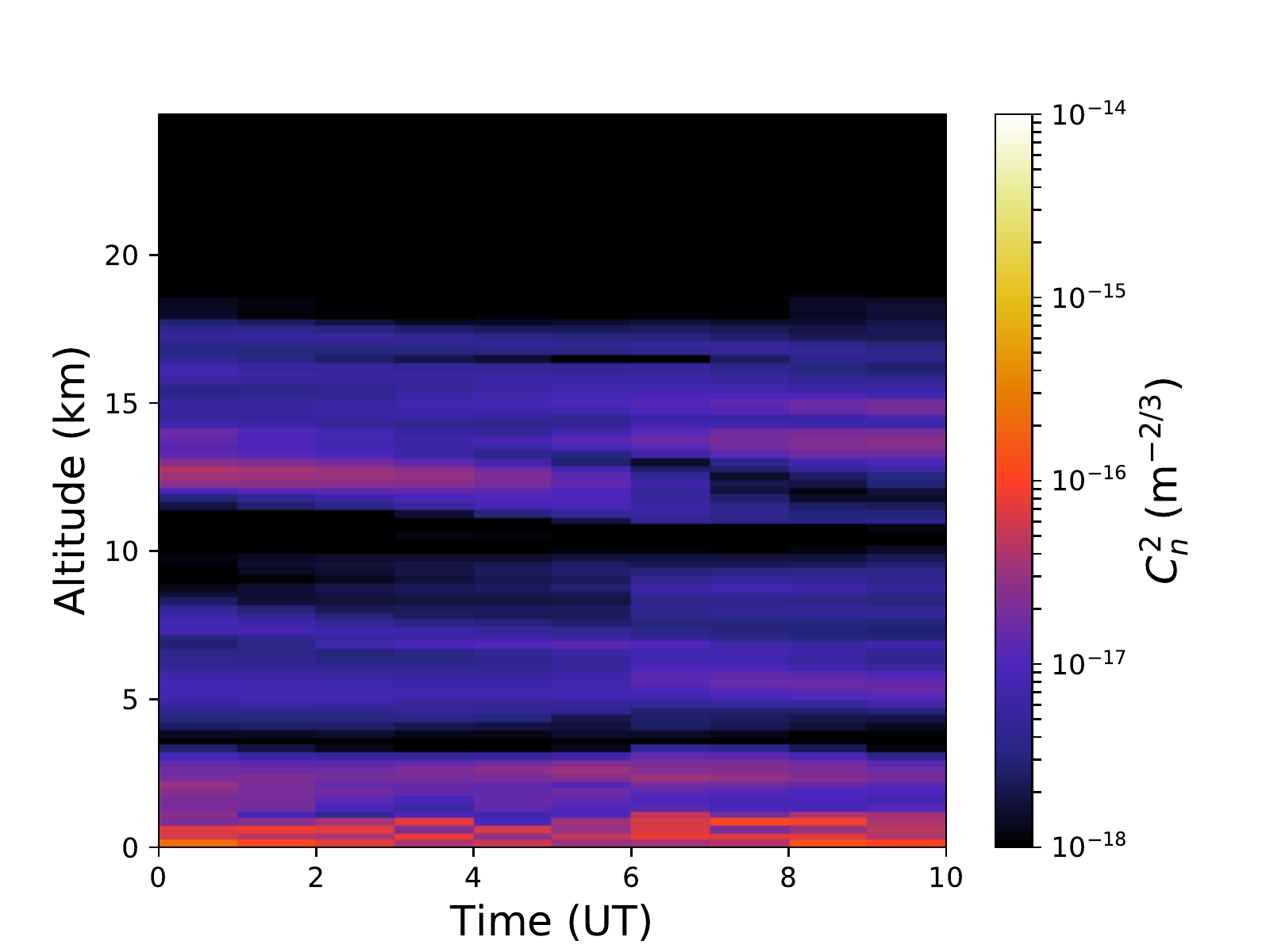} \\	
	\includegraphics[width=0.23\textwidth,trim={2cm 0 1cm 0}]{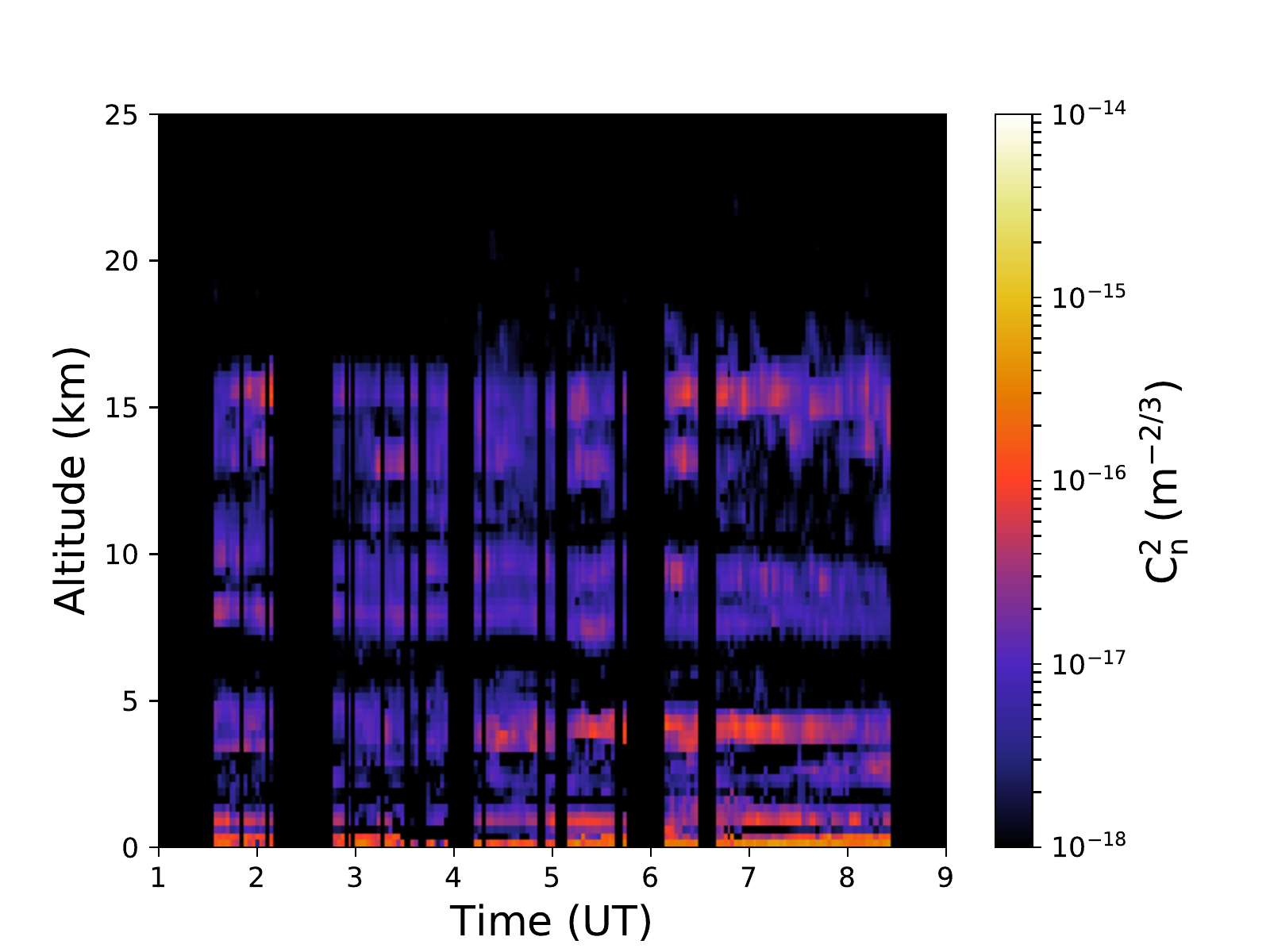} &
    	\includegraphics[width=0.23\textwidth,trim={2cm 0 1cm 0}]{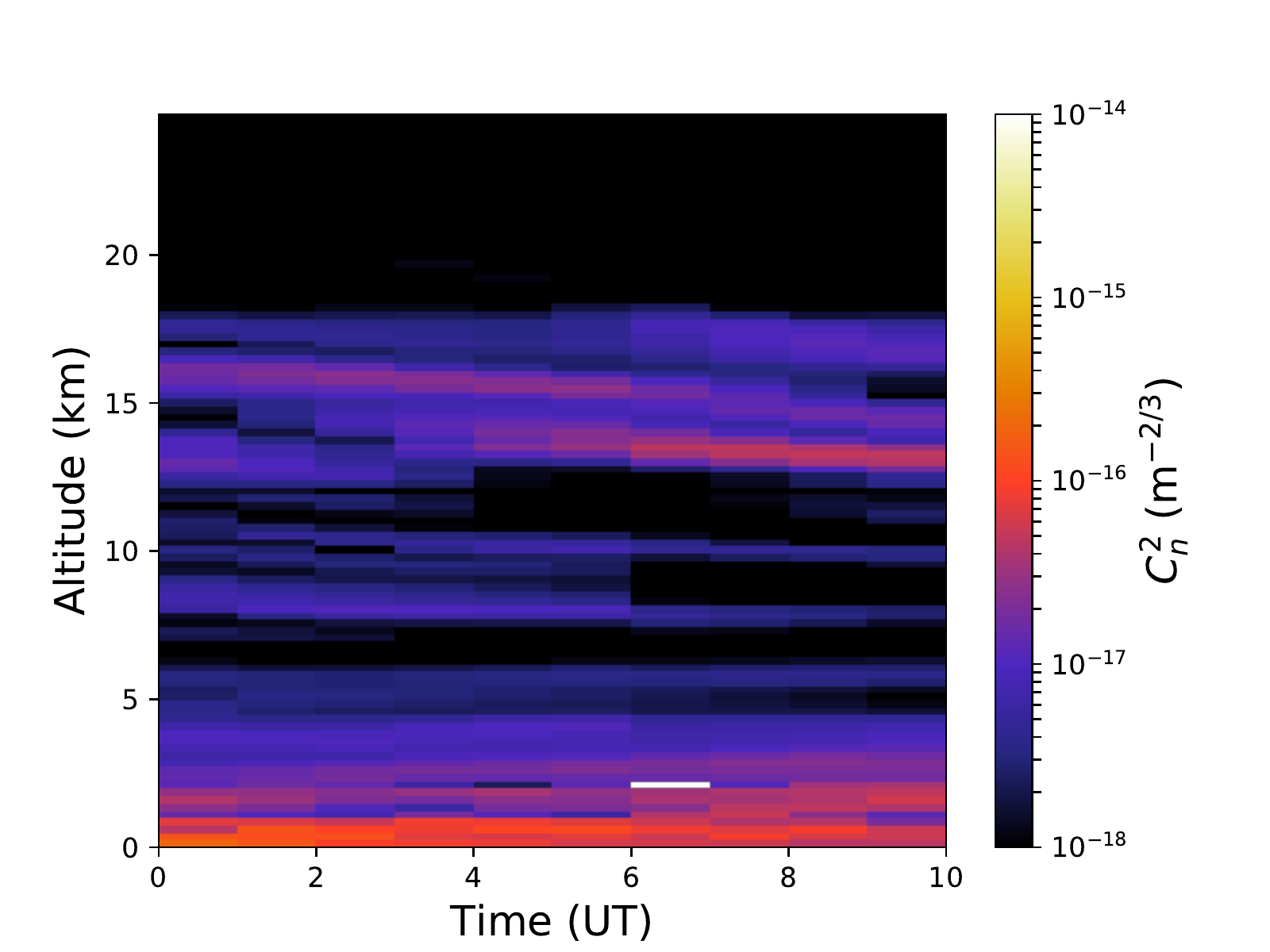} &
    	\includegraphics[width=0.23\textwidth,trim={2cm 0 1cm 0}]{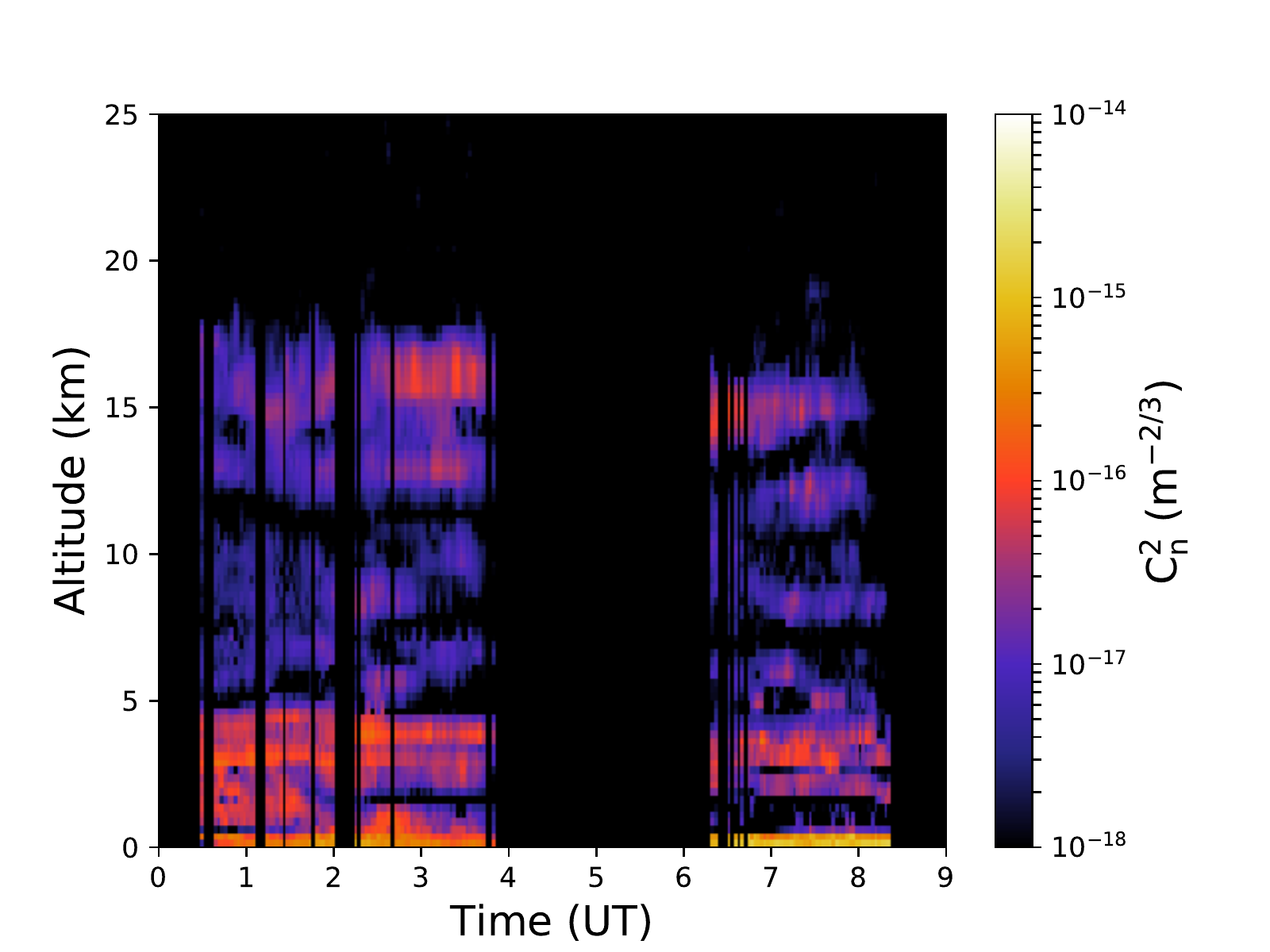} &
    	\includegraphics[width=0.23\textwidth,trim={2cm 0 1cm 0}]{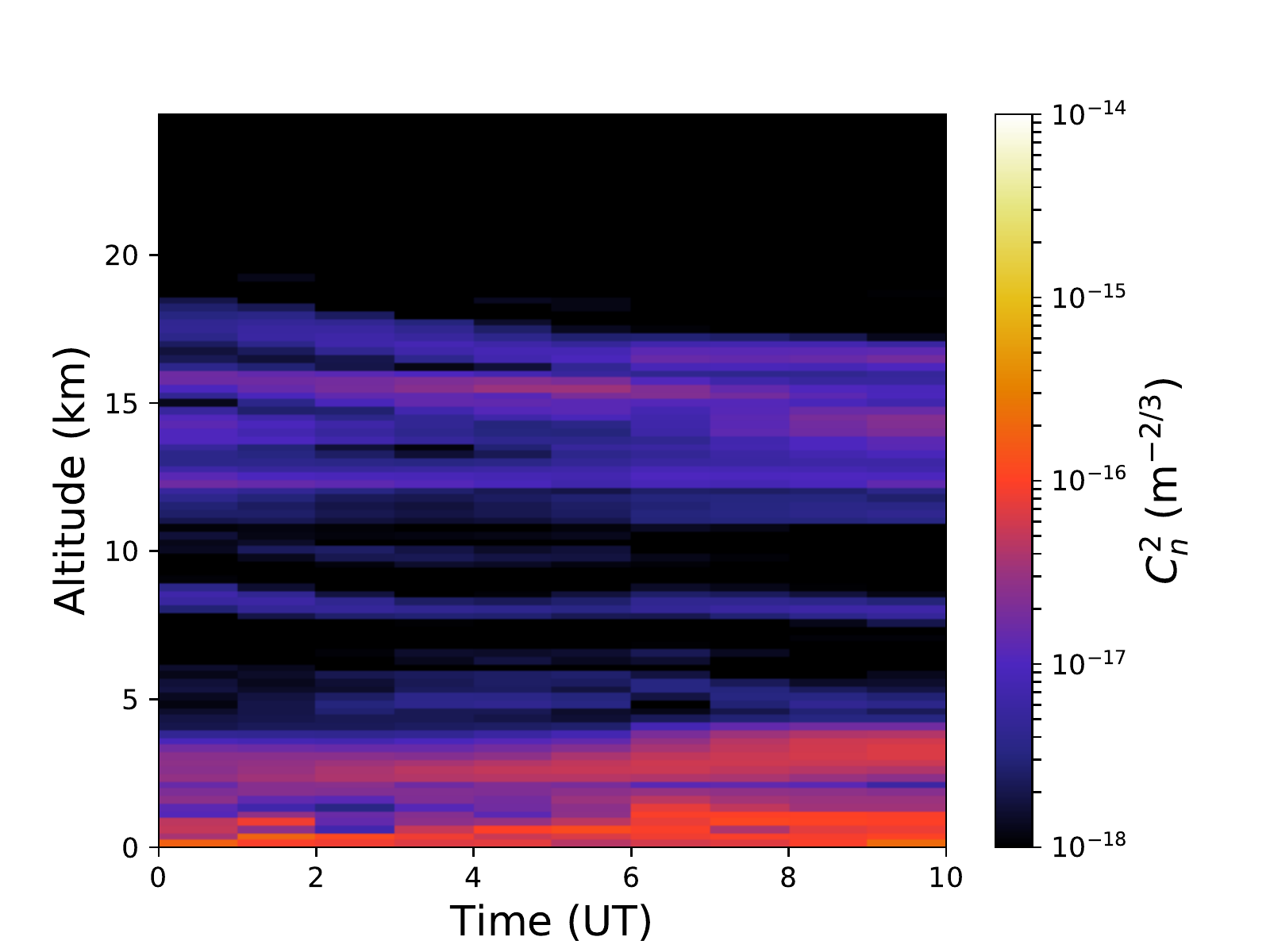} \\	
	
\end{array}$
\caption{Example vertical profiles as measured by the stereo-SCIDAR (green) and estimated by the ECMWF GCM model (red). The profiles shown are the median for an individual night of observation. The coloured region shows the interquartile range. These profiles are from the nights beginning 14th - 17th December 2017 and 13th - 15th, 18th-19th January 2018.}
\label{fig:seqProfiles6}
\end{figure*}

\section{Nightly median profiles}
\label{sect:nightMedian}
\begin{figure*}
\centering
$\begin{array}{ccc}
	\includegraphics[width=0.3\textwidth,trim={1cm 0 0cm 0}]{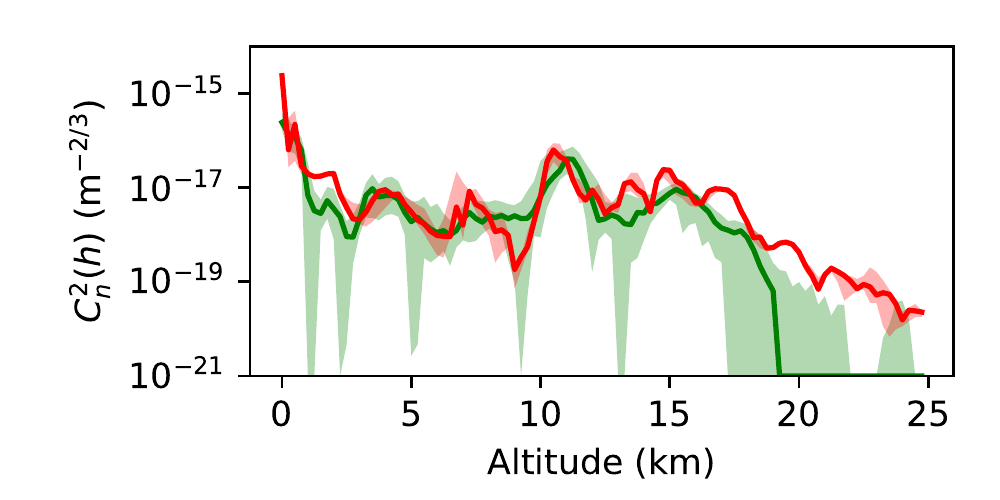} &
    \includegraphics[width=0.3\textwidth,trim={1cm 0 0cm 0}]{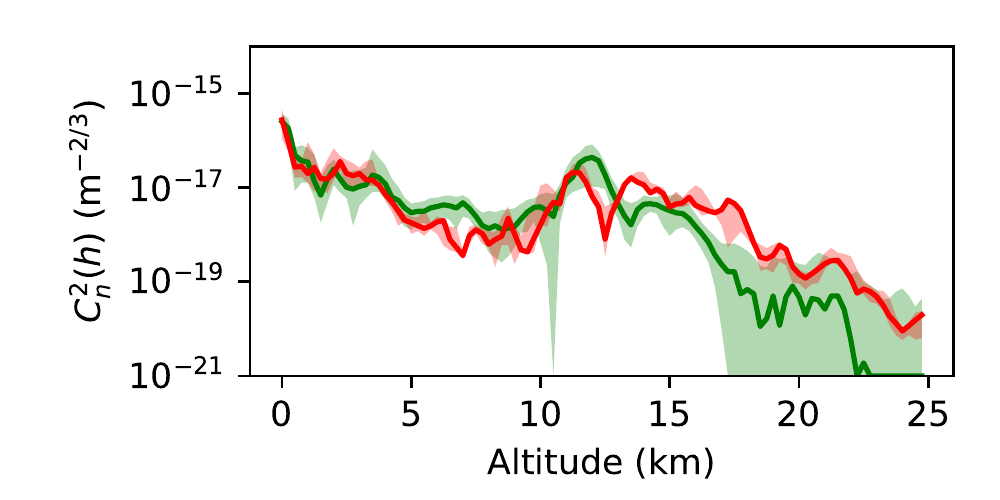} &
    \includegraphics[width=0.3\textwidth,trim={1cm 0 0cm 0}]{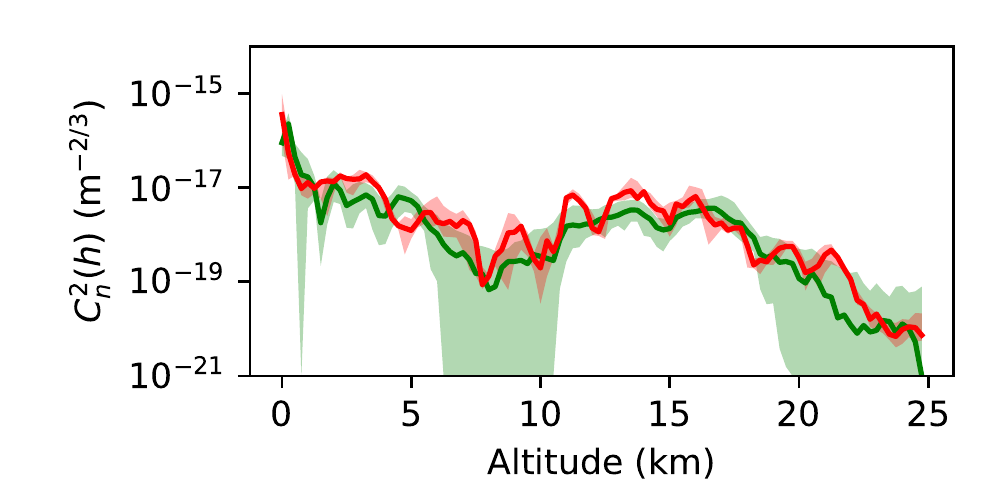}\\
    \includegraphics[width=0.3\textwidth,trim={1cm 0 0cm 0}]{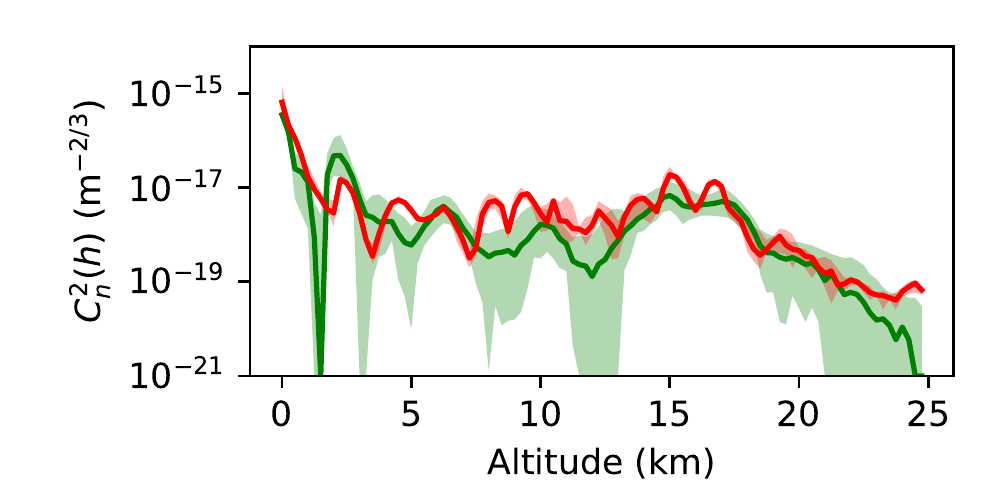} &
    
    \includegraphics[width=0.3\textwidth,trim={1cm 0 0cm 0}]{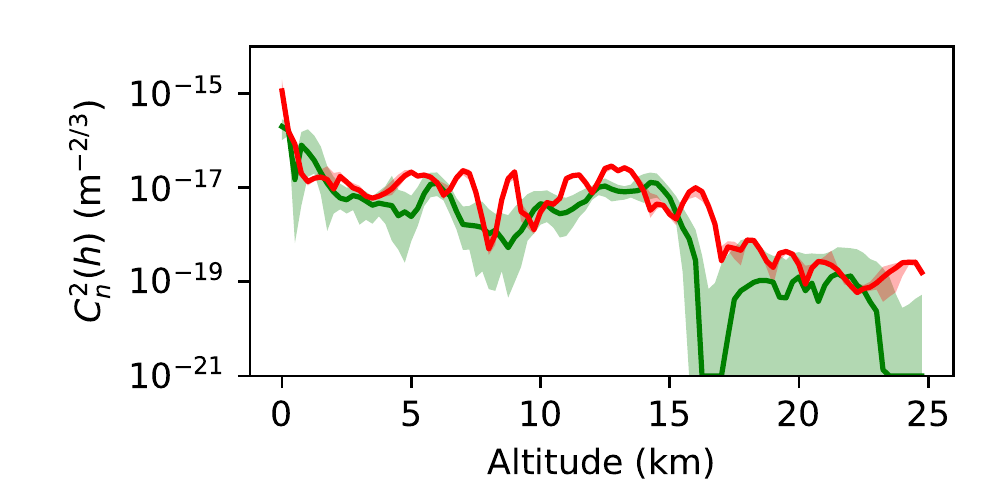} &
    \includegraphics[width=0.3\textwidth,trim={1cm 0 0cm 0}]{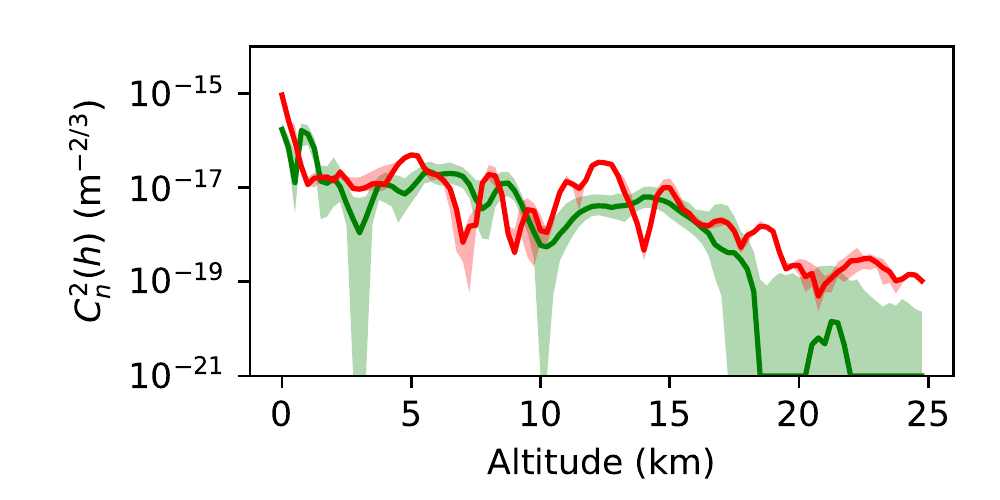}\\
    \includegraphics[width=0.3\textwidth,trim={1cm 0 0cm 0}]{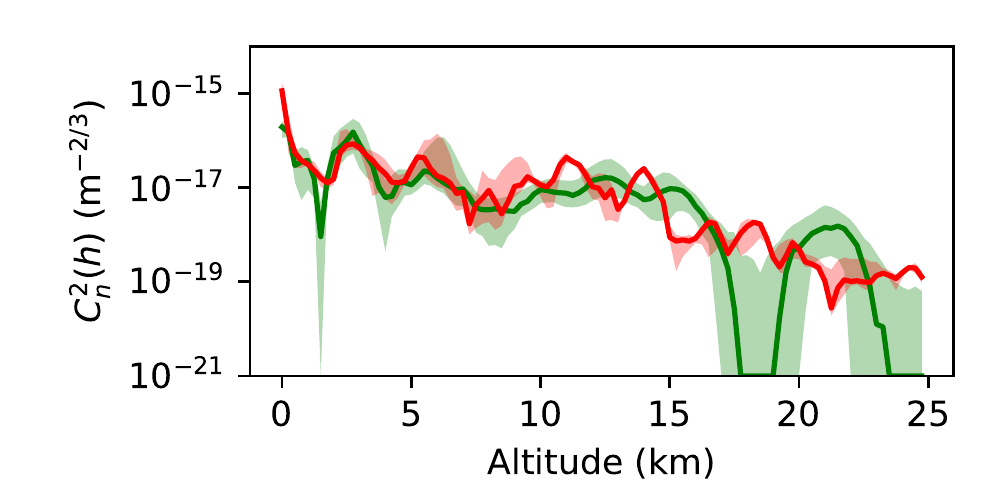} &
    \includegraphics[width=0.3\textwidth,trim={1cm 0 0cm 0}]{images/medianScidarEcmwf_PAR_20160726.pdf} &
    \includegraphics[width=0.3\textwidth,trim={1cm 0 0cm 0}]{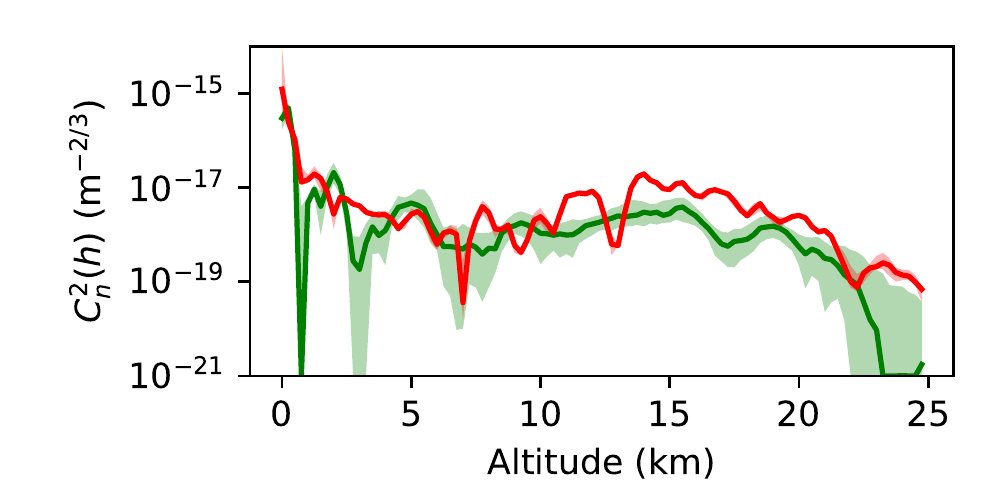}\\
    
    \includegraphics[width=0.3\textwidth,trim={1cm 0 0cm 0}]{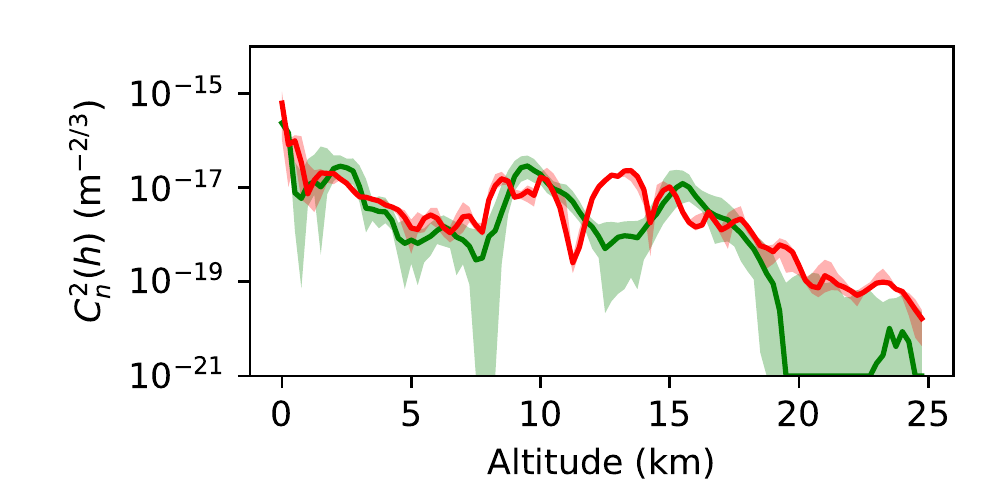} &
	\includegraphics[width=0.3\textwidth,trim={1cm 0 0cm 0}]{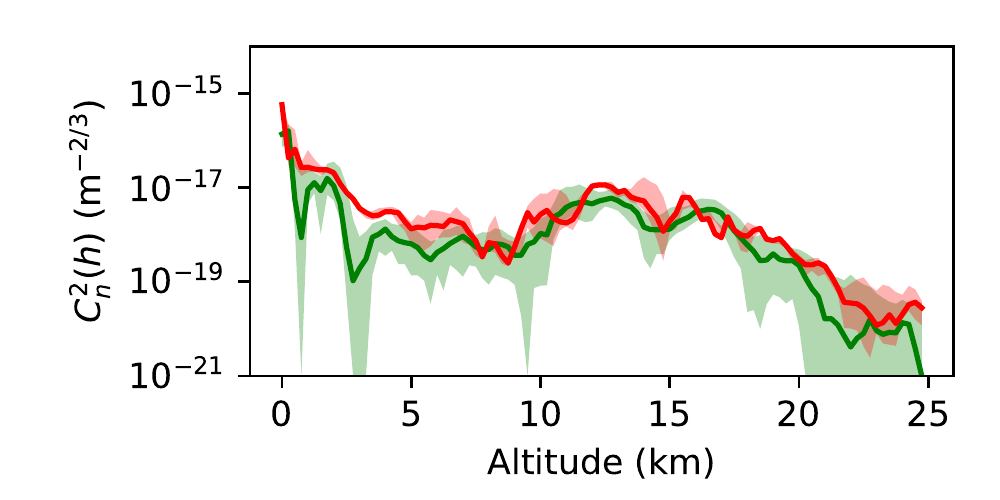} &
	\includegraphics[width=0.3\textwidth,trim={1cm 0 0cm 0}]{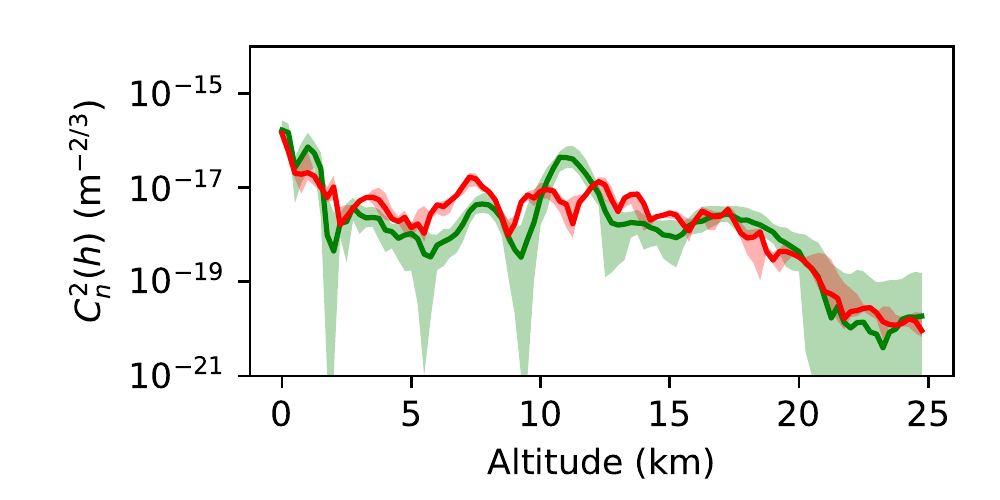} \\
    \includegraphics[width=0.3\textwidth,trim={1cm 0 0cm 0}]{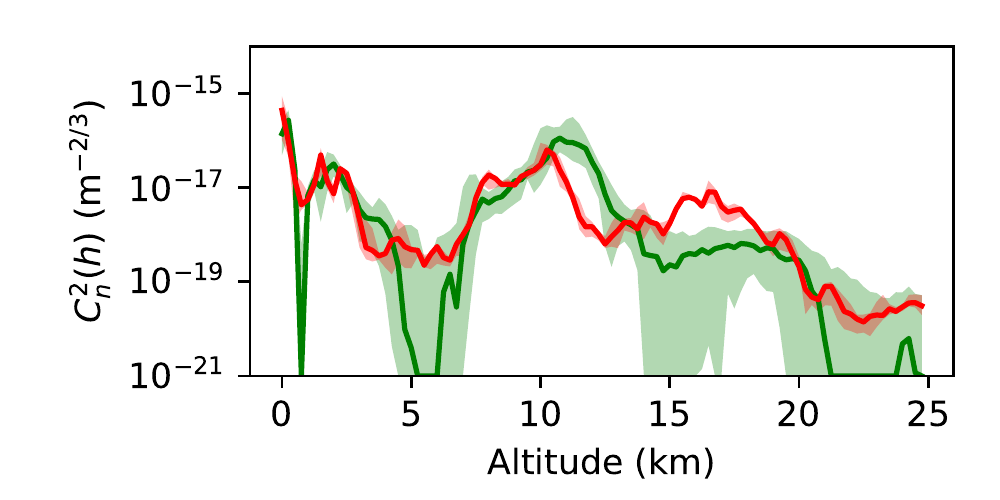} &
    
    \includegraphics[width=0.3\textwidth,trim={1cm 0 0cm 0}]{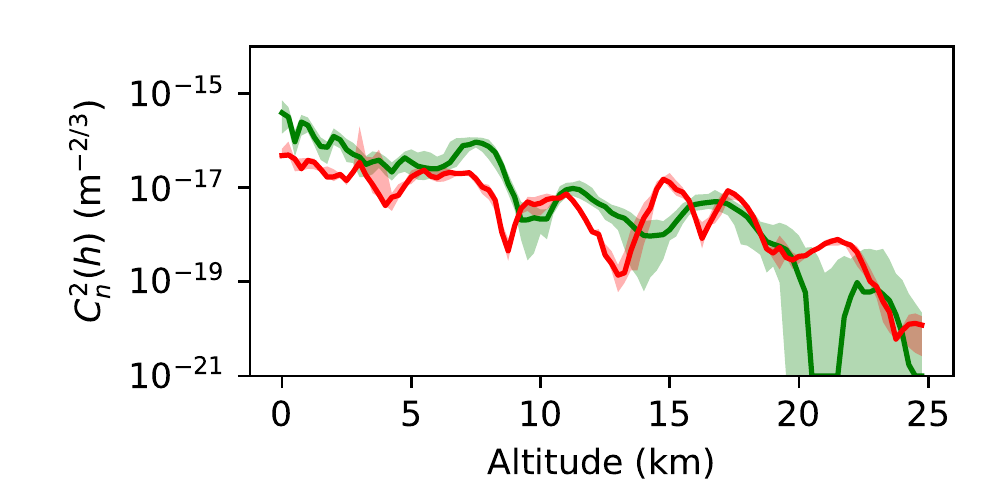} &
    \includegraphics[width=0.3\textwidth,trim={1cm 0 0cm 0}]{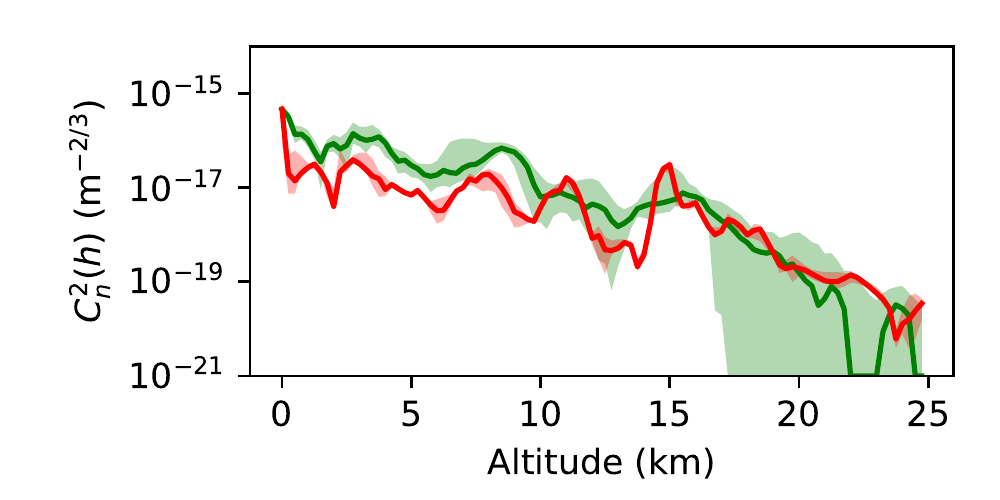} \\
    \includegraphics[width=0.3\textwidth,trim={1cm 0 0cm 0}]{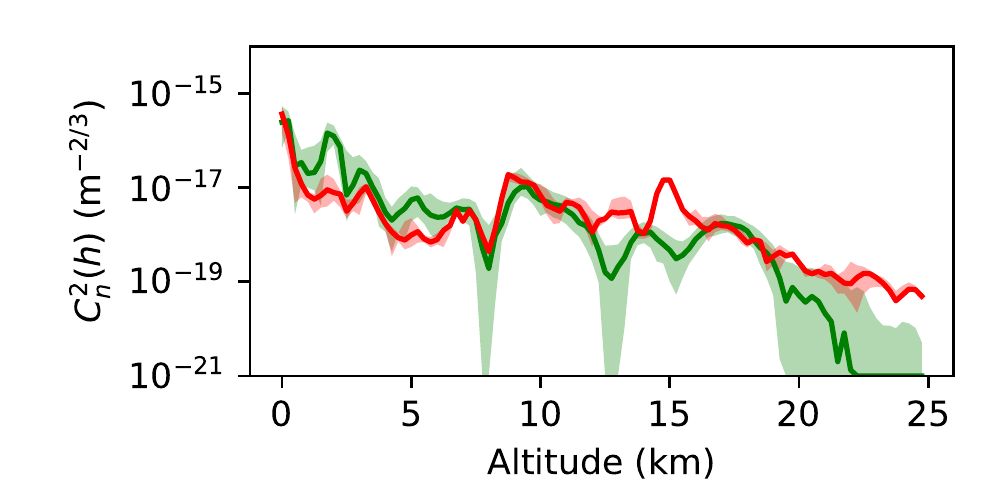} &
\end{array}$
\caption{Example vertical profiles as measured by the stereo-SCIDAR (green) and estimated by the ECMWF GCM model (red). The profiles shown are the median for an individual night of observation. The coloured region shows the interquartile range. These profiles are from the nights beginning 26th - 29th April, 22nd-26th July, 30th-31st October, 1st-2nd November, 10th-12th December 2016.}
\label{fig:allProfiles1}
\end{figure*}

\begin{figure*}
\centering
$\begin{array}{ccc}
    \includegraphics[width=0.3\textwidth,trim={1cm 0 0cm 0}]{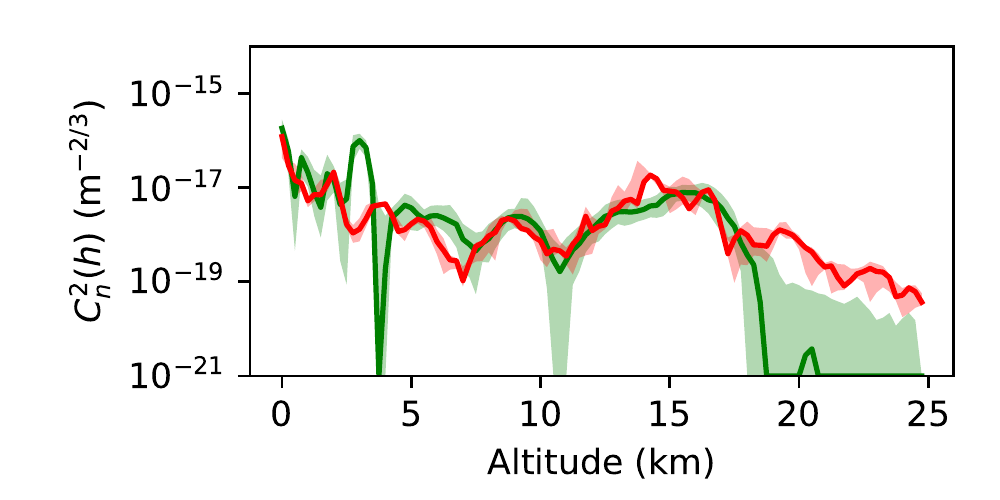} &
	\includegraphics[width=0.3\textwidth,trim={1cm 0 0cm 0}]{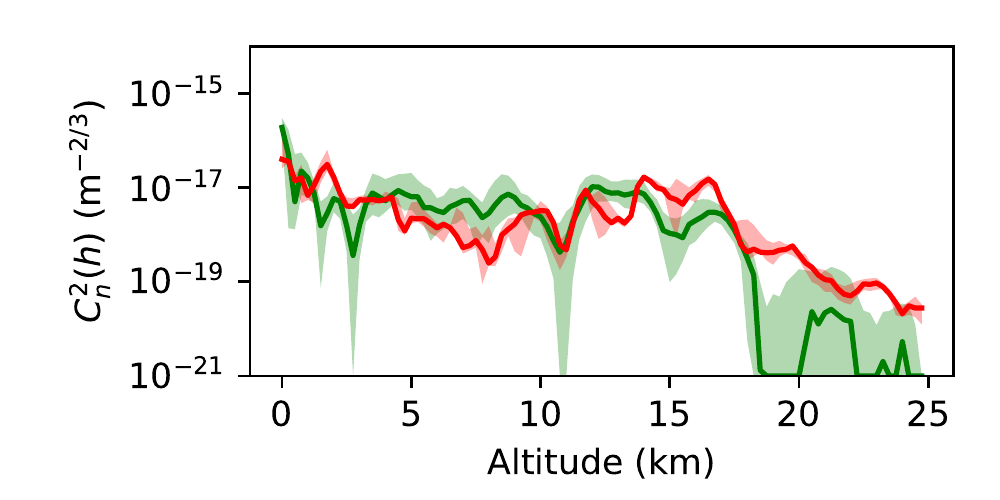} &
	\includegraphics[width=0.3\textwidth,trim={1cm 0 0cm 0}]{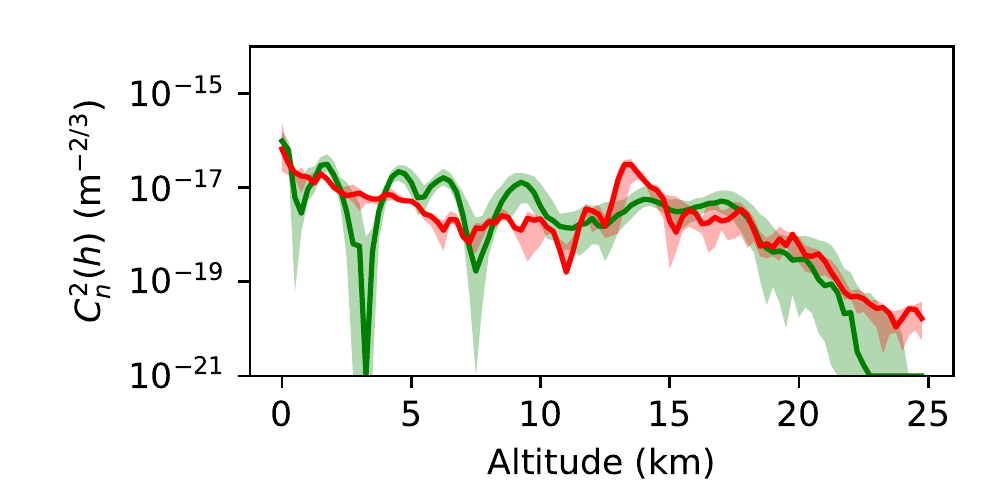} \\
	
    \includegraphics[width=0.3\textwidth,trim={1cm 0 0cm 0}]{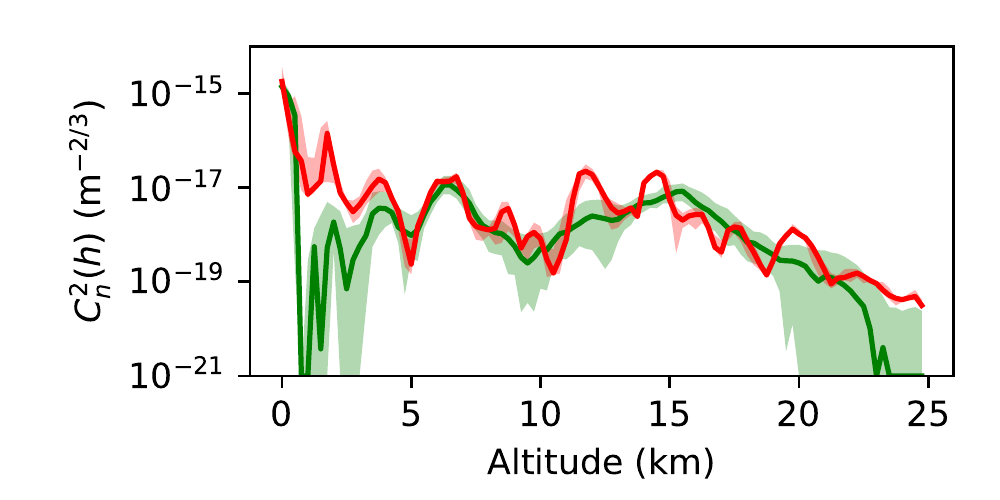} &
    \includegraphics[width=0.3\textwidth,trim={1cm 0 0cm 0}]{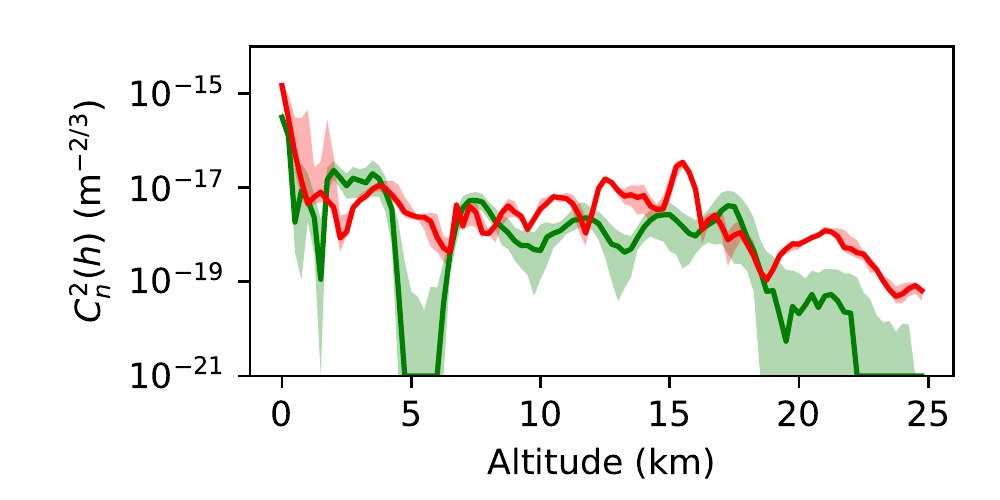} &
    \includegraphics[width=0.3\textwidth,trim={1cm 0 0cm 0}]{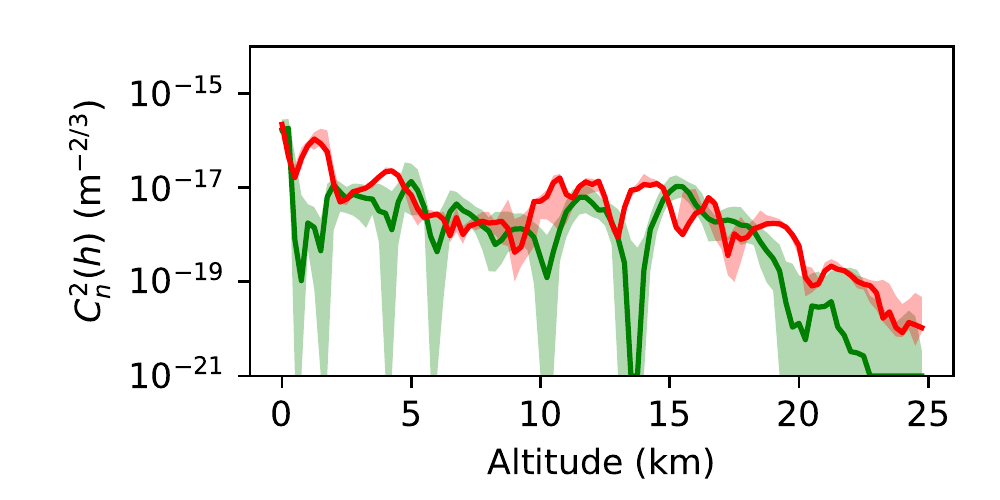} \\
    \includegraphics[width=0.3\textwidth,trim={1cm 0 0cm 0}]{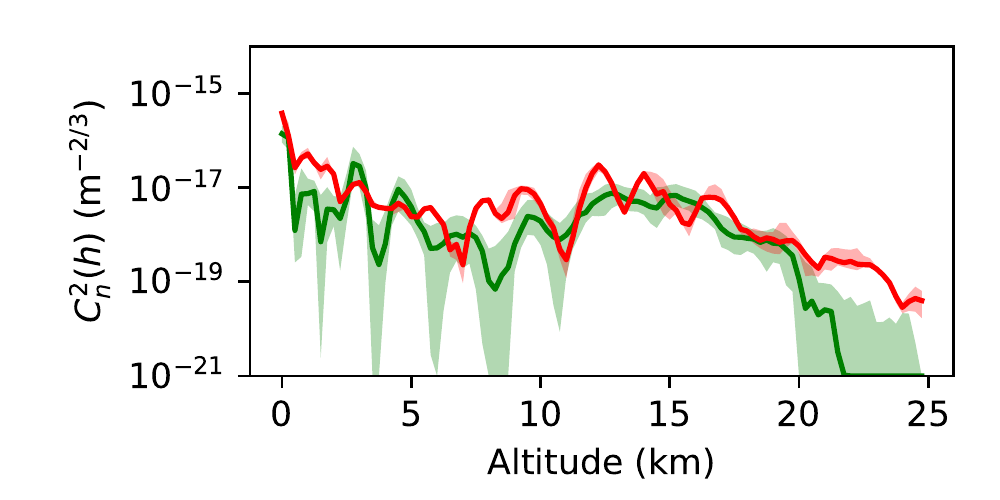} &
    \includegraphics[width=0.3\textwidth,trim={1cm 0 0cm 0}]{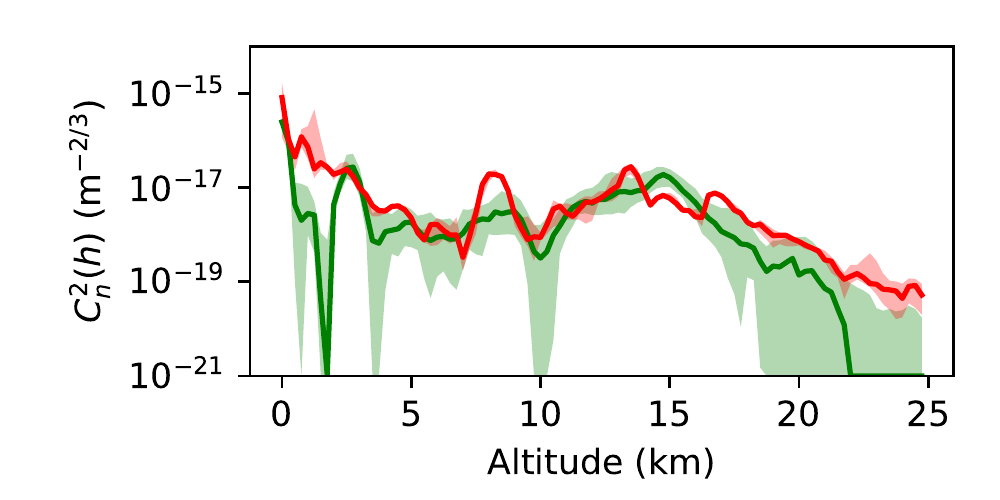} &
    \includegraphics[width=0.3\textwidth,trim={1cm 0 0cm 0}]{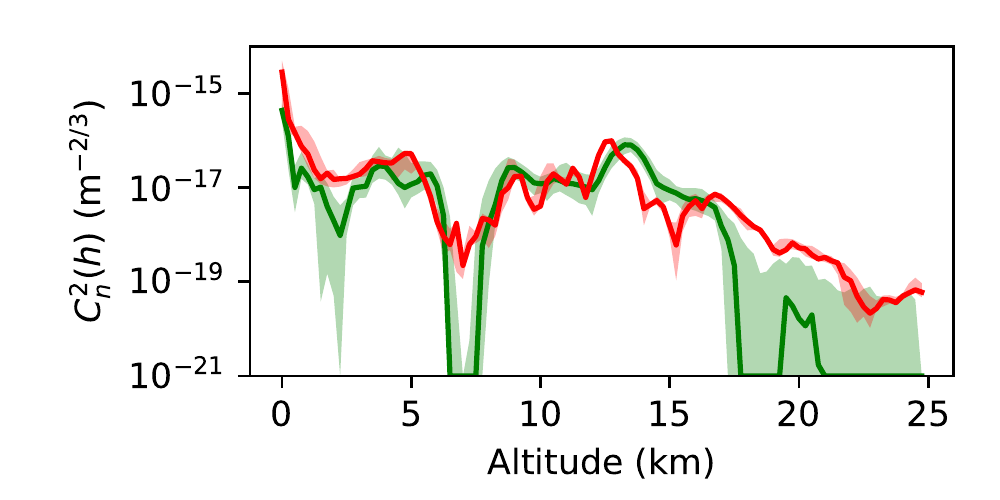} \\
    \includegraphics[width=0.3\textwidth,trim={1cm 0 0cm 0}]{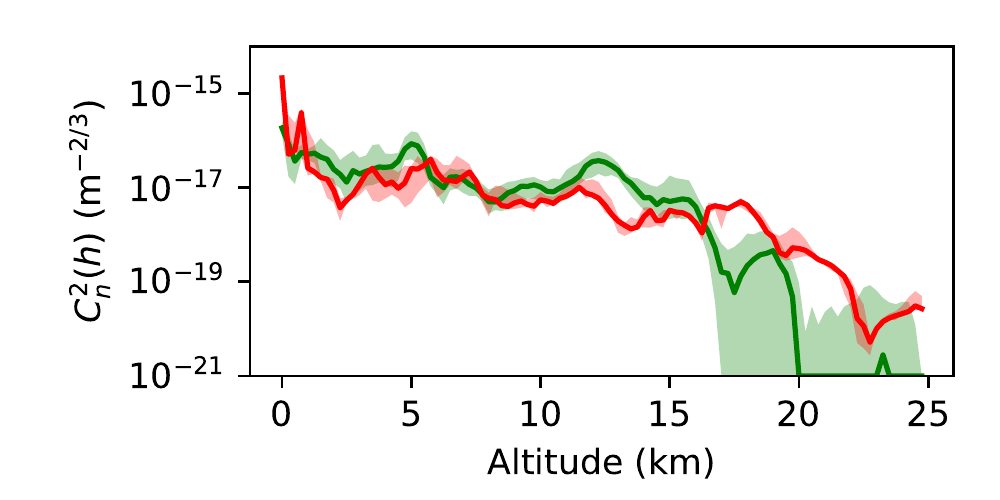} &
    
    \includegraphics[width=0.3\textwidth,trim={1cm 0 0cm 0}]{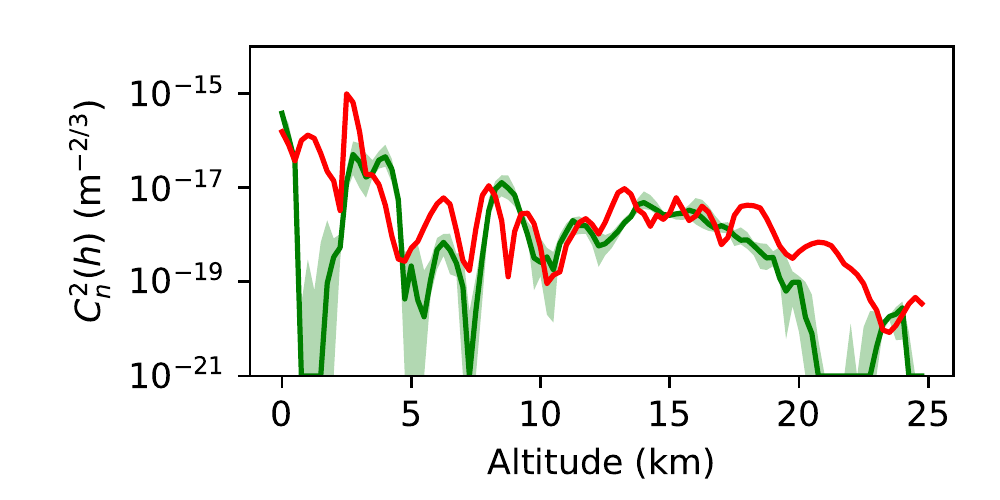} &
    \includegraphics[width=0.3\textwidth,trim={1cm 0 0cm 0}]{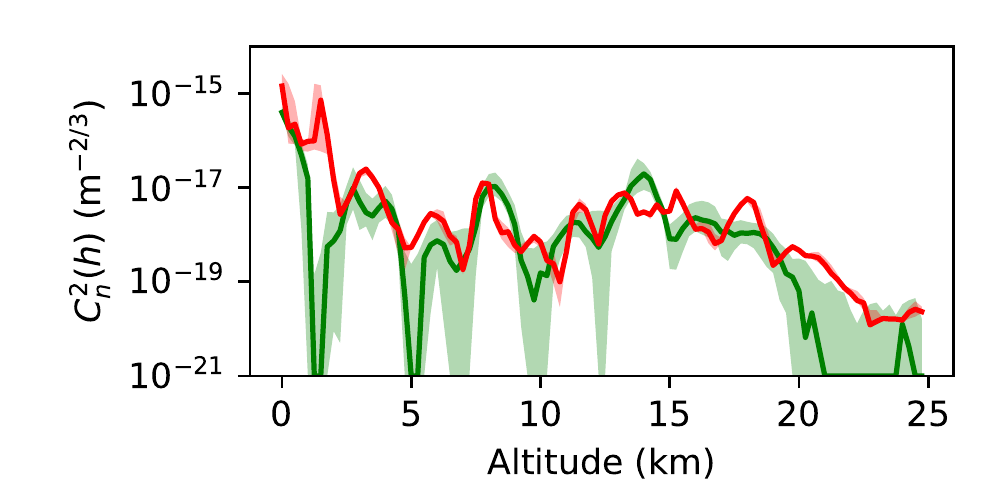} \\
    \includegraphics[width=0.3\textwidth,trim={1cm 0 0cm 0}]{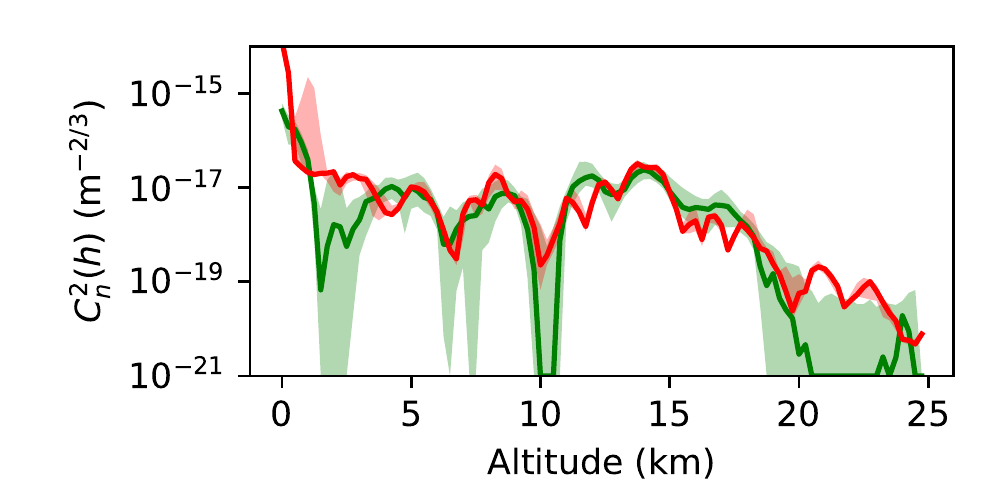} &
    \includegraphics[width=0.3\textwidth,trim={1cm 0 0cm 0}]{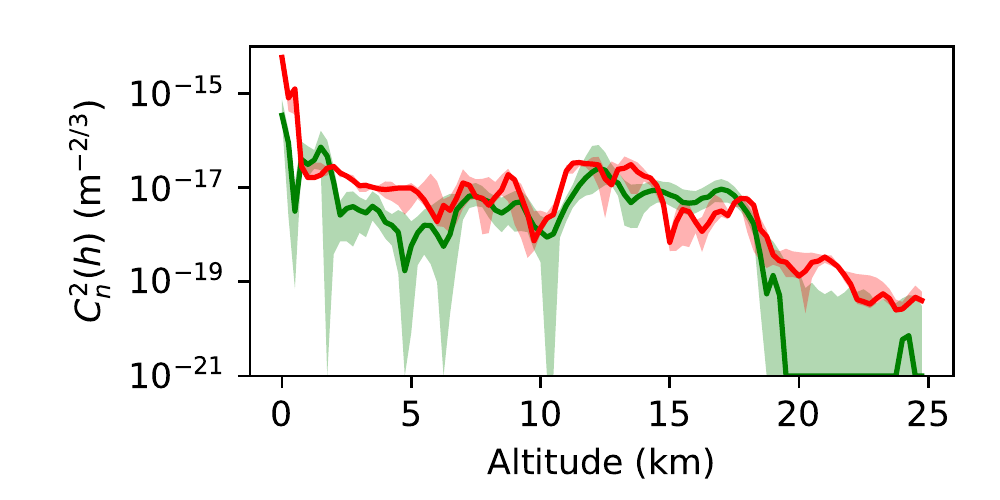} &
    \includegraphics[width=0.3\textwidth,trim={1cm 0 0cm 0}]{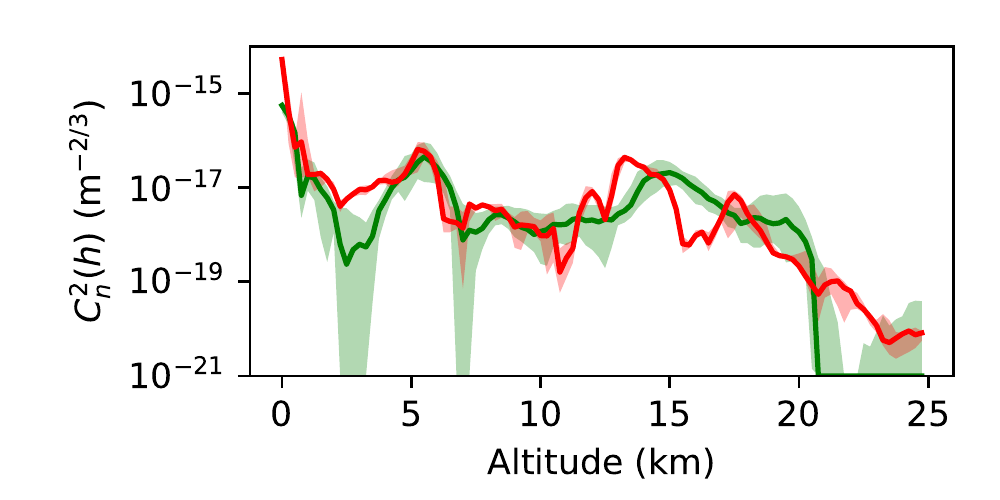} \\         
    \includegraphics[width=0.3\textwidth,trim={1cm 0 0cm 0}]{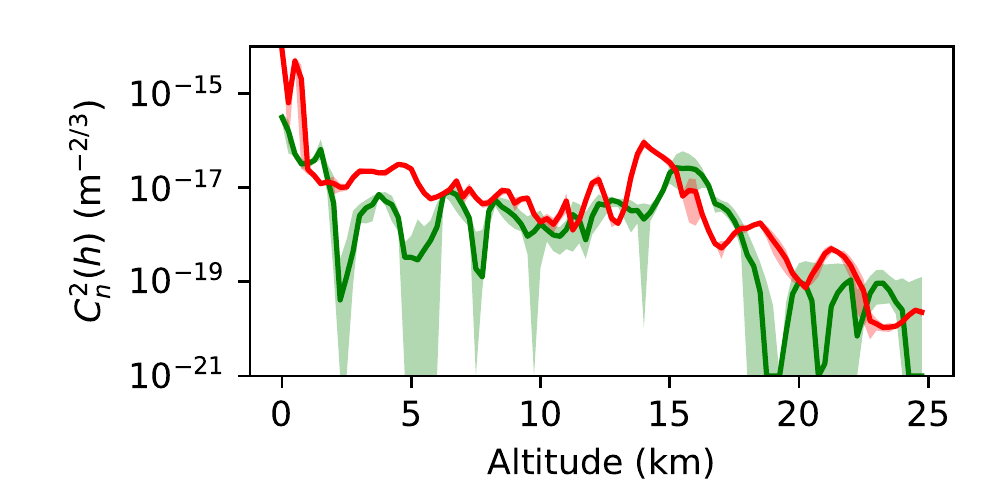} &
    
    \includegraphics[width=0.3\textwidth,trim={1cm 0 0cm 0}]{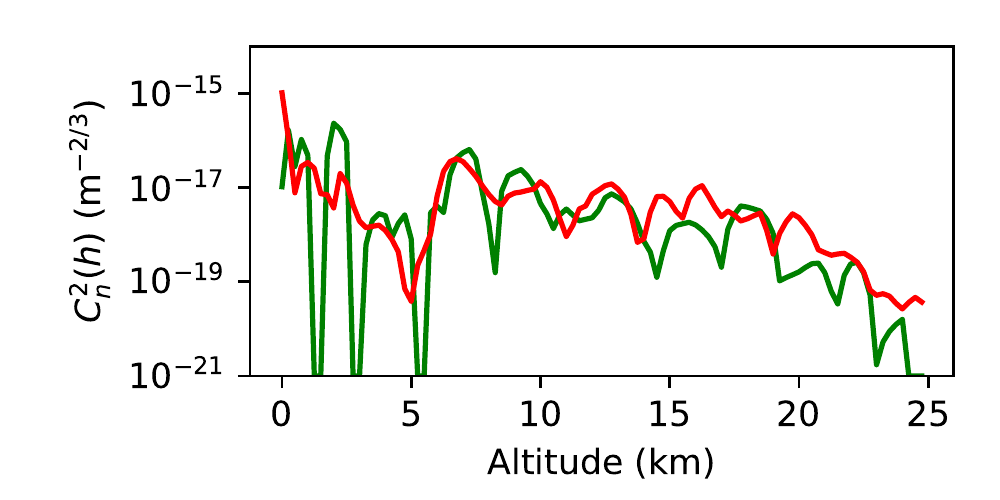} &
    \includegraphics[width=0.3\textwidth,trim={1cm 0 0cm 0}]{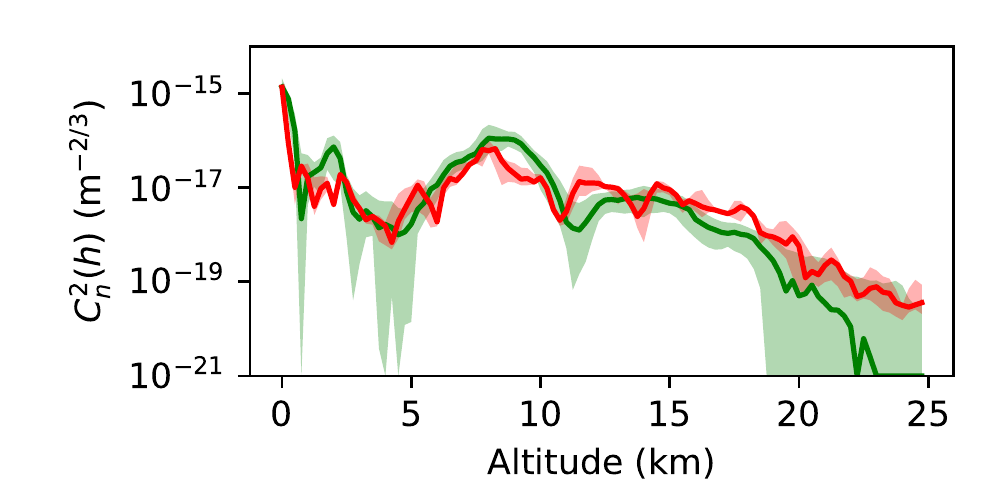} \\         
    \includegraphics[width=0.3\textwidth,trim={1cm 0 0cm 0}]{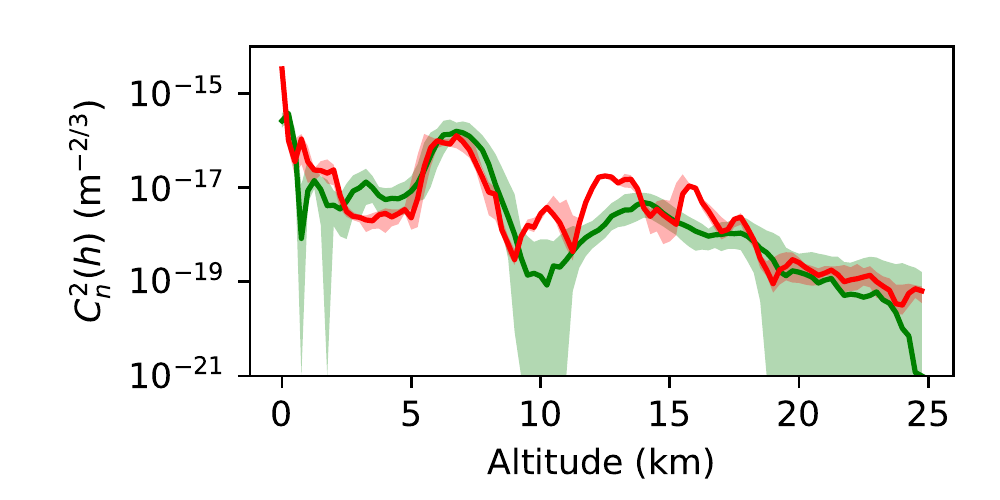} &
    \includegraphics[width=0.3\textwidth,trim={1cm 0 0cm 0}]{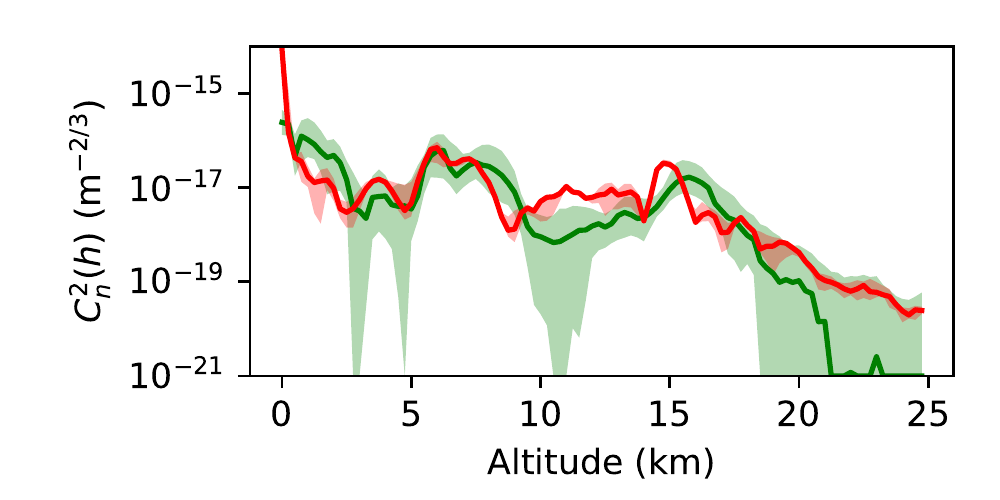} &

\end{array}$
\caption{Example vertical profiles as measured by the stereo-SCIDAR (green) and estimated by the ECMWF GCM model (red). The profiles shown are the median for an individual night of observation. The coloured region shows the interquartile range. These profiles are from the nights beginning 7th - 9th March, 12th -18th April, 4th - 9th May and 7th - 10th June 2017.}
\label{fig:allProfiles2}
\end{figure*}

\begin{figure*}
\centering
$\begin{array}{ccc}
    \includegraphics[width=0.3\textwidth,trim={1cm 0 0cm 0}]{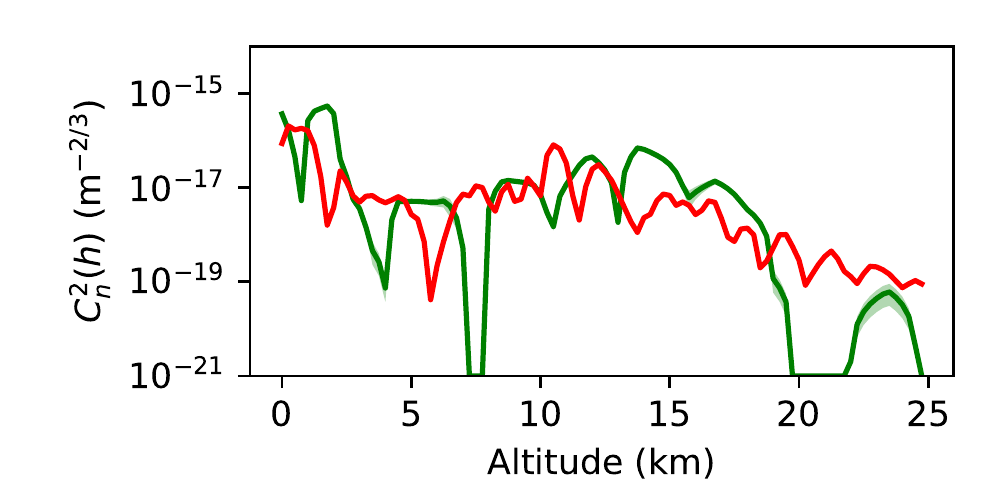} &
    \includegraphics[width=0.3\textwidth,trim={1cm 0 0cm 0}]{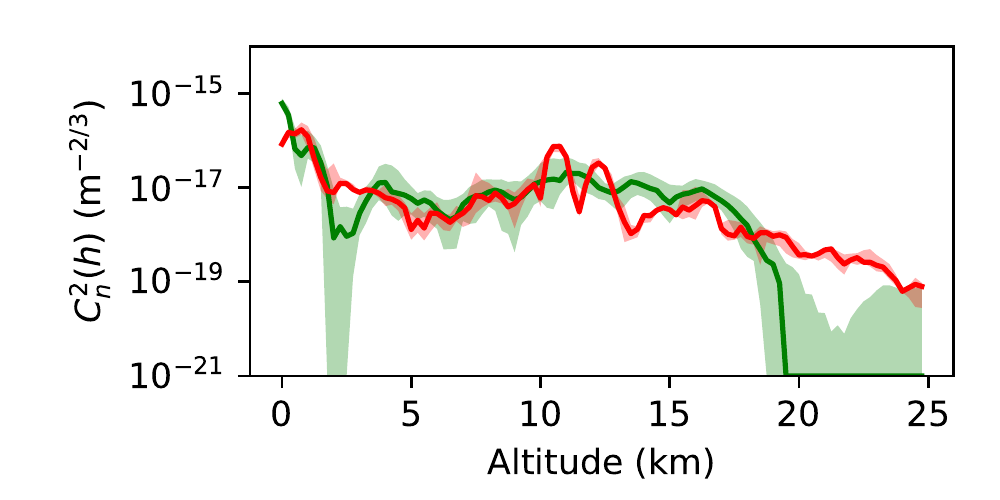} &
    \includegraphics[width=0.3\textwidth,trim={1cm 0 0cm 0}]{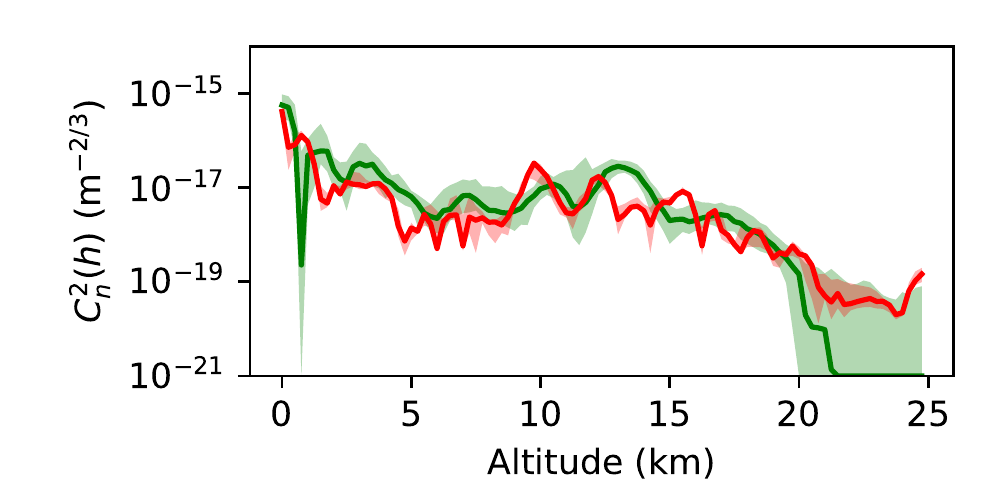} \\
    \includegraphics[width=0.3\textwidth,trim={1cm 0 0cm 0}]{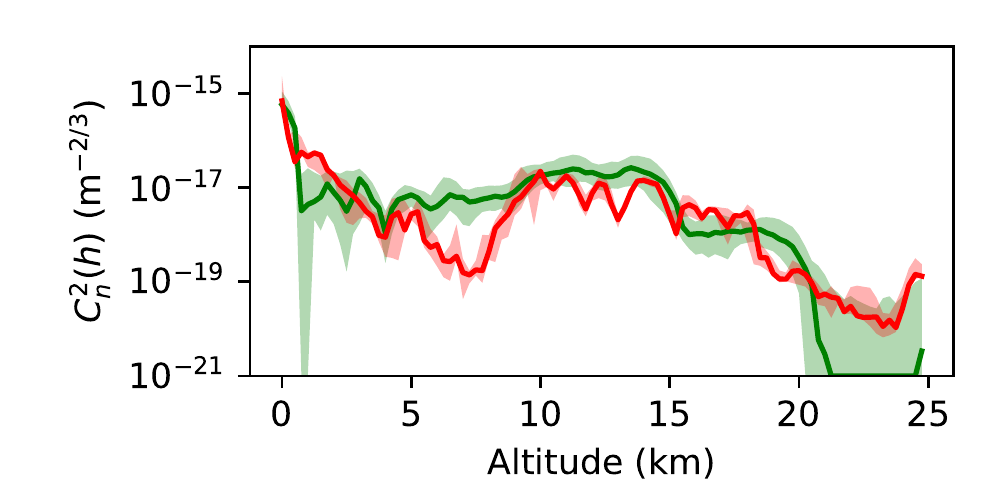} &
    \includegraphics[width=0.3\textwidth,trim={1cm 0 0cm 0}]{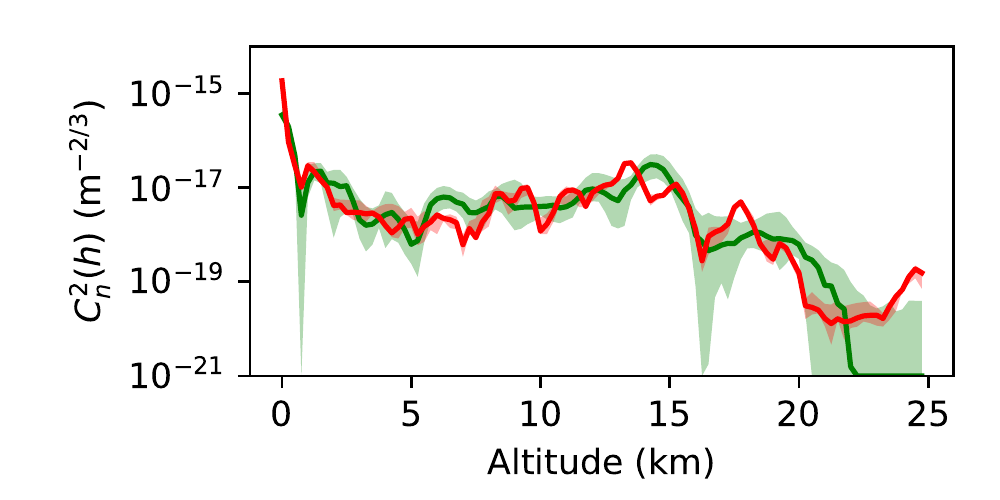} &
    \includegraphics[width=0.3\textwidth,trim={1cm 0 0cm 0}]{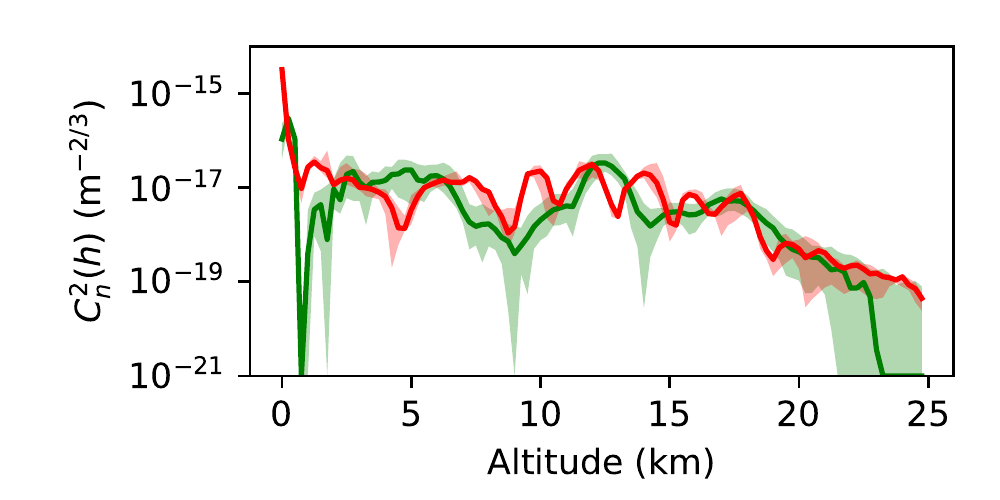} \\
    \includegraphics[width=0.3\textwidth,trim={1cm 0 0cm 0}]{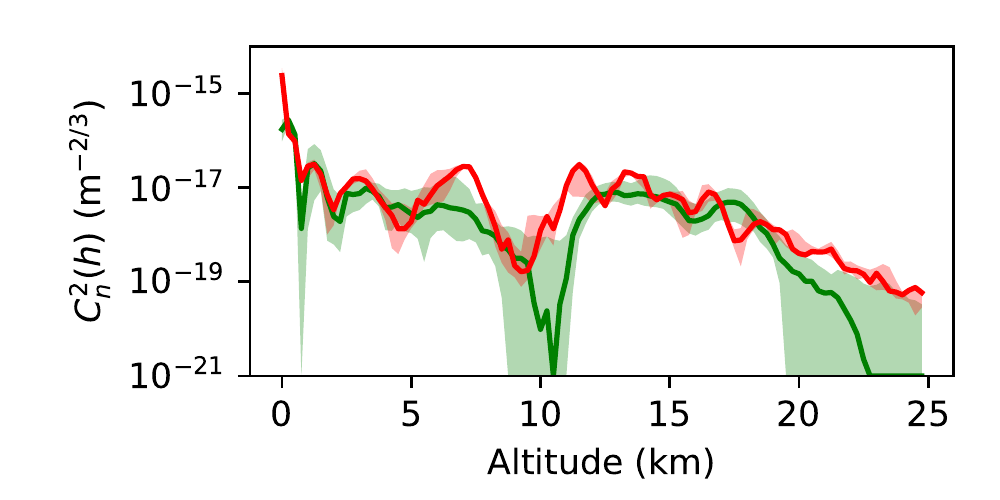} &
    \includegraphics[width=0.3\textwidth,trim={1cm 0 0cm 0}]{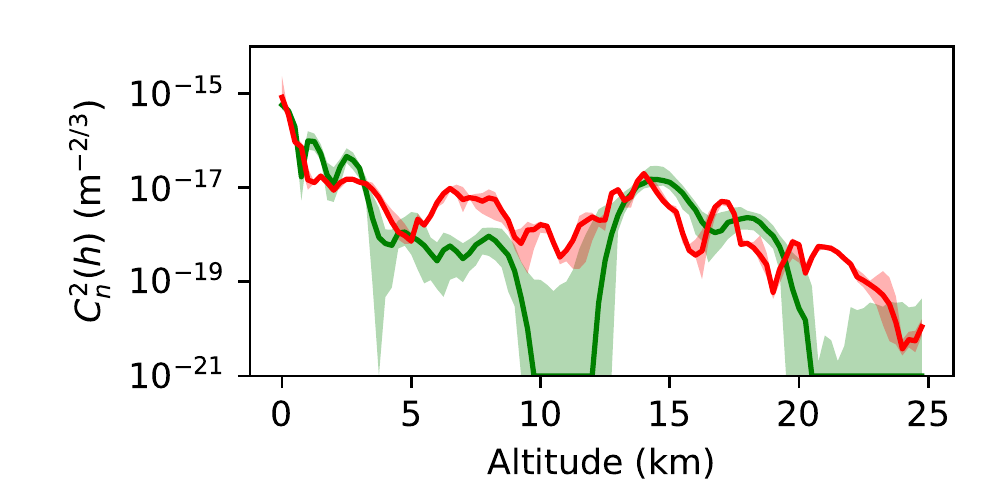} &
    
    \includegraphics[width=0.3\textwidth,trim={1cm 0 0cm 0}]{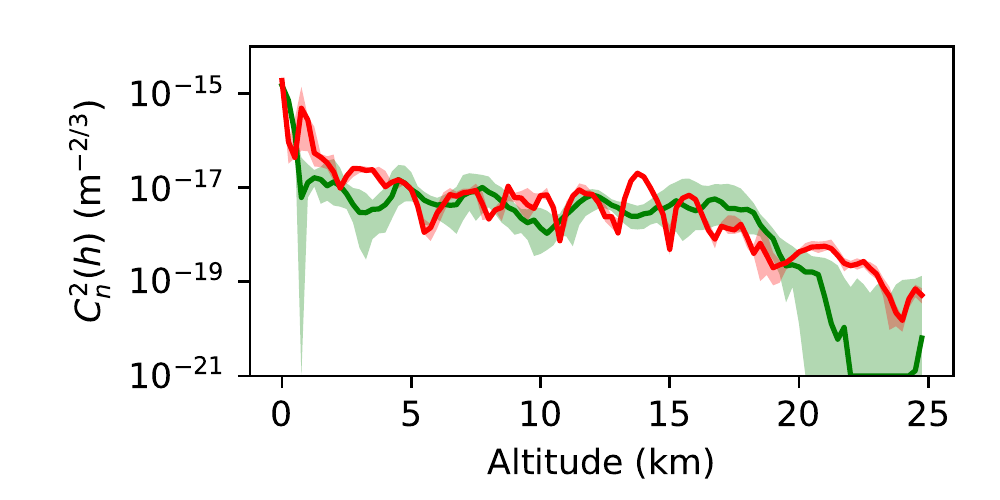} \\
    \includegraphics[width=0.3\textwidth,trim={1cm 0 0cm 0}]{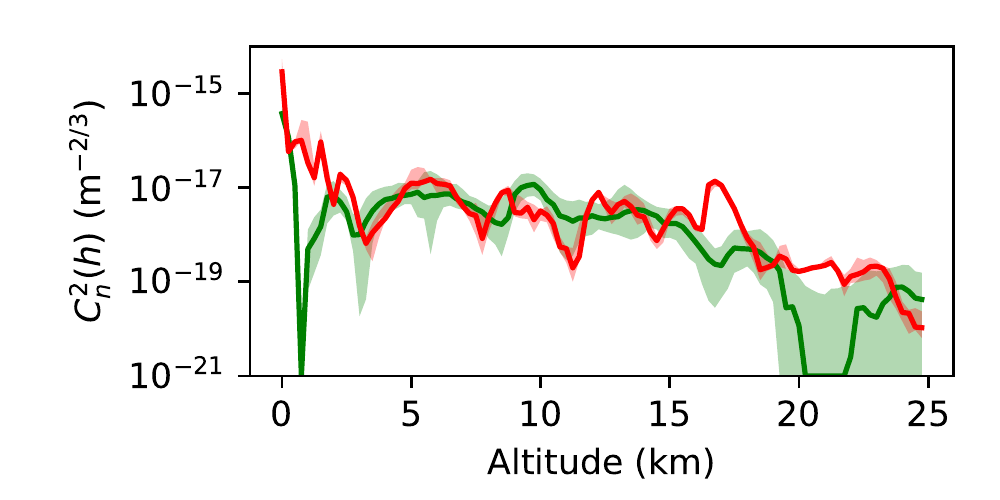} &
    \includegraphics[width=0.3\textwidth,trim={1cm 0 0cm 0}]{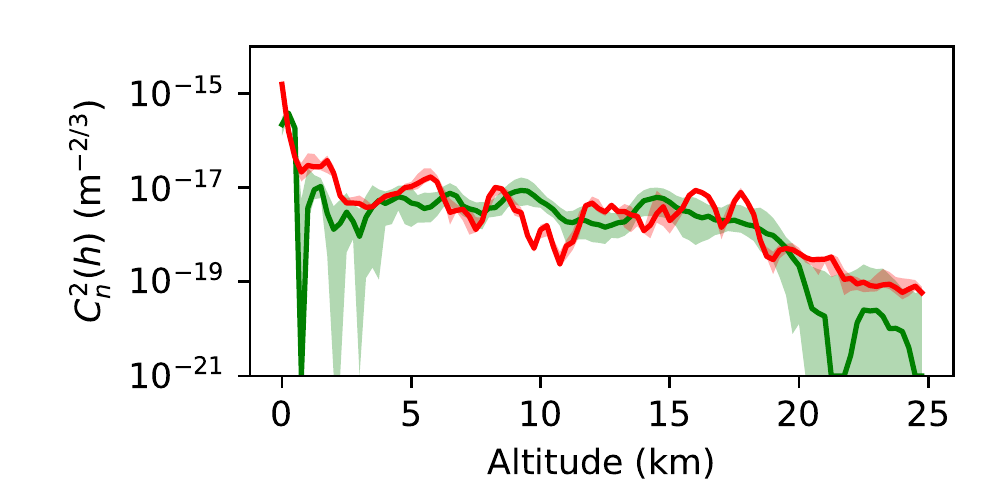} &
    \includegraphics[width=0.3\textwidth,trim={1cm 0 0cm 0}]{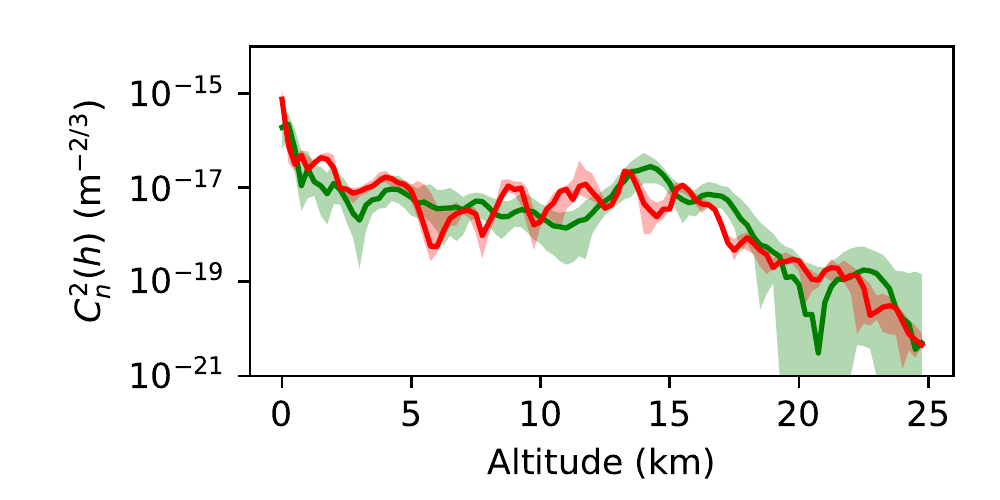} \\
    \includegraphics[width=0.3\textwidth,trim={1cm 0 0cm 0}]{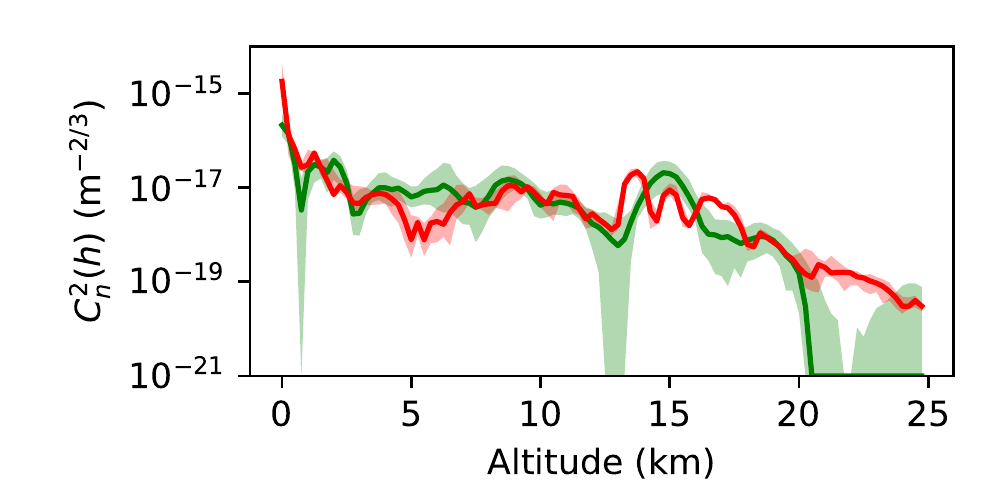} &
    \includegraphics[width=0.3\textwidth,trim={1cm 0 0cm 0}]{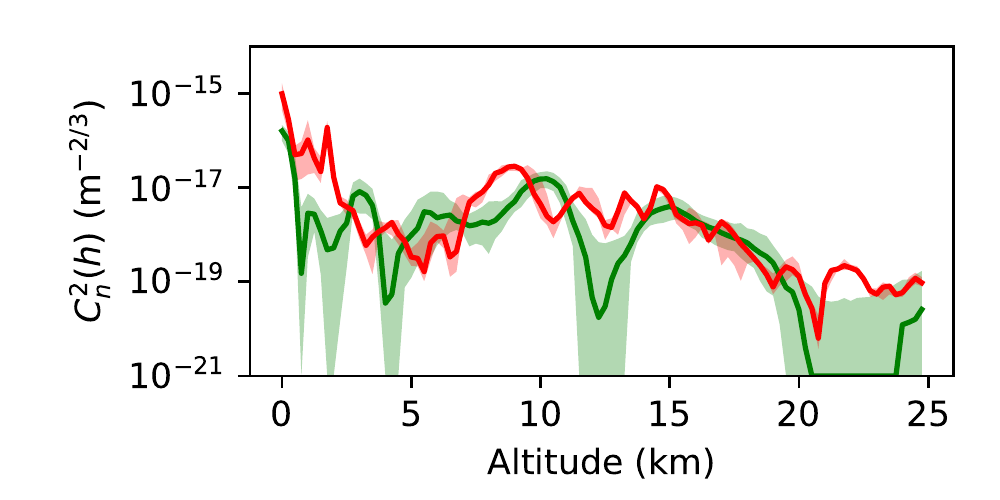} &
    
    \includegraphics[width=0.3\textwidth,trim={1cm 0 0cm 0}]{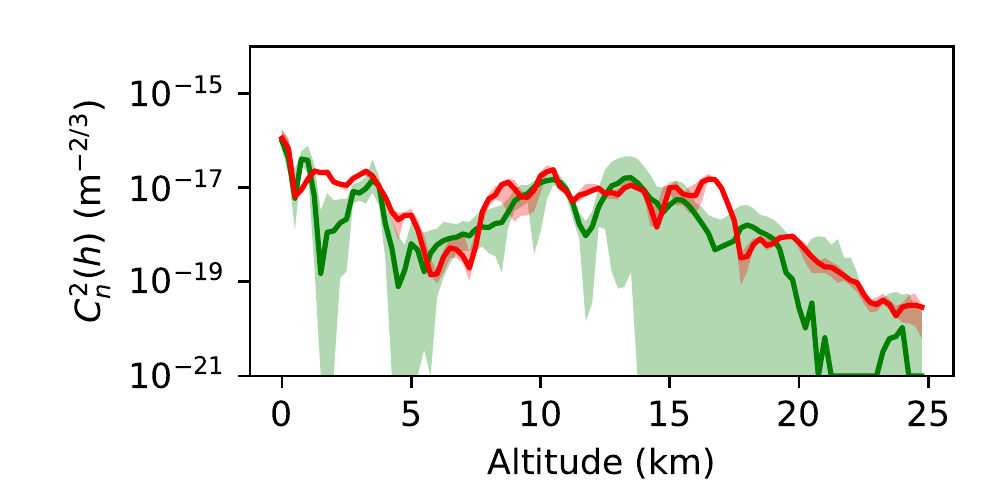} \\
    \includegraphics[width=0.3\textwidth,trim={1cm 0 0cm 0}]{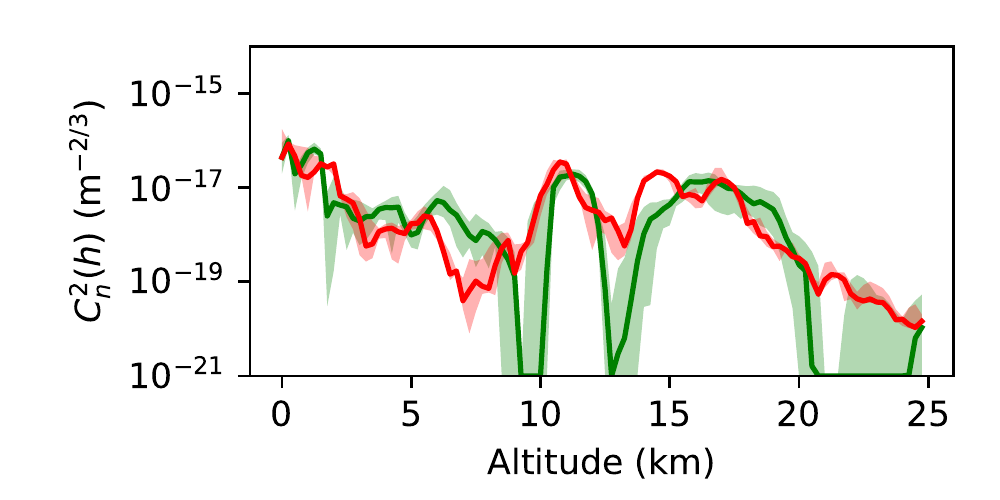} &
    \includegraphics[width=0.3\textwidth,trim={1cm 0 0cm 0}]{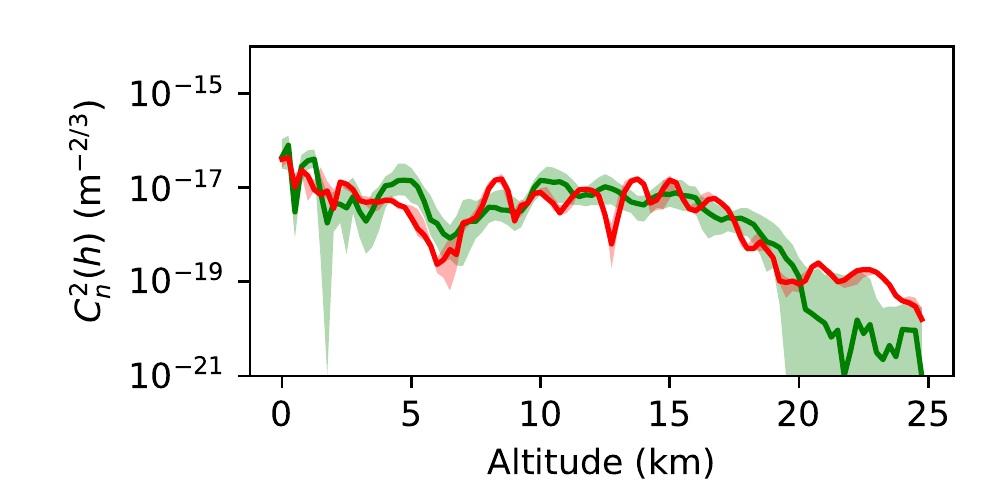} &
    \includegraphics[width=0.3\textwidth,trim={1cm 0 0cm 0}]{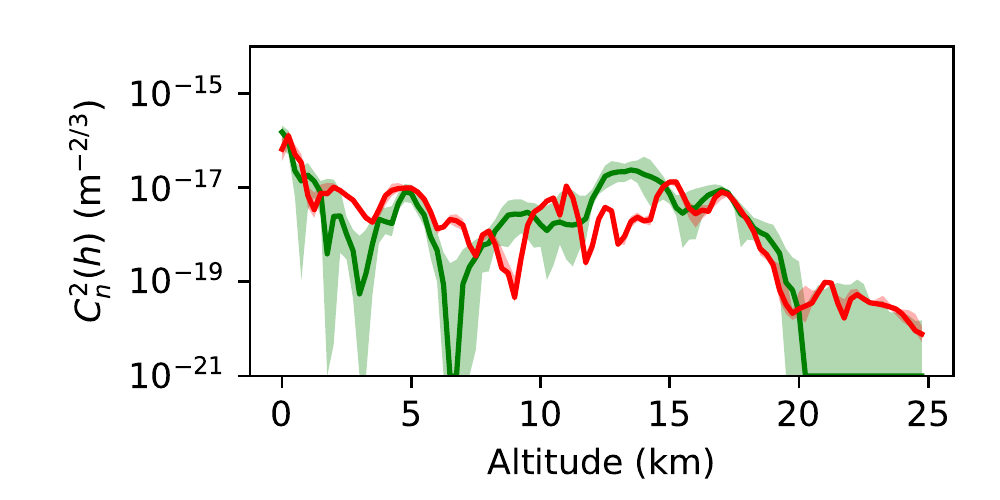} \\
    \includegraphics[width=0.3\textwidth,trim={1cm 0 0cm 0}]{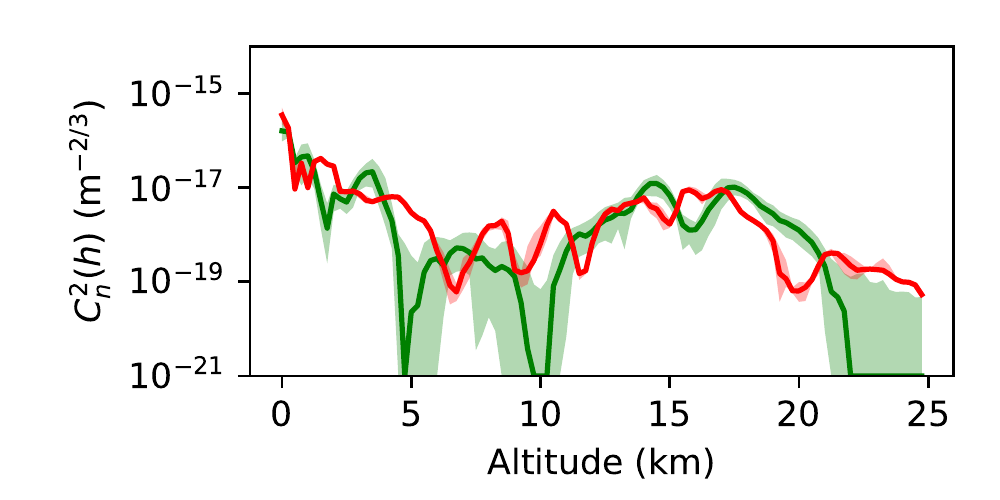} &
    \includegraphics[width=0.3\textwidth,trim={1cm 0 0cm 0}]{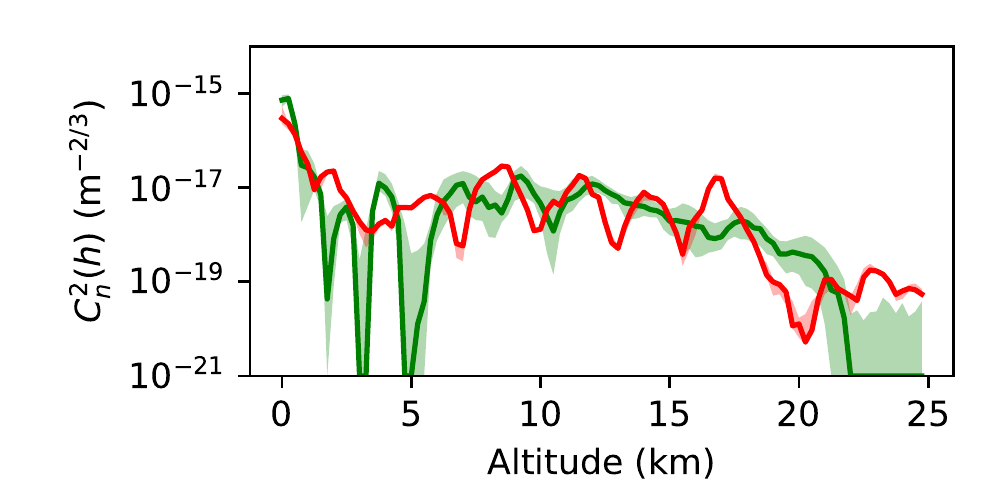} &

\end{array}$
\caption{Example vertical profiles as measured by the stereo-SCIDAR (green) and estimated by the ECMWF GCM model (red). The profiles shown are the median for an individual night of observation. The coloured region shows the interquartile range. These profiles are from the nights beginning 2nd - 9th July, 3rd - 8th August and 4th - 9th November 2017.}
\label{fig:allProfiles3}
\end{figure*}

\begin{figure*}
\centering
$\begin{array}{ccc}

    \includegraphics[width=0.3\textwidth,trim={1cm 0 0cm 0}]{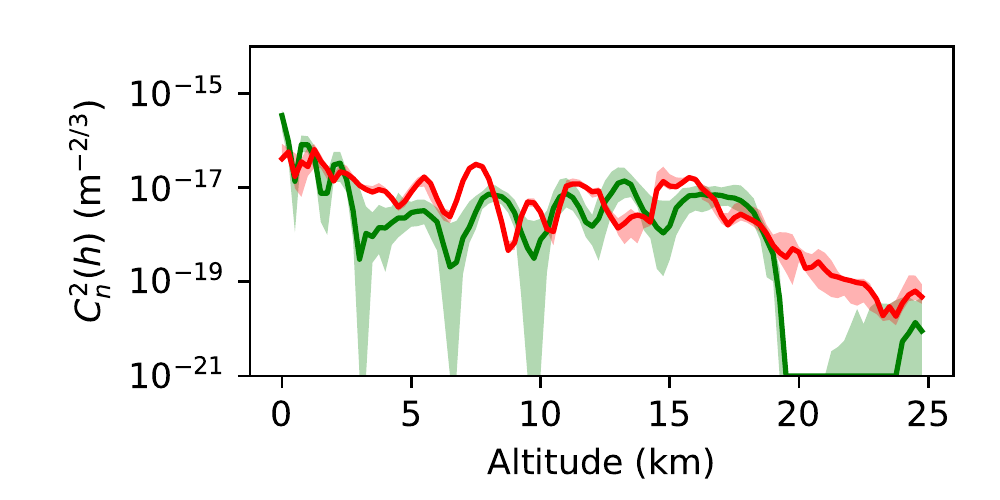} &
    \includegraphics[width=0.3\textwidth,trim={1cm 0 0cm 0}]{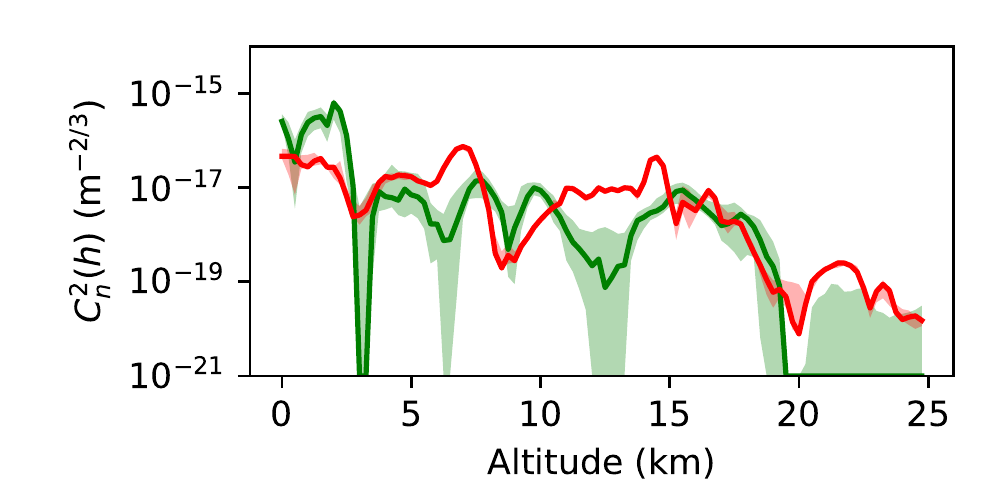} &
    \includegraphics[width=0.3\textwidth,trim={1cm 0 0cm 0}]{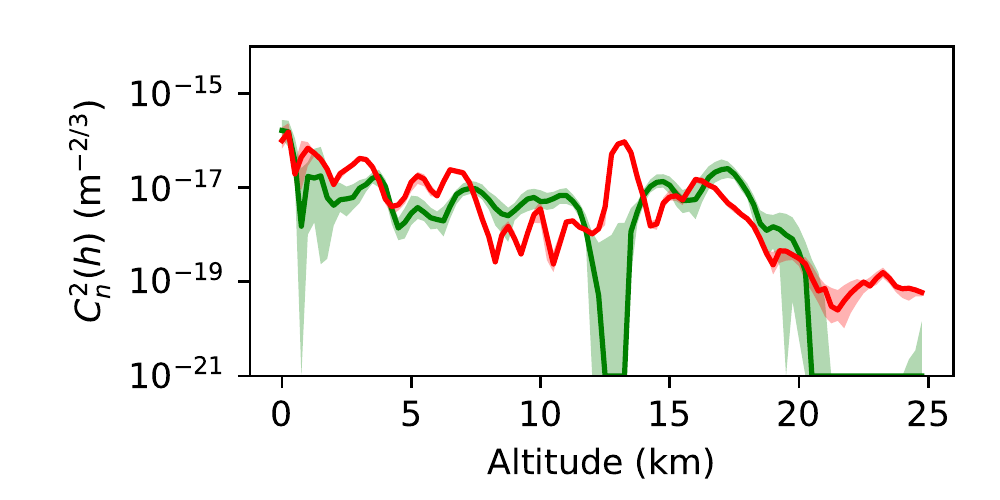} \\

    \includegraphics[width=0.3\textwidth,trim={1cm 0 0cm 0}]{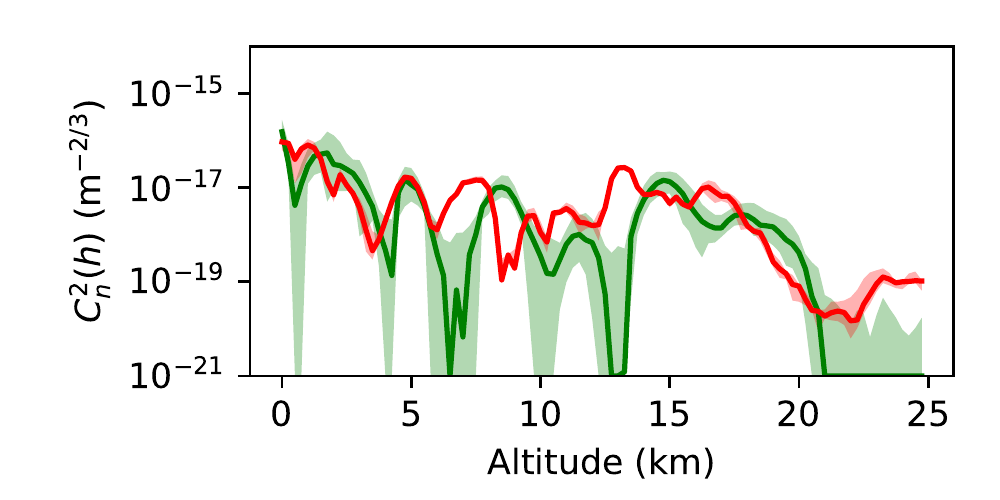} &
    \includegraphics[width=0.3\textwidth,trim={1cm 0 0cm 0}]{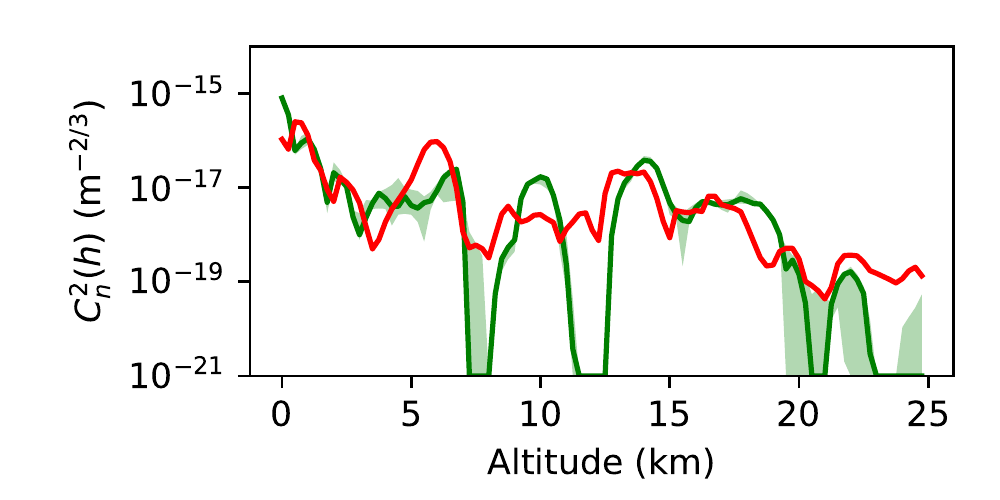} &
    \includegraphics[width=0.3\textwidth,trim={1cm 0 0cm 0}]{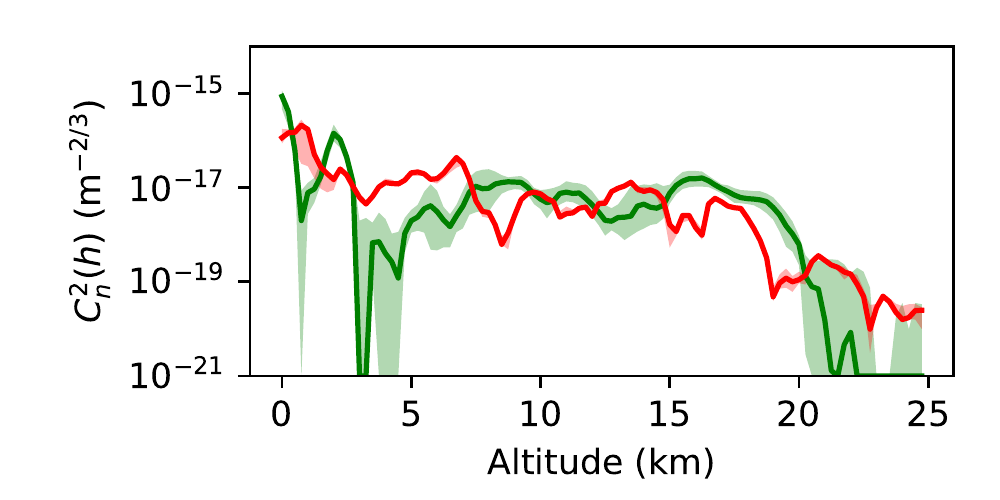} \\
    \includegraphics[width=0.3\textwidth,trim={1cm 0 0cm 0}]{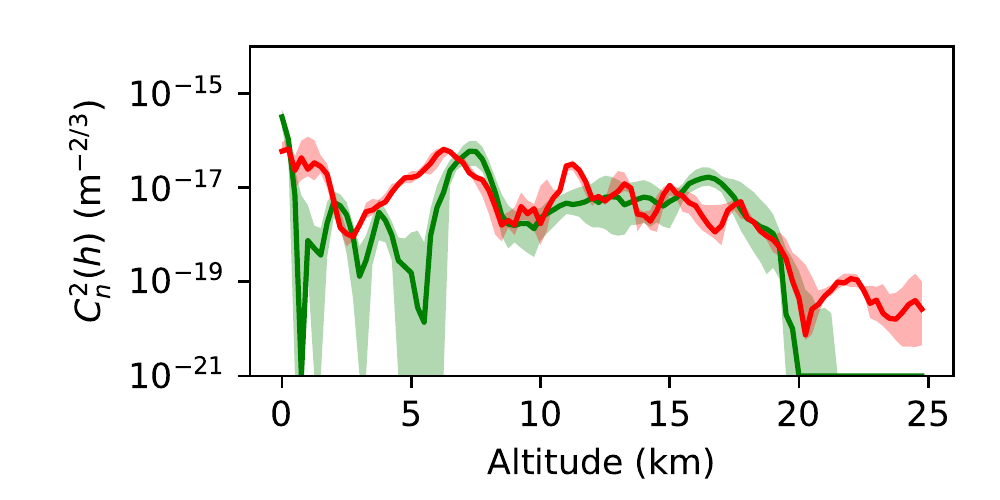} &

    \includegraphics[width=0.3\textwidth,trim={1cm 0 0cm 0}]{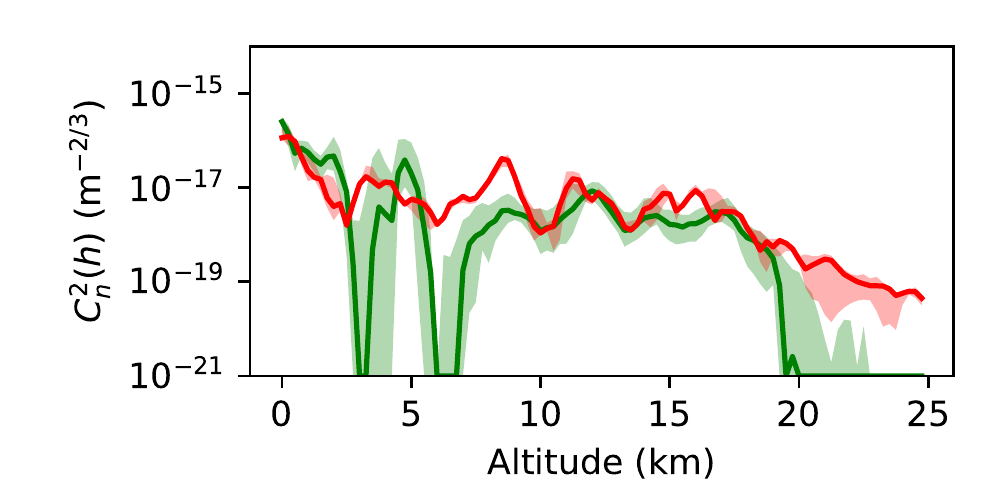} &
    \includegraphics[width=0.3\textwidth,trim={1cm 0 0cm 0}]{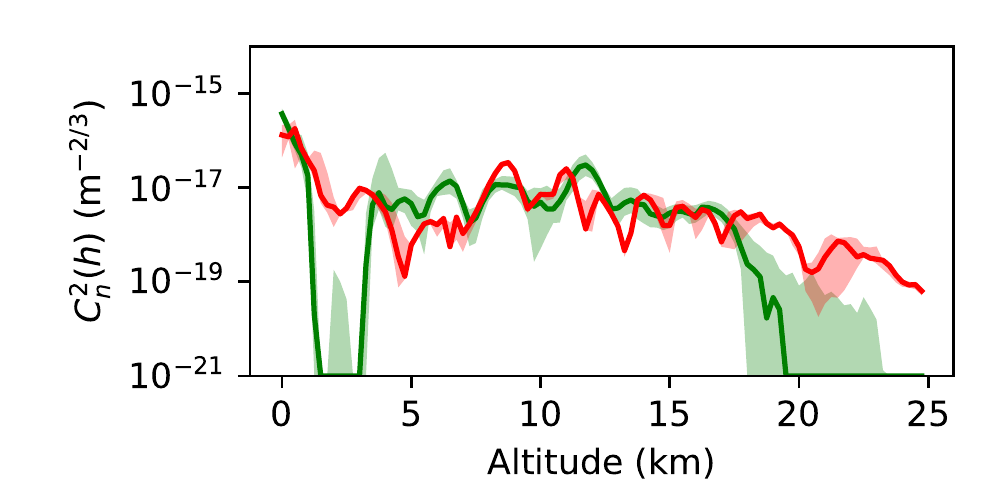} \\
    \includegraphics[width=0.3\textwidth,trim={1cm 0 0cm 0}]{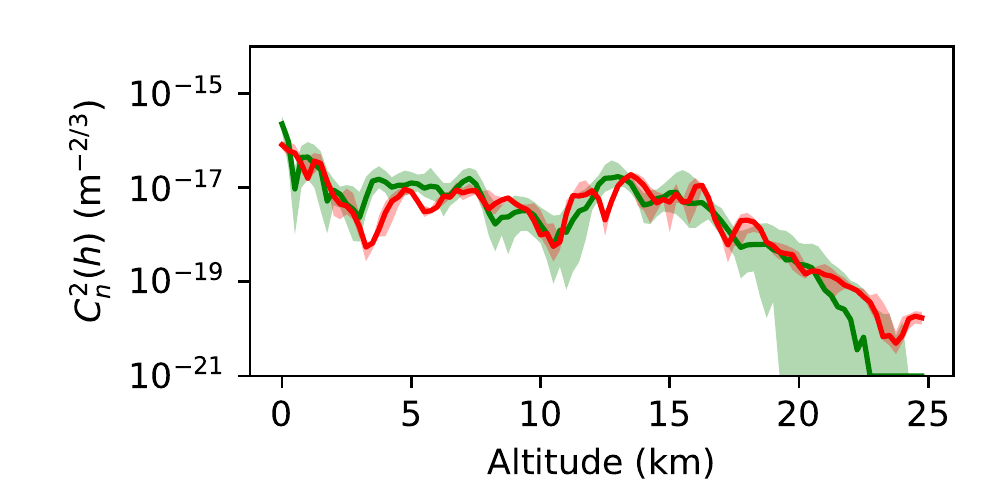} &
    \includegraphics[width=0.3\textwidth,trim={1cm 0 0cm 0}]{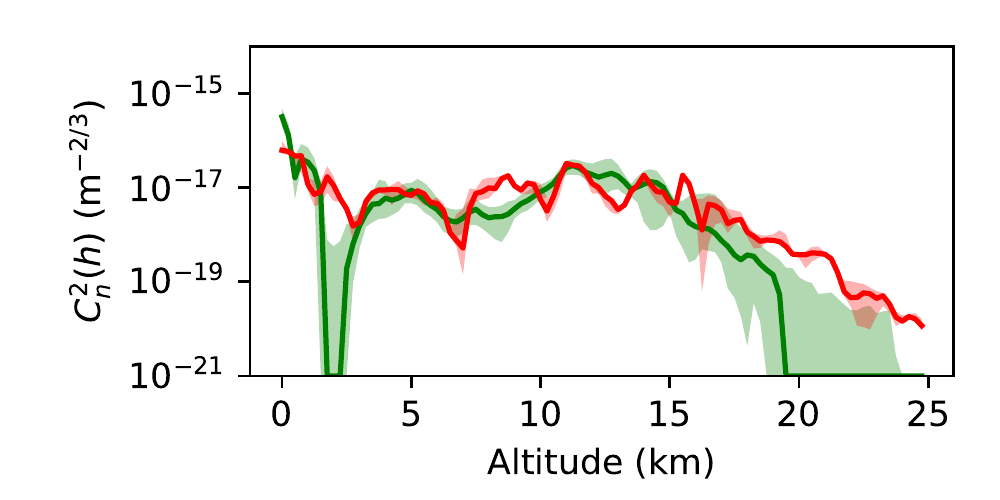} &
    \includegraphics[width=0.3\textwidth,trim={1cm 0 0cm 0}]{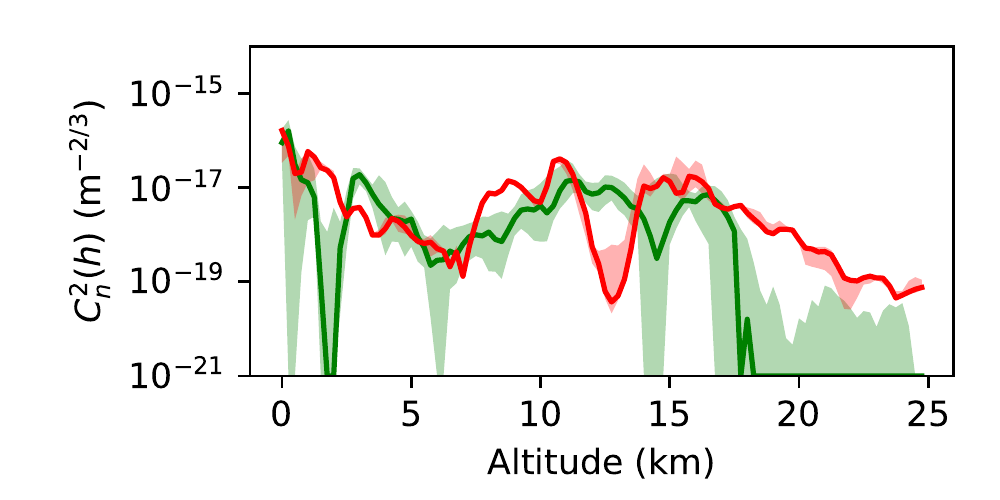} \\  

        \includegraphics[width=0.3\textwidth,trim={1cm 0 0cm 0}]{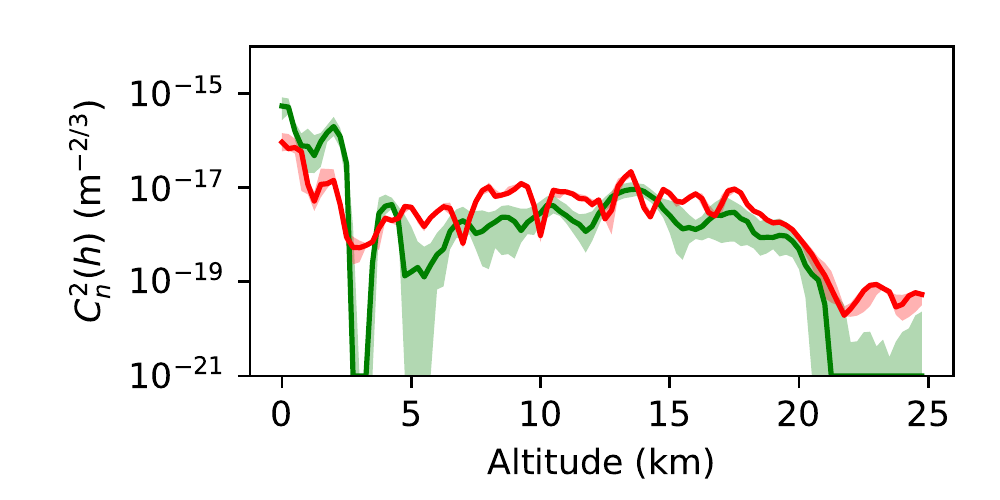} &
    \includegraphics[width=0.3\textwidth,trim={1cm 0 0cm 0}]{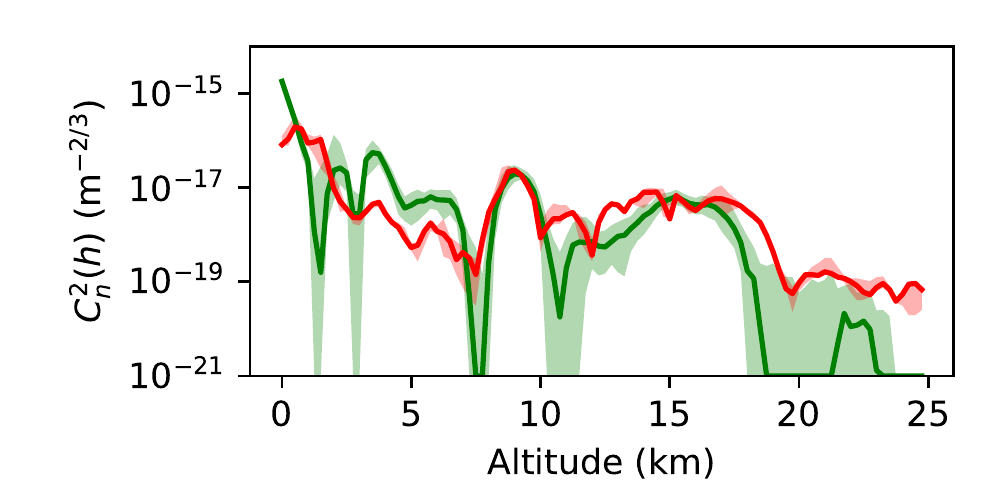} &
    \includegraphics[width=0.3\textwidth,trim={1cm 0 0cm 0}]{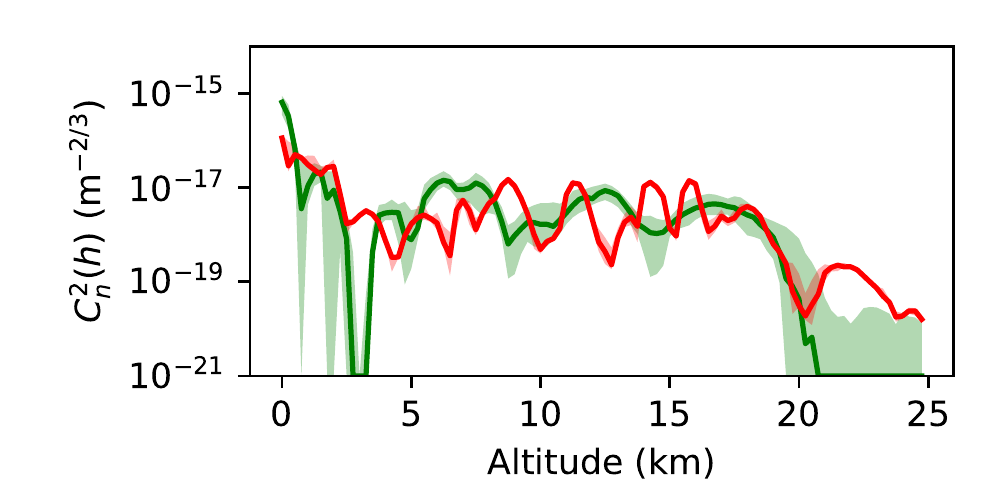} \\
    \includegraphics[width=0.3\textwidth,trim={1cm 0 0cm 0}]{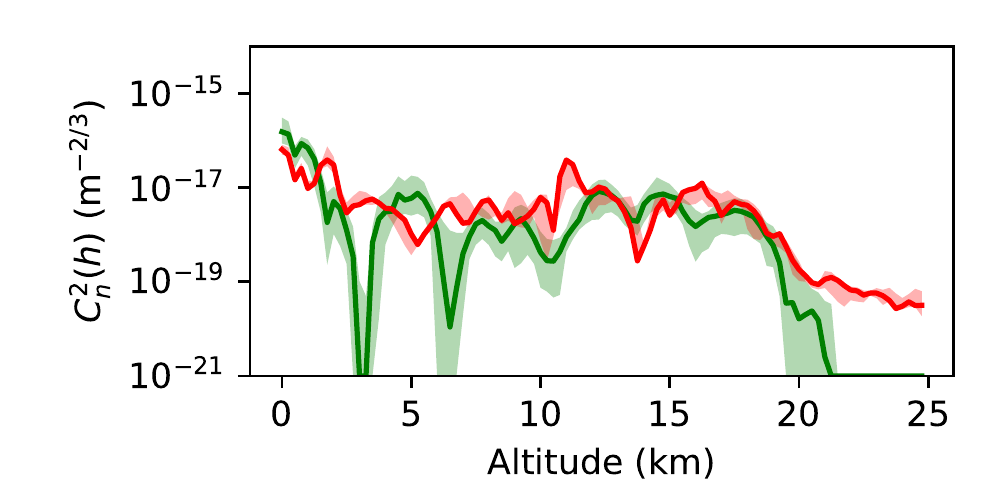} &
    \includegraphics[width=0.3\textwidth,trim={1cm 0 0cm 0}]{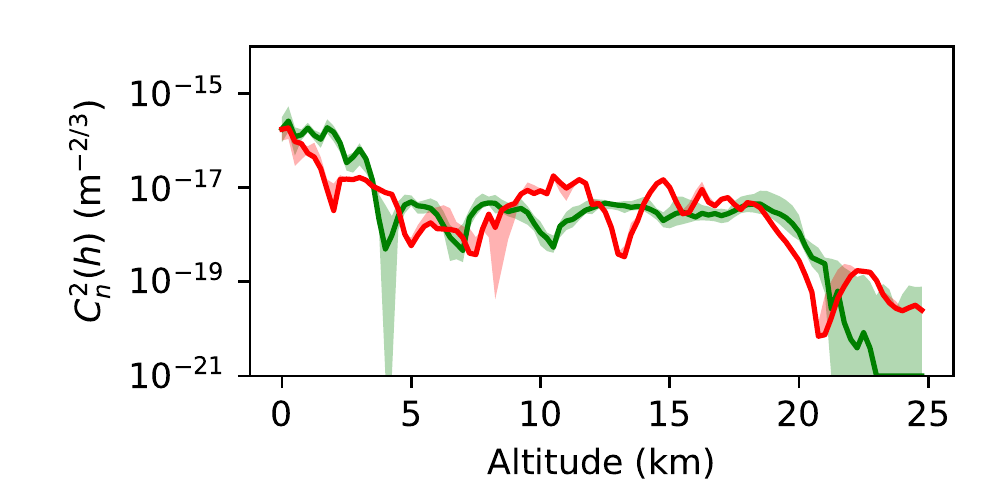} &
    \includegraphics[width=0.3\textwidth,trim={1cm 0 0cm 0}]{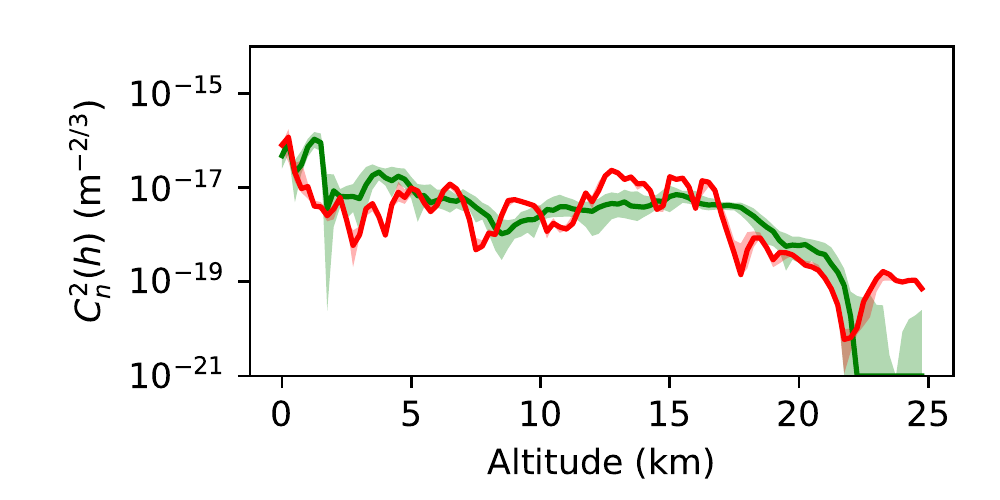} \\  

\end{array}$
\caption{Example vertical profiles as measured by the stereo-SCIDAR (green) and estimated by the ECMWF GCM model (red). The profiles shown are the median for an individual night of observation. The coloured region shows the interquartile range. These profiles are from the nights beginning 18th - 19th November, 29th - 30th November, 1st - 2nd December, 5th December, 8th - 18th December 2017.}
\label{fig:allProfiles4}
\end{figure*}

\begin{figure*}
\centering
$\begin{array}{ccc}
    \includegraphics[width=0.3\textwidth,trim={1cm 0 0cm 0}]{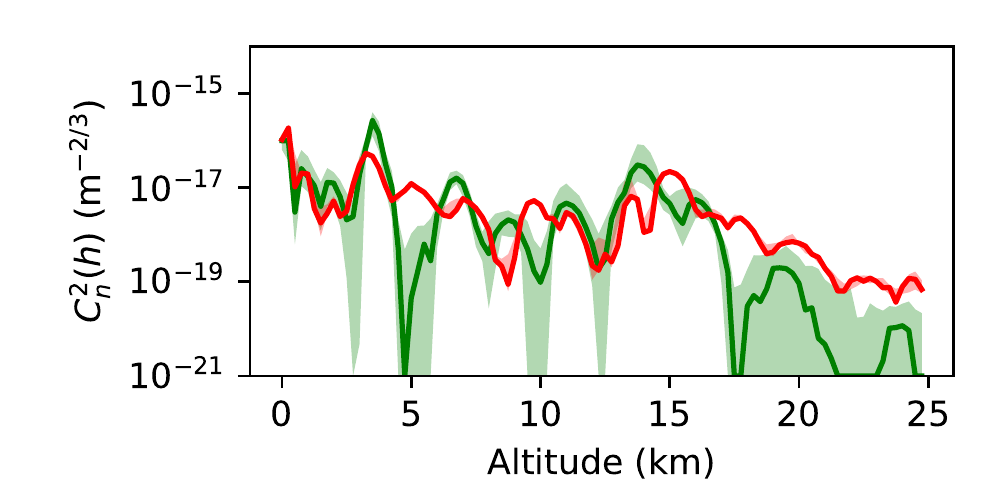} &
    \includegraphics[width=0.3\textwidth,trim={1cm 0 0cm 0}]{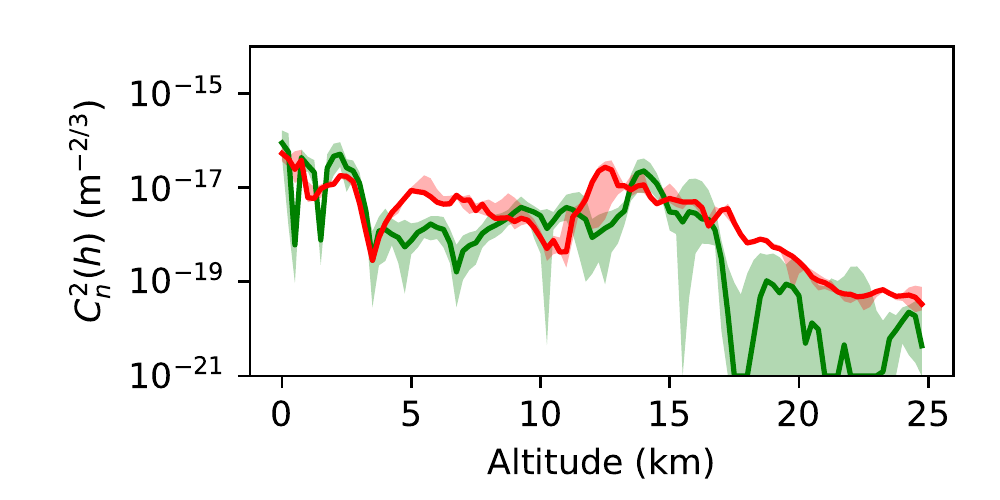} &
    \includegraphics[width=0.3\textwidth,trim={1cm 0 0cm 0}]{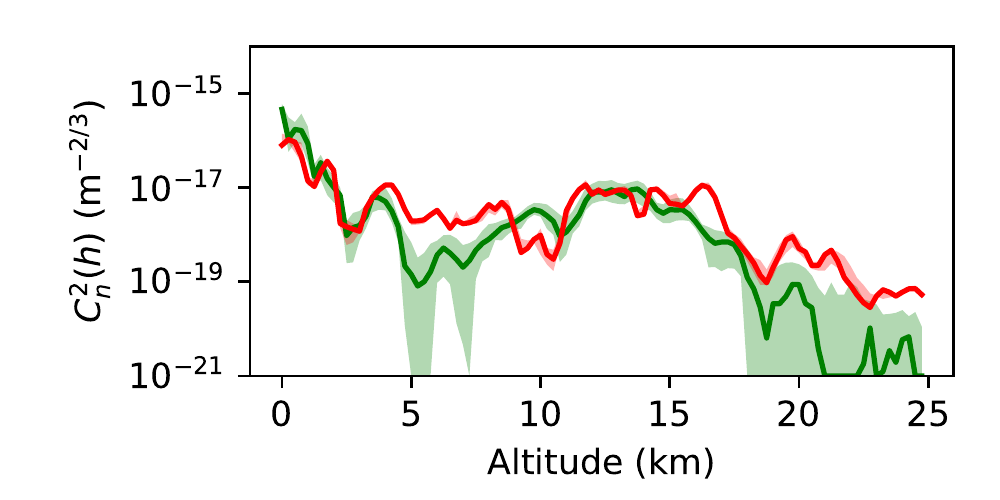} \\
    \includegraphics[width=0.3\textwidth,trim={1cm 0 0cm 0}]{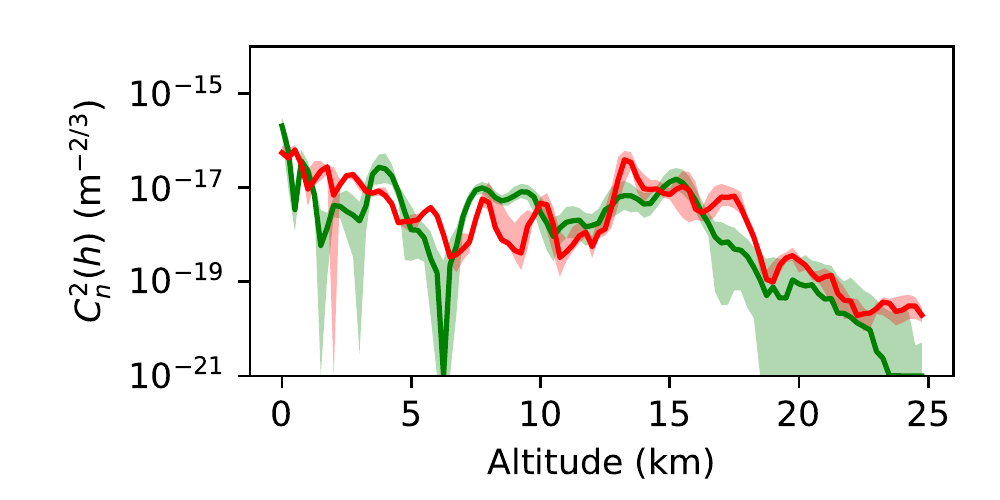} &
    \includegraphics[width=0.3\textwidth,trim={1cm 0 0cm 0}]{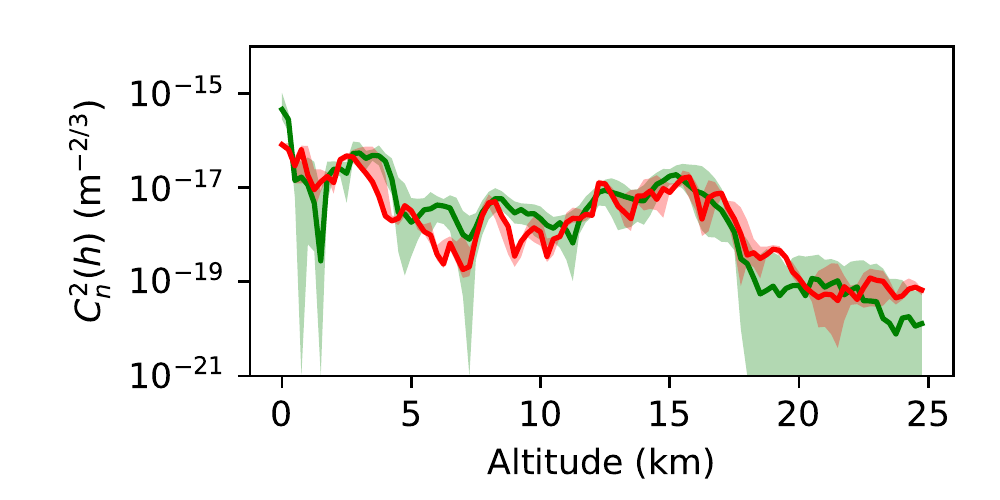} &
\end{array}$
\caption{Example vertical profiles as measured by the stereo-SCIDAR (green) and estimated by the ECMWF GCM model (red). The profiles shown are the median for an individual night of observation. The coloured region shows the interquartile range. These profiles are from the nights beginning 13th - 15th January, 18th - 19th January 2018.}
\label{fig:allProfiles5}
\end{figure*}

Figures~\ref{fig:allProfiles1}, \ref{fig:allProfiles2}, \ref{fig:allProfiles3}, \ref{fig:allProfiles4} and \ref{fig:allProfiles5} show the nightly median for all the stereo-SCIDAR nights at Cerro Paranal in 2016. The ECMWF turbulence forecast is also shown.

\label{lastpage}

\end{document}